\documentclass{LMCS}

\def\dOi{10(4:11)2014}
\lmcsheading%
{\dOi}
{1--58}
{}
{}
{Mar.~10, 2014}
{Dec.~17, 2014}
{}

\ACMCCS{[{\bf Mathematics of computing}]: 
	Probability and statistics---Stochastic processes---Markov processes; 
	Design and analysis of algorithms---Mathematical
        optimization---Continuous optimization---Linear Programming; 
	[{\bf Software and its engineering}]: Software creation and
        management---Software verification and validation---Formal
        software verification}

\usepackage{amsmath}
\usepackage{amsfonts}
\usepackage{amssymb}
\usepackage{latexsym}
\usepackage{stmaryrd}
\usepackage{hyperref}
\usepackage{wrapfig}

\usepackage{pgf}
\usepackage{tikz}
\usetikzlibrary{automata,arrows,intersections,positioning,calc,shapes}

\usepackage[noend]{algorithmic}
\usepackage{algorithm}

\usepackage{cite}

\allowdisplaybreaks

\newcommand{\mystackrel}[2]{%
  \mathrel{\vbox{\offinterlineskip\ialign{%
    \hfil##\hfil\cr
    $\scriptstyle#1$\cr
    \noalign{\kern.2ex}
    $#2$\cr
}}}}

\newenvironment{exampleStart}[1][]{\begin{exa}[#1]}{\hfill$\lozenge$\end{exa}}
\newenvironment{exampleCont}{\begin{exa}}{\hfill$\lozenge$\end{exa}}

\newenvironment{myproof}[0]{\proof}{\qed}
\newenvironment{myoutline}{\proof[Proof outline]}{\qed}

\newenvironment{definition}{\begin{defi}}{\end{defi}}
\newenvironment{corollary}{\begin{cor}}{\end{cor}}
\newenvironment{proposition}{\begin{prop}}{\end{prop}}
\newenvironment{theorem}{\begin{thm}}{\end{thm}}
\newenvironment{lemma}{\begin{lem}}{\end{lem}}

\theoremstyle{plain}
\newtheorem{result}[thm]{Result}

\newcommand{\bigO}[1]{\mathcal{O}(#1)}

\renewcommand{\epsilon}{\varepsilon}

\newcommand{\reals}{\mathbb{R}}
\newcommand{\posreals}{\reals^{\geq 0}}
\newcommand{\strictposreals}{\reals^{> 0}}
\newcommand{\nat}{\mathbb{N}}

\newcommand{\setcond}[2]{\{\, #1 \mid #2 \,\}}
\newcommand{\setnocond}[1]{\{#1\}}
\newcommand{\setcardinality}[1]{|#1|}

\newcommand{\sd}{\mu} 
\newcommand{\gd}{\rho} 
\newcommand{\dirac}[1]{\delta_{#1}}

\newcommand{\probeval}[2]{#1(#2)}

\newcommand{\functionGenericArgument}{\,\cdot\,}

\newcommand{\functioneval}[2]{#1(#2)}

\newcommand{\Disc}[1]{\mathrm{Disc}(#1)}
\newcommand{\SubDisc}[1]{\mathrm{SubDisc}(#1)}
\newcommand{\Supp}[1]{\mathrm{Supp}(#1)}

\newcommand{\family}[2]{{{\setnocond{#1}}_{#2}}}

\newcommand{\aut}[1][A]{\mathcal{#1}}
\newcommand{\mdp}[1][M]{\mathcal{#1}}
\newcommand{\stateSet}{S}
\newcommand{\actionSet}{\Sigma}
\newcommand{\internalActionSet}{\mathtt{H}}
\newcommand{\externalActionSet}{\mathtt{E}}
\newcommand{\startState}[1][s]{\bar{#1}}
\newcommand{\transitionRelation}{\mathit{D}}
\newcommand{\transitionProbability}{\mathit{P}}
\newcommand{\rewardFunction}{r}
\newcommand{\expectedReward}[2][]{\mathbb{E}_{#1}^{#2}}
\newcommand{\probPolicyFrag}[2][\policy]{\transitionProbability^{#1}(#2)}
\newcommand{\parComp}{\parallel}

\newcommand{\NetworkTBetaMuRel}[4]{G(#1,#2,{#4})}
\newcommand{\LPproblemTBetaMuRel}[4]{{\hyperWeakCombinedTransition{#1}{#2}{\liftrelord[{#4}]#3}}}
\newcommand{\minCostLPproblemTBetaMuRel}[4]{{\min_{\costf}\hyperWeakCombinedTransition{#1}{#2}{\liftrelord[{#4}]#3}}}

\newcommand{\minCostLPValue}{\mathfrak{C}}
\newcommand{\givenCost}[1][c]{\mathbf{#1}}

\newcommand{\transitionsWithLabel}[2][]{\transitionRelation_{#1}(#2)}

\newcommand{\project}[2]{\mathrm{proj}_{#1}(#2)}

\newcommand{\hidden}{\tau}
\newcommand{\apparent}[1]{\nu_{#1}}

\newcommand{\tr}{\mathit{tr}}
\newcommand{\source}[1]{\mathit{src}(#1)}
\newcommand{\action}[1]{\mathit{act}(#1)}

\newcommand{\prefix}{\leqslant}
\newcommand{\frags}[1]{\mathit{frags}(#1)}
\newcommand{\finiteFrags}[1]{\mathit{frags}^{*}(#1)}
\newcommand{\length}[1]{\lvert #1 \rvert}
\newcommand{\cone}[1]{C_{#1}}
\newcommand{\trace}[1]{\mathit{trace}(#1)}
\newcommand{\last}[1]{\mathit{last}(#1)}
\newcommand{\first}[1]{\mathit{first}(#1)}
\newcommand{\emptytrace}{\epsilon}
\newcommand{\fragRestriction}[2]{#1{\downharpoonright}#2}

\newcommand{\combined}{\mathrm{C}}
\newcommand{\strongTransition}[3]{{#1 \mystackrel{#2}{\longrightarrow} #3}}
\newcommand{\strongCombinedTransition}[3]{{#1 \mystackrel{#2}{\longrightarrow}_{\combined} #3}}

\newcommand{\weakCombinedTransition}[3]{{#1 \mystackrel{#2}{\Longrightarrow}_{\combined} #3}}

\newcommand{\hyperWeakTransition}[3]{{#1 \mystackrel{#2}{\Longrightarrow} #3}}
\newcommand{\hyperWeakCombinedTransition}[3]{{#1 \mystackrel{#2}{\Longrightarrow}_{\combined} #3}}

\newcommand{\relord}[1][\relsymbol]{\mathcal{#1}}
\newcommand{\rel}[1][\relsymbol]{\mathrel{\relord[#1]}}

\newcommand{\idrelord}[1][\idrelsymbol]{\relord[#1]}

\newcommand{\wbrelord}[1][\wbrelsymbol]{\relord[#1]}
\newcommand{\wbrel}[1][\wbrelsymbol]{\rel[#1]}

\newcommand{\costrelord}[1][\costrelsymbol]{\relord[#1]}
\newcommand{\costrel}[1][\costrelsymbol]{\rel[#1]}
\newcommand{\liftrelord}[1][\relord]{\mathcal{L}(#1)}
\newcommand{\liftrel}[1][\relord]{\mathrel{\liftrelord[#1]}}

\newcommand{\relationComposition}{\mathbin{\circ}}

\newcommand{\borderTerm}{border}
\newcommand{\borderStateSetOrd}[1][\relord]{\mathcal{B}(#1)}
\newcommand{\borderStateSet}[1][\relord]{\borderStateSetOrd[#1]}

\newcommand{\sched}{\sigma}
\newcommand{\schedeval}[2]{#1(#2)}
\newcommand{\policy}{\pi}
\newcommand{\policyeval}[2]{#1(#2)}

\newcommand{\equivclass}[0]{\mathcal{E}}
\newcommand{\relclass}[2]{[#1]_{#2}}
\newcommand{\relimage}[2]{#2(#1)}
\newcommand{\quotienting}[2]{#1/#2}
\newcommand{\partitionset}[2][\relord]{\quotienting{#2}{#1}}

\newcommand{\costf}{c}
\newcommand{\cost}[2][]{\costf_{#1}(#2)}
\newcommand{\gen}[1]{\mathrm{gen}(#1)}

\newcommand{\costAut}[2][\aut]{(#1,#2)}
\newcommand{\costAutPed}[3][\aut]{\costAut[#1_{#3}]{#2_{#3}}}

\newcommand{\strongBisim}{\sim}
\newcommand{\strongProbBisim}{\sim_{p}}
\newcommand{\strongCostBisim}{\sim}
\newcommand{\strongCostProbBisim}{\sim_{p}}
\newcommand{\strongCostBisimMinorCost}{\lesssim}
\newcommand{\strongCostProbBisimMinorCost}{\lesssim_{p}}
\newcommand{\weakBisim}{\approx_{p}}
\newcommand{\weakCostBisim}{\approx_{p}}
\newcommand{\weakCostBisimMinorCost}{\lessapprox_{p}}

\newcommand{\incomingflow}{\vec{f}}
\newcommand{\netsource}{{\vartriangle}}
\newcommand{\netsink}{{\blacktriangledown}}

\newcommand{\acronym}[1]{\ensuremath{\textsl{#1}}}
\newcommand{\PA}{\acronym{PA}}
\newcommand{\CPA}{\acronym{CPA}}

\newcommand{\MDP}{\acronym{MDP}}

\newcommand{\proc}[1]{\textnormal{\scshape#1}}
\newcommand{\MinorCostProc}{\proc{MinorCost}}
\newcommand{\QuotientProc}{\proc{Quotient}}
\newcommand{\FindSplitProc}{\proc{FindSplit}}
\newcommand{\progHeader}[1]{#1}

\newcommand{\icc}{\mathit{ICC}}
\newcommand{\wccHopsRadiusTransmprob}[3]{\mathit{WCC}(#1,#2,#3)}
\newcommand{\sendMessage}[1]{\mathit{s}_{#1}}
\newcommand{\receiveMessage}[1]{\mathit{r}_{#1}}
\newcommand{\hop}[1]{\mathit{t}_{#1}}
\newcommand{\msgSet}{\mathit{Msg}}

\begin{document}

\title[Cost Preserving Bisimulations for Probabilistic Automata]{Cost Preserving Bisimulations\\for Probabilistic Automata\rsuper*}

\author[A.~Turrini]{Andrea Turrini\rsuper a}
\address{{\lsuper a}State Key Laboratory of Computer Science, Institute of Software, Chinese Academy of Sciences, Beijing, China}
\email{turrini@ios.ac.cn}

\author[H.~Hermanns]{Holger Hermanns\rsuper b}
\address{{\lsuper b}Saarland University -- Computer Science, Saarbr\"{u}cken, Germany}
\email{hermanns@cs.uni-saarland.de}


\titlecomment{{\lsuper*}An extended abstract appeared in CONCUR 2013}

\begin{abstract}
\noindent
Probabilistic automata constitute a versatile and elegant model for concurrent probabilistic systems. They are equipped with a compositional theory supporting abstraction, enabled by weak probabilistic bisimulation serving as the reference notion for summarising the effect of abstraction.

This paper considers probabilistic automata augmented with costs. 
It extends the notions of weak transitions in probabilistic automata in such a way that the costs incurred along a weak transition are captured. 
This gives rise to cost-preserving and cost-bounding variations of weak probabilistic bisimilarity, for which we establish compositionality properties with respect to parallel composition. 
Furthermore, polynomial-time decision algorithms are proposed, that can be effectively used to compute reward-bounding abstractions of Markov decision processes in a compositional manner.
\end{abstract}

\maketitle

\section{Introduction}
\label{sec:introduction}

Markov Decision Processes (\MDP{}s) are mathematical models widely used in operations research, automated planning, decision support systems and related fields. 
In the concurrent systems context, they appear in the form of Probabilistic Automata (\PA{}s)~\cite{Seg95}. 
\PA{}s form the backbone model of successful model checkers such as PRISM~\cite{HKNP06} enabling the analysis of randomised concurrent systems.

In probabilistic automata, probabilistic experiments can be performed inside a transition. 
This embodies a clear separation between probability and nondeterminism, and is represented by transitions of the form $\strongTransition{s}{a}{\sd}$, where $s$ is a state, $a$ is an action label, and $\sd$ is a probability distribution on states. 
Labelled transition systems are instances of this model family, obtained by restricting to Dirac distributions (assigning full probability to single states). 
Thus, foundational concepts and results of standard concurrency theory are retained in full and extend smoothly to the model of probabilistic automata. 
This includes notions of strong, branching and weak probabilistic bisimilarity~\cite{Seg95}. 

As one of the classical concurrency theory manifestations, weak probabilistic bisimilarity is a congruence relation for parallel composition on \PA{}s. 
In other contexts, this has enabled powerful compositional minimisation approaches to combat the state space explosion problem in explicit state verification approaches~\cite{CGMTZ96,KKZJ07,HK00}. 
This is rooted in the availability of effective minimisation algorithms for weak bisimulation implemented in tools like CADP~\cite{CGHLMS10}, MRMC~\cite{KZHHJ11} or~sigref~\cite{WHHSB06}. 
In the \PA{} context, this avenue has not been explored, mainly because for a long time only an exponential decision algorithm for weak probabilistic bisimilarity was known~\cite{CS02}, and it was unclear how to turn the decision algorithm into a minimisation algorithm. 
Lately, these two problems have been successfully attacked: 
A polynomial time algorithm for deciding weak probabilistic bisimilarity~\cite{HT12} has been devised, and has been embedded into a minimisation algorithm~\cite{EHSTZ13}, producing in polynomial time the minimal canonical representation with respect to weak probabilistic bisimilarity for any given \PA{}. 
Therefore, compositional minimisation can now be followed also in the context of \PA{}s and \MDP{}s.

\MDP{} models are usually decorated with cost or reward structures, with the intention to minimise costs or maximise rewards along the model execution. 
Likewise, in tools like PRISM, \PA{}s appear augmented with cost or reward structures. 
It is hence a natural question how costs can be embedded into the approach discussed above, and this is what the paper is about: 
We propose Cost Probabilistic Automata (\CPA{}s), a model where \emph{cost} is any kind of quantity associated with the transitions of the automata, and we aim to minimise the cost. 
For instance, we can consider as the cost of a transition the power needed to transmit a message, the time spent in the computation modelled by the transition, the (monetary) risk associated with an action, the expense of some work, and so on. 

We then turn our attention to strong and weak probabilistic bisimulation that accounts for costs. 
Costs for weak transitions are interpreted in line with the vast body of literature on \MDP{}s. 
As a strict option, we require weak transition costs to be matched exactly for bisimilar states, inducing \emph{cost-preserving weak probabilistic bisimulation}. 
As a weaker alternative, we ask them to be bounded from one \CPA{} to the other, leading to the notion of \emph{minor cost weak probabilistic bisimulation}.

When establishing the base properties expected from these kind of definitions, especially transitivity of minor cost weak probabilistic bisimulation turns out to be quite intricate to prove. 
We also show that both relations are compositional: 
Cost-preserving weak probabilistic bisimulation is a congruence with respect to parallel composition, and minor cost weak probabilistic bisimulation is a precongruence.

Furthermore, we provide polynomial time algorithms for all the cost related relations discussed, and present an application of minor cost weak probabilistic bisimulation to a multi-hop wireless communication scenario where the cost structure represents transmission power which in turn depends on physical distances.

The algorithmic advancement is rooted in an alternative interpretation of weak transition costs, which agrees with the original one with respect to cost expectations, but provides us with the technical assets to establish a link to the polynomial time algorithm for \PA{} weak probabilistic bisimilarity. 
At the core of that algorithm is a polynomial number of linear programming (LP) problems, each of them checking the existence of a specific weak transition, and this is what we manage to bridge to also in the cost setting.

\noindent \textbf{Organisation of the paper.}  
After introducing preliminaries in Section~\ref{sec:Preliminaries} and probabilistic automata background in Section~\ref{sec:probabilisticAutomata}, we present cost probabilistic automata in Section~\ref{sec:costProbabilisticAutomata}, the strong and weak cost-preserving bisimilarities in Section~\ref{sec:costPreservingBisims}, and the strong and weak cost-bounding bisimilarities in Section~\ref{sec:costBoundingBisims} where we study their properties, and exemplify the usefulness of minor cost weak bisimulation by means of a wireless channel example. 
We then revisit the LP problem formulation behind weak probabilistic bisimilarity in Section~\ref{sec:weakTransitionAsLPP} so as to arrive at polynomial-time algorithms for all bisimilarities we introduced. 
Related work and possible extensions are discussed  in Section~\ref{sec:discussion} and we conclude the paper in Section~\ref{sec:conclusion} with some remarks. 
To keep the presentation of the paper clear, we moved all non-trivial proofs to the appendix.

Parts of this paper are based on a conference publication in CONCUR 2013~\cite{HT13}. 

\section{Mathematical Preliminaries}
\label{sec:Preliminaries}
This section recalls the basic mathematical preliminaries together with the notational conventions we adhere to in this work.
Given a function $f \colon \reals \times \reals \to \reals$, we say that $f$ is 
\begin{itemize}
\item 
	\emph{symmetric} if, for each $x,y \in \reals$, it holds $\functioneval{f}{x,y} = \functioneval{f}{y,x}$;
\item
	\emph{zero-preserving} if $\functioneval{f}{0,0} = 0$;
\item
	\emph{distributive} over convex combination if, for any finite sets $I,J \subseteq \nat$, each $\family{x_{i} \in \reals}{i \in I}$, and each $\family{y_{j} \in \reals}{j \in J}$, each $\family{p_{i} \in \strictposreals}{i \in I}$, and each $\family{q_{j} \in \strictposreals}{j \in J}$, the following holds:
	\begin{itemize}
	\item 
		$\functioneval{f}{x, y} = \sum_{j \in J} q_{j} \cdot \functioneval{f}{x, y_{j}}$, and
	\item
		$\functioneval{f}{x, y} = \sum_{i \in I} p_{i} \cdot \functioneval{f}{x_{i}, y}$,
	\end{itemize}
	where $x = \sum_{i \in I} p_{i} \cdot x_{i}$ and $y = \sum_{j \in J} q_{j} \cdot y_{j}$;
\item
	\emph{monotone increasing} if, for each $x,x',y,y' \in \reals$ with $x < x'$ and $y < y'$, the following holds:
	\begin{itemize}
	\item 
		$\functioneval{f}{x,y} < \functioneval{f}{x',y}$,
	\item
		$\functioneval{f}{x,y} < \functioneval{f}{x,y'}$, and 
	\item
		$\functioneval{f}{x,y} < \functioneval{f}{x',y'}$.
	\end{itemize}
\end{itemize}

\noindent For a set $X$, denote by $\Disc{X}$ the set of discrete probability distributions over $X$, and by $\SubDisc{X}$ the set of discrete sub-probability distributions over $X$. 
Given $\gd \in \SubDisc{X}$ and $Y \subseteq X$, we write $\probeval{\gd}{Y}$ for $\sum_{y \in Y} \probeval{\gd}{y}$.
Given $\gd \in \SubDisc{X}$, we denote by $\Supp{\gd}$ the set $\setcond{x \in X}{\probeval{\gd}{x} >0}$, by $\probeval{\gd}{\bot}$ the value $1 - \probeval{\gd}{X}$ where $\bot \notin X$, and by $\dirac{x}$, where $x \in X \cup \setnocond{\bot}$, the \emph{Dirac} distribution such that $\probeval{\gd}{y} = 1$ for $y = x$, $0$ otherwise.
For a sub-probability distribution $\gd$, we also write $\gd = \setcond{(x, p_x)}{x \in X}$ where $p_{x}$ is the probability of $x$.
Given $\gd_{x} \in \SubDisc{X}$ and $\gd_{y} \in \SubDisc{Y}$, we denote by $\gd_{x} \times \gd_{y}$ the sub-probability distribution over $X \times Y$ defined by $\probeval{\gd_{x} \times \gd_{y}}{u, v} = \probeval{\gd_{x}}{u} \cdot \probeval{\gd_{y}}{v}$ for each $(u, v) \in X \times Y$.
Given a finite set $I$ of indexes, a family $\family{p_{i} \in \strictposreals}{i \in I}$ such that $\sum_{i \in I} p_{i} = 1$, and a family $\family{\gd_{i} \in \SubDisc{X}}{i \in I}$, we say that $\gd$ is the \emph{convex combination} of $\family{\gd_{i}}{i \in I}$ according to $\family{p_{i}}{i \in I}$, denoted by $\sum_{i \in I} p_{i} \cdot \gd_{i}$, if for each $x \in X$, $\probeval{\gd}{x} = \sum_{i \in I} p_{i} \cdot \probeval{\gd_{i}}{x}$.

Given a relation $\relord \subseteq X \times X$, we say that $\relord$ is a preorder if it is reflexive and transitive. 
We say that $\relord$ is an equivalence relation if it is a symmetric preorder.
Given an equivalence relation $\relord$ on $X$, we denote by $\partitionset{X}$ the set of equivalence classes induced by $\relord$ and, for $x \in X$, by $\relclass{x}{\relord}$ the class $\equivclass \in \partitionset{X}$ such that $x \in \equivclass$.
We denote by $\idrelord$ the \emph{identity relation}, i.e., the equivalence relation having $\relclass{x}{\idrelord} = \setnocond{x}$ for each $x \in X$.

Given the relations $\relord_{1} \subseteq X \times Y$ and $\relord_{2} \subseteq Y \times Z$, the \emph{composition} of $\relord_{1}$ 
and $\relord_{2}$, denoted by $\relord_{1} \relationComposition \relord_{2}$, is the relation $\relord \subseteq X \times Z$ defined as 
$\relord = \setcond{(x,z)}{\exists y \in Y. x \rel_{1} y \rel_{2} z}$.
If $\relord_{1}$ and $\relord_{2}$ are equivalence relations on $X \cup Y$ and $Y \cup Z$, respectively, then $\relord_{1} \relationComposition \relord_{2}$ is the equivalence relation $\relord $ on $X \cup Z$ defined as the symmetric and transitive closure of $\setcond{(x,z)}{\exists y \in Y. x \rel_{1} y \rel_{2} z} \cup \setcond{(x,x') \in X \times X}{x \rel_{1} x'} \cup \setcond{(z,z') \in Z \times Z}{z \rel_{2} z'}$.

Given the relations $\relord_{1} \subseteq W \times X$ and $\relord_{2} \subseteq Y \times Z$, the \emph{cross-product} of $\relord_{1}$ and $\relord_{2}$, denoted by $\relord_{1} \times \relord_{2}$, is the relation $\relord \subseteq (W \times Y) \times (X \times Z)$ such that $(w,y) \rel (x,z)$ if and only if $w \rel_{1} x$ and $y \rel_{2} z$.

The lifting $\liftrelord$~\cite{JL91} of a relation $\relord \subseteq X \times Y$ is defined as: 
For $\gd_{X} \in \Disc{X}$ and $\gd_{Y} \in \Disc{Y}$, $\gd_{X} \liftrel \gd_{Y}$ holds if there exists a \emph{weighting function} $w \colon X \times Y \to [0,1]$ such that
\begin{enumerate}
\item 
	$w(x,y) > 0$ implies $x \rel y$,
\item 
	$\sum_{y \in Y} w(x,y) = \gd_{X}(x)$, and
\item 
	$\sum_{x \in X} w(x,y) = \gd_{Y}(y)$.
\end{enumerate}
The lifting of relations has some interesting properties:
\begin{enumerate}
\item 
\label{item:lifting_statesAndDirac}
	$x \rel y$ if and only if $\dirac{x} \liftrel \dirac{y}$.
\item 
\label{item:lifting_presevedEmptySet}
	$\relord = \emptyset$ if and only if $\liftrelord = \emptyset$.
\item 
\label{item:lifting_inclusion}
	If $\relord \subseteq \relord[S]$, then $\liftrelord \subseteq \liftrelord[{\relord[S]}]$.
\item 
\label{item:lifting_reflexivity}
	If $\relord$ is reflexive, then $\liftrelord$ is reflexive.
\item 
\label{item:lifting_symmetry}
	If $\relord$ is symmetric, then $\liftrelord$ is symmetric.
\item 
\label{item:lifting_transitivity}
	If $\relord$ is transitive, then $\liftrelord$ is transitive.
\item 
\label{item:lifting_genericTransitivity}
	If $\gd_{x} \liftrel \gd_{y}$ and $\gd_{y} \liftrel[{\relord[S]}] \gd_{z}$, then $\gd_{x} \liftrel[{\rel \relationComposition \rel[S]}] \gd_{z}$.
\item 
\label{item:lifting_composition}
	If $\gd_{x} \liftrel \gd_{y}$, then $\gd_{x} \times \gd_{z} \liftrel[\relord \times \idrelord] \gd_{y} \times \gd_{z}$.
\item
\label{item:lifting_convexCombination}
	Given a finite set $I$ of indexes, a family $\family{p_{i} \in \strictposreals}{i \in I}$ such that $\sum_{i \in I} p_{i} = 1$, a family $\family{\gd_{x,i} \in \Disc{X}}{i \in I}$, and a family $\family{\gd_{y,i} \in \Disc{Y}}{i \in I}$, if $\gd_{x,i} \liftrel \gd_{y,i}$ for each $i \in I$, then $\sum_{i \in I} p_{i} \cdot \gd_{x,i} \liftrel \sum_{i \in I} p_{i} \cdot \gd_{y,i}$.
\end{enumerate}

\section{Probabilistic Automata}
\label{sec:probabilisticAutomata}
We now recall the definition of probabilistic automata as proposed by Segala in~\cite{Seg95} as \emph{simple} probabilistic automata. 
We then review strong and weak bisimilarities on \PA{}s together with their properties. 
We follow the notation used in~\cite{Seg06}.

\begin{definition}
A \emph{Probabilistic Automaton} (\PA{}) $\aut$ is a tuple $(\stateSet, \startState, \actionSet, \transitionRelation)$, where $\stateSet$ is a countable set of \emph{states}, $\startState \in \stateSet$ is the \emph{start state}, $\actionSet$ is a countable set of \emph{actions}, and $\transitionRelation \subseteq \stateSet \times \actionSet \times \Disc{\stateSet}$ is a \emph{probabilistic transition relation}.
\end{definition}

The set $\actionSet$ is divided in two sets $\internalActionSet$ and $\externalActionSet$ of internal (hidden) and external actions, respectively;
we let $s$,$t$,$u$,$v$, and their variants with indices range over $\stateSet$; 
$a$, $b$ range over actions; 
and $\hidden$ range over internal actions.
We also denote the generic elements of a probabilistic automaton $\aut$ by $\stateSet$, $\startState$, $\actionSet$, $\transitionRelation$, and we propagate primes and indices when necessary. 
Thus, for example, the probabilistic automaton $\aut'_{i}$ has states $\stateSet'_{i}$, start state $\startState'_{i}$, actions $\actionSet'_{i}$ and 
transition relation $\transitionRelation'_{i}$. 

A transition $\tr = (s, a, \sd) \in \transitionRelation$, also denoted by $\strongTransition{s}{a}{\sd}$, is said to \emph{leave} from state $s$, to be \emph{labelled} by $a$, and to \emph{lead} to the \emph{target} distribution $\sd$, also denoted by $\sd_{\tr}$.
We denote by $\source{\tr}$ the \emph{source} state $s$ and by $\action{\tr}$ the \emph{action} $a$.
We also say that $s$ enables action $a$, that action $a$ is enabled from $s$, and that $(s,a,\sd)$ is enabled from $s$.
Finally, we let $\transitionsWithLabel{a} = \setcond{\tr \in \transitionRelation}{\action{\tr} = a}$ be the set of transitions with label $a$.

If we restrict the nondeterminism in each state so that at the state enables at most one transition per action, we obtain the Markov decision process model~\cite{Bel57,How07,Put05}.
This model is widely used in operations research and artificial intelligence literature to represent systems exhibiting both probabilistic and nondeterministic behaviours.
Usually, the actual action labelling a transition is ignored while in the \PA{} setting, actions are used for synchronisation on parallel composition, as we will see below.
\begin{definition}
\label{def:mdp}
A \emph{Markov Decision Process} (\MDP{}) $\mdp$ is a tuple $(\stateSet, \startState, \actionSet, \transitionProbability,\rewardFunction)$ that can be considered as a variation of a \PA{} with a functional transition relation $\transitionProbability \colon \stateSet \times \actionSet \to \Disc{\stateSet}$. 
\end{definition}
Since an \MDP{} is a special case of \PA{}, we adopt the same terminology and notation for both models.
Given a state $s$, we denote by $\actionSet(s)$ the set of actions enabled by $s$, i.e., $\actionSet(s) = \setcond{a \in \actionSet}{\text{$\functioneval{\transitionProbability}{s,a}$ is defined}}$.

In this paper we consider only finite models, i.e., \PA{}s or \MDP{}s such that $\stateSet$ and $\transitionRelation$ (or $\transitionProbability$) are finite. 
Moreover, we assume that every state can be reached from the start state.

\begin{exampleStart}[A wireless communication channel]
\label{ex:wcc}
As an example of a \PA{}, consider a reliable wireless communication channel used to transmit messages belonging to the set $\msgSet$ from a sender to a receiver.
\begin{figure}[t]
	\centering
	\begin{tikzpicture}[->,>=stealth',shorten >=1pt,auto]
	\scriptsize

	\node (ss) at (0,0.75) {\normalsize$\startState$};
	\node (h0) at ($(ss) + (-3.5,-0.75)$) {\normalsize$h^{m}_{0}$};
	\node (h1) at ($(ss) + (-1.75,-0.75)$) {\normalsize$h^{m}_{1}$};
	\node (h2) at ($(ss) + (0,-0.75)$) {\normalsize$h^{m}_{2}$};
	\node (hnm1) at ($(ss) + (1.75,-0.75)$) {\normalsize$h^{m}_{n-1}$};
	\node (hn) at ($(ss) + (3.5,-0.75)$) {\normalsize$h^{m}_{n}$};
	
	\node (wcc) at ($(ss) + (0,-2)$) {\normalsize$\wccHopsRadiusTransmprob{n}{r}{p}$};
	
	\draw ($(ss.north) + (0,0.3)$) to (ss);
	\draw (ss) to [bend right=10] node [above] {$\sendMessage{m}$} (h0.north east);
	\draw[name path=t1h0h1] (h0.east) to node [above] {$\hop{r}$} node [below, very near end] {$p$} (h1);
	\draw[name path=t1h0h0] (h0.east) .. controls ($(h0.east) + (1.25,-0.75)$) and ($(h0) + (0,-0.75)$) .. node[very near end] {$1-p$} (h0);
	\path[name path=t1] (h0.east) circle (0.5);
	\draw[name intersections={of=t1h0h1 and t1, by=i112},
		name intersections={of=t1h0h0 and t1},
		-,shorten >=0pt]
	(i112) to[bend left] (intersection-2);
	\draw[name path=t2h1h2] (h1.east) to node [above] {$\hop{r}$} node [below, very near end] {$p$} (h2);
	\draw[name path=t2h1h1] (h1.east) .. controls ($(h1.east) + (1.25,-0.75)$) and ($(h1) + (0,-0.75)$) .. node[very near end] {$1-p$} (h1);
	\path[name path=t2] (h1.east) circle (0.5);
	\draw[name intersections={of=t2h1h2 and t2, by=i223},
		name intersections={of=t2h1h1 and t2},
		-,shorten >=0pt]
	(i223) to[bend left] (intersection-2);
	\draw[dotted,-] (h2) to (hnm1);
	\draw[name path=tnm1hnm1hn] (hnm1.east) to node [above] {$\hop{r}$} node [below, very near end] {$p$} (hn);
	\draw[name path=tnm1hnm1hnm1] (hnm1.east) .. controls ($(hnm1.east) + (1.25,-0.75)$) and ($(hnm1) + (0,-0.75)$) .. node[very near end] {$1-p$} (hnm1);
	\path[name path=tnm1] (hnm1.east) circle (0.5);
	\draw[name intersections={of=tnm1hnm1hn and tnm1, by=inm1nm1n},
		name intersections={of=tnm1hnm1hnm1 and tnm1},
		-,shorten >=0pt]
	(inm1nm1n) to[bend left] (intersection-1);
	\draw (hn.north west) to [bend right=10] node [above] {$\receiveMessage{m}$} (ss);

	\node (ess) at (-4,-3) {\normalsize$\startState$};
	\node (eh0) at ($(ess) + (-1.75,-0.75)$) {\normalsize$h^{m}_{0}$};
	\node (eh1) at ($(ess) + (0,-0.75)$) {\normalsize$h^{m}_{1}$};
	\node (eh2) at ($(ess) + (1.75,-0.75)$) {\normalsize$h^{m}_{2}$};
	
	\node (ewcc) at ($(ess) + (0,-2)$) {\normalsize$\wccHopsRadiusTransmprob{2}{5}{\frac{3}{4}}$};
	
	\draw ($(ess.north) + (0,0.3)$) to (ess);
	\draw (ess) to [bend right=10] node [above, near end] {$\sendMessage{m}$} (eh0.north east);
	\draw[name path=t1h0h1] (eh0.east) to node [above] {$\hop{5}$} node [below, very near end] {$\frac{3}{4}$} (eh1);
	\draw[name path=t1h0h0] (eh0.east) .. controls ($(eh0.east) + (1.25,-0.75)$) and ($(eh0) + (0,-0.75)$) .. node[very near end] {$\frac{1}{4}$} (eh0);
	\path[name path=t1] (eh0.east) circle (0.5);
	\draw[name intersections={of=t1h0h1 and t1, by=e112},
		name intersections={of=t1h0h0 and t1},
		-,shorten >=0pt]
	(e112) to[bend left] (intersection-2);
	\draw[name path=t2h1h2] (eh1.east) to node [above] {$\hop{5}$} node [below, very near end] {$\frac{3}{4}$} (eh2);
	\draw[name path=t2h1h1] (eh1.east) .. controls ($(eh1.east) + (1.25,-0.75)$) and ($(eh1) + (0,-0.75)$) .. node[very near end] {$\frac{1}{4}$} (eh1);
	\path[name path=t2] (eh1.east) circle (0.5);
	\draw[name intersections={of=t2h1h2 and t2, by=e223},
		name intersections={of=t2h1h1 and t2},
		-,shorten >=0pt]
	(e223) to[bend left] (intersection-2);
	\draw (eh2.north west) to [bend right=10] node [above, near start] {$\receiveMessage{m}$} (ess);

	\node (iss) at (4,-3) {\normalsize$\startState$};
	\node (ih0) at ($(iss) + (0,-1.25)$) {\normalsize$h^{m}_{0}$};
	
	\node (iwcc) at ($(iss) + (0,-2)$) {\normalsize$\icc = \wccHopsRadiusTransmprob{0}{\infty}{1}$};
	
	\draw ($(iss.north) + (0,0.3)$) to (iss);
	\draw (iss) to [bend right=30] node [left, near end] {$\sendMessage{m}$} (ih0);
	\draw (ih0) to [bend right=30] node [right, near start] {$\receiveMessage{m}$} (iss);
	\end{tikzpicture}
	\caption{The wireless communication channel \PA{} $\wccHopsRadiusTransmprob{n}{r}{p}$, its concrete instance $\wccHopsRadiusTransmprob{2}{5}{\frac{3}{4}}$, and the ideal communication channel \PA{} $\icc = \wccHopsRadiusTransmprob{0}{\infty}{1}$}
	\label{fig:wirelessCC}
\end{figure}
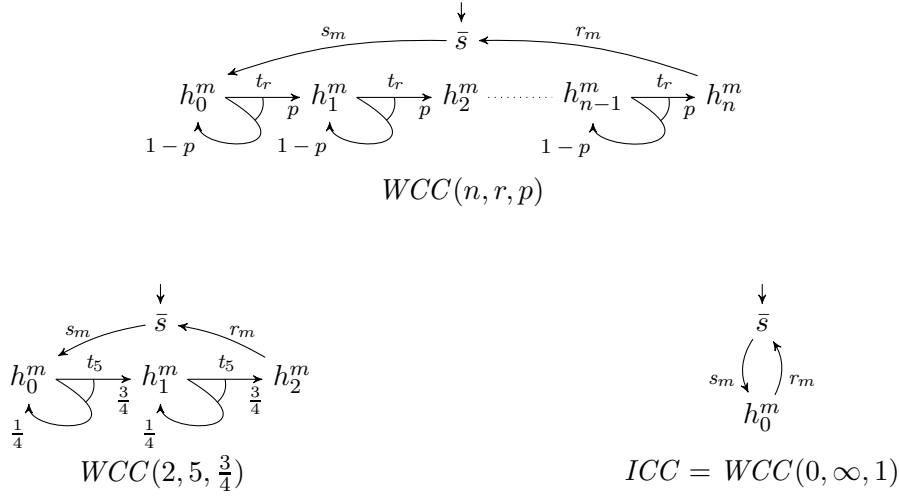
The wireless implementation of the communication channel is depicted in Figure~\ref{fig:wirelessCC}: 
We graphically mark the start state $\startState$ of the automaton with an incoming arrow without source state; 
such arrow does not represent a transition of the automaton as other arrows do.

The \PA{} $\wccHopsRadiusTransmprob{n}{r}{p}$ models a communication channel that requires $n$ intermediate nodes (hops) to reach the receiver where the probability to transmit correctly the message from each node to the successor is $p$.
Each intermediate node has a transmission radius $r$, and this parameter will become useful when determining the transmission cost in terms of power consumed, power that we aim to reduce. 
In this \PA{}, the message $m$ to transmit is obtained from the sender via the external $\sendMessage{m}$ action and it is delivered to the receiver by using the external action $\receiveMessage{m}$.
Internal action $\hop{r}$ models the transmission of the message $m$ from one node to the successor distant at most $r$, the transmission radius.

As a concrete example of wireless communication channel, we consider the case with two intermediate hops, a transmission radius of $5$, and a success transmission probability of $0.75$; 
the corresponding automaton is the \PA{} $\wccHopsRadiusTransmprob{2}{5}{\frac{3}{4}}$ shown in the bottom left-hand part of Figure~\ref{fig:wirelessCC}.

The ideal communication channel is modelled by the \PA{} $\wccHopsRadiusTransmprob{0}{\infty}{1}$, that is, the automaton that does not require intermediate nodes:
The corresponding automaton is the \PA{} $\icc$ shown in the bottom right-hand part of Figure~\ref{fig:wirelessCC}.
Obviously, $\icc$ models a reliable communication channel since the message is delivered with probability $1$ just after having received it.
\end{exampleStart}

The following definition of parallel composition is just an equivalent rewriting of the definition provided in~\cite{Seg06}.
\begin{definition}
\label{def:paParallelComposition}
	Given two \PA{}s $\aut_{1}$ and $\aut_{2}$, we say that $\aut_{1}$ and $\aut_{2}$ are \emph{compatible} if $\actionSet_{1} \cap \internalActionSet_{2} = \emptyset = \internalActionSet_{1} \cap \actionSet_{2}$.
	
	Given two compatible \PA{}s $\aut_{1}$ and $\aut_{2}$, the \emph{parallel composition} of $\aut_{1}$ and $\aut_{2}$, denoted by $\aut_{1} \parComp \aut_{2}$, is the probabilistic automaton $\aut = (\stateSet, \startState, \actionSet, \transitionRelation)$ where 
	\begin{itemize}
	\item 
		$\stateSet = \stateSet_{1} \times \stateSet_{2}$,
	\item
		$\startState = (\startState_{1}, \startState_{2})$,
	\item
		$\actionSet = \externalActionSet \cup \internalActionSet$ where $\externalActionSet = \externalActionSet_{1} \cup \externalActionSet_{2}$ and $\internalActionSet = \internalActionSet_{1} \cup \internalActionSet_{2}$, and 
	\item
		$((s_{1},s_{2}), a, \sd_{1} \times \sd_{2}) \in \transitionRelation$ if and only if
		\begin{itemize}
		\item
			whenever $a \in \actionSet_{1} \cap \actionSet_{2}$, $(s_{1}, a, \sd_{1}) \in \transitionRelation_{1}$ and $(s_{2}, a, \sd_{2}) \in \transitionRelation_{2}$, 
		\item
			whenever $a \in \actionSet_{1} \setminus \actionSet_{2}$, $(s_{1}, a, \sd_{1}) \in \transitionRelation_{1}$ and $\sd_{2} = \dirac{s_{2}}$, and 
		\item
			whenever $a \in \actionSet_{2} \setminus \actionSet_{1}$, $(s_{2}, a, \sd_{2}) \in \transitionRelation_{2}$ and $\sd_{1} = \dirac{s_{1}}$. 
		\end{itemize}
	\end{itemize}
	For $a \in \actionSet_{1} \setminus \actionSet_{2}$, we denote by $(s_{2}, \apparent{a}, \dirac{s_{2}})$ the \emph{apparent} internal transition corresponding to not performing any transition from $s_{2}$ in the composed transition, and similarly for $a \in \actionSet_{2} \setminus \actionSet_{1}$.
\end{definition}
For two compatible \PA{}s $\aut_{1}$ and $\aut_{2}$ and their parallel composition $\aut_{1} \parComp \aut_{2}$, we refer to $\aut_{1}$ and $\aut_{2}$ as the component automata and to $\aut_{1} \parComp \aut_{2}$ as the composed automaton.

The composition of two compatible \MDP{}s is not necessarily an \MDP{} because the composed transition relation might become non-functional. 
Thus, \MDP{}s are not closed under parallel composition, in contrast to \PA{}s. 

\begin{figure}[t]
	\centering
	\begin{tikzpicture}[->,>=stealth',shorten >=1pt,auto]
	\scriptsize
	
	\node (aut0) at (0,0) {\normalsize$\aut_{0}$};
	\node (s0) at ($(aut0) + (0,4.75)$) {\normalsize$\startState_{0}$};
	\node (t0) at ($(s0) + (-1,-1.25)$) {\normalsize$t_{0}$};
	\node (u0) at ($(s0) + (1,-1.25)$) {\normalsize$u_{0}$};
	\node (v0) at ($(t0) + (0,-1.25)$) {\normalsize$v_{0}$};
	\node (w0) at ($(u0) + (0,-1.25)$) {\normalsize$w_{0}$};
	\node (x0) at ($(v0) + (-0.75,-1.25)$) {\normalsize$x_{0}$};
	\node (y0) at ($(v0) + (0.75,-1.25)$) {\normalsize$y_{0}$};
	\node (z0) at ($(w0) + (0,-1.25)$) {\normalsize$z_{0}$};
	
	\draw ($(s0.north) + (0,0.3)$) to (s0);
	\draw (s0) to (t0);
	\draw (s0) to (u0);
	\draw (t0) to node[left] {$a$} (v0);
	\draw (u0) to node[left] {$a$} (w0);
	\draw (v0) to (x0);
	\draw (v0) to (y0);
	\draw (w0) to (z0);

	\node (aut1) at (6,0) {\normalsize$\aut_{1}$};
	\node (s1) at ($(aut1) + (0,4.75)$) {\normalsize$\startState_{1}$};
	\node (t1) at ($(s1) + (-1,-1.25)$) {\normalsize$t_{1}$};
	\node (u1) at ($(s1) + (1,-1.25)$) {\normalsize$u_{1}$};
	\node (v1) at ($(t1) + (0,-1.25)$) {\normalsize$v_{1}$};
	\node (w1) at ($(u1) + (0,-1.25)$) {\normalsize$w_{1}$};
	\node (x1) at ($(v1) + (0,-1.25)$) {\normalsize$x_{1}$};
	
	\draw ($(s1.north) + (0,0.3)$) to (s1);
	\draw (s1) to (t1);
	\draw (s1) to (u1);
	\draw (t1) to node[left] {$a$} (v1);
	\draw (u1) to node[left] {$a$} (w1);
	\draw (v1) to (x1);
\end{tikzpicture}
\vskip5mm
\begin{tikzpicture}[->,>=stealth',shorten >=1pt,auto]
	\scriptsize
		
	\node (aut01) at (0,0) {\normalsize$\aut_{0} \parComp \aut_{1}$};
	\node (s0s1) at ($(aut01) + (1.25,7.5)$) {\normalsize$(\startState_{0},\startState_{1})$};
	\node (t0s1) at ($(s0s1) + (-6,-1.25)$) {\normalsize$(t_{0},s_{1})$};
	\node (u0s1) at ($(s0s1) + (-2,-1.25)$) {\normalsize$(u_{0},s_{1})$};
	\node (s0t1) at ($(s0s1) + (2,-1.25)$) {\normalsize$(s_{0},t_{1})$};
	\node (s0u1) at ($(s0s1) + (5,-1.25)$) {\normalsize$(s_{0},u_{1})$};
	\node (t0t1) at ($(t0s1) + (0,-1.25)$) {\normalsize$(t_{0},t_{1})$};
	\node (t0u1) at ($(u0s1) + (0,-1.25)$) {\normalsize$(t_{0},u_{1})$};
	\node (u0t1) at ($(s0t1) + (0,-1.25)$) {\normalsize$(u_{0},t_{1})$};
	\node (u0u1) at ($(s0u1) + (0,-1.25)$) {\normalsize$(u_{0},u_{1})$};
	\node (v0v1) at ($(t0t1) + (0,-1.25)$) {\normalsize$(v_{0},v_{1})$};
	\node (v0w1) at ($(t0u1) + (0,-1.25)$) {\normalsize$(v_{0},w_{1})$};
	\node (w0v1) at ($(u0t1) + (0,-1.25)$) {\normalsize$(w_{0},v_{1})$};
	\node (w0w1) at ($(u0u1) + (0,-1.25)$) {\normalsize$(w_{0},w_{1})$};
	\node (x0v1) at ($(v0v1) + (-1.5,-1.25)$) {\normalsize$(x_{0},v_{1})$};
	\node (y0v1) at ($(v0v1) + (0,-1.25)$) {\normalsize$(y_{0},v_{1})$};
	\node (v0x1) at ($(v0v1) + (1.5,-1.25)$) {\normalsize$(v_{0},x_{1})$};
	\node (x0x1) at ($(y0v1) + (-1,-1.25)$) {\normalsize$(x_{0},x_{1})$};
	\node (y0x1) at ($(y0v1) + (1,-1.25)$) {\normalsize$(y_{0},x_{1})$};
	\node (x0w1) at ($(v0w1) + (-1,-2.5)$) {\normalsize$(x_{0},w_{1})$};
	\node (y0w1) at ($(v0w1) + (1,-2.5)$) {\normalsize$(y_{0},w_{1})$};
	\node (z0v1) at ($(w0v1) + (-1,-1.25)$) {\normalsize$(z_{0},v_{1})$};
	\node (w0x1) at ($(w0v1) + (1,-1.25)$) {\normalsize$(w_{0},x_{1})$};
	\node (z0x1) at ($(z0v1) + (1,-1.25)$) {\normalsize$(z_{0},x_{1})$};
	\node (z0w1) at ($(w0w1) + (0,-2.5)$) {\normalsize$(z_{0},w_{1})$};
	
	\draw ($(s0s1.north) + (0,0.3)$) to (s0s1);
	\draw (s0s1) to (t0s1);
	\draw (s0s1) to (u0s1);
	\draw (s0s1) to (s0t1);
	\draw (s0s1) to (s0u1);
	\draw (t0s1) to (t0t1);
	\draw (t0s1) to (t0u1);
	\draw (u0s1) to (u0t1);
	\draw (u0s1) to (u0u1);
	\draw (s0t1) to (t0t1);
	\draw (s0t1) to (u0t1);
	\draw (s0u1) to (t0u1);
	\draw (s0u1) to (u0u1);
	\draw (t0t1) to node[left] {$a$} (v0v1);
	\draw (t0u1) to node[left] {$a$} (v0w1);
	\draw (u0t1) to node[left] {$a$} (w0v1);
	\draw (u0u1) to node[left] {$a$} (w0w1);
	\draw (v0v1) to (x0v1);
	\draw (v0v1) to (y0v1);
	\draw (v0v1) to (v0x1);
	\draw (x0v1) to (x0x1);
	\draw (y0v1) to (y0x1);
	\draw (v0x1) to (x0x1);
	\draw (v0x1) to (y0x1);
	\draw (v0w1) to (x0w1);
	\draw (v0w1) to (y0w1);
	\draw (w0v1) to (z0v1);
	\draw (w0v1) to (w0x1);
	\draw (z0v1) to (z0x1);
	\draw (w0x1) to (z0x1);
	\draw (w0w1) to (z0w1);
	\end{tikzpicture}
	\caption{Example of parallel composition (fragment generated from the composed state $(\startState_{0},\startState_{1})$)}
	\label{fig:parallelComposition}
\end{figure}
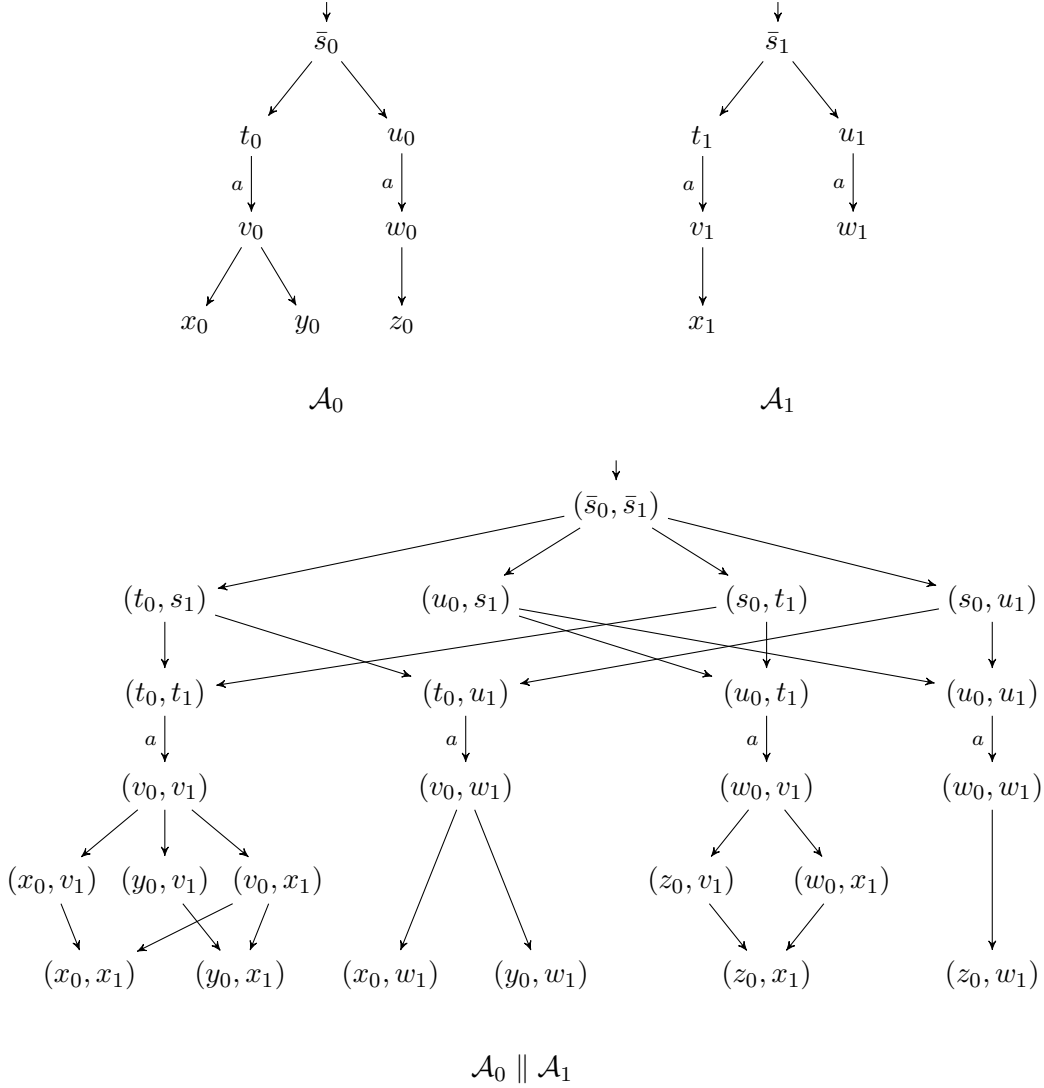
\begin{exampleStart}
	As an example of parallel composition, consider the two automata $\aut_{0}$ and $\aut_{1}$ depicted in Figure~\ref{fig:parallelComposition}.
	To keep the picture clear, we omitted the probability values since all transitions lead to a Dirac measure, as well as the $\hidden$ label on internal transitions.
	The automaton $\aut_{0} \parComp \aut_{1}$ is the fragment of the parallel composition of $\aut_{0}$ and $\aut_{1}$ reachable from the composed start state $(\startState_{0}, \startState_{1})$.
\end{exampleStart}

\subsection{Strong Probabilistic Bisimulation}
Strong probabilistic bisimilarity is the base notion for our considerations~\cite{Seg95,Seg06}. 
It uses the concept of combined transitions, defined as follows.  
Given a \PA{} $\aut$, a finite set $I$ of indexes, a family $\family{p_{i} \in \strictposreals}{i \in I}$ such that $\sum_{i \in I} p_{i} = 1$, and a family $\family{\strongTransition{s}{a}{\sd_{i}} \in \transitionRelation}{i \in I}$, we say that $s$ enables the \emph{strong combined transition} $(s,a,\sd)$, denoted by $\strongCombinedTransition{s}{a}{\sd}$, if $\sd = \sum_{i \in I} p_{i} \cdot \sd_{i}$.

\begin{definition}
\label{def:strongProbBisim}
	Let $\aut_{1}$, $\aut_{2}$ be two \PA{}s.  
	An equivalence relation $\rel$ on the disjoint union $\stateSet_{1} \uplus \stateSet_{2}$ is a \emph{strong probabilistic bisimulation} if, for each pair of states $s,t \in \stateSet_{1} \uplus \stateSet_{2}$ such that $s \rel t$, if $\strongTransition{s}{a}{\sd_{s}}$ for some probability distribution $\sd_{s}$, then there exists $\sd_{t}$ such that $\strongCombinedTransition{t}{a}{\sd_{t}}$ and $\sd_{s} \liftrel \sd_{t}$.

	We say that $\aut_{1}$ and $\aut_{2}$ are strong probabilistic bisimilar if there exists a strong probabilistic bisimulation $\relord$ on $\stateSet_{1} \uplus \stateSet_{2}$ such that $\startState_{1} \rel \startState_{2}$ and we say that two states $s_{1}$ and $s_{2}$ are strong probabilistic bisimilar if $s_{1} \rel s_{2}$. 
	We denote strong probabilistic bisimilarity by $\strongProbBisim$.
\end{definition}
If $\strongCombinedTransition{}{a}{}$ is replaced by $\strongTransition{}{a}{}$ in the above definition, then \emph{strong bisimilarity} (denoted~$\strongBisim$) results, as opposed to strong \emph{probabilistic} bisimilarity as we define it here.  
So, the difference between these two relations is that the state $t$ has to match the transition enabled by $s$ by using an ordinary transition in strong bisimulation, and a combined transition in strong probabilistic bisimulation. 
In the remainder of the paper, we may refer to $s$ as the challenger, to $t$ as the defender, and to the condition on transitions as the step condition.

It is known~\cite{Seg95} that both strong and strong probabilistic bisimilarities are equivalence relations on the set of all \PA{}s, and that they are the coarsest relations satisfying their respective bisimulation definitions. 
Furthermore, they are preserved by parallel composition, thus they are congruence relations with respect to parallel composition (and other algebraic operators such as nondeterministic choice and sequential composition).

\subsection{Weak Transitions} 

In the setting of labelled transition systems, weak transitions are used to abstract from internal computations~\cite{Mil89}. 
Intuitively, an internal weak transition is formed by an arbitrary long sequence of internal transitions, and an external weak transition is formed by an external transition preceded and followed by arbitrary long sequences of internal transitions. 
To lift this idea to the setting of probabilistic automata is a little intricate owed to the fact that transitions branch into distributions, and one thus has to work with tree-like objects instead of sequences, as detailed in the sequel.
 
An \emph{execution fragment} of a \PA{} $\aut$ is a finite or infinite sequence of alternating states and actions $\alpha = s_{0} a_{1} s_{1} a_{2} s_{2} \dots$ starting from a state $s_{0}$, also denoted by $\first{\alpha}$, and, if the sequence is finite, ending with a state denoted by $\last{\alpha}$, such that for each $i > 0$ there exists a transition $(s_{i-1}, a_{i}, \sd_{i}) \in \transitionRelation$ such that $\probeval{\sd_{i}}{s_{i}}> 0$.
The \emph{length} of $\alpha$, denoted by $\length{\alpha}$, is the number of occurrences of actions in $\alpha$.
If $\alpha$ is infinite, then $\length{\alpha} = \infty$.
Denote by $\frags{\aut}$ the set of execution fragments of $\aut$ and by $\finiteFrags{\aut}$ the set of finite execution fragments of $\aut$.
An execution fragment $\alpha$ is a \emph{prefix} of an execution fragment $\alpha'$, denoted by $\alpha \prefix \alpha'$, if the sequence $\alpha$ is a prefix of the sequence $\alpha'$.
The \emph{trace} $\trace{\alpha}$ of $\alpha$ is the sub-sequence of external actions of $\alpha$; 
we denote by $\emptytrace$ the empty trace and we define $\trace{a} = a$ for $a \in \externalActionSet$ and $\trace{a} = \emptytrace$ for $a \in \internalActionSet$. 

A \emph{scheduler} for a \PA{} $\aut$ is a function $\sched \colon \finiteFrags{\aut} \to \SubDisc{\transitionRelation}$ such that for each $\alpha \in \finiteFrags{\aut}$, $\schedeval{\sched}{\alpha} \in \SubDisc{\setcond{\tr \in \transitionRelation}{\source{\tr} = \last{\alpha}}}$.
Given a scheduler $\sched$ and a finite execution fragment $\alpha$, the distribution $\schedeval{\sched}{\alpha}$ describes how transitions are chosen to move on from $\last{\alpha}$.
A scheduler $\sched$ and a state $s$ induce a probability distribution $\sd_{\sched,s}$ over execution fragments as follows.
The basic measurable events are the cones of finite execution fragments, where the cone of $\alpha$, denoted by $\cone{\alpha}$, is the set $\setcond{\alpha' \in \frags{\aut}}{\alpha \prefix \alpha'}$.
The probability $\sd_{\sched,s}$ of a cone $\cone{\alpha}$ is defined recursively as follows:
\[
	\probeval{\sd_{\sched,s}}{\cone{\alpha}} =
	\begin{cases}
		0 & \text{if $\alpha = t$ for a state $t \neq s$,} \\
		1 & \text{if $\alpha = s$,} \\
		\probeval{\sd_{\sched,s}}{\cone{\alpha'}} \cdot \sum_{\tr \in \transitionsWithLabel{a}} \probeval{\schedeval{\sched}{\alpha'}}{\tr} \cdot  \probeval{\sd_{\tr}}{t}
			& \text{if $\alpha = \alpha'a t$.}
	\end{cases}
\]
Standard measure theoretical arguments ensure that $\sd_{\sched,s}$ extends uniquely to the $\sigma$-field generated by cones.
We call the resulting measure $\sd_{\sched,s}$ a \emph{probabilistic execution fragment} of $\aut$ and we say that it is generated by $\sched$ from $s$.
Given a finite execution fragment $\alpha$, we define $\probeval{\sd_{\sched,s}}{\alpha}$ as $\probeval{\sd_{\sched,s}}{\alpha} = \probeval{\sd_{\sched,s}}{\cone{\alpha}} \cdot \probeval{\schedeval{\sched}{\alpha}}{\bot}$, where $\probeval{\schedeval{\sched}{\alpha}}{\bot}$ is the probability of terminating the computation after $\alpha$ has occurred.

\begin{definition}
\label{def:weakCombinedTransition}
	We say that there is a \emph{weak combined transition} from $s \in \stateSet$ to $\sd \in \Disc{\stateSet}$ labelled by $a \in \actionSet$, denoted by $\weakCombinedTransition{s}{a}{\sd}$, if there exists a scheduler $\sched$ such that the following holds for the induced probabilistic execution fragment $\sd_{\sched,s}$:
	\begin{enumerate}
	\item 
		$\probeval{\sd_{\sched,s}}{\finiteFrags{\aut}} = 1$;
	\item 
		for each $\alpha \in \finiteFrags{\aut}$, if $\probeval{\sd_{\sched,s}}{\alpha} > 0$ then $\trace{\alpha} = \trace{a}$;
	\item 
		for each state $t$, $\probeval{\sd_{\sched,s}}{\setcond{\alpha \in \finiteFrags{\aut}}{\last{\alpha} = t}} = \probeval{\sd}{t}$.
	\end{enumerate}
	In this case, we say that the weak combined transition $\weakCombinedTransition{s}{a}{\sd}$ is induced by $\sched$.
\end{definition}

Although the definition of weak combined transitions is admittedly intricate, it is just the obvious extension of weak transitions on labelled transition systems to the setting with probabilities.  
We refer to Segala~\cite{Seg06} for more details on weak combined transitions.

\begin{exampleCont}
	Consider again the \PA{} $\wccHopsRadiusTransmprob{2}{5}{\frac{3}{4}}$, depicted in Figure~\ref{fig:wirelessCC}, and the weak combined transition $\weakCombinedTransition{h^{m}_{0}}{\hidden}{\dirac{h^{m}_{1}}}$.
	To simplify the notation, let us denote by $\hidden$ the internal action $t_{5}$ of $\wccHopsRadiusTransmprob{2}{5}{\frac{3}{4}}$.
	In order to show that $\weakCombinedTransition{h^{m}_{0}}{\hidden}{\dirac{h^{m}_{1}}}$ is actually a weak combined transition of $\wccHopsRadiusTransmprob{2}{5}{\frac{3}{4}}$, we have to exhibit a scheduler $\sched$ inducing it.
	It is easy to verify that $\sched$ defined as: 
	$\schedeval{\sched}{\alpha} = \dirac{\strongTransition{h^{m}_{0}}{\hidden}{\gd_{0}}}$ if $\last{\alpha} = h^{m}_{0}$, $\dirac{\bot}$ otherwise, where $\gd_{0} = \setnocond{(h^{m}_{1}, \frac{3}{4}), (h^{m}_{0}, \frac{1}{4})}$, induces the transition $\weakCombinedTransition{h^{m}_{0}}{\hidden}{\dirac{h^{m}_{1}}}$.
	Consider, for instance, the probability of stopping in $h^{m}_{1}$, i.e., the sum of the probability of each finite execution fragment ending with $h^{m}_{1}$, i.e., execution fragments of the form $(h^{m}_{0} \hidden)^{n+1} h^{m}_{1} (\hidden h^{m}_{1})^{l}$ where $l, n \in \nat$;
	it is easy to derive that for $n \in \nat$, $\probeval{\sd_{\sched,h^{m}_{0}}}{(h^{m}_{0} \hidden)^{n+1} h^{m}_{1}} = 1 \cdot (1 \cdot \frac{1}{4})^{n} \cdot \frac{3}{4} \cdot 1 = (\frac{1}{4})^{n} \cdot \frac{3}{4}$ and that for $l, n \in \nat$, $\probeval{\sd_{\sched,h^{m}_{0}}}{(h^{m}_{0} \hidden)^{n+1} h^{m}_{1} (\hidden h^{m}_{1})^{l+1}} = 1 \cdot (1 \cdot \frac{1}{4}) ^{n} \cdot \frac{3}{4} \cdot (0 \cdot \frac{1}{4})^{l+1} \cdot 1 = 0$.
	Note that the factor $0$ appearing in $\probeval{\sd_{\sched,h^{m}_{0}}}{(h^{m}_{0} \hidden)^{n+1} h^{m}_{1} (\hidden h^{m}_{1})^{l+1}}$ comes from the fact that for each $\alpha$ such that $\last{\alpha} = h^{m}_{1}$, $\probeval{\schedeval{\sched}{\alpha}}{\strongTransition{h^{m}}{\hidden}{\gd_{1}}} = 0$, where $\gd_{1} = \setnocond{(h^{m}_{2}, \frac{3}{4}), (h^{m}_{1}, \frac{1}{4})}$.
	Hence we have that $\probeval{\sd_{\sched, h^{m}_{0}}}{\setcond{\alpha \in \finiteFrags{\aut}}{\last{\alpha} = h^{m}_{1}}} = \probeval{\sd_{\sched,h^{m}_{0}}}{\setcond{(h^{m}_{0} \hidden)^{n+1} h^{m}_{1}}{n \in \nat}} + \probeval{\sd_{\sched,h^{m}_{0}}}{\setcond{(h^{m}_{0} \hidden)^{n+1} h^{m}_{1} (\hidden h^{m}_{1})^{l+1}}{l, n \in \nat}} = \sum_{n \in \nat} (\frac{1}{4})^{n} \cdot \frac{3}{4} + 0 = \frac{3}{4} \cdot \frac{1}{1-\frac{1}{4}} = 1 = \probeval{\dirac{h^{m}_{1}}}{h^{m}_{1}}$.
\end{exampleCont}

We say that there is a \emph{hyper-transition} from $\gd \in \Disc{\stateSet}$ to $\sd \in \Disc{\stateSet}$ labelled by $a \in \actionSet$, denoted by $\hyperWeakCombinedTransition{\gd}{a}{\sd}$, if there exists a family of weak combined transitions $\family{\weakCombinedTransition{s}{a}{\sd_{s}}}{s \in \Supp{\gd}}$ such that $\sd = \sum_{s \in \Supp{\gd}} \probeval{\gd}{s} \cdot \sd_{s}$.
Given $\strongTransition{s}{a}{\gd}$ and $\hyperWeakCombinedTransition{\gd}{\hidden}{\sd}$, we denote by $\weakCombinedTransition{\strongTransition{s}{a}{\gd}}{\hidden}{\sd}$ the weak combined transition $\weakCombinedTransition{s}{a}{\sd}$ obtained by concatenating $\strongTransition{s}{a}{\gd}$ and $\hyperWeakCombinedTransition{\gd}{\hidden}{\sd}$ (cf.~\cite[Proposition~3.6]{LSV07}).

\subsection{Weak Transition Compositions} 
\label{ssec:weakTranComposition}

Since we are working in a compositional setting, it will become important to discuss how weak transitions are composed via a parallel composition of \PA{}s, respectively in what sense a weak transition of the composed system can be decomposed into component weak transitions.

Given two automata $\aut_{0}$ and $\aut_{1}$, it is possible to construct a weak combined transition $\weakCombinedTransition{(s_{0}, s_{1})}{a}{\sd_{0} \times \sd_{1}}$ for the composed automaton $\aut_{0} \parComp \aut_{1}$ given two weak combined transitions $\weakCombinedTransition{s_{i}}{a_{i}}{\sd_{i}}$ of the component automata, provided that actions $a_{0}$ and $a_{1}$ are either the same external action, or $a_{i} = a$ and $a_{1-i} = \hidden$.
The construction of the composed weak combined transition is quite easy and we illustrate it on the two automata $\aut_{0}$ and $\aut_{1}$ and their parallel composition $\aut_{0} \parComp \aut_{1}$ shown in Figure~\ref{fig:parallelComposition}.

\begin{exampleStart}
\label{ex:composedTransition}
	As weak combined transitions of $\aut_{0}$ and $\aut_{1}$, consider the weak combined transitions $\tr_{0} = \weakCombinedTransition{\startState_{0}}{a}{\sd_{0}}$ where $\sd_{0} = \setnocond{(x_{0}, \frac{1}{4}), (y_{0}, \frac{1}{4}), (z_{0}, \frac{1}{2})}$ and $\tr_{1} = \weakCombinedTransition{\startState_{1}}{a}{\sd_{1}}$ where $\sd_{1} = \setnocond{(x_{1}, \frac{1}{2}), (w_{1}, \frac{1}{2})}$.

	The expected composed weak combined transition of $\aut_{1} \parComp \aut_{2}$ is $\weakCombinedTransition{(\startState_{0}, \startState_{1})}{a}{\sd_{01}}$ where the measure $\sd_{01}$ assigns value $\frac{1}{8}$ to the states $(x_{0},x_{1})$, $(y_{0},x_{1})$, $(x_{0},w_{1})$, and $(y_{0},w_{1})$ and value $\frac{1}{4}$ to the states $(z_{0},x_{1})$ and $(z_{0},w_{1})$. 

	It is easy to verify that both $\tr_{1}$ and $\tr_{2}$ are induced by the scheduler that chooses uniformly the transitions enabled by each state.
	More precisely, $\tr_{0}$ is induced by the scheduler $\sched_{0}$ defined as follows:
	\[
		\schedeval{\sched_{0}}{\alpha} = 
		\begin{cases}
			\setnocond{(\strongTransition{\startState_{0}}{\hidden}{\dirac{t_{0}}}, \frac{1}{2}), (\strongTransition{\startState_{0}}{\hidden}{\dirac{u_{0}}}, \frac{1}{2})}& \text{if $\alpha = \startState_{0}$,} \\
			\setnocond{(\strongTransition{t_{0}}{a}{\dirac{v_{0}}},1)} & \text{if $\alpha = \startState_{0} \hidden t_{0}$,} \\
			\setnocond{(\strongTransition{u_{0}}{a}{\dirac{w_{0}}},1)} & \text{if $\alpha = \startState_{0} \hidden u_{0}$,} \\
			\setnocond{(\strongTransition{v_{0}}{\hidden}{\dirac{x_{0}}}, \frac{1}{2}), (\strongTransition{v_{0}}{\hidden}{\dirac{y_{0}}}, \frac{1}{2})}& \text{if $\alpha = \startState_{0} \hidden t_{0} a v_{0}$,} \\
			\setnocond{(\strongTransition{w_{0}}{\hidden}{\dirac{z_{0}}},1)} & \text{if $\alpha = \startState_{0} \hidden u_{0} a w_{0}$, and} \\
			\dirac{\bot} & \text{otherwise.}
		\end{cases}
	\]
	Similarly, $\tr_{1}$ is induced by the scheduler $\sched_{1}$ defined as follows:
	\[
		\schedeval{\sched_{1}}{\alpha} = 
		\begin{cases}
			\setnocond{(\strongTransition{\startState_{1}}{\hidden}{\dirac{t_{1}}}, \frac{1}{2}), (\strongTransition{\startState_{1}}{\hidden}{\dirac{u_{1}}}, \frac{1}{2})}& \text{if $\alpha = \startState_{1}$,} \\
			\setnocond{(\strongTransition{t_{1}}{a}{\dirac{v_{1}}},1)} & \text{if $\alpha = \startState_{1} \hidden t_{1}$,} \\
			\setnocond{(\strongTransition{u_{1}}{a}{\dirac{w_{1}}},1)} & \text{if $\alpha = \startState_{1} \hidden u_{1}$,} \\
			\setnocond{(\strongTransition{v_{1}}{\hidden}{\dirac{x_{1}}}, 1)}& \text{if $\alpha = \startState_{1} \hidden t_{1} a v_{1}$, and} \\
			\dirac{\bot} & \text{otherwise.}
		\end{cases}
	\]
\end{exampleStart}

For $i=0,1$, let $\sched_{i}$ be the scheduler inducing $\weakCombinedTransition{s_{i}}{a_{i}}{\sd_{i}}$.
Suppose that $a_{0}$ and $a_{1}$ are the same external action $a$.
From $(s_{0}, s_{1})$, extend each obtained execution fragment $\alpha_{0}$ by scheduling the transition $\strongTransition{(v_{0},s_{1})}{\hidden}{\sd_{0} \times \dirac{s_{1}}}$ with probability $\probeval{\schedeval{\sched_{0}}{\alpha_{0}}}{\strongTransition{v_{0}}{\hidden}{\sd_{0}}}$ until no more internal transitions can be performed according to $\sched_{0}$ (here we write $\schedeval{\sched_{0}}{\alpha_{0}}$ to mean $\schedeval{\sched_{0}}{\beta_{0}}$ where $\beta_{0}$ is $\alpha_{0}$ where the second component of each state has been dropped). 

\begin{exampleCont}
	According to the above procedure, the scheduler $\sched_{01}$ that is expected to induce $\weakCombinedTransition{(\startState_{0}, \startState_{1})}{a}{\sd_{01}}$ performs the following choice:
	\begin{center}
		$\schedeval{\sched_{01}}{(\startState_{0},\startState_{1})} = \setnocond{(\strongTransition{(\startState_{0},\startState_{1})}{\hidden}{\dirac{(t_{0},\startState_{1})}}, \frac{1}{2}), (\strongTransition{(\startState_{0},\startState_{1})}{\hidden}{\dirac{(u_{0},\startState_{1})}}, \frac{1}{2})}$.
	\end{center}
\end{exampleCont}

When no more internal transitions can be performed according to $\sched_{0}$, extend each execution fragment $\alpha_{0}$ with $\last{\alpha_{0}} = (t_{0}, s_{1})$ with the execution fragment $\alpha_{1}$ obtained by scheduling the transitions $\strongTransition{(t_{0},v_{1})}{\hidden}{\dirac{t_{0}} \times \sd_{1}}$ with probability $\probeval{\schedeval{\sched_{1}}{\alpha_{1}}}{\strongTransition{v_{1}}{\hidden}{\sd_{1}}}$ until no more internal transitions can be performed according to $\sched_{1}$.

\begin{exampleCont}
	The resulting choices of the scheduler $\sched_{01}$ after this extension are the following:
	\[
		\schedeval{\sched_{01}}{\alpha} = 
		\begin{cases}
			\setnocond{(\strongTransition{(t_{0},\startState_{1})}{\hidden}{\dirac{(t_{0},t_{1})}}, \frac{1}{2}), (\strongTransition{(t_{0},\startState_{1})}{\hidden}{\dirac{(t_{0},u_{1})}}, \frac{1}{2})}& \text{if $\alpha = (\startState_{0},\startState_{1}) \hidden (t_{0},\startState_{1})$,} \\
			\setnocond{(\strongTransition{(u_{0},\startState_{1})}{\hidden}{\dirac{(u_{0},t_{1})}}, \frac{1}{2}), (\strongTransition{(u_{0},\startState_{1})}{\hidden}{\dirac{(u_{0},u_{1})}}, \frac{1}{2})}& \text{if $\alpha = (\startState_{0},\startState_{1}) \hidden (u_{0},\startState_{1})$.}
		\end{cases}
	\]
\end{exampleCont}

For each execution fragment $\alpha_{0} \alpha_{1}$, let $(v_{0}, v_{1})$ be $\last{\alpha_{0} \alpha_{1}}$ and choose the transition $\strongTransition{(v_{0},v_{1})}{a}{\sd_{0} \times \sd_{1}}$ with probability equal to $\probeval{\schedeval{\sched_{0}}{\alpha_{0}}}{\strongTransition{v_{0}}{a}{\sd_{0}}} \cdot \probeval{\schedeval{\sched_{1}}{\alpha_{1}}}{\strongTransition{v_{1}}{a}{\sd_{1}}}$.

\begin{exampleCont}
	The resulting choices of the scheduler $\sched_{01}$ after this extension are the following:
	\[
		\schedeval{\sched_{01}}{\alpha} = 
		\begin{cases}
			\setnocond{(\strongTransition{(t_{0},t_{1})}{a}{\dirac{(v_{0},v_{1})}}, 1)}& \text{if $\alpha = (\startState_{0},\startState_{1}) \hidden (t_{0},\startState_{1}) \hidden (t_{0},t_{1})$,} \\
			\setnocond{(\strongTransition{(t_{0},u_{1})}{a}{\dirac{(v_{0},w_{1})}}, 1)}& \text{if $\alpha = (\startState_{0},\startState_{1}) \hidden (t_{0},\startState_{1}) \hidden (t_{0},u_{1})$,} \\
			\setnocond{(\strongTransition{(u_{0},t_{1})}{a}{\dirac{(w_{0},v_{1})}}, 1)}& \text{if $\alpha = (\startState_{0},\startState_{1}) \hidden (u_{0},\startState_{1}) \hidden (u_{0},t_{1})$,} \\
			\setnocond{(\strongTransition{(u_{0},u_{1})}{a}{\dirac{(w_{0},w_{1})}}, 1)}& \text{if $\alpha = (\startState_{0},\startState_{1}) \hidden (u_{0},\startState_{1}) \hidden (u_{0},u_{1})$.} 
		\end{cases}
	\]
\end{exampleCont}

Let $\alpha_{0} \alpha_{1} a (u_{0}, u_{1})$ be one of the resulting execution fragments;
extend $\alpha_{0} \alpha_{1} a (u_{0}, u_{1})$ with the execution fragment $\alpha_{0a}$ obtained scheduling the transition $\strongTransition{(v_{0},u_{1})}{\hidden}{\sd_{0} \times \dirac{s_{1}}}$ with probability $\probeval{\schedeval{\sched_{0}}{\alpha_{0} a (u_{0}, u_{1}) \alpha_{0a}}}{\strongTransition{v_{0}}{\hidden}{\sd_{0}}}$  until no more internal transitions can be performed according to $\sched_{0}$.

\begin{exampleCont}
	The resulting choices of the scheduler $\sched_{01}$ after this extension are the following:
	\[
		\schedeval{\sched_{01}}{\alpha} =
		\left\{
		\kern-5pt
		\begin{array}{ll}
			\multicolumn{2}{l}{\setnocond{(\strongTransition{(v_{0},v_{1})}{\hidden}{\dirac{(x_{0},v_{1})}}, \frac{1}{2}), (\strongTransition{(v_{0},v_{1})}{\hidden}{\dirac{(y_{0},v_{1})}}, \frac{1}{2})}} \\
			& \text{if $\alpha = (\startState_{0},\startState_{1}) \hidden (t_{0},\startState_{1}) \hidden (t_{0},t_{1}) a (v_{0},v_{1})$,} \\
			\multicolumn{2}{l}{\setnocond{(\strongTransition{(v_{0},w_{1})}{\hidden}{\dirac{(x_{0},w_{1})}}, \frac{1}{2}), (\strongTransition{(v_{0},w_{1})}{\hidden}{\dirac{(y_{0},w_{1})}}, \frac{1}{2})}} \\
			& \text{if $\alpha = (\startState_{0},\startState_{1}) \hidden (t_{0},\startState_{1}) \hidden (t_{0},u_{1}) a (v_{0},w_{1})$,} \\
			\setnocond{(\strongTransition{(w_{0},v_{1})}{\hidden}{\dirac{(z_{0},v_{1})}}, 1)} & \text{if $\alpha = (\startState_{0},\startState_{1}) \hidden (u_{0},\startState_{1}) \hidden (u_{0},t_{1}) a (w_{0},v_{1})$,} \\
			\setnocond{(\strongTransition{(w_{0},w_{1})}{\hidden}{\dirac{(z_{0},w_{1})}}, 1)} & \text{if $\alpha = (\startState_{0},\startState_{1}) \hidden (u_{0},\startState_{1}) \hidden (u_{0},u_{1}) a (w_{0},w_{1})$.}
		\end{array}
		\right.
	\]
\end{exampleCont}

When no more internal transitions can be performed according to $\sched_{0}$, extend each execution fragment $\alpha_{0} \alpha_{1} a (u_{0}, u_{1}) \alpha_{0a}$ with $\last{\alpha_{0a}} = (x_{0}, u_{1})$ with the execution fragment $\alpha_{1a}$ obtained by scheduling the transitions $\strongTransition{(x_{0}, v_{1})}{\hidden}{\dirac{x_{0}} \times \sd_{1}}$ with probability equal to $\probeval{\schedeval{\sched_{1}}{\alpha_{1} a (u_{0}, u_{1}) \alpha_{1a}}}{\strongTransition{v_{1}}{\hidden}{\sd_{1}}}$ until no more internal transitions can be performed according to $\sched_{1}$.
Since there may be finite execution fragments for which the resulting scheduler is still undefined, extend the scheduler by mapping such execution fragments to $\dirac{\bot}$.

\begin{exampleCont}
	The resulting choices of the scheduler $\sched_{01}$ after this extension are the following:
	\[
		\schedeval{\sched_{01}}{\alpha} =
		\begin{cases}
			\setnocond{(\strongTransition{(x_{0},v_{1})}{\hidden}{\dirac{(x_{0},x_{1})}}, 1)} & \text{if $\alpha = (\startState_{0},\startState_{1}) \hidden (t_{0},\startState_{1}) \hidden (t_{0},t_{1}) a (v_{0},v_{1}) \hidden (x_{0}, v_{1})$,} \\
			\setnocond{(\strongTransition{(y_{0},v_{1})}{\hidden}{\dirac{(y_{0},x_{1})}}, 1)} & \text{if $\alpha = (\startState_{0},\startState_{1}) \hidden (t_{0},\startState_{1}) \hidden (t_{0},t_{1}) a (v_{0},v_{1}) \hidden (y_{0}, v_{1})$,} \\
			\setnocond{(\strongTransition{(z_{0},v_{1})}{\hidden}{\dirac{(z_{0},x_{1})}}, 1)} & \text{if $\alpha = (\startState_{0},\startState_{1}) \hidden (u_{0},\startState_{1}) \hidden (u_{0},t_{1}) a (w_{0},v_{1}) \hidden (z_{0}, v_{1})$,} \\
			\dirac{\bot} & \text{otherwise.}
		\end{cases}
	\]
	It is routine to verify that the scheduler $\sched_{01}$ induces $\weakCombinedTransition{(\startState_{0},\startState_{1})}{a}{\sd_{01}}$ as desired.
\end{exampleCont}

If $a_{0}$ and $a_{1}$ are not the same external action $a$, then the construction is similar, except for the scheduling of the transitions with label $a$ (if external) and the extension with the subsequent internal transitions.
It can be shown that this construction actually leads to the weak combined transition $\weakCombinedTransition{(s_{0}, s_{1})}{a}{\sd_{0} \times \sd_{1}}$.

The inverse operation, that is, the decomposition of a weak combined transition of a composed automaton into weak combined transitions of the component automata, is possible as well.
In fact,~\cite[Section~4.3.2]{Seg95} shows how identify two weak combined transitions,
one for each component automaton, corresponding to the view that each
component automaton has of the composed weak combined transition.
\begin{definition}
	Given two compatible \PA{}s $\aut_{0}$ and $\aut_{1}$, let $\aut$ be their parallel composition $\aut_{0} \parComp \aut_{1}$ and consider a weak combined transition $\weakCombinedTransition{s}{a}{\sd}$ with $s = (s_{0}, s_{1})$.
	The \emph{projection} $\project{i}{\weakCombinedTransition{s}{a}{\sd}}$ of $\weakCombinedTransition{s}{a}{\sd}$ on the component automaton $\aut_{i}$ is the weak combined transition $\weakCombinedTransition{s_{i}}{a_{i}}{\sd_{i}}$ where $a_{i} = a$ if $a \in \actionSet_{i}$, $a_{i} = \hidden$ otherwise, and for each $t_{i} \in \stateSet_{i}$, $\probeval{\sd_{i}}{t_{i}} = \sum_{t_{1-i} \in \stateSet_{1-i}} \probeval{\sd}{t_{0}, t_{1}}$.
\end{definition}
Note that the action $a_{i}$ labelling $\weakCombinedTransition{s_{i}}{a_{i}}{\sd_{i}}$ depends on how $\aut_{i}$ considers the action $a$: 
if $a$ is an action of $\aut_{i}$ (independently on whether it is internal or external), then $a_{i} = a$; 
otherwise, it means that $a$ is an action only of $\aut_{1-i}$, so the contribution of $\aut_{i}$ to $\weakCombinedTransition{s}{a}{\sd}$ involves possibly only internal transitions, thus we use $\hidden$ as label for $\weakCombinedTransition{s_{i}}{a_{i}}{\sd_{i}}$.

The technical construction of $\weakCombinedTransition{s_{i}}{a}{\sd_{i}}$ is rather involved (cf.~\cite[Section~4.3.2]{Seg95}), since it requires to manage correctly the probabilistic choices of the scheduler. 
Intuitively, for obtaining the projection on the component automaton $\aut_{i}$, each execution fragment and the probabilistic execution fragment underlying the weak combined transition of the composed automaton is compressed by removing the pairs of actions and states corresponding to only a transition from the other component automaton $\aut_{1-i}$, i.e., the composed transition involves an apparent transition for $\aut_{i}$.

\subsection{Weak Probabilistic Bisimulation} 
\label{ssec:weakProbBisimulation}

The above definition of weak combined transitions (Definition~\ref{def:weakCombinedTransition}) naturally lead us to the definition of the weak counterpart of strong probabilistic bisimilarity, namely weak probabilistic bisimilarity~\cite{Seg95,Seg06}. 

\begin{definition}
	Let $\aut_{1}$, $\aut_{2}$ be two \PA{}s.  
	An equivalence relation $\rel$ on the disjoint union $\stateSet_{1} \uplus \stateSet_{2}$ is a \emph{weak probabilistic bisimulation} if, for each pair of states $s,t \in \stateSet_{1} \uplus \stateSet_{2}$ such that $s \rel t$, if $\strongTransition{s}{a}{\sd_{s}}$ for some probability distribution $\sd_{s}$, then there exists $\sd_{t}$ such that $\weakCombinedTransition{t}{a}{\sd_{t}}$ and $\sd_{s} \liftrel \sd_{t}$.

	We say that $\aut_{1}$ and $\aut_{2}$ are weak probabilistic bisimilar if there exists a weak probabilistic bisimulation $\relord$ on $\stateSet_{1} \uplus \stateSet_{2}$ such that $\startState_{1} \rel \startState_{2}$ and we say that two states $s_{1}$ and $s_{2}$ are weak probabilistic bisimilar if $s_{1} \rel s_{2}$. 
	We denote weak probabilistic bisimilarity by $\weakBisim$.
\end{definition}

As happens for the strong case, it is known~\cite{Seg95} that weak probabilistic bisimilarity is an equivalence relation on the set of all \PA{}s, and that it is the coarsest relation satisfying its bisimulation definition. 
Furthermore, it is preserved by parallel composition, thus it is a congruence relation with respect to parallel composition.

\begin{exampleCont}
	Consider  any instance $\wccHopsRadiusTransmprob{n}{r}{p}$ and the ideal communication channel $\icc$.
	It is quite easy to verify that $\icc \weakBisim \wccHopsRadiusTransmprob{n}{r}{p}$ for each $n \in \nat$, $r \in \posreals$, and $p \in (0,1]$, where the relation $\relord$ justifying $\icc \weakBisim \wccHopsRadiusTransmprob{n}{r}{p}$ has for each $m \in \msgSet$ one class containing all $h^{m}_{i}$ states and another class containing start states.
	This means, by transitivity of $\weakBisim$, that $\wccHopsRadiusTransmprob{n}{r}{p} \weakBisim \wccHopsRadiusTransmprob{n'}{r'}{p'}$ for each possible value of $n, n' \in \nat$, $r, r' \in \posreals$, and $p,p' \in (0,1]$.
\end{exampleCont}

There exists also a notion of weak bisimulation on probabilistic automata, obtained by restricting the step condition of the weak probabilistic bisimulation to use only Dirac schedulers. 
This echoes the difference between strong and strong probabilistic bisimilarity.
A Dirac scheduler is a scheduler $\sched$ such that for each execution fragment $\alpha$, $\schedeval{\sched}{\alpha}$ is a Dirac distribution. 
The main problem with this weak bisimulation is that it is not transitive~\cite{Den05}, opposed to weak probabilistic bisimilarity. 

\section{Cost Probabilistic Automata}
\label{sec:costProbabilisticAutomata}

We are now ready to discuss the cost augmented probabilistic automata model that will be in our focus. 
As already hinted at in Example~\ref{ex:wcc}, we consider as \emph{cost} any kind of quantity associated with the transitions of the automaton $\aut$. 
We aim to minimise these costs.
We model the cost of the transitions by a function $\costf$ that assigns to each transition a non-negative real value.
\begin{definition}
	A \emph{cost probabilistic automaton} (\CPA{}) is a pair $\costAut{\costf}$ where $\aut = (\stateSet, \startState, \actionSet, \transitionRelation)$ is a probabilistic automaton and $\costf$, the \emph{transition cost function}, is a total function $\costf \colon \transitionRelation \to \posreals$.
\end{definition}
The above definition follows (and generalises) the standard definition of \emph{reward structure} we find for \MDP{}s (cf.~\cite[Section~2.1.3]{Put05}):
\begin{definition}
	A \emph{Markov decision process with rewards} is a tuple $(\stateSet, \startState, \actionSet, \transitionProbability, \rewardFunction)$ where $(\stateSet, \startState, \actionSet, \transitionProbability)$ is an \MDP{} and $\rewardFunction \colon \stateSet \times \actionSet \to \reals$ is a \emph{reward function} or \emph{structure}.
\end{definition}
In this paper we consider only non-negative rewards, i.e., it is assumed as $\rewardFunction(s,a) \geq 0$ for each $(s,a) \in \stateSet \times \actionSet$. 
We usually interpret them as transition \emph{costs}.

\subsection{Strong Combined Transition Cost}

The extension of costs from a single transition to a convex combination of transitions is straightforward: 
It is canonical to consider as cost the weighted sum of the costs of the transitions which are being combined. 
This corresponds directly to the expected reward criterion we find in the operations research literature~\cite{How07}.
\begin{definition}
	Given a \CPA{} $\costAut{\costf}$, a finite set $I$ of indexes, a family $\family{p_{i} \in \strictposreals}{i \in I}$ such that $\sum_{i \in I} p_{i} = 1$, and a family $\family{\strongTransition{s}{a}{\sd_{i}} \in \transitionRelation}{i \in I}$,
	let $\strongCombinedTransition{s}{a}{\sd}$ be the resulting strong combined transition.
	Then, the cost of $\strongCombinedTransition{s}{a}{\sd}$ is defined as 
	\[
		\cost{\strongCombinedTransition{s}{a}{\sd}} = \sum_{i \in I} p_{i} \cdot \cost{\strongTransition{s}{a}{\sd_{i}}}\text{.}
	\]
\end{definition}
This definition will be used in the definition of cost-preserving strong probabilistic bisimilarity.

\subsection{Weak Combined Transition Cost}

While there is a canonical way to generalise transition costs to strong combined transitions, it is not so obvious how to faithfully extend this to weak combined transitions: 
There are several ways of extending the cost from a single transition to a sequence of transitions, and hence to a weak combined transition, and we elaborate on this in the sequel.  
A prominent possibility is to consider the weighted sum of the costs of all involved finite execution fragments. 
This approach matches the standard interpretation in the operations research literature for expected reward criteria~\cite{Put05,How07} for \MDP{}s.

\begin{definition}
\label{def:rewardAsUsualDefinition}
	Given an \MDP{} $\aut[M] = (\stateSet, \startState, \actionSet, \transitionProbability, \rewardFunction)$, a finite execution fragment $\alpha = s_{1} a_{1} \dots s_{n} a_{n} s_{n+1} \in \finiteFrags{\aut[M]}$, and a policy $\policy$, let $\fragRestriction{\alpha}{i} = s_{1} a_{1} \dots a_{i-1} s_{i}$ be the $i$-prefix of $\alpha$, $\functioneval{\rewardFunction}{\alpha} = \sum_{i=1}^{n} \functioneval{\rewardFunction}{s_{i},a_{i}}$, and $\probPolicyFrag{\alpha} = \probeval{\dirac{\startState}}{s_{1}} \cdot \prod_{i=1}^{n} \probeval{\policyeval{\policy}{\fragRestriction{\alpha}{i}}}{a_{i}} \cdot \probeval{\probeval{\transitionProbability}{s_{i},a_{i}}}{s_{i+1}}$.
  
	Then the \emph{expected total reward with horizon $N$ under policy $\policy$} is defined as $\expectedReward[N]{\policy} = \sum_{\alpha \in \setcond{\alpha \in \finiteFrags{\aut[M]}}{\length{\alpha} = N}} \functioneval{\rewardFunction}{\alpha} \cdot \probPolicyFrag{\alpha}$.
\end{definition}
Since probabilistic automata are a conservative extension of \MDP{}, we extend this notion to weak transition costs by taking into account the resolution of the nondeterminism as induced by a given scheduler.
This approach is similar to a radial characterisation of the cost: 
If we imagine the target probability distribution as the border of the execution, then each finite execution fragment is like a ray leaving the source state $s$ and reaching such border where the execution stops. 
The cost of reaching the border is then the weighted sum $\sum_{\alpha \in \finiteFrags{\aut}} \cost[\sched]{\alpha} \cdot \probeval{\sd_{\sched,s}}{\alpha}$ of the cost $\cost[\sched]{\alpha}$ of each ray $\alpha$ that is given by the sum of the cost of each transition part of the ray weighted by the probability of such transition.

\begin{definition}
\label{def:costAsUsualDefinition}
	Given a \CPA{} $\costAut{\costf}$, a state $s$, an action $a$, a probability distribution $\sd$, and a scheduler $\sched$ inducing the weak combined transition $\weakCombinedTransition{s}{a}{\sd}$, we define the cost $\cost[\sched]{\weakCombinedTransition{s}{a}{\sd}}$ of the weak combined transition $\weakCombinedTransition{s}{a}{\sd}$ as 
	\[
		\cost[\sched]{\weakCombinedTransition{s}{a}{\sd}} = \sum_{\alpha \in \finiteFrags{\aut}} \cost[\sched]{\alpha} \cdot \probeval{\sd_{\sched,s}}{\alpha}
	\]
	where 
	$\cost[\sched]{\alpha} = \cost[\sched]{\alpha'} + \sum_{\tr \in \transitionsWithLabel{a}} \cost{\tr} \cdot \schedeval{\widehat{\sched}}{\alpha', t, a, \tr}$ if $\alpha = \alpha'a t$, $0$ otherwise, and 
	where $\widehat{\sched} \colon \finiteFrags{\aut} \times \stateSet \times \actionSet \times \transitionRelation \to \posreals$ is defined as:
	\[
		\schedeval{\widehat{\sched}}{\alpha, t, a, \tr} = 
		\begin{cases}
			\dfrac{\probeval{\schedeval{\sched}{\alpha}}{\tr} \cdot \probeval{\sd_{\tr}}{t}}{\sum_{\tr \in \transitionsWithLabel{a}} \probeval{\schedeval{\sched}{\alpha}}{\tr} \cdot \probeval{\sd_{\tr}}{t}} & \text{if $\sum_{\tr \in \transitionsWithLabel{a}} \probeval{\schedeval{\sched}{\alpha}}{\tr} \cdot \probeval{\sd_{\tr}}{t} > 0$,} \\
			0 & \text{otherwise.}
		\end{cases}
	\]
\end{definition}
\noindent When the scheduler $\sched$ is clear from the context, we just write $\cost{\weakCombinedTransition{s}{a}{\sd}}$.

In the above definition, the function $\schedeval{\widehat{\sched}}{\alpha, t, a, \tr}$ is used to normalise the contribution of each scheduled transition to the cost of the resulting finite execution fragment, so that the probabilistic effects of the choice of the scheduler and the transition are correctly managed only by $\probeval{\sd_{\sched,s}}{\alpha}$.
We remark that here $\probeval{\sd_{\sched,s}}{\alpha} > 0$ implies that $\trace{\alpha} = \trace{a}$ and that $\probeval{\probeval{\sd_{\sched,s}}{\alpha}}{\bot} > 0$, i.e., the cost of $\alpha$ is considered in the sum only when the the computation stops.
This ensures also the correctness of the definition, since it is not possible to account multiple times the same probability values. 
This is a particular consequence of the fact that each finite execution fragment is a measurable event, as explained in~\cite[Example~3.1]{LSV07}.

When we restrict Definition~\ref{def:costAsUsualDefinition} to \MDP{}s, it coincides with Definition~\ref{def:rewardAsUsualDefinition}, so the definition of cost of the weak combined transition induced by a scheduler $\sched$ is a conservative extension of the definition of expected total reward with horizon $N$ under policy $\policy$:
\begin{proposition}
\label{pro:costAsExpectedRewardInMDP}
	Given an \MDP{} $\aut[M] = (\stateSet, \startState, \actionSet, \transitionProbability)$ and a policy $\policy$, 
	let $\aut$ be the \PA{} $(\stateSet, \startState, \actionSet, \transitionRelation)$ where $\transitionRelation = \setcond{(s, a, \probeval{\transitionProbability}{s,a})}{s \in \stateSet, a \in \actionSet(s)}$. 
	For each $N \in \nat$, $\tr = (s, a, \sd) \in \transitionRelation$ and $\alpha \in \finiteFrags{\aut[M]}$, let $\cost{\tr} = \functioneval{\rewardFunction}{s,a}$, 
	$\probeval{\schedeval{\sched}{\alpha}}{\tr} =\probeval{\policyeval{\policy}{\alpha}}{a}$ if $\length{\alpha} < N$, $0$ otherwise, 
	and $\weakCombinedTransition{\startState}{\hidden}{\sd}$ be the weak combined transition of $\aut$ induced by the scheduler $\sched$ when all actions are considered as internal. 
	Then, it holds that for each  $N \in \nat$, 
	\[
		\expectedReward[N]{\policy} = \cost[\sched]{\weakCombinedTransition{\startState}{\hidden}{\sd}}\text{.}
	\]
\end{proposition}
\begin{myoutline}
	The proof is based on a simple manipulation of the definition of expected total reward with horizon $N$ under policy $\policy$.
\end{myoutline}

\begin{exampleCont}
	Consider the \CPA{} $\costAut[\wccHopsRadiusTransmprob{2}{5}{\frac{3}{4}}]{\costf}$ where $\costf$ assigns cost $25$ to each transition labelled by the internal action $\hop{5}$; 
	the weak combined transition $\weakCombinedTransition{h^{m}_{0}}{\hop{5}}{\dirac{h^{m}_{2}}}$ can be seen as the concatenation of the two transitions $\weakCombinedTransition{h^{m}_{0}}{\hop{5}}{\dirac{h^{m}_{1}}}$ and $\weakCombinedTransition{h^{m}_{1}}{\hop{5}}{\dirac{h^{m}_{2}}}$. 
	It is routine to check that each $\weakCombinedTransition{h^{m}_{0}}{\hop{r}}{\dirac{h^{m}_{1}}}$ (and similarly for $\weakCombinedTransition{h^{m}_{1}}{\hop{5}}{\dirac{h^{m}_{2}}}$) is induced by the scheduler $\sched_{0}$ 
	such that $\schedeval{\sched_{0}}{\alpha} = \dirac{\tr^{m}_{0}}$ if $\last{\alpha} = h^{m}_{0}$, $\dirac{\bot}$ otherwise,
	where $\tr^{m}_{0} = \strongTransition{h^{m}_{0}}{\hop{5}}{\setnocond{(h^{m}_{1}, \frac{3}{4}), (h^{m}_{0}, \frac{1}{4})}}$.
	Now, consider the finite execution fragment $\alpha = (h^{m}_{0} \hop{5})^{n+1} h^{m}_{1}$: 
	According to Definition~\ref{def:costAsUsualDefinition}, it has cost $\cost[\sched_{0}]{\alpha} = (n+1) \cdot 25$.
	The probability $\probeval{\sd_{\sched_{0},h^{m}_{0}}}{\alpha}$ of $\alpha$ is $\frac{1}{4}^{n} \cdot \frac{3}{4}$ while the probability of each $\alpha' \in \finiteFrags{\wccHopsRadiusTransmprob{2}{5}{\frac{3}{4}}} \setminus \setcond{(h^{m}_{0} \hop{5})^{n+1} h^{m}_{1}}{n \in \nat}$ is $0$, thus the cost of the transition $\weakCombinedTransition{h^{m}_{0}}{\hop{5}}{\dirac{h^{m}_{1}}}$ as induced by $\sched_{0}$ is $\cost[\sched_{0}]{\weakCombinedTransition{h^{m}_{0}}{\hop{5}}{\dirac{h^{m}_{1}}}} = \sum_{n \in \nat} (n+1) \cdot 25 \cdot \frac{1}{4}^{n} \cdot \frac{3}{4} = 25 \cdot \frac{3}{4} \cdot \sum_{n \in \nat} (n+1) \cdot \frac{1}{4}^{n} = \frac{25 \cdot \frac{3}{4}}{\frac{1}{4}} \cdot \sum_{n \in \nat} (n+1) \cdot \frac{1}{4}^{n+1} = \frac{25 \cdot \frac{3}{4}}{\frac{1}{4}} \cdot \frac{\frac{1}{4}}{(1-\frac{1}{4})^{2}} = \frac{25}{\frac{3}{4}}$, hence $\weakCombinedTransition{h^{m}_{0}}{\hop{5}}{\dirac{h^{m}_{2}}}$ has cost $2 \cdot \frac{25}{\frac{3}{4}}$.
	
	By using a similar approach, it is easy to generalise the above result to the \CPA{} $\costAut[\wccHopsRadiusTransmprob{n}{r}{p}]{\costf}$, where $\costf$ assigns cost $\givenCost[r]$ to each transition labelled by the internal action $\hop{r}$, and the weak combined transition $\weakCombinedTransition{h^{m}_{0}}{\hop{r}}{\dirac{h^{m}_{n}}}$: 
	the resulting cost for such transition is $n \cdot \frac{\givenCost[r]}{p}$.
\end{exampleCont}

According to Definition~\ref{def:costAsUsualDefinition}, we consider a weak transition cost as a radial characterisation of the cost: 
The cost of reaching the border, where the execution stops, is the weighted sum of the cost of each ray $\alpha$ that is given by the sum of the cost of each transition part of the ray weighted by the probability of such transition.

The same execution border can be reached also by an increasing sequence of concentric balls whose centre is the source state $s$ and whose radii are the number of performed steps.  
In this case the overall cost is the cost of performing the first step from the start state $s$ to the radius $1$ ball plus the cost of the steps from radius $1$ to radius $2$ ball plus the cost of the steps from radius $2$ to radius $3$ ball, and so on, where the cost of each single transition is weighted by the probability of performing it. 
This construction can be formalised as follows:
\begin{definition}
\label{def:costAsWorkingForTheProof}
	Given a \CPA{} $\costAut{\costf}$, a state $s$, an action $a$, a probability distribution $\sd$, and a scheduler $\sched$ inducing the weak combined transition $\weakCombinedTransition{s}{a}{\sd}$, we define the cost $\cost[\sched]{\weakCombinedTransition{s}{a}{\sd}}$ of $\weakCombinedTransition{s}{a}{\sd}$ as 
	\[
		\cost[\sched]{\weakCombinedTransition{s}{a}{\sd}} = \sum_{\alpha \in \finiteFrags{\aut}} \cost[\sched,s]{\alpha}
	\]
	where 
	\[
		\cost[\sched,s]{\alpha} =
		\begin{cases}
			\probeval{\sd_{\sched,s}}{\cone{\alpha'}} \cdot \sum_{\tr \in \transitionsWithLabel{b}} \cost{\tr} \cdot \probeval{\schedeval{\sched}{\alpha'}}{\tr} \cdot \probeval{\sd_{\tr}}{t} & \text{if $\alpha = \alpha'b t$,} \\
			0 & \text{otherwise}
		\end{cases}
	\]
\end{definition}
As it can be expected, Definition~\ref{def:costAsWorkingForTheProof} is an equivalent characterisation of Definition~\ref{def:costAsUsualDefinition}:
\begin{proposition}
\label{pro:equivalenceOfCostDefinitions}
	Given a \CPA{} $\costAut{\costf}$, a state $s$, an action $a$, a probability distribution $\sd$, and a scheduler $\sched$ inducing the weak combined transition $\weakCombinedTransition{s}{a}{\sd}$, it holds that 
	\[
		\costf^{\mathit{Ray}}_{\sched}(\weakCombinedTransition{s}{a}{\sd}) = \costf^{\mathit{Ball}}_{\sched}(\weakCombinedTransition{s}{a}{\sd})
	\]
	where costs $\costf^{\mathit{Ray}}_{\sched}$ and $\costf^{\mathit{Ball}}_{\sched}$ are defined according to Definition~\ref{def:costAsUsualDefinition} and~\ref{def:costAsWorkingForTheProof}, respectively.
\end{proposition}
\begin{myoutline}
	The proof is based on a manipulation of the definition of the cost $\costf^{\mathit{Ray}}_{\sched}$ and of the probability of a cone, together with the fact that $\sched$ induces a weak combined transition.
\end{myoutline}
The concentric ball characterisation will be taken up later in Section~\ref{sec:transitioncosts} where we discuss the algorithmic aspects of \CPA{}s.

\subsection{Parallel Composition on Cost Probabilistic Automata}

We now extend the parallel composition operator to the cost setting. 
As expected, it is based on the underlying probabilistic automata definition. 
However, defining the cost of transitions in the parallel composition is not entirely obvious since we have to carefully define the cost of the resulting transitions.

\begin{definition}
	Given two \CPA{}s $\costAutPed{\costf}{1}$ and $\costAutPed{\costf}{2}$, we say that $\costAutPed{\costf}{1}$ and $\costAutPed{\costf}{2}$ are \emph{compatible} if the underlying \PA{}s $\aut_{1}$ and $\aut_{2}$ are compatible.
\end{definition}
	
\begin{definition}
	Given two compatible \CPA{}s $\costAutPed{\costf}{1}$ and $\costAutPed{\costf}{2}$, consider the parallel composition $\aut_{1} \parComp \aut_{2} = (\stateSet, \startState, \actionSet, \transitionRelation)$ of $\aut_{1}$ and $\aut_{2}$. 
	Let $\costf \colon \transitionRelation \to \posreals$ be total;
	we say that $\costf$ is \emph{cost-preserving} if there exists a symmetric, zero-preserving, distributive (over convex combination) and monotone increasing function $f \colon \reals \times \reals \to \reals$ such that, for each $((s_{1},s_{2}), a, \sd_{1} \times \sd_{2}) \in \transitionRelation$, 
	\[
		\cost{(s_{1},s_{2}), a, \sd_{1} \times \sd_{2}} = 
		\begin{cases}
			\functioneval{f}{\cost[1]{s_{1}, a, \sd_{1}}, \cost[2]{s_{2}, a, \sd_{2}}} & \text{if $a \in \actionSet_{1} \cap \actionSet_{2}$,} \\
			\functioneval{f}{\cost[1]{s_{1}, a, \sd_{1}}, \cost[2]{s_{2}, \apparent{a}, \dirac{s_{2}}}} & \text{if $a \in \actionSet_{1} \setminus \actionSet_{2}$,}\\
			\functioneval{f}{\cost[1]{s_{1}, \apparent{a}, \dirac{s_{1}}}, \cost[2]{s_{2}, a, \sd_{2}}} & \text{if $a \in \actionSet_{2} \setminus \actionSet_{1}$, and}\\
			0 & \text{otherwise,}
		\end{cases}
	\]
	where we extend $\costf_{i}$ by $\cost[i]{s_{i}, \apparent{a}, \dirac{s_{i}}} = 0$ for each apparent transition $(s_{i}, \apparent{a}, \dirac{s_{i}})$, $i=1,2$.
	We say that $f$ is the \emph{generator} of $\costf$ and we denote it by $\gen{\costf}$.
\end{definition}
The generator function $f$ describes how to compute the cost of a composed transition. 
A common instance of the function $f$ is addition ($+$), since often the cost of the composed transition is just the sum of the cost of the component transitions. 
The generator function generalises this to arbitrary functions provided they are symmetric, zero-preserving, distributive (over convex combination), and monotone increasing. 
These are the properties needed to establish compositionality of the bisimulations we are going to introduce in the remainder of the section. 
The generator function allows us to obtain more general results for parallel composition than if restricting to addition.

\begin{definition}
	Given two compatible \CPA{}s $\costAutPed{\costf}{1}$ and $\costAutPed{\costf}{2}$ and a cost-preserving function $\costf$, the \emph{parallel composition} of $\costAutPed{\costf}{1}$ and $\costAutPed{\costf}{2}$, denoted by $\costAutPed{\costf}{1} \parComp_{\costf} \costAutPed{\costf}{2}$, is the cost probabilistic automaton $\costAut{\costf} = (\aut_{1} \parComp \aut_{2}, \costf)$.
\end{definition}

\section{Cost Preserving Bisimulations}
\label{sec:costPreservingBisims}

The previous section has discussed how costs can be lifted to strong combined, respectively weak combined transitions. 
This is now incorporated into the probabilistic bisimulations of \PA{}s.

\subsection{Strong Cost Preserving Bisimulations}
\label{ssec:strongCostPreservingBisim}

Since a \CPA{} is an ordinary \PA{} enriched with a cost function, one might consider a naive lifting of \PA{} strong (probabilistic) bisimulation, where two \CPA{}s are strong (probabilistic) bisimilar if the underlying \PA{}s are. 
However this definition obviously falls too short, since it may relate states with different cost behaviours.
For this reason and following~\cite{GDG03} we define a refined notion of bisimulation where each transition of the challenging state has to be matched by the defender state by enabling a strong (combined) transition, as in ordinary strong (probabilistic) bisimulation, and, in addition, the costs of challenging and defending transitions must agree. 
In other words, the extension of strong bisimulation and strong probabilistic bisimulation from \PA{}s to \CPA{}s enforces that additional to the original conditions, a defender must match the transition costs proposed by the challenger.
\begin{definition}
\label{def:strongCostBisimEq}
	Given two \CPA{}s $\costAutPed{\costf}{1}$ and $\costAutPed{\costf}{2}$, an equivalence relation $\relord$ on the disjoint union $\stateSet_{1} \uplus \stateSet_{2}$ is a \emph{strong probabilistic cost-preserving bisimulation} if for each pair of states $s,t \in \stateSet_{1} \uplus \stateSet_{2}$ such that $s \rel t$, if $\strongTransition{s}{a}{\sd_{s}}$, then there exists $\sd_{t}$ such that $\strongCombinedTransition{t}{a}{\sd_{t}}$, $\sd_{s} \liftrel \sd_{t}$, and $\cost[d]{\strongCombinedTransition{t}{a}{\sd_{t}}} = \cost[c]{\strongTransition{s}{a}{\sd_{s}}}$ where $\costf_{d}$ and $\costf_{c}$ are the cost functions of the defender and the challenger \CPA{}, respectively.

	Two \CPA{}s $\costAutPed{\costf}{1}$ and $\costAutPed{\costf}{2}$ are strong probabilistic cost-preserving bisimilar if there exists a strong probabilistic cost-preserving bisimulation $\relord$ on $\stateSet_{1} \uplus \stateSet_{2}$ such that $\startState_{1} \rel \startState_{2}$.  
	We denote  strong probabilistic cost-preserving bisimilarity by $\strongCostProbBisim$.
\end{definition}
By using this definition of bisimulation, we have that states enabling transitions with different cost are no more bisimilar, since they do not respect cost constraints.

As in the pure probabilistic setting (Definition~\ref{def:strongProbBisim}), the above definition can be twisted to give rise to a \emph{strong cost-preserving bisimilarity}, denoted~$\strongCostBisim$, by replacing all occurrences of $\strongCombinedTransition{}{a}{}$ by $\strongTransition{}{a}{}$.  
It is then trivial to show that two strong cost-preserving bisimilar \CPA{}s are also strong probabilistic cost-preserving bisimilar.

\subsection{Properties of Strong Cost Preserving Bisimilarities.}

The relations we have defined above inherit all the relevant properties from their cost-insensitive counterparts.

\begin{proposition}
\label{pro:strongCostBisimEqImpliesStrongCostProbBisimEq}
	Given two \CPA{}s $\costAutPed{\costf}{1}$ and $\costAutPed{\costf}{2}$, if $\costAutPed{\costf}{1} \strongCostBisim \costAutPed{\costf}{2}$, then $\costAutPed{\costf}{1} \strongCostProbBisim \costAutPed{\costf}{2}$.
\end{proposition}
\begin{myproof}
	The result is immediate, since each transition is also a combined transition; 
	this implies that the relation $\relord$ justifying $\costAutPed{\costf}{1} \strongCostBisim \costAutPed{\costf}{2}$ is also a strong probabilistic cost-preserving bisimulation, thus $\costAutPed{\costf}{1} \strongCostProbBisim \costAutPed{\costf}{2}$.
\end{myproof}

Another important property of both strong and strong probabilistic cost-preserving bisimilarities is that they are equivalence relations on the set of \CPA{}s. 

\begin{proposition}
\label{pro:strongCostBisimEqIsEquivalenceRelation}
	Strong and strong probabilistic cost-preserving bisimilarities are equivalence relations on the set of \CPA{}s.
\end{proposition}
\begin{myoutline}
	Reflexivity and symmetry are trivial as they follow immediately from reflexivity and symmetry of the underlying equivalence relations.
	Transitivity is based on the equivalence relation witnessing the transitivity of the underlying bisimulation for \PA{}s: with a simple manipulation of combined transitions, it is shown that the costs are preserved.
\end{myoutline}

Strong probabilistic cost-preserving bisimilarity is the coarsest strong cost-preserving probabilistic bisimulation, and similarly for strong cost-preserving  bisimilarity.
As one may expect, both strong and strong probabilistic cost-preserving bisimilarities are preserved by parallel composition.
\begin{proposition}
\label{pro:strongCostBisimEqIsCompositional}
	Given two \CPA{}s $\costAutPed{\costf}{1}$ and $\costAutPed{\costf}{2}$, if $\costAutPed{\costf}{1} \strongCostProbBisim \costAutPed{\costf}{2}$, then for each $\costAutPed{\costf}{3}$ compatible with both $\costAutPed{\costf}{1}$ and $\costAutPed{\costf}{2}$ and each pair of cost-preserving functions $\costf_{l}$ and $\costf_{r}$ with $\gen{\costf_{l}} = \gen{\costf_{r}}$, $\costAutPed{\costf}{1} \parComp_{\costf_{l}} \costAutPed{\costf}{3} \strongCostProbBisim \costAutPed{\costf}{2} \parComp_{\costf_{r}} \costAutPed{\costf}{3}$, and similarly for $\strongCostBisim$.
\end{proposition}
\begin{myoutline}
	The proof is based on the relation justifying the compositionality of the underlying bisimulation between \PA{}s; 
	the preservation of the cost of the transitions is based on the properties of the cost-preserving functions $\costf_{l}$ and $\costf_{r}$ with $\gen{\costf_{l}} = \gen{\costf_{r}}$.
\end{myoutline}

Since we have shown that both strong and strong probabilistic cost-preserving bisimilarities are equivalence relations, and are preserved by parallel composition, they are congruences.

In the previous proposition, as well as in the following Propositions~\ref{pro:weakCostBisimEqIsCompositional},~\ref{pro:strongMinCostSimIsCompositional} and~\ref{pro:weakMinCostSimIsCompositional}, we make use of the condition $\gen{\costf_{l}} = \gen{\costf_{r}}$ about the generators of the cost functions for the parallel composition, so we assume the same generator to be used on both sides.
This condition is needed to transfer the cost of a composed transition in one automaton to an equivalent composed transition in the other automaton. 

For simplicity, let us consider $\strongCostBisim$ and an action $a \in \actionSet_{1} \cap \actionSet_{3}$.
Given $\aut_{1} \strongBisim \aut_{2}$, we know that $\aut_{1} \parComp \aut_{3} \strongBisim \aut_{2} \parComp \aut_{3}$ is justified by the relation $\relord = \relord_{12} \times \idrelord_{3}$ where $\relord_{12}$ is the strong bisimulation between $\aut_{1}$ and $\aut_{2}$ and $\idrelord_{3}$ is the identity relation on $\stateSet_{3}$.
Given $(s_{1},s_{3}) \rel (s_{2},s_{3})$ and a challenging transition $\strongTransition{(s_{1},s_{3})}{a}{\sd_{1} \times \sd_{3}}$, the defender transition is $\strongTransition{(s_{2},s_{3})}{a}{\sd_{2} \times \sd_{3}}$; 
$\strongTransition{s_{2}}{a}{\sd_{2}}$ is the defender transition when $s_{2}$ is challenged by $s_{1}$ in the step condition of $\relord_{12}$ with the transition $\strongTransition{s_{1}}{a}{\sd_{1}}$.

The construction for the strong cost-preserving bisimulation is essentially the same;
the only difference is that we have also to check the equality of the cost.
To prove the equality, by definition we know that $\cost[l]{\strongTransition{(s_{1},s_{3})}{a}{\sd_{1} \times \sd_{3}}} = \functioneval{\gen{\costf_{l}}}{\functioneval{\costf_{1}}{\strongTransition{s_{1}}{a}{\sd_{1}}}, \functioneval{\costf_{3}}{\strongTransition{s_{3}}{a}{\sd_{3}}}}$ and that $\cost[r]{\strongTransition{(s_{2},s_{3})}{a}{\sd_{2} \times \sd_{3}}} = \functioneval{\gen{\costf_{r}}}{\functioneval{\costf_{2}}{\strongTransition{s_{2}}{a}{\sd_{2}}}, \functioneval{\costf_{3}}{\strongTransition{s_{3}}{a}{\sd_{3}}}}$.
Moreover, from $\costAutPed{\costf}{1} \strongCostBisim \costAutPed{\costf}{2}$, we know that $\functioneval{\costf_{1}}{\strongTransition{s_{1}}{a}{\sd_{1}}} = \functioneval{\costf_{2}}{\strongTransition{s_{2}}{a}{\sd_{2}}}$.

This means that $\cost[l]{\strongTransition{(s_{1},s_{3})}{a}{\sd_{1} \times \sd_{3}}} = \functioneval{\gen{\costf_{l}}}{\functioneval{\costf_{1}}{\strongTransition{s_{1}}{a}{\sd_{1}}}, \functioneval{\costf_{3}}{\strongTransition{s_{3}}{a}{\sd_{3}}}}$ is actually equal to $\functioneval{\gen{\costf_{l}}}{\functioneval{\costf_{2}}{\strongTransition{s_{2}}{a}{\sd_{2}}}, \functioneval{\costf_{3}}{\strongTransition{s_{3}}{a}{\sd_{3}}}}$.
The condition $\gen{\costf_{l}} = \gen{\costf_{r}}$ allows us to derive $\functioneval{\gen{\costf_{l}}}{\functioneval{\costf_{2}}{\strongTransition{s_{2}}{a}{\sd_{2}}}, \functioneval{\costf_{3}}{\strongTransition{s_{3}}{a}{\sd_{3}}}} = \functioneval{\gen{\costf_{r}}}{\functioneval{\costf_{2}}{\strongTransition{s_{2}}{a}{\sd_{2}}}, \functioneval{\costf_{3}}{\strongTransition{s_{3}}{a}{\sd_{3}}}} = \cost[r]{\strongTransition{(s_{2},s_{3})}{a}{\sd_{2} \times \sd_{3}}}$, as needed.
If we omit such condition, then there is no way to establish the equality of the costs of the two composed transitions.

\subsection{Weak Cost Preserving Bisimulations}

We now discuss options for weak bisimulations on \CPA{}s, so as to ignore internal computations as long as these do not change the visible behaviour of the system.
As we have mentioned towards the end of Section~\ref{ssec:weakProbBisimulation}, weak bisimilarity on \PA{}s is not transitive, and therefore we base our considerations solely on weak probabilistic bisimilarity.

We follow the intuition  of the strong cost-preserving setting discussed in Section~\ref{ssec:strongCostPreservingBisim} and introduce a refined notion of weak probabilistic bisimulation where each transition $\strongTransition{s}{a}{\sd_{s}}$ of the challenging state $s$ has to be matched by the defender state $t$ by enabling a weak combined transition $\weakCombinedTransition{t}{a}{\sd_{t}}$ such that $\sd_{s} \liftrel \sd_{t}$ as in ordinary weak probabilistic bisimulation, and, in addition, the costs of challenging and defending transitions must agree.
\begin{definition}
\label{def:weakCostBisimEq}
	Given two \CPA{}s $\costAutPed{\costf}{1}$ and $\costAutPed{\costf}{2}$, an equivalence relation $\relord$ on the disjoint union $\stateSet_{1} \uplus \stateSet_{2}$ is a \emph{weak probabilistic cost-preserving bisimulation} if for each pair of states $s,t \in \stateSet_{1} \uplus \stateSet_{2}$ such that $s \rel t$, if $\strongTransition{s}{a}{\sd_{s}}$, then there exists $\sd_{t}$ such that $\weakCombinedTransition{t}{a}{\sd_{t}}$, $\sd_{s} \liftrel \sd_{t}$, and $\cost[d]{\weakCombinedTransition{t}{a}{\sd_{t}}} = \cost[c]{\strongTransition{s}{a}{\sd_{s}}}$ where $\costf_{d}$ and $\costf_{c}$ are the cost functions of the defender and the challenger \CPA{}, respectively.

	Two \CPA{}s $\costAutPed{\costf}{1}$ and $\costAutPed{\costf}{2}$ are weak probabilistic cost-preserving bisimilar if there exists a weak probabilistic cost-preserving bisimulation $\relord$ on $\stateSet_{1} \uplus \stateSet_{2}$ such that $\startState_{1} \rel \startState_{2}$.  We denote weak probabilistic cost-preserving bisimilarity by $\weakCostBisim$.
\end{definition}

\subsection{Properties of Weak Cost-Preserving Probabilistic Bisimilarity.} 

A first property is that strong probabilistic cost-preserving bisimilarity is a special case of weak cost-preserving probabilistic bisimilarity. 
This result is straightforward, as every strong combined transition is also a weak combined transition.

\begin{proposition}
\label{pro:strongCostProbBisimImpliesWeakCostBisim}
	Given two \CPA{}s $\costAutPed{\costf}{1}$ and $\costAutPed{\costf}{2}$, if $\costAutPed{\costf}{1} \strongCostProbBisim \costAutPed{\costf}{2}$, then $\costAutPed{\costf}{1} \weakCostBisim \costAutPed{\costf}{2}$.
\end{proposition}

Another desirable property is that weak probabilistic cost-preserving bisimilarity is an equivalence relation on the set of \CPA{}s. 
\begin{proposition}
\label{pro:weakCostBisimEqIsEquivalenceRelation}
	Weak probabilistic cost-preserving bisimilarity is an equivalence relation on the set of \CPA{}s.
\end{proposition}
\begin{myoutline}
	Similarly to the proof of Proposition~\ref{pro:strongCostBisimEqIsEquivalenceRelation}, reflexivity and symmetry are trivial as they follow immediately from reflexivity and symmetry of the underlying equivalence relations.
	Transitivity is based on the equivalence relation witnessing the transitivity of the underlying bisimulation for \PA{}s: 
	with a simple manipulation of the matching combined transitions, it is shown that the costs according to Definition~\ref{def:costAsWorkingForTheProof} are preserved.
\end{myoutline}

In order to show that weak probabilistic cost-preserving bisimilarity is preserved by parallel composition, we state a preliminary result concerning the cost of weak combined transitions of the parallel composition of two \PA{}s.
\begin{lemma}
\label{lem:costWeakTransPCfunctionOfComponentWTs}
	Given two compatible \CPA{}s $\costAutPed{\costf}{1}$ and $\costAutPed{\costf}{2}$ and a cost-preserving function $\costf$, let $\weakCombinedTransition{(s_{1}, s_{2})}{a}{\sd_{1} \times \sd_{2}}$ be a weak combined transition of $\costAutPed{\costf}{1} \parComp_{\costf} \costAutPed{\costf}{2}$.
	Then, $\cost{\weakCombinedTransition{(s_{1}, s_{2})}{a}{\sd_{1} \times \sd_{2}}} = \functioneval{\gen{\costf}}{\cost[1]{\weakCombinedTransition{s_{1}}{a}{\sd_{1}}}, \cost[2]{\weakCombinedTransition{s_{2}}{a}{\sd_{2}}}}$, where for $i=1,2$, $\weakCombinedTransition{s_{i}}{a}{\sd_{i}} = \project{i}{\weakCombinedTransition{(s_{1}, s_{2})}{a}{\sd_{1} \times \sd_{2}}}$.
\end{lemma}
\begin{myoutline}
	The result follows from Definition~\ref{def:costAsWorkingForTheProof}, the properties of $\gen{\costf}$, and simple manipulation of the terms of summations.
\end{myoutline}

This lemma puts us in the position to show that weak probabilistic cost-preserving bisimulation is preserved by parallel composition.

\begin{proposition}
\label{pro:weakCostBisimEqIsCompositional}
	Given two \CPA{}s $\costAutPed{\costf}{1}$ and $\costAutPed{\costf}{2}$, if $\costAutPed{\costf}{1} \weakCostBisim \costAutPed{\costf}{2}$, then for each $\costAutPed{\costf}{3}$ compatible with both $\costAutPed{\costf}{1}$ and $\costAutPed{\costf}{2}$ and each pair of cost-preserving functions $\costf_{l}$ and $\costf_{r}$ with $\gen{\costf_{l}} = \gen{\costf_{r}}$, $\costAutPed{\costf}{1} \parComp_{\costf_{l}} \costAutPed{\costf}{3} \weakCostBisim \costAutPed{\costf}{2} \parComp_{\costf_{r}} \costAutPed{\costf}{3}$.
\end{proposition}
\begin{myoutline}
	The result is based on the relation justifying the compositionality of the underlying weak probabilistic bisimulation on \PA{}s;
	the equality of the costs is shown by Lemma~\ref{lem:costWeakTransPCfunctionOfComponentWTs}.
\end{myoutline}

Since we have shown that weak probabilistic cost-preserving bisimulation is an equivalence relation and that it is preserved by parallel composition, it is a congruence.

\section{Cost Bounding Bisimulations}
\label{sec:costBoundingBisims}

The definitions of strong, strong probabilistic, and especially weak probabilistic cost-preserv\-ing bisimulation allow us to relate different-size \CPA{}s that have the same behaviour and the same cost.
Since we are working in a setting where our aim is to minimise the cost while preserving the behaviour of a \CPA{} $\costAut{\costf}$, we will now discuss how to relax the cost equality in the bisimulation definitions so that one automaton has a cost that is at most the cost of the other one. We first consider the strong case, and then discuss the weak setting. 

\subsection{Strong Cost Bounding Bisimulations}

We first consider an extension of the strong (probabilistic) cost-preserving bisimulation. 
The central idea is to require that the cost of the defender transition is at most the cost of the challenger transition.

\begin{definition}
\label{def:strongCostBisimMinorCost}
	Given two \CPA{}s $\costAutPed{\costf}{1}$ and $\costAutPed{\costf}{2}$, an equivalence relation $\relord$ on the disjoint union $\stateSet_{1} \uplus \stateSet_{2}$ is a \emph{minor cost strong probabilistic bisimulation} from $\costAutPed{\costf}{1}$ to $\costAutPed{\costf}{2}$ if $\relord$ is a strong probabilistic bisimulation for $\aut_{1}$ and $\aut_{2}$ and for each $\strongTransition{s_{2}}{a}{\sd_{2}} \in \transitionRelation_{2}$ and each $s_{1} \in \stateSet_{1}$ such that $s_{2} \rel s_{1}$, there exists $\sd_{1}$ such that $\strongCombinedTransition{s_{1}}{a}{\sd_{1}}$, $\sd_{2} \liftrel \sd_{1}$, and $\cost[1]{\strongCombinedTransition{s_{1}}{a}{\sd_{1}}} \leq \cost[2]{\strongTransition{s_{2}}{a}{\sd_{2}}}$.

	We say that $\costAutPed{\costf}{1}$ is minor cost strong probabilistic bisimilar to $\costAutPed{\costf}{2}$ if there exists a minor cost strong probabilistic bisimulation $\relord$ such that $\startState_{2} \rel \startState_{1}$. 
	We denote  minor cost strong probabilistic bisimilarity from $\costAutPed{\costf}{1}$ to $\costAutPed{\costf}{2}$ by $\costAutPed{\costf}{1} \strongCostProbBisimMinorCost \costAutPed{\costf}{2}$ and we say that $\costAutPed{\costf}{1}$ is in minor cost strong probabilistic bisimilarity with $\costAutPed{\costf}{2}$.
\end{definition}

Similar to Definitions~\ref{def:strongProbBisim} and~\ref{def:strongCostBisimEq}, the above definition can be twisted to give rise to a \emph{minor cost strong bisimilarity} (denoted~$\strongCostBisimMinorCost$), obtained by replacing all occurrences of $\strongCombinedTransition{}{a}{}$ by $\strongTransition{}{a}{}$.

\subsubsection{Properties of Strong Cost Bounding Bisimilarities}

We now discuss the core properties of the above relations. 

As in the cost-preserving case, it is straightforward to show that two minor cost strong bisimilar \CPA{}s are also minor cost strong probabilistic bisimilar.

\begin{proposition}
\label{pro:strongCostBisimMinorCostImpliesStrongCostProbBisimMinorCost}
	Given two \CPA{}s $\costAutPed{\costf}{1}$ and $\costAutPed{\costf}{2}$, if $\costAutPed{\costf}{1} \strongCostBisimMinorCost \costAutPed{\costf}{2}$, then $\costAutPed{\costf}{1} \strongCostProbBisimMinorCost \costAutPed{\costf}{2}$.
\end{proposition}
\begin{myproof}
	The result is immediate, since each transition is also a combined transition.
\end{myproof}

Both minor cost strong and strong probabilistic bisimilarities are reflexive and transitive relations, but not symmetric, of course. 
Thus they are preorders.
\begin{proposition}
\label{pro:strongCostBisimMinorCostIsPreorder}
	Minor cost strong and strong probabilistic bisimilarities are preorders on the set of \CPA{}s.
\end{proposition}
\begin{myoutline}
	Similarly to the proof of Proposition~\ref{pro:strongCostBisimEqIsEquivalenceRelation}, reflexivity is trivial as it follows immediately from reflexivity of the underlying equivalence relations.
	Transitivity is based on the relation witnessing the transitivity of the underlying bisimulation for \PA{}s: 
	with a simple manipulation of the matching transitions, it is shown that the costs are correctly bounded.
\end{myoutline}

Minor cost strong probabilistic bisimilarity is the coarsest minor cost strong probabilistic bisimulation, and similarly for minor cost strong  bisimilarity. 
As we will discuss below, both minor cost strong and strong probabilistic bisimulations are preserved by parallel composition.

\begin{proposition}
\label{pro:strongMinCostSimIsCompositional}
	Given two \CPA{}s $\costAutPed{\costf}{1}$ and $\costAutPed{\costf}{2}$, if $\costAutPed{\costf}{1} \strongCostProbBisimMinorCost \costAutPed{\costf}{2}$, then for each $\costAutPed{\costf}{3}$ compatible with both $\costAutPed{\costf}{1}$ and $\costAutPed{\costf}{2}$ and each pair of cost-preserving functions $\costf_{l}$ and $\costf_{r}$ with $\gen{\costf_{l}} = \gen{\costf_{r}}$, $\costAutPed{\costf}{1} \parComp_{\costf_{l}} \costAutPed{\costf}{3} \strongCostProbBisimMinorCost \costAutPed{\costf}{2} \parComp_{\costf_{r}} \costAutPed{\costf}{3}$, and similarly for $\strongCostBisimMinorCost$.
\end{proposition}
\begin{myoutline}
	The proof is based on the relation justifying the compositionality of the underlying bisimulation between \PA{}s; 
	as for Proposition~\ref{pro:strongCostBisimEqIsCompositional}, the bounding of the cost of the transitions is based on the properties of the cost-preserving functions $\costf_{l}$ and $\costf_{r}$ with $\gen{\costf_{l}} = \gen{\costf_{r}}$.
\end{myoutline}

Since we have shown that both minor cost strong and strong probabilistic bisimulations are preorders and they are preserved by parallel composition, they are precongruences.

\subsection{Weak Cost Bounding Bisimulations}

Also in the weak setting, we will relax the cost equality condition from Definition~\ref{def:weakCostBisimEq} by instead requiring that the cost of the defender matching transition is at most the cost of the challenger transition.  
Despite the simplicity of this idea, the formal definition is quite involved since we have to consider properly the cost of internal transitions.

To shed some light on this, consider an automaton $\aut_{1}$ performing three internal steps $\strongTransition{\startState_{1}}{\hidden}{\dirac{t_{1}}}$, $\strongTransition{t_{1}}{\hidden}{\dirac{u_{1}}}$, and $\strongTransition{u_{1}}{\hidden}{\dirac{v_{1}}}$ where each step has cost $5$ followed by an external step $\strongTransition{v_{1}}{a}{\dirac{x_{1}}}$ with cost $2$ and an automaton $\aut_{2}$ that performs four steps $\strongTransition{\startState_{2}}{\hidden}{\dirac{t_{2}}}$, $\strongTransition{t_{2}}{\hidden}{\dirac{u_{2}}}$, $\strongTransition{u_{2}}{\hidden}{\dirac{v_{2}}}$, and $\strongTransition{v_{2}}{\hidden}{\dirac{w_{2}}}$ each with cost $3$ followed by an external step $\strongTransition{w_{2}}{a}{\dirac{x_{2}}}$ with cost $2$. 
Graphically, the two automata appear as follows, where we put the action and the cost as label of the transitions.
The length of the arrows is proportional to the cost of the transition.
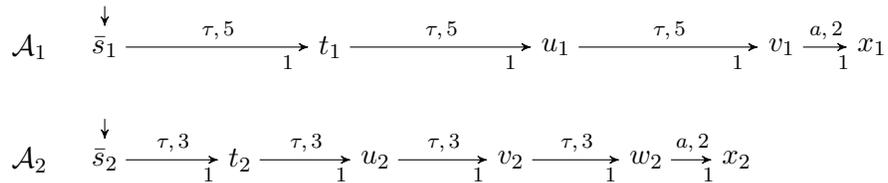
\begin{center}
	\begin{tikzpicture}[->,>=stealth',shorten >=1pt,auto]
	\path[use as bounding box] (-0.5,1) rectangle (11.6,-2.25);
	\scriptsize

	\node (aut1) at (0,0) {\normalsize$\aut_{1}$};
	\node (ss1) at ($(aut1) + (1,0)$) {\normalsize$\startState_{1}$};
	\node (t1) at ($(ss1) + (3,0)$) {\normalsize$t_{1}$};
	\node (u1) at ($(t1) + (3,0)$) {\normalsize$u_{1}$};
	\node (v1) at ($(u1) + (3,0)$) {\normalsize$v_{1}$};
	\node (x1) at ($(v1) + (1.2,0)$) {\normalsize$x_{1}$};

	\draw ($(ss1.north) + (0,0.3)$) to (ss1);
	\draw (ss1) to node[above] {$\hidden,5$} node[below, very near end] {$1$} (t1);
	\draw (t1) to node[above] {$\hidden,5$} node[below, very near end] {$1$} (u1);
	\draw (u1) to node[above] {$\hidden,5$} node[below, very near end] {$1$} (v1);
	\draw (v1) to node[above] {$a,2$} node[below, very near end] {$1$} (x1);
	
	\node (aut2) at (0,-1.5) {\normalsize$\aut_{2}$};
	\node (ss2) at ($(aut2) + (1,0)$) {\normalsize$\startState_{2}$};
	\node (t2) at ($(ss2) + (1.8,0)$) {\normalsize$t_{2}$};
	\node (u2) at ($(t2) + (1.8,0)$) {\normalsize$u_{2}$};
	\node (v2) at ($(u2) + (1.8,0)$) {\normalsize$v_{2}$};
	\node (w2) at ($(v2) + (1.8,0)$) {\normalsize$w_{2}$};
	\node (x2) at ($(w2) + (1.2,0)$) {\normalsize$x_{2}$};

	\draw ($(ss2.north) + (0,0.3)$) to (ss2);
	\draw (ss2) to node[above] {$\hidden,3$} node[below, very near end] {$1$} (t2);
	\draw (t2) to node[above] {$\hidden,3$} node[below, very near end] {$1$} (u2);
	\draw (u2) to node[above] {$\hidden,3$} node[below, very near end] {$1$} (v2);
	\draw (v2) to node[above] {$\hidden,3$} node[below, very near end] {$1$} (w2);
	\draw (w2) to node[above] {$a,2$} node[below, very near end] {$1$} (x2);
	
	\end{tikzpicture}
\end{center}

An external observer is able to recognise that the behaviour of $\aut_{1}$  is more expensive than the one of $\aut_{2}$ since the overall cost is $17$ for the former, $14$ for the latter.
However, from a state-based bisimulation point of view, $\aut_{2}$ is not always cheaper than $\aut_{1}$:
Let $\setnocond{\setnocond{\startState_{1}, \startState_{2}}, \setnocond{t_{1}, t_{2}}, \setnocond{u_{1}, u_{2}}, \setnocond{v_{1}, v_{2}, w_{2}}, \setnocond{x_{1}, x_{2}}}$ be the equivalence classes of $\relord$; 
it is easy to verify that $\relord$ is a weak probabilistic bisimulation between $\aut_{1}$ and $\aut_{2}$:
When $\aut_{1}$ performs $\strongTransition{\startState_{1}}{\hidden}{\dirac{t_{1}}}$ with cost $5$, $\aut_{2}$ replies with $\strongTransition{\startState_{2}}{\hidden}{\dirac{t_{2}}}$ with cost $3 \leq 5$ and $t_{1} \rel t_{2}$. 
Note that $\aut_{2}$ can not perform the subsequent transition $\strongTransition{t_{2}}{\hidden}{\dirac{u_{2}}}$ since in this case the overall cost would be $6 \nleq 5$.
The same happens for transitions $\strongTransition{t_{1}}{\hidden}{\dirac{u_{1}}}$ and $\strongTransition{u_{1}}{\hidden}{\dirac{v_{1}}}$ that are matched by $\strongTransition{t_{2}}{\hidden}{\dirac{u_{2}}}$ and $\strongTransition{u_{2}}{\hidden}{\dirac{v_{2}}}$, respectively.
Since $\aut_{1}$ now performs $\strongTransition{v_{1}}{a}{\dirac{x_{1}}}$ with cost $2$, $v_{2}$ is not able to match this transition with a cost at most $2$: 
In order to match the transition, $\aut_{2}$ has to perform both transitions $\strongTransition{v_{2}}{\hidden}{\dirac{w_{2}}}$ and $\strongTransition{w_{2}}{a}{\dirac{x_{2}}}$ whose cost is $5 \nleq 2$.

These considerations indicate that internal challenger transitions should not be considered separately but as a whole, so in order to abstract away from costs of single challenger internal transitions while preserving the overall cost, we consider for the challenger the cost of reaching the \borderTerm{} states, i.e., states where the automaton performs an external action or exhibits a different behaviour by changing the current class as induced by the weak bisimulation relation.
\begin{definition}
\label{def:borderState}
	Given a \PA{} $\aut$ and an equivalence relation $\relord$ over $\stateSet$, we say that a state $s$ is a \emph{\borderTerm{} state} if there exists $\strongTransition{s}{a}{\sd} \in \transitionRelation$ such that either $\probeval{\sd}{\relclass{s}{\relord}} < 1$ or $a \in \externalActionSet$. 
	
	We denote the set of all \borderTerm{} states with respect to $\relord$ by $\borderStateSetOrd$.
\end{definition}

\begin{definition}
\label{def:weakCostBisimMinorCost}
	Let $\costAutPed{\costf}{1}$ and $\costAutPed{\costf}{2}$ be two \CPA{}s. 
	Let $\wbrelord$ be an equivalence relation on the disjoint union $\stateSet_{1} \uplus \stateSet_{2}$ and $\costrelord \subseteq \wbrelord \cap (\stateSet_{2} \times \stateSet_{1})$ such that for each $s_{2} \in \stateSet_{2}$ there exists $s_{1} \in \stateSet_{1}$ such that $s_{2} \costrel s_{1}$. 
	Then we say that $(\wbrelord, \costrelord)$ is a \emph{minor cost weak probabilistic bisimulation} from $\costAutPed{\costf}{1}$ to $\costAutPed{\costf}{2}$ if $\wbrelord$ is a weak probabilistic bisimulation for $\aut_{1}$ and $\aut_{2}$ and for each $\strongTransition{s_{2}}{a}{\sd_{2}} \in \transitionRelation_{2}$ and each $s_{1} \in \stateSet_{1}$ such that $s_{2} \costrel s_{1}$, 
	\begin{enumerate}
	\item \label{def:weakCostBisimMinorCost:stepConditionOnBorderStates}
		if there exists $\gd_{2} \in \Disc{\borderStateSetOrd[\wbrelord] \cap \stateSet_{2}}$ such that $\hyperWeakCombinedTransition{\sd_{2}}{\hidden}{\gd_{2}}$, then there exists $\gd_{1} \in \Disc{\borderStateSetOrd[\wbrelord] \cap \stateSet_{1}}$ such that 
		\begin{enumerate}
			\item $\weakCombinedTransition{s_{1}}{a}{\gd_{1}}$, 
			\item $\gd_{2} \liftrel[\costrelord] \gd_{1}$, 
			\item $\cost[1]{\weakCombinedTransition{s_{1}}{a}{\gd_{1}}} \leq \cost[2]{\weakCombinedTransition{\strongTransition{s_{2}}{a}{\sd_{2}}}{\hidden}{\gd_{2}}}$, and 
			\item $\min \setcond{\cost[2]{\hyperWeakCombinedTransition{\sd_{2}}{\hidden}{\gd}}}{\gd \in \Disc{\borderStateSet[\wbrelord] \cap \stateSet_{2}}} = \cost[2]{\hyperWeakCombinedTransition{\sd_{2}}{\hidden}{\gd_{2}}}$; or
		\end{enumerate}
	\item \label{def:weakCostBisimMinorCost:stepConditionOnInternalStates}
		if there does not exist $\gd_{2} \in \Disc{\borderStateSetOrd[\wbrelord] \cap \stateSet_{2}}$ such that $\hyperWeakCombinedTransition{\sd_{2}}{\hidden}{\gd_{2}}$, then there exists $\sd_{1} \in \Disc{\stateSet_{1}}$ such that $\weakCombinedTransition{s_{1}}{a}{\sd_{1}}$, $\sd_{2} \liftrel[\costrelord] \sd_{1}$, and $\cost[1]{\weakCombinedTransition{s_{1}}{a}{\sd_{1}}} \leq \cost[2]{\strongTransition{s_{2}}{a}{\sd_{2}}}$.
	\end{enumerate}

	\noindent We say that $\costAutPed{\costf}{1}$ is minor cost weak probabilistic bisimilar to $\costAutPed{\costf}{2}$ if there exists a minor cost weak probabilistic bisimulation $(\wbrelord, \costrelord)$ such that $\startState_{2} \costrel \startState_{1}$. 
	We denote minor cost weak probabilistic bisimilarity from $\costAutPed{\costf}{1}$ to $\costAutPed{\costf}{2}$ by $\costAutPed{\costf}{1} \weakCostBisimMinorCost \costAutPed{\costf}{2}$ and we say that $\costAutPed{\costf}{1}$ is in minor cost weak probabilistic bisimilarity with $\costAutPed{\costf}{2}$.
\end{definition}

\subsection{Properties of Minor Cost Weak Probabilistic Bisimilarity}

A first property is that minor cost strong probabilistic bisimilarity is a special case of minor cost weak probabilistic bisimilarity. 
This result is rather easy, as every strong combined transition is also a weak combined transition.

\begin{proposition}
\label{pro:strongMinCostProbBisimImpliesWeakMinCostBisim}
	Given two \CPA{}s $\costAutPed{\costf}{1}$ and $\costAutPed{\costf}{2}$, if $\costAutPed{\costf}{1} \strongCostProbBisimMinorCost \costAutPed{\costf}{2}$, then $\costAutPed{\costf}{1} \weakCostBisimMinorCost \costAutPed{\costf}{2}$.
\end{proposition}
\begin{myoutline}
	The proof is based on the relation $\wbrelord$ justifying $\costAutPed{\costf}{1} \strongCostProbBisimMinorCost \costAutPed{\costf}{2}$; 
	the relation $\costrelord$ is constructed as $\wbrelord \cap (\stateSet_{2} \times \stateSet_{1})$ and the pair $(\wbrelord, \costrelord)$ shows $\costAutPed{\costf}{1} \weakCostBisimMinorCost \costAutPed{\costf}{2}$.
\end{myoutline}

The proof idea can be reused to show that weak probabilistic cost-preserving bisimilarity is a special case of minor cost weak probabilistic bisimilarity.
\begin{proposition}
\label{pro:weakCostBisimImpliesWeakMinCostBisim}
	Given two \CPA{}s $\costAutPed{\costf}{1}$ and $\costAutPed{\costf}{2}$, if $\costAutPed{\costf}{1} \weakCostBisim \costAutPed{\costf}{2}$, then $\costAutPed{\costf}{1} \weakCostBisimMinorCost \costAutPed{\costf}{2}$.
\end{proposition}
\begin{myproof}
	The proof is a literal recapitulation of the proof of Proposition~\ref{pro:strongMinCostProbBisimImpliesWeakMinCostBisim}; 
	the only difference being that when we match $\strongTransition{t_{2}}{b}{\gamma_{2}}$, instead of using the strong combined transition $\strongCombinedTransition{t_{1}}{b}{\gamma_{1}}$, we use the corresponding weak combined transition $\weakCombinedTransition{t_{1}}{b}{\gamma_{1}}$ that has cost $\cost[1]{\strongCombinedTransition{t_{1}}{b}{\gamma_{1}}} = \cost[2]{\strongTransition{t_{2}}{b}{\gamma_{2}}}$. 
	So it is immediate to derive that $\cost[1]{\weakCombinedTransition{s_{1}}{a}{\gd_{1}}} \leq \cost[2]{\weakCombinedTransition{\strongTransition{s_{2}}{a}{\sd_{2}}}{\hidden}{\gd_{2}}}$ (as it is indeed $\cost[1]{\weakCombinedTransition{s_{1}}{a}{\gd_{1}}} = \cost[2]{\weakCombinedTransition{\strongTransition{s_{2}}{a}{\sd_{2}}}{\hidden}{\gd_{2}}}$).
	Similarly for the challenging $\strongTransition{s_{2}}{a}{\sd_{2}}$ when $\sd_{2}$ can not be extended to reach the border.
\end{myproof}

As for the strong case, we have that  minor cost weak probabilistic bisimilarity is reflexive and transitive, thus it is a preorder. 
The proof of transitivity is not trivial.
\begin{proposition}
\label{pro:weakMinCostSimIsPreorder}
	Minor cost weak probabilistic bisimilarity is a preorder on the set of \CPA{}s.
\end{proposition}
\begin{myoutline}
	The proof is rather involved and it is based on relations $\wbrelord_{31} = \wbrelord_{32} \relationComposition \wbrelord_{21}$ and $\costrelord_{31} = \costrelord_{32} \relationComposition \costrelord_{21}$ obtained as composition of the relations $(\wbrelord_{21}, \costrelord_{21})$ and $(\wbrelord_{32}, \costrelord_{32})$ justifying $\costAutPed{\costf}{1} \weakCostBisimMinorCost \costAutPed{\costf}{2}$ and $\costAutPed{\costf}{2} \weakCostBisimMinorCost \costAutPed{\costf}{3}$, respectively.
	By manipulating the definition of weak combined transition, the properties of $(\wbrelord_{21}, \costrelord_{21})$ and $(\wbrelord_{32}, \costrelord_{32})$ allow us to show that $(\wbrelord_{31},  \costrelord_{31})$ is a witness for $\costAutPed{\costf}{1} \weakCostBisimMinorCost \costAutPed{\costf}{3}$.
\end{myoutline}

Minor cost weak probabilistic bisimilarity is the coarsest minor cost weak probabilistic bisimulation. 
And as desired, minor cost weak probabilistic bisimulation is preserved by parallel composition.
	
\begin{proposition}
\label{pro:weakMinCostSimIsCompositional}
	Given two \CPA{}s $\costAutPed{\costf}{1}$ and $\costAutPed{\costf}{2}$, if $\costAutPed{\costf}{1} \weakCostBisimMinorCost \costAutPed{\costf}{2}$, then for each $\costAutPed{\costf}{3}$ compatible with both $\costAutPed{\costf}{1}$ and $\costAutPed{\costf}{2}$ and each pair of cost-preserving functions $\costf_{l}$ and $\costf_{r}$ with $\gen{\costf_{l}} = \gen{\costf_{r}}$, $\costAutPed{\costf}{1} \parComp_{\costf_{l}} \costAutPed{\costf}{3} \weakCostBisimMinorCost \costAutPed{\costf}{2} \parComp_{\costf_{r}} \costAutPed{\costf}{3}$.
\end{proposition}
\begin{myoutline}
	The proof is based on the pair of relations $\wbrelord_{p} = \wbrelord \times \idrelord$ and $\costrelord_{p} = \costrelord \times \idrelord$ obtained as cross-product of the relations $(\wbrelord, \costrelord)$ and $(\idrelord, \idrelord)$ justifying $\costAutPed{\costf}{1} \weakCostBisimMinorCost \costAutPed{\costf}{2}$ and $\costAutPed{\costf}{3} \weakCostBisimMinorCost \costAutPed{\costf}{3}$, respectively.
	By manipulating the definition of weak combined transition, the properties of $(\wbrelord, \costrelord)$ and of $\costf_{l}$ and $\costf_{r}$ allow us to show that $(\wbrelord_{p},  \costrelord_{p})$ is a witness for $\costAutPed{\costf}{1} \parComp_{\costf_{l}} \costAutPed{\costf}{3} \weakCostBisimMinorCost \costAutPed{\costf}{2} \parComp_{\costf_{r}} \costAutPed{\costf}{3}$.
\end{myoutline}

Since we have shown that minor cost weak probabilistic bisimilarity is a preorder and it is preserved by parallel composition, it is a precongruence.

\subsection{The Cost of the Wireless Communication Channel}
\label{ssec:WCCExample}

We now apply the minor cost weak probabilistic bisimulation to the reliable wireless communication channel introduced in Section~\ref{sec:Preliminaries} and depicted in Figure~\ref{fig:wirelessCC}, page~\pageref{fig:wirelessCC}.
As cost, we consider the function $\costf$ that assigns cost $1$ to transitions labelled by $\sendMessage{m}$ or $\receiveMessage{m}$ and cost $r^{2}$ to transitions labelled by $\hop{r}$.
We use value $1$ to represent a constant power consumption relative to sending/receiving message actions and value $r^{2}$ to model the energy, quadratic on the transmission radius, required to transmit a message via wireless.

As a concrete example, consider the two instances $\aut_{23} = \wccHopsRadiusTransmprob{2}{3}{\frac{1}{2}}$ and $\aut_{32} = \wccHopsRadiusTransmprob{3}{2}{\frac{1}{2}}$ of the wireless communication channel connecting sender and receiver that are at distance $6$. 
To avoid name collisions, we rename the states $h^{m}_{j}$ of $\wccHopsRadiusTransmprob{3}{2}{\frac{1}{2}}$ to $k^{m}_{j}$ for $0 \leq j \leq 3$.
It is easy to verify that the equivalence relation $\wbrelord$ whose classes are $\setnocond{\startState_{23},\startState_{32}}$ and $\setcond{h^{m}_{i}, k^{m}_{j}}{0 \leq i \leq 2, 0 \leq j \leq 3}$ for each $m \in \msgSet$ justifies $\aut_{23} \weakBisim \aut_{32}$, so consider the two \CPA{}s $\costAut[\aut_{23}]{\costf}$ and $\costAut[\aut_{32}]{\costf}$.
We suspect that $\costAut[\aut_{32}]{\costf} \weakCostBisimMinorCost \costAut[\aut_{23}]{\costf}$, but not the reverse, since intuitively $\costAut[\aut_{23}]{\costf}$ has overall cost $26$ for sending and receiving a single message while $\costAut[\aut_{32}]{\costf}$ has overall cost $38$. 
In order to show $\costAut[\aut_{32}]{\costf} \weakCostBisimMinorCost \costAut[\aut_{23}]{\costf}$, we have to find a suitable relation $\costrelord$ that, together with $\wbrelord$, satisfies the conditions of Definition~\ref{def:weakCostBisimMinorCost}.
A suitable relation is $\costrelord = \setnocond{(\startState_{23}, \startState_{32})} \cup \bigcup_{m \in \msgSet} \setcond{(h^{m}_{i}, k^{m}_{3})}{0 \leq i \leq 2}$: 
Consider the pair $(\startState_{23}, \startState_{32})$ and the only available transition $\strongTransition{\startState_{23}}{\sendMessage{m}}{\dirac{h^{m}_{0}}}$. 
Since $\borderStateSet[\wbrelord] = \setnocond{\startState_{23}, \startState_{32}} \cup \setcond{h^{m}_{2}, k^{m}_{3}}{m \in \msgSet}$, the only possible $\gd_{23} \in \Disc{\borderStateSet[\wbrelord] \cap \stateSet_{23}}$ such that $\hyperWeakTransition{\dirac{h^{m}_{0}}}{\hidden}{\gd_{23}}$ is $\gd_{23} = \dirac{h^{m}_{2}}$.
In order to match such transition, $\startState_{32}$ enables the weak transition $\weakCombinedTransition{\startState_{32}}{\sendMessage{m}}{\dirac{k^{m}_{3}}}$ that satisfies $\dirac{h^{m}_{2}} \liftrel[\costrelord] \dirac{k^{m}_{3}}$.
The last condition we have to verify is that $\cost{\weakCombinedTransition{\startState_{32}}{\sendMessage{m}}{\dirac{k^{m}_{3}}}} \leq \cost{\weakCombinedTransition{\strongTransition{\startState_{23}}{\sendMessage{m}}{\dirac{h^{m}_{0}}}}{\hidden}{\dirac{h^{m}_{2}}}}$; 
this constraint is satisfied since $\cost{\weakCombinedTransition{\startState_{32}}{\sendMessage{m}}{\dirac{k^{m}_{3}}}} = 25$ while $\cost{\weakCombinedTransition{\strongTransition{\startState_{23}}{\sendMessage{m}}{\dirac{h^{m}_{0}}}}{\hidden}{\dirac{h^{m}_{2}}}} = 37$.
It is routine to check the remaining pairs of states, thus $\costAut[\aut_{32}]{\costf} \weakCostBisimMinorCost \costAut[\aut_{23}]{\costf}$.

Now, assume $\costAut[\aut_{23}]{\costf} \weakCostBisimMinorCost \costAut[\aut_{32}]{\costf}$: 
By definition, it must hold that $\startState_{32} \costrel \startState_{23}$, so consider the transition $\strongTransition{\startState_{32}}{s_{m}}{\dirac{k^{m}_{0}}}$.
For sure $k^{m}_{3}$ and $h^{m}_{2}$ are border states, as well as $\startState_{32}$ and $\startState_{23}$.
Moreover, $\startState_{32}$ and $\startState_{23}$ can not be related by $\wbrelord$ to any other state as they are the only states performing  $s_{m}$.
Suppose that these are the only border states; 
this implies that $\strongTransition{\startState_{32}}{s_{m}}{\dirac{k^{m}_{0}}}$ has to be extended to $\hyperWeakCombinedTransition{\strongTransition{\startState_{32}}{s_{m}}{\dirac{k^{m}_{0}}}}{\hidden}{\dirac{k^{m}_{3}}}$ whose cost is $25$.
The only possibility for $\startState_{23}$ to match such transition while respecting the cost constraint is to perform the weak combined transition $\weakCombinedTransition{\startState_{23}}{s_{m}}{\dirac{h^{m}_{i}}}$ with $i=0$ or $i=1$ and $k^{m}_{3} \costrel h^{m}_{i}$.
Note that we can not use $\weakCombinedTransition{\startState_{23}}{s_{m}}{\dirac{h^{m}_{2}}}$ since its cost is $37 \nleq 25$.
Independently on the chosen $i$, since $k^{m}_{3} \costrel h^{m}_{i}$ and $\strongTransition{k^{m}_{3}}{r_{m}}{\dirac{\startState_{32}}}$, $h^{m}_{i}$ has to perform the weak combined transition $\weakCombinedTransition{h^{m}_{i}}{r_{m}}{\dirac{\startState_{23}}}$ whose cost is $1 + 18 \cdot (2-i) \nleq 1$, so the condition is not satisfied.
By applying the same approach to the case where we consider other states as border states, we can derive a similar failure, thus there does not exist any suitable cost relation $\costrelord$ with $\startState_{32} \costrel \startState_{23}$, hence $\costAut[\aut_{23}]{\costf} \not\weakCostBisimMinorCost \costAut[\aut_{32}]{\costf}$.

\section{Decision Algorithms for Bisimulations via Linear Programming}
\label{sec:weakTransitionAsLPP}

In the previous sections we have discussed foundational properties of the cost probabilistic bisimulation variations. 
This section develops polynomial time decision algorithms for them. 
We focus on the intricacies faced when deciding the weak relations, algorithms for the strong relations are  derived later.  

To start with we revisit the ideas underlying the equivalence of weak transitions and linear programming problems, as developed in~\cite{HT12}, and then extend this to the cost setting.  
At its core, and inspired by network flow problems, is the observation that one can view a transition $\weakCombinedTransition{t}{\hidden}{\sd_{t}}$ of the \PA{} $\aut$ as a \emph{flow} where the initial probability mass $\dirac{t}$ flows and splits along internal transitions according to \emph{(i)} the transition target distributions and \emph{(ii)} the scheduler resolutions of the nondeterminism occurring along the weak transition. 
Similarly, for $a \neq \hidden$, i.e., $a \in \externalActionSet$, one can view $\weakCombinedTransition{t}{a}{\sd_{t}}$ as a \emph{flow} flowing along internal transitions and exactly one transition with label $a$ for each stream, again splitting in accordance with the transition target distributions and the scheduler resolutions of the nondeterminism.

From this observation one can derive an LP problem $\LPproblemTBetaMuRel{t}{a}{\sd}{\relord[E]}$, proposed in~\cite{HT12}, used to validate or refute the existence of a weak combined transition $\weakCombinedTransition{t}{a}{\sd_{t}}$ such that $\sd \liftrel[{\relord[E]}] \sd_{t}$. 
Here it is assumed that $\relord[E]$ is an equivalence relation on $\stateSet$; 
but we can extend it to an arbitrary relation $\relord \subseteq \stateSet \times \stateSet$ as follows: 
Checking that there exists $\sd_{t}$ such that $\weakCombinedTransition{t}{a}{\sd_{t}}$ and $\sd \liftrel \sd_{t}$ is equivalent, by properties of $\liftrel[\functionGenericArgument]$, to finding distributions $\sd_{t}$ and $\sd'_{t}$ such that $\weakCombinedTransition{t}{a}{\sd_{t}}$, $\sd_{t} \liftrel[\idrelord] \sd'_{t}$, and $\sd \liftrel \sd'_{t}$, where $\idrelord$ is the identity relation on $\stateSet$.  
Since verifying $\sd \liftrel \sd'_{t}$ is itself equivalent~\cite[Lemma~5.1]{BEMC00} to solving a maximum flow problem, such a flow problem can be merged with the $\LPproblemTBetaMuRel{t}{a}{\sd'_{t}}{\idrelord}$ LP problem. 
This abstracts from the actual distribution $\sd'_{t}$, so as to extend it to a binary relation $\relord$, as we formalise in the sequel.

For a \PA{} $\aut = (\stateSet, \startState, \actionSet, \transitionRelation)$ and $\relord \subseteq \stateSet \times \stateSet$, for $a \in \externalActionSet$, the network $\NetworkTBetaMuRel{t}{a}{\sd}{\rel} = (V,E)$ has the set of vertices $V =\setnocond{\netsource,\netsink} \cup \stateSet \cup \stateSet^{\tr} \cup \stateSet_{a} \cup \stateSet^{\tr}_{a} \cup \stateSet_{\relord}$ where
\begin{align*}
	\stateSet^{\tr} & {} = \setcond{v^{\tr}}{v \in \stateSet, \tr = \strongTransition{v}{b}{\rho} \in \transitionRelation, b \in \setnocond{a, \hidden}}\text{,}\\
	\stateSet_{a} & {} = \setcond{v_{a}}{v \in \stateSet}\text{,}\\
	\stateSet^{\tr}_{a} & {} = \setcond{v^{\tr}_{a}}{v^{\tr} \in \stateSet^{\tr}}\text{, and}\\
	\stateSet_{\relord} & {} = \setcond{s_{\relord}}{s \in \stateSet}
\end{align*}
and the set of arcs is
\begin{eqnarray*}
	E &=& \setnocond{(\netsource,t)} \cup \setcond{(v_{a},u_{\relord}),
	(u_{\relord},\netsink)}{u,v \in \stateSet, v \rel u} \\ &\cup& \setcond{(v,v^{\tr}), (v^{\tr}, v'), (v_{a},v^{\tr}_{a}), (v^{\tr}_{a},v'_{a})}{\tr = \strongTransition{v}{\hidden}{\rho} \in \transitionRelation, v' \in \Supp{\rho}}\\ &\cup& \setcond{(v,v^{\tr}_{a}), (v^{\tr}_{a}, v'_{a})}{\tr = \strongTransition{v}{a}{\rho} \in \transitionRelation, v' \in \Supp{\rho}}.
\end{eqnarray*}

When instead $a \in \internalActionSet$, the definition is simpler: $V = \setnocond{\netsource,\netsink} \cup \stateSet \cup
\stateSet^{\tr} \cup \stateSet_{\relord}$ and 
\begin{eqnarray*}
	E &=& \setnocond{(\netsource,t)} \cup \setcond{(v,u_{\relord}),
	(u_{\relord},\netsink)}{u,v \in \stateSet, v \rel u} \\ &\cup& \setcond{(v,v^{\tr}), (v^{\tr}, v')}{\tr = \strongTransition{v}{\hidden}{\rho} \in \transitionRelation, v' \in \Supp{\rho}}.
\end{eqnarray*}

\begin{exampleCont}
\label{ex:network}
	As an example of the construction of the network, consider the automaton $\wccHopsRadiusTransmprob{2}{5}{\frac{3}{4}}$ depicted in Figure~\ref{fig:wirelessCC}, the state $h^{m}_{1}$, the action $\receiveMessage{m}$, and the equivalence relation $\relord$ on states whose induced classes are $\setnocond{\startState}$ and $\setnocond{h^{m}_{0}, h^{m}_{1}, h^{m}_{2}}$ for each message $m$.
	Denote the transitions of the automaton $\wccHopsRadiusTransmprob{2}{5}{\frac{3}{4}}$ as follows: 
	$\tr_{s} = \strongTransition{\startState}{\sendMessage{m}}{\dirac{h^{m}_{0}}}$,
	$\tr_{0} = \strongTransition{h^{m}_{0}}{\hop{5}}{\setnocond{(h^{m}_{0}, \frac{1}{4}), (h^{m}_{1}, \frac{3}{4})}}$,
	$\tr_{1} = \strongTransition{h^{m}_{1}}{\hop{5}}{\setnocond{(h^{m}_{1}, \frac{1}{4}), (h^{m}_{2}, \frac{3}{4})}}$, and 
	$\tr_{2} = \strongTransition{h^{m}_{2}}{\receiveMessage{m}}{\dirac{\startState}}$.
	The network $\NetworkTBetaMuRel{h^{m}_{1}}{\receiveMessage{m}}{\dirac{\startState}}{\relord}$ is as follows:
	\begin{center}
	\resizebox{\textwidth}{!}
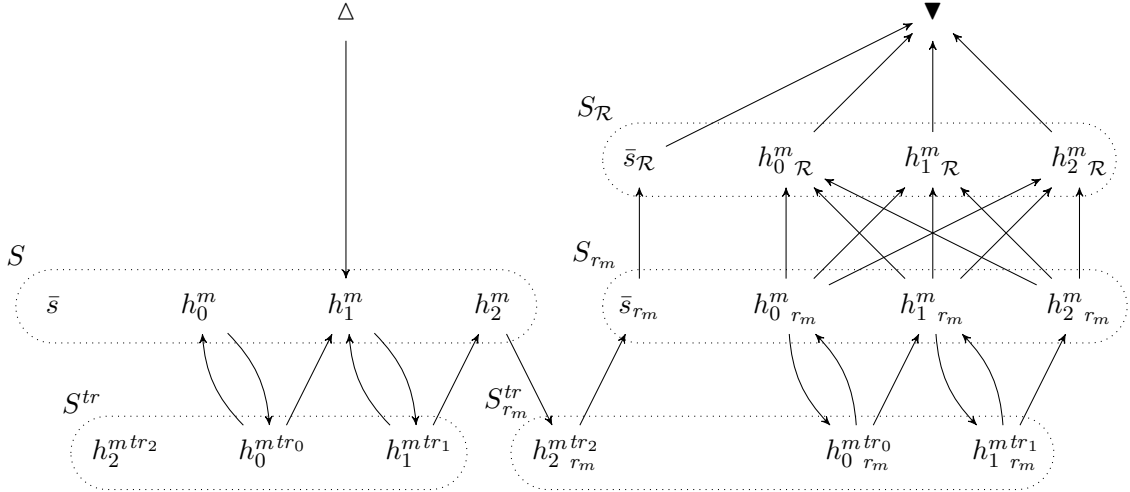
{%
	\begin{tikzpicture}[->,>=stealth',auto]
	\path[use as bounding box] (-4.8,-2) rectangle (10.7,5.5);

	\node (source) at (0,5) {$\vphantom{{h^{m}_{1}}^{\tr_{2}}_{\receiveMessage{m}}} \netsource$};
	\node (h1) at ($(source) + (0,-4)$) {$\vphantom{{h^{m}_{1}}^{\tr_{2}}_{\receiveMessage{m}}} h^{m}_{1}$};
	\node (h1t) at ($(h1) + (1,-2)$) {$\vphantom{{h^{m}_{1}}^{\tr_{2}}_{\receiveMessage{m}}} {h^{m}_{1}}^{\tr_{1}}$};
	\node (h0) at ($(h1) + (-2,0)$) {$\vphantom{{h^{m}_{1}}^{\tr_{2}}_{\receiveMessage{m}}} h^{m}_{0}$};
	\node (h0t) at ($(h0) + (1,-2)$) {$\vphantom{{h^{m}_{1}}^{\tr_{2}}_{\receiveMessage{m}}} {h^{m}_{0}}^{\tr_{0}}$};
	\node (h2) at ($(h1) + (2,0)$) {$\vphantom{{h^{m}_{1}}^{\tr_{2}}_{\receiveMessage{m}}} h^{m}_{2}$};
	\node (h2rt) at ($(h2) + (1,-2)$) {$\vphantom{{h^{m}_{1}}^{\tr_{2}}_{\receiveMessage{m}}} {h^{m}_{2}}^{\tr_{2}}_{\receiveMessage{m}}$};
	\node (h2t) at ($(h0) + (-1,-2)$) {$\vphantom{{h^{m}_{1}}^{\tr_{2}}_{\receiveMessage{m}}} {h^{m}_{2}}^{\tr_{2}}$};
	\node (s) at ($(h0) + (-2,0)$) {$\vphantom{{h^{m}_{1}}^{\tr_{2}}_{\receiveMessage{m}}} \startState$};
	\node (sr) at ($(h2) + (2,0)$) {$\vphantom{{h^{m}_{1}}^{\tr_{2}}_{\receiveMessage{m}}} \startState_{\receiveMessage{m}}$};
	\node (srel) at ($(sr) + (0,2)$) {$\vphantom{{h^{m}_{1}}^{\tr_{2}}_{\receiveMessage{m}}} \startState_{\relord}$};
	\node (sink) at ($(srel) + (4,2)$) {$\vphantom{{h^{m}_{1}}^{\tr_{2}}_{\receiveMessage{m}}} \netsink$};
	\node (h0rel) at ($(sink) + (-2,-2)$) {$\vphantom{{h^{m}_{1}}^{\tr_{2}}_{\receiveMessage{m}}} {h^{m}_{0}}_{\relord}$};
	\node (h1rel) at ($(h0rel) + (2,0)$) {$\vphantom{{h^{m}_{1}}^{\tr_{2}}_{\receiveMessage{m}}} {h^{m}_{1}}_{\relord}$};
	\node (h2rel) at ($(h1rel) + (2,0)$) {$\vphantom{{h^{m}_{1}}^{\tr_{2}}_{\receiveMessage{m}}} {h^{m}_{2}}_{\relord}$};
	\node (h2r) at ($(h2rel) + (0,-2)$) {$\vphantom{{h^{m}_{1}}^{\tr_{2}}_{\receiveMessage{m}}} {h^{m}_{2}}_{\receiveMessage{m}}$};
	\node (h1r) at ($(h1rel) + (0,-2)$) {$\vphantom{{h^{m}_{1}}^{\tr_{2}}_{\receiveMessage{m}}} {h^{m}_{1}}_{\receiveMessage{m}}$};
	\node (h1rt) at ($(h1r) + (1,-2)$) {$\vphantom{{h^{m}_{1}}^{\tr_{2}}_{\receiveMessage{m}}} {h^{m}_{1}}^{\tr_{1}}_{\receiveMessage{m}}$};
	\node (h0r) at ($(h0rel) + (0,-2)$) {$\vphantom{{h^{m}_{1}}^{\tr_{2}}_{\receiveMessage{m}}} {h^{m}_{0}}_{\receiveMessage{m}}$};
	\node (h0rt) at ($(h0r) + (1,-2)$) {$\vphantom{{h^{m}_{1}}^{\tr_{2}}_{\receiveMessage{m}}} {h^{m}_{0}}^{\tr_{0}}_{\receiveMessage{m}}$};

	\draw[dotted, rounded corners=5mm] ($(s) - (0.5,0.5)$) rectangle ($(h2) + (0.6,0.5)$);
	\node (S) at ($(s) + (-0.5,0.7)$) {$\stateSet$};

	\draw[dotted, rounded corners=5mm] ($(sr) - (0.5,0.5)$) rectangle ($(h2r) + (0.65,0.5)$);
	\node (Sr) at ($(sr) + (-0.6,0.7)$) {$\stateSet_{\receiveMessage{m}}$};

	\draw[dotted, rounded corners=5mm] ($(srel) - (0.5,0.5)$) rectangle ($(h2rel) + (0.5,0.5)$);
	\node (Srel) at ($(srel) + (-0.6,0.7)$) {$\stateSet_{\relord}$};

	\draw[dotted, rounded corners=5mm] ($(h2t) - (0.75,0.5)$) rectangle ($(h1t) + (0.65,0.5)$);
	\node (St) at ($(h2t) + (-0.6,0.7)$) {$\stateSet^{\tr}$};

	\draw[dotted, rounded corners=5mm] ($(h2rt) - (0.75,0.5)$) rectangle ($(h1rt) + (0.65,0.5)$);
	\node (Srt) at ($(h2rt) + (-0.8,0.7)$) {$\stateSet^{\tr}_{\receiveMessage{m}}$};

	\draw (source) to (h1);
	\draw[bend left = 20] (h0) to (h0t);
	\draw[bend left = 20] (h0t) to (h0);
	\draw (h0t) to (h1);
	\draw[bend left = 20] (h1) to (h1t);
	\draw[bend left = 20] (h1t) to (h1);
	\draw (h1t) to (h2);
	\draw (h2) to (h2rt);
	\draw (h2rt) to (sr);
	\draw (sr) to (srel);
	\draw (srel) to (sink);
	\draw (h0r) to (h0rel);
	\draw (h0r) to (h1rel);
	\draw (h0r) to (h2rel);
	\draw (h0rel) to (sink);
	\draw[bend right = 20] (h0r) to (h0rt);
	\draw[bend right = 20] (h0rt) to (h0r);
	\draw (h0rt) to (h1r);
	\draw (h1r) to (h0rel);
	\draw (h1r) to (h1rel);
	\draw (h1r) to (h2rel);
	\draw (h1rel) to (sink);
	\draw[bend right = 20] (h1r) to (h1rt);
	\draw[bend right = 20] (h1rt) to (h1r);
	\draw (h1rt) to (h2r);
	\draw (h2r) to (h0rel);
	\draw (h2r) to (h1rel);
	\draw (h2r) to (h2rel);
	\draw (h2rel) to (sink);
	\end{tikzpicture}
	}
	\end{center}

	In the network we have also highlighted the different sets of vertices obtained from the states and the transitions of the automaton, by surrounding them with dotted lines.  
\end{exampleCont}

As in~\cite{HT12}, this network $\NetworkTBetaMuRel{t}{a}{\sd}{\relord}$ and the associated maximum flow problem can not be used directly to encode a weak combined transition since it is not possible to force the flow to split proportional to the transition probability distributions.  
Instead an ordinary LP problem can be derived from the network, which is enriched with additional constraints called \emph{balancing factors}. 
A balancing factor models a probabilistic choice and ensures a balance between flows that leave a vertex so as to respect the probability values in a probabilistic choice, i.e., when leaving a vertex $v \in \stateSet^{\tr} \cup \stateSet^{\tr}_{a}$.

\begin{definition}[cf.~{\cite[Definition~6]{HT12}}]
	Given a \PA{} $\aut$, $\relord \subseteq \stateSet \times \stateSet$, $\sd \in \Disc{\stateSet}$, and $t \in \stateSet$, for $a \in \externalActionSet$ we define the $\LPproblemTBetaMuRel{t}{a}{\sd}{\relord}$ LP problem associated to the network graph $(V,E) = \NetworkTBetaMuRel{t}{a}{\sd}{\relord}$ as follows:
	\[
	\begin{array}{lll}
			\multicolumn{2}{l}{\max \sum_{(x,y) \in E} -f_{x,y}} \\
		\multicolumn{2}{l}{\text{under constraints}} \\ 	
			f_{u,v} \geq 0 & \multicolumn{2}{l}{\qquad \qquad \text{for each $(u,v) \in E$}}\\
			f_{\netsource,t} = 1 \qquad & \\
			f_{v_{\relord},\netsink} = \probeval{\sd}{v} & \multicolumn{2}{l}{\qquad \qquad \text{for each $v \in \stateSet_{\relord}$}} \\
			\multicolumn{2}{l}{\sum_{u \in \setcond{x}{(x,v) \in E}} f_{u,v} - \sum_{u \in \setcond{y}{(v,y) \in E}} f_{v,u} = 0 \qquad} & \hskip2.2mm\text{for each $v \in V \setminus \setnocond{\netsource,\netsink}$} \\
			f_{v^{\tr},v'} - \probeval{\rho}{v'} f_{v,v^{\tr}} = 0 & 
			\multicolumn{2}{l}{\qquad \qquad \text{for each $\tr = \strongTransition{v}{\hidden}{\rho} \in \transitionRelation$ and $v' \in \Supp{\rho}$}}\\
			f_{v^{\tr}_{a},v'_{a}} - \probeval{\rho}{v'} f_{v_{a},v^{\tr}_{a}} = 0 & \multicolumn{2}{l}{\qquad \qquad \text{for each $\tr = \strongTransition{v}{\hidden}{\rho} \in \transitionRelation$ and $v' \in \Supp{\rho}$}}\\
			f_{v^{\tr}_{a},v'_{a}} - \probeval{\rho}{v'} f_{v,v^{\tr}_{a}} = 0 & \multicolumn{2}{l}{\qquad \qquad \text{for each $\tr = \strongTransition{v}{a}{\rho} \in \transitionRelation$ and $v' \in \Supp{\rho}$}}
	\end{array}
	\]
\end{definition}

When $a \in \internalActionSet$, the LP problem $\LPproblemTBetaMuRel{t}{\hidden}{\sd}{\relord}$ associated to $\NetworkTBetaMuRel{t}{\hidden}{\sd}{\relord}$ is defined as above without the last two groups of constraints.
\begin{exampleCont}
\label{ex:lpProblem}
	Consider again the automaton $\wccHopsRadiusTransmprob{2}{5}{\frac{3}{4}}$ depicted in Figure~\ref{fig:wirelessCC}, the state $h^{m}_{1}$, the action $\receiveMessage{m}$, and the equivalence relation $\relord$ on states whose induced classes are $\setnocond{\startState}$ and $\setnocond{h^{m}_{0}, h^{m}_{1}, h^{m}_{2}}$ for each message $m$.
	We have seen in the Example~\ref{ex:network} the network $\NetworkTBetaMuRel{h^{m}_{1}}{\receiveMessage{m}}{\dirac{\startState}}{\relord}$. 
	Consider the probability measure $\dirac{\startState}$ and denote the transitions of the automaton $\wccHopsRadiusTransmprob{2}{5}{\frac{3}{4}}$ as follows: 
	$\tr_{s} = \strongTransition{\startState}{\sendMessage{m}}{\dirac{h^{m}_{0}}}$,
	$\tr_{0} = \strongTransition{h^{m}_{0}}{\hop{5}}{\setnocond{(h^{m}_{0}, \frac{1}{4}), (h^{m}_{1}, \frac{3}{4})}}$,
	$\tr_{1} = \strongTransition{h^{m}_{1}}{\hop{5}}{\setnocond{(h^{m}_{1}, \frac{1}{4}), (h^{m}_{2}, \frac{3}{4})}}$, and 
	$\tr_{2} = \strongTransition{h^{m}_{2}}{\receiveMessage{m}}{\dirac{\startState}}$.

	Besides the constraints for the non-negativity of the variables, the $\LPproblemTBetaMuRel{h^{m}_{1}}{\receiveMessage{m}}{\dirac{\startState}}{\relord}$ LP problem associated to the network $\NetworkTBetaMuRel{h^{m}_{1}}{\receiveMessage{m}}{\dirac{\startState}}{\relord}$ has the following constraints:
	\begin{itemize}
	\item
		initial flow and challenging probabilities:
		\[
			\begin{array}{lclclclcl}
				f_{\netsource,h^{m}_{1}} = 1 & \quad & f_{\startState_{\relord},\netsink} = 1 & \quad &
				f_{{h^{m}_{0}}_{\relord},\netsink} = 0 & \quad & f_{{h^{m}_{2}}_{\relord},\netsink} = 0 & \quad & f_{{h^{m}_{2}}_{\relord},\netsink} = 0 
			\end{array}
		\]
	\item 
		conservation of the flow for vertices in $\stateSet$:
		\[
			\begin{array}{lcl}
				f_{{h^{m}_{0}}^{\tr_{0}}, h^{m}_{0}} - f_{h^{m}_{0}, {h^{m}_{0}}^{\tr_{0}}} = 0 & \quad & f_{\netsource, h^{m}_{1}} + f_{{h^{m}_{0}}^{\tr_{0}}, h^{m}_{1}} + f_{{h^{m}_{1}}^{\tr_{1}}, h^{m}_{1}} - f_{h^{m}_{1}, {h^{m}_{1}}^{\tr_{1}}} = 0 \\
				f_{{h^{m}_{1}}^{\tr_{1}}, h^{m}_{2}} - f_{h^{m}_{2}, {h^{m}_{2}}^{\tr_{2}}_{\receiveMessage{m}}} = 0
			\end{array}
		\]
	\item 
		conservation of the flow for vertices in $\stateSet^{\tr}$:
		\[
			\begin{array}{lcl}
				f_{h^{m}_{0}, {h^{m}_{0}}^{\tr_{0}}} - f_{{h^{m}_{0}}^{\tr_{0}}, h^{m}_{0}} - f_{{h^{m}_{0}}^{\tr_{0}}, h^{m}_{1}} = 0 & \quad & f_{h^{m}_{1}, {h^{m}_{1}}^{\tr_{1}}} - f_{{h^{m}_{1}}^{\tr_{1}}, h^{m}_{1}} - f_{{h^{m}_{1}}^{\tr_{1}}, h^{m}_{2}} = 0
			\end{array}
		\]
	\item 
		conservation of the flow for vertices in $\stateSet_{\receiveMessage{m}}$:
		\[
			\begin{array}{l}
				f_{{h^{m}_{2}}^{\tr_{2}}_{\receiveMessage{m}}, \startState_{\receiveMessage{m}}} - f_{\startState_{\receiveMessage{m}}, \startState_{\relord}} = 0 \\
				f_{{h^{m}_{0}}^{\tr_{0}}_{\receiveMessage{m}}, {h^{m}_{0}}_{\receiveMessage{m}}} - f_{{h^{m}_{0}}_{\receiveMessage{m}}, {h^{m}_{0}}^{\tr_{0}}_{\receiveMessage{m}}} - f_{{h^{m}_{0}}_{\receiveMessage{m}}, {h^{m}_{0}}_{\relord}} - f_{{h^{m}_{0}}_{\receiveMessage{m}}, {h^{m}_{0}}_{\relord}} - f_{{h^{m}_{0}}_{\receiveMessage{m}}, {h^{m}_{2}}_{\relord}} = 0 \\
				f_{{h^{m}_{0}}^{\tr_{0}}_{\receiveMessage{m}}, {h^{m}_{1}}_{\receiveMessage{m}}} + f_{{h^{m}_{1}}^{\tr_{1}}_{\receiveMessage{m}}, {h^{m}_{1}}_{\receiveMessage{m}}} - f_{{h^{m}_{1}}_{\receiveMessage{m}}, {h^{m}_{1}}^{\tr_{1}}_{\receiveMessage{m}}} - f_{{h^{m}_{1}}_{\receiveMessage{m}}, {h^{m}_{0}}_{\relord}} - f_{{h^{m}_{1}}_{\receiveMessage{m}}, {h^{m}_{1}}_{\relord}} - f_{{h^{m}_{1}}_{\receiveMessage{m}}, {h^{m}_{2}}_{\relord}} = 0 \\
				f_{{h^{m}_{1}}^{\tr_{1}}_{\receiveMessage{m}}, {h^{m}_{2}}_{\receiveMessage{m}}} - f_{{h^{m}_{2}}_{\receiveMessage{m}}, {h^{m}_{0}}_{\relord}} - f_{{h^{m}_{2}}_{\receiveMessage{m}}, {h^{m}_{0}}_{\relord}} - f_{{h^{m}_{2}}_{\receiveMessage{m}}, {h^{m}_{2}}_{\relord}} = 0			
			\end{array}
		\]
	\item 
		conservation of the flow for vertices in $\stateSet^{\tr}_{\receiveMessage{m}}$:
		\[
			\begin{array}{l}
				f_{{h^{m}_{0}}_{\receiveMessage{m}}, {h^{m}_{0}}^{\tr_{0}}_{\receiveMessage{m}}} - f_{{h^{m}_{0}}^{\tr_{0}}_{\receiveMessage{m}}, {h^{m}_{0}}_{\receiveMessage{m}}} - f_{{h^{m}_{0}}^{\tr_{0}}_{\receiveMessage{m}}, {h^{m}_{1}}_{\receiveMessage{m}}} = 0 \\
				f_{{h^{m}_{1}}_{\receiveMessage{m}}, {h^{m}_{1}}^{\tr_{1}}_{\receiveMessage{m}}} - f_{{h^{m}_{1}}^{\tr_{1}}_{\receiveMessage{m}}, {h^{m}_{1}}_{\receiveMessage{m}}} - f_{{h^{m}_{1}}^{\tr_{1}}_{\receiveMessage{m}}, {h^{m}_{2}}_{\receiveMessage{m}}} = 0 \\
				f_{h^{m}_{2}, {h^{m}_{2}}^{\tr_{2}}_{\receiveMessage{m}}} - f_{{h^{m}_{2}}^{\tr_{2}}_{\receiveMessage{m}}, \startState_{\receiveMessage{m}}} = 0
			\end{array}
		\]
	\item
		conservation of the flow for vertices in $\stateSet_{\relord}$:
		\[
			\begin{array}{l}
				f_{\startState_{\receiveMessage{m}}, \startState_{\relord}} - f_{\startState_{\relord}, \netsink} = 0 \\
				f_{{h^{m}_{0}}_{\receiveMessage{m}}, {h^{m}_{0}}_{\relord}} + f_{{h^{m}_{1}}_{\receiveMessage{m}}, {h^{m}_{0}}_{\relord}} + f_{{h^{m}_{2}}_{\receiveMessage{m}}, {h^{m}_{0}}_{\relord}} - f_{{h^{m}_{0}}_{\relord}, \netsink} = 0 \\
				f_{{h^{m}_{0}}_{\receiveMessage{m}}, {h^{m}_{1}}_{\relord}} + f_{{h^{m}_{1}}_{\receiveMessage{m}}, {h^{m}_{1}}_{\relord}} + f_{{h^{m}_{2}}_{\receiveMessage{m}}, {h^{m}_{1}}_{\relord}} - f_{{h^{m}_{1}}_{\relord}, \netsink} = 0 \\
				f_{{h^{m}_{0}}_{\receiveMessage{m}}, {h^{m}_{2}}_{\relord}} + f_{{h^{m}_{1}}_{\receiveMessage{m}}, {h^{m}_{2}}_{\relord}} + f_{{h^{m}_{2}}_{\receiveMessage{m}}, {h^{m}_{2}}_{\relord}} - f_{{h^{m}_{2}}_{\relord}, \netsink} = 0
			\end{array}
		\]
	\item
		balancing constraints:
		\[
			\begin{array}{lcl}
				f_{{h^{m}_{0}}^{\tr_{0}}, h^{m}_{0}} - \frac{1}{4} \cdot f_{h^{m}_{0}, {h^{m}_{0}}^{\tr_{0}}} = 0 & \quad & f_{{h^{m}_{0}}^{\tr_{0}}, h^{m}_{1}} - \frac{3}{4} \cdot f_{h^{m}_{0}, {h^{m}_{0}}^{\tr_{0}}} = 0 \\
				f_{{h^{m}_{1}}^{\tr_{1}}, h^{m}_{1}} - \frac{1}{4} \cdot f_{h^{m}_{1}, {h^{m}_{1}}^{\tr_{1}}} = 0 & \quad & f_{{h^{m}_{1}}^{\tr_{1}}, h^{m}_{2}} - \frac{3}{4} \cdot f_{h^{m}_{1}, {h^{m}_{1}}^{\tr_{1}}} = 0 \\
				f_{{h^{m}_{2}}^{\tr_{2}}_{\receiveMessage{m}}, \startState_{\receiveMessage{m}}} - 1 \cdot f_{h^{m}_{2}, {h^{m}_{2}}^{\tr_{2}}_{\receiveMessage{m}}} = 0\\
				f_{{h^{m}_{0}}^{\tr_{0}}_{\receiveMessage{m}}, {h^{m}_{0}}_{\receiveMessage{m}}} - \frac{1}{4} \cdot f_{{h^{m}_{0}}_{\receiveMessage{m}}, {h^{m}_{0}}^{\tr_{0}}_{\receiveMessage{m}}} = 0 & \quad & f_{{h^{m}_{0}}^{\tr_{0}}_{\receiveMessage{m}}, {h^{m}_{1}}_{\receiveMessage{m}}} - \frac{3}{4} \cdot f_{{h^{m}_{0}}_{\receiveMessage{m}}, {h^{m}_{0}}^{\tr_{0}}_{\receiveMessage{m}}} = 0 \\
				f_{{h^{m}_{1}}^{\tr_{1}}_{\receiveMessage{m}}, {h^{m}_{1}}_{\receiveMessage{m}}} - \frac{1}{4} \cdot f_{{h^{m}_{1}}_{\receiveMessage{m}}, {h^{m}_{1}}^{\tr_{1}}_{\receiveMessage{m}}} = 0 & \quad & f_{{h^{m}_{1}}^{\tr_{1}}_{\receiveMessage{m}}, {h^{m}_{2}}_{\receiveMessage{m}}} - \frac{3}{4} \cdot f_{{h^{m}_{1}}_{\receiveMessage{m}}, {h^{m}_{1}}^{\tr_{1}}_{\receiveMessage{m}}} = 0 
			\end{array}
		\]
	\end{itemize}
	A solution that maximises the objective function assigns value $0$ to all variables except for the following variables:
	\[
		\begin{array}{lclclcl}
			f_{\netsource, h^{m}_{1}} = 1 & \quad & f_{h^{m}_{1}, {h^{m}_{1}}^{\tr_{1}}} = \frac{4}{3} & \quad & f_{{h^{m}_{1}}^{\tr_{1}}, h^{m}_{1}} = \frac{1}{3} & \quad & f_{{h^{m}_{1}}^{\tr_{1}}, h^{m}_{2}} = 1 \\
			f_{h^{m}_{2}, {h^{m}_{2}}^{\tr_{2}}_{\receiveMessage{m}}} = 1 & \quad & f_{{h^{m}_{2}}^{\tr_{2}}_{\receiveMessage{m}}, \startState_{\receiveMessage{m}}} = 1 & \quad & f_{\startState_{\receiveMessage{m}}, \startState_{\relord}} = 1 & \quad & f_{\startState_{\relord}, \netsink} = 1
		\end{array}
	\]
	It is not uncommon to have variables with value greater than $1$, as happens for $f_{h^{m}_{1}, {h^{m}_{1}}^{\tr_{1}}}$, in particular when such variables correspond to edges in a cycle.
\end{exampleCont}

The LP problem $\LPproblemTBetaMuRel{t}{a}{\sd}{\relord}$ is equivalent to a weak combined transition modulo $\relord$, in the sense that any feasible solution of the LP problem is enough to establish the transition (cf.~\cite[Theorem~8]{HT12}).  
So the objective function has no impact, and this gives us room to use for instance some $\min \sum_{(x,y) \in E} c_{x,y} \cdot f_{x,y}$ as objective function. 
In this way, a weak transition can also be seen as a minimum cost flow problem plus balancing constraints.  
In the sequel we explore how to use the objective function to compute and minimise the cost of performing a weak combined transition.

\subsection{Incorporating Transition Costs}
\label{sec:transitioncosts}

In order to extend our computational approach to costs we revisit the concentric ball characterisation of weak transition cost from Definition~\ref{def:costAsWorkingForTheProof}.  
This is worthwhile when the weak combined transition is induced by a \emph{determinate} scheduler~\cite{CS02}, that is, a scheduler $\sched$ such that for each pair of finite execution fragments $\alpha$, $\alpha'$, if $\trace{\alpha} = \trace{\alpha'}$ and $\last{\alpha} = \last{\alpha'}$, then $\schedeval{\sched}{\alpha} = \schedeval{\sched}{\alpha'}$.  
Under these schedulers, the resolution of the nondeterminism is the same for all finite execution fragments having the same trace and the same final state, so we can rearrange addends and factors in Definition~\ref{def:costAsWorkingForTheProof} in order to express the cost of the weak combined transition as the sum of the cost of each transition weighted by the sum of the probabilities of performing it regardless of the ball we are considering.  
Since there is a strict relation between probabilities of reaching a state and the flow entering the corresponding vertex (cf.~\cite[Corollary~2]{HT12}), the overall resulting cost is the sum of the cost of each transition $\tr = \strongTransition{u}{b}{\gd}$ multiplied by the flow from $u$ to $u^{\tr}$.  
This consideration is the base for the following definition, where we encode the transition costs in the LP problem as coefficients of the objective function.

\begin{definition}
\label{def:minCostLPProblem}
	Given a \CPA{} $\costAut{\costf}$, a binary relation $\relord$ on $\stateSet$, a probability distribution $\sd \in \Disc{\stateSet}$, and a state $t \in \stateSet$, for action $a \in \externalActionSet$ we define the min-cost LP problem $\minCostLPproblemTBetaMuRel{t}{a}{\sd}{\relord}$ associated to the network $\NetworkTBetaMuRel{t}{a}{\sd}{\relord}$ as follows.

	\[
		\begin{array}{lll}
			\multicolumn{2}{l}{\min \sum_{(x,y) \in E} \cost[f]{x,y} \cdot f_{x,y}} \\
			\multicolumn{2}{l}{\text{under constraints}} \\ 	
			f_{u,v} \geq 0 & \multicolumn{2}{l}{\qquad \qquad \text{for each $(u,v) \in E$}}\\
			f_{\netsource,t} = 1 \qquad & \\
			f_{v_{\relord},\netsink} = \probeval{\sd}{v} & \multicolumn{2}{l}{\qquad \qquad \text{for each $v \in \stateSet_{\relord}$}}\\
			\multicolumn{2}{l}{\sum_{u \in \setcond{x}{(x,v) \in E}} f_{u,v} - \sum_{u \in \setcond{y}{(v,y) \in E}} f_{v,u} = 0 \qquad} & \hskip2.2mm\text{for each $v \in V \setminus \setnocond{\netsource,\netsink}$} \\
			f_{v^{\tr},v'} - \probeval{\rho}{v'} f_{v,v^{\tr}} = 0 & \multicolumn{2}{l}{\qquad \qquad \text{for each $\tr = \strongTransition{v}{\hidden}{\rho} \in \transitionRelation$ and $v' \in \Supp{\rho}$}} \\
			f_{v^{\tr}_{a},v'_{a}} - \probeval{\rho}{v'} f_{v_{a},v^{\tr}_{a}} = 0 & \multicolumn{2}{l}{\qquad \qquad \text{for each $\tr = \strongTransition{v}{\hidden}{\rho} \in \transitionRelation$ and $v' \in \Supp{\rho}$}}\\
			f_{v^{\tr}_{a},v'_{a}} - \probeval{\rho}{v'} f_{v,v^{\tr}_{a}} = 0 & \multicolumn{2}{l}{\qquad \qquad \text{for each $\tr = \strongTransition{v}{a}{\rho} \in \transitionRelation$ and $v' \in \Supp{\rho}$}}
		\end{array}
	\]
	where $\costf_{f} \colon E \to \posreals$ is a total function defined as follows:
	\[
		\cost[f]{x,y} =
		\begin{cases}
			\cost{\tr} & \text{if $\tr = \strongTransition{v}{\hidden}{\gd}$, $x = v$, $y = v^{\tr}$,}\\
			\cost{\tr} & \text{if $\tr = \strongTransition{v}{\hidden}{\gd}$, $x = v_{a}$, $y = v^{\tr}_{a}$,}\\
			\cost{\tr} & \text{if $\tr = \strongTransition{v}{a}{\gd}$, $x = v$, $y = v^{\tr}_{a}$, }\\
			0 & \text{otherwise.}
		\end{cases}
	\]
	If $\minCostLPproblemTBetaMuRel{t}{a}{\sd}{\relord}$ has an optimal solution $f^{o}$, then we denote by $\minCostLPValue$ the minimum cost $\minCostLPValue = \sum_{(x,y) \in E} \cost[f]{x,y} \cdot f^{o}_{x,y}$.
\end{definition}
When $a \in \internalActionSet$, the min-cost LP problem $\minCostLPproblemTBetaMuRel{t}{\hidden}{\sd}{\relord}$ associated to the network $\NetworkTBetaMuRel{t}{\hidden}{\sd}{\relord}$ is defined as above without the last two groups of constraints.

A first obvious result is that $\minCostLPproblemTBetaMuRel{t}{a}{\sd}{\relord}$ is feasible if and only if $\LPproblemTBetaMuRel{t}{a}{\sd}{\relord}$ is feasible, since the only difference between the two problems is the objective function that does not affect the feasibility of an LP problem:
\begin{proposition}
\label{pro:minCostLPProblemIsFeasibleIFFLPProblemIsFeasible}
	Given a \CPA{} $\costAut{\costf}$, $\relord \subseteq \stateSet \times \stateSet$, $a \in \actionSet$, $\sd \in \Disc{\stateSet}$, and $t \in \stateSet$, the minimisation LP problem $\minCostLPproblemTBetaMuRel{t}{a}{\sd}{\relord}$ has a feasible solution $f^{*}$ if and only if $f^{*}$ is a feasible solution of the LP problem $\LPproblemTBetaMuRel{t}{a}{\sd}{\relord}$.
\end{proposition}
\begin{myproof}
	The equivalence holds since $\minCostLPproblemTBetaMuRel{t}{a}{\sd}{\rel}$ and $\LPproblemTBetaMuRel{t}{a}{\sd}{\rel}$ have the same set of constraints.
\end{myproof}

Similarly, as generating and checking the existence of a valid solution of the LP problem $\LPproblemTBetaMuRel{t}{a}{\sd}{\relord}$ is polynomial in $N = \max \setnocond{\setcardinality{\stateSet},\setcardinality{\transitionRelation}}$ (cf.~\cite[Theorem~7]{HT12}), the same holds for $\minCostLPproblemTBetaMuRel{t}{a}{\sd}{\relord}$:
\begin{corollary}
\label{cor:MinCostLPProblemIsPolynomial}
	Given a \CPA{} $\costAut{\costf}$, $\relord \subseteq \stateSet \times \stateSet$, $a \in \actionSet$, $\sd \in \Disc{\stateSet}$, and $t \in \stateSet$, 
	generating and checking the existence of a valid solution of the minimisation LP problem $\minCostLPproblemTBetaMuRel{t}{a}{\sd}{\relord}$ is polynomial in $N = \max \setnocond{\setcardinality{\stateSet},\setcardinality{\transitionRelation}}$.
\end{corollary}
\begin{myproof}
	The result follows immediately from Proposition~\ref{pro:minCostLPProblemIsFeasibleIFFLPProblemIsFeasible} and ~\cite[Theorem~7]{HT12}.
\end{myproof}

Since $\LPproblemTBetaMuRel{t}{a}{\sd}{\relord}$ is feasible if and only if there exists a scheduler $\sched$ that induces $\weakCombinedTransition{t}{a}{\sd_{t}}$ such that $\sd \liftrel \sd_{t}$, we may expect a similar result regarding costs, that is, $\minCostLPproblemTBetaMuRel{t}{a}{\sd}{\relord}$ is feasible with optimal value $\minCostLPValue$ if and only if there exists a scheduler $\sched$ that induces $\weakCombinedTransition{t}{a}{\sd_{t}}$ such that $\sd \liftrel \sd_{t}$ and $\cost{\weakCombinedTransition{t}{a}{\sd_{t}}} = \minCostLPValue$.
Note that in general it is not possible to obtain such a result: 
There can be different ways to resolve nondeterminism, i.e., different schedulers, that induce the same weak combined transition but with different costs. 
Thus we can not talk about \emph{the cost} of a weak combined transition, but of the cost of the weak combined transition \emph{as induced by the scheduler $\sched$}.
For instance, consider the automaton $\aut$ depicted on the right whose transitions are $\tr_{1} = \strongTransition{\startState}{a}{\dirac{t}}$, $\tr_{2} = \strongTransition{\startState}{\hidden}{\dirac{v}}$, and $\tr_{3} = \strongTransition{v}{a}{\dirac{t}}$, each one with cost $1$.

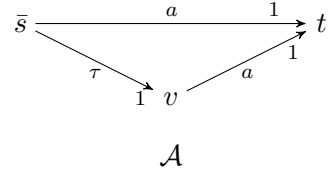
\begin{wrapfigure}[8]{r}{50mm}
\centering
\begin{tikzpicture}[->,>=stealth',shorten >=1pt,auto]
	\scriptsize
	\path[use as bounding box] (0,0) rectangle (4,-1.75);
	
	\node (ss) at (0,0) {\normalsize$\startState$};
	\node (t) at (4,0) {\normalsize$t$};
	\node (v) at (2,-1) {\normalsize$v$};
	\node (aut) at (2, -1.75) {\normalsize$\aut$};
	
	\draw (ss) to node[above] {$a$} node[above, very near end] {$1$} (t);
	\draw (ss) to node[below] {$\hidden$} node[below, very near end] {$1$} (v);
	\draw (v) to node[below] {$a$} node[below, very near end] {$1$} (t);
\end{tikzpicture}
\caption{An automaton}
\label{fig:automataDifferenWeakTransitionCost}
\end{wrapfigure}
It is straightforward to check that the scheduler $\sched_{1}$ such that $\schedeval{\sched_{1}}{\startState} = \dirac{\tr_{1}}$ and $\schedeval{\sched_{1}}{\alpha} = \dirac{\bot}$ for each finite execution fragment $\alpha \neq \startState$ induces the weak combined transition $\weakCombinedTransition{\startState}{a}{\dirac{t}}$ whose cost is $1$; 
the same transition is induced also by the scheduler $\sched_{2}$ defined as $\schedeval{\sched_{2}}{\startState} = \dirac{\tr_{2}}$, $\schedeval{\sched_{2}}{\startState \hidden v} = \dirac{\tr_{3}}$, and $\schedeval{\sched_{2}}{\alpha} = \dirac{\bot}$ for each other finite execution fragment $\alpha$.
However the cost as induced by $\sched_{2}$ is $\cost[\sched_{2}]{\weakCombinedTransition{\startState}{a}{\dirac{t}}} = 2 \neq 1 = \cost[\sched_{1}]{\weakCombinedTransition{\startState}{a}{\dirac{t}}}$;
it is easy to show that $1 \leq \cost[\sched]{\weakCombinedTransition{\startState}{a}{\dirac{t}}} \leq 2$ for each scheduler $\sched$ inducing $\weakCombinedTransition{\startState}{a}{\dirac{t}}$.
Note that there are uncountably many such schedulers, each one corresponding to a different resolution of the choice between $\tr_{1} = \strongTransition{\startState}{a}{\dirac{t}}$ and $\tr_{2} = \strongTransition{\startState}{\hidden}{\dirac{v}}$: 
In general, we can denote such choice as the distribution $\setnocond{(\tr_{1}, p), (\tr_{2}, 1-p)}$ where $p \in [0,1]$.

The cost given by a scheduler and the value of the objective function of the corresponding LP problem are however related: 
\begin{theorem}
\label{thm:minCostLPequivalentToCostWeakTransitionLifting}
	Given a \CPA{} $\costAut{\costf}$, $\relord \subseteq \stateSet \times \stateSet$, $a \in \actionSet$, $\sd \in \Disc{\stateSet}$, and $t \in \stateSet$, consider the $\minCostLPproblemTBetaMuRel{t}{a}{\sd}{\relord}$ LP problem.
	The following implications hold:
	\begin{enumerate}
		\item
			If there exists a scheduler $\sched$ for $\aut$ that induces $\weakCombinedTransition{t}{a}{\sd_{t}}$ such that $\sd \liftrel \sd_{t}$, then $\minCostLPproblemTBetaMuRel{t}{a}{\sd}{\relord}$ has an optimal solution $f^{o}$ such that $\minCostLPValue \leq \cost{\weakCombinedTransition{t}{a}{\sd_{t}}}$.
		\item 
			If $\minCostLPproblemTBetaMuRel{t}{a}{\sd}{\relord}$ has an optimal solution $f^{o}$, then there exists a scheduler $\sched$ for $\aut$ that induces $\weakCombinedTransition{t}{a}{\sd_{t}}$ such that $\sd \liftrel \sd_{t}$ and $\cost{\weakCombinedTransition{t}{a}{\sd_{t}}} = \minCostLPValue$.
	\end{enumerate}
\end{theorem}
\begin{myoutline}
	The proof is mainly based on the proof of~\cite[Theorem~8]{HT12}; 
	the relations between the optimal value and the cost of the weak combined transition are shown by manipulating the summations in the objective function of the LP problem and of the cost of the weak combined transition, together with the fact that the flow incoming a vertex $v$ in $\stateSet \cup \stateSet_{a}$ equals the sum of the probabilities of the cones of finite execution fragment ending with the state $v$.
\end{myoutline}

As immediate corollaries we have that the cost given by the optimal solution of the $\minCostLPproblemTBetaMuRel{t}{a}{\sd}{\relord}$ LP problem corresponds to the minimum cost induced by any scheduler inducing $\weakCombinedTransition{t}{a}{\sd_{t}}$ and that finding such minimum is polynomial. 
\begin{corollary}
\label{cor:minCostLPequalToMinCostWeakTransitionLifting}
	Given a \CPA{} $\costAut{\costf}$, $\relord \subseteq \stateSet \times \stateSet$, $a \in \actionSet$, $\sd \in \Disc{\stateSet}$, and $t \in \stateSet$ such that there exists $\weakCombinedTransition{t}{a}{\sd_{t}}$ with $\sd \liftrel \sd_{t}$, 
	the LP problem $\minCostLPproblemTBetaMuRel{t}{a}{\sd}{\relord}$ has minimum cost $\minCostLPValue = \min\setcond{\cost[\sched]{\weakCombinedTransition{t}{a}{\sd_{t}}}}{\text{$\sched$ induces $\weakCombinedTransition{t}{a}{\sd_{t}}$ such that $\sd \liftrel \sd_{t}$}}$.
\end{corollary}
\begin{myproof}
	Let $\sched$ be a scheduler that induces the transition $\weakCombinedTransition{t}{a}{\sd_{t}}$ with $\sd \liftrel \sd_{t}$ such that $\cost[\sched]{\weakCombinedTransition{t}{a}{\sd_{t}}} = \min\setcond{\cost[\sched]{\weakCombinedTransition{t}{a}{\sd_{t}}}}{\text{$\sched$ induces $\weakCombinedTransition{t}{a}{\sd_{t}}$}}$.
	By Theorem~\ref{thm:minCostLPequivalentToCostWeakTransitionLifting}, we can derive that $\minCostLPproblemTBetaMuRel{t}{a}{\sd}{\rel}$ has an optimal solution $f^{o}$ such that $\minCostLPValue = \sum_{(x,y) \in E} \cost[f]{(x,y)} \cdot f^{o}_{x,y} \leq \cost[\sched]{\weakCombinedTransition{t}{a}{\sd_{t}}}$, hence $\minCostLPValue \leq \min\setcond{\cost[\sched]{\weakCombinedTransition{t}{a}{\sd_{t}}}}{\text{$\sched$ induces $\weakCombinedTransition{t}{a}{\sd_{t}}$ such that $\sd \liftrel \sd_{t}$}}$.

	Suppose, for the sake of contradiction, that there exists a scheduler $\sched$ inducing $\weakCombinedTransition{t}{a}{\sd_{t}}$ such that $\sd \liftrel \sd_{t}$ such that $\cost[\sched]{\weakCombinedTransition{t}{a}{\sd_{t}}} < \minCostLPValue$.
	Hence, by Theorem~\ref{thm:minCostLPequivalentToCostWeakTransitionLifting}, we have that $\minCostLPproblemTBetaMuRel{t}{a}{\sd}{\rel}$ has a solution $f^{*}$, that is induced by the scheduler $\sched$, such that $\sum_{(x,y) \in E} \cost[f]{(x,y)} \cdot f^{*}_{x,y} \leq \cost[\sched]{\weakCombinedTransition{t}{a}{\sd_{t}}} < \minCostLPValue$, but this contradicts the fact that $f^{o}$ is optimal.
	Thus for each scheduler $\sched'$, $\cost[\sched']{\weakCombinedTransition{t}{a}{\sd_{t}}} \geq \minCostLPValue $.
	Theorem~\ref{thm:minCostLPequivalentToCostWeakTransitionLifting} implies also that there exists a scheduler $\sched'$ inducing the transition $\weakCombinedTransition{t}{a}{\sd_{t}}$ such that $\sd \liftrel \sd_{t}$ and that $\cost[\sched']{\weakCombinedTransition{t}{a}{\sd_{t}}} = \minCostLPValue$, thus $\minCostLPValue = \min\setcond{\cost[\sched]{\weakCombinedTransition{t}{a}{\sd_{t}}}}{\text{$\sched$ induces $\weakCombinedTransition{t}{a}{\sd_{t}}$ such that $\sd \liftrel \sd_{t}$}}$.
\end{myproof}

\begin{corollary}
\label{cor:minCostWeakTransitionIsPolynomial}
	Given a \CPA{} $\costAut{\costf}$, $\relord \subseteq \stateSet \times \stateSet$, $a \in \actionSet$, $\sd \in \Disc{\stateSet}$, and $t \in \stateSet$, finding $\min\setcond{\cost[\sched]{\weakCombinedTransition{t}{a}{\sd_{t}}}}{\text{$\sched$ induces $\weakCombinedTransition{t}{a}{\sd_{t}}$ such that $\sd \liftrel \sd_{t}$}}$ is polynomial in $N = \max \setnocond{\setcardinality{\stateSet},\setcardinality{\transitionRelation}}$.
\end{corollary}
\begin{myproof}
	The result follows immediately from Corollaries~\ref{cor:minCostLPequalToMinCostWeakTransitionLifting} and~\ref{cor:MinCostLPProblemIsPolynomial}.
\end{myproof}

Extending the above results to hyper-transitions of the \CPA{} $\costAut{\costf}$ is straightforward, since we can consider each hyper-transition $\hyperWeakCombinedTransition{\gd}{a}{\sd}$ as the weak combined transition $\weakCombinedTransition{h}{a}{\sd}$ in the \CPA{} $\costAut[\aut']{\costf'}$ that is $\costAut{\costf}$ enriched with the fresh state $h$ and the transition $\strongTransition{h}{\hidden}{\gd}$ whose cost is set to $0$.

\subsection{Deciding Cost Bisimulations}
\label{ssec:decisionProcedure}

We now show how we can decide in polynomial time the cost bisimulations we have presented for \CPA{} in Section~\ref{sec:costProbabilisticAutomata}. 
We commence our discussion with the most intricate relation, minor cost weak probabilistic bisimulation, and then move on to the simpler weak probabilistic cost-preserving bisimulation. 
Only after that we consider the strong probabilistic cost relations and finally the strong cost relations.

\subsubsection{Deciding Minor Cost Weak Probabilistic Bisimulation}
\label{ssec:decisionProcedureMCWPB}

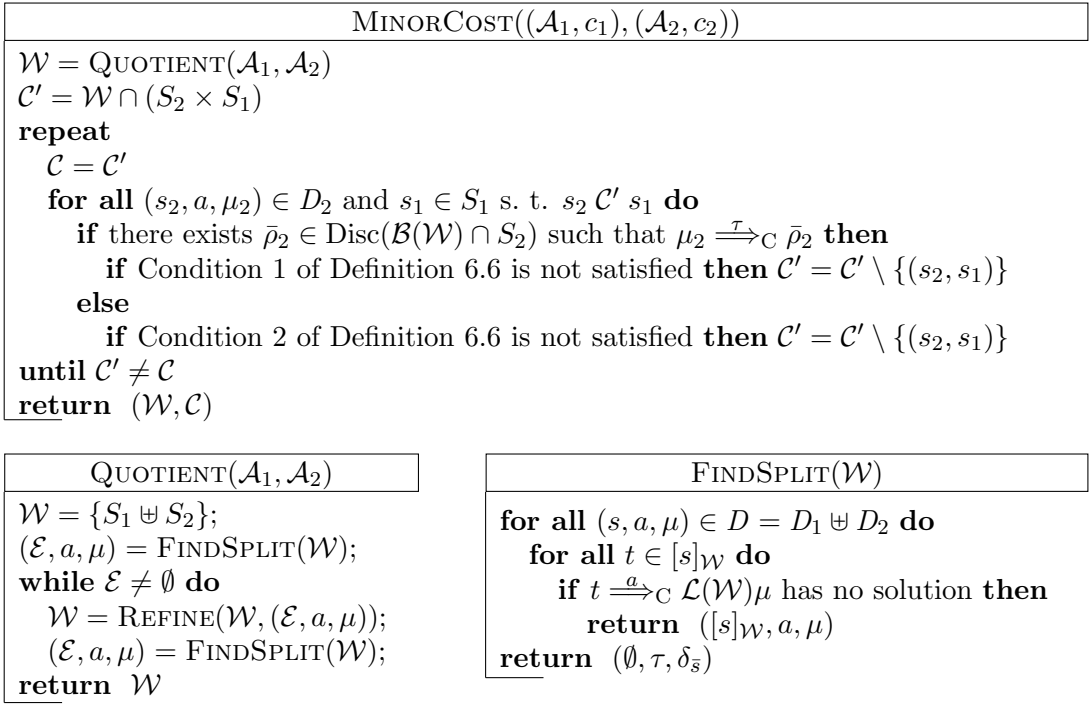
\begin{figure}[t]
\centering
	\begin{tikzpicture}
		\node[draw, text height=2ex, minimum width=144mm, inner sep=0pt, minimum height=3ex] (headingMC) at (0,0) {\progHeader{$\MinorCostProc(\costAutPed{\costf}{1}, \costAutPed{\costf}{2})$}};
		\node[below=-0.1ex of headingMC] {%
		\hskip-4mm
		\begin{minipage}[h]{144mm}
			\begin{algorithmic}
				\STATE{$\wbrelord = \QuotientProc(\aut_{1},\aut_{2})$}
				\STATE{$\costrelord' = \wbrelord \cap (\stateSet_{2} \times \stateSet_{1}$)}
				\REPEAT
					\STATE{$\costrelord = \costrelord'$}
					\FORALL{$(s_{2},a,\sd_{2}) \in \transitionRelation_{2}$ and $s_{1} \in \stateSet_{1}$ s.~t. $s_{2} \costrel' s_{1}$}
						\IF{there exists $\bar{\gd}_{2} \in \Disc{\borderStateSetOrd[\wbrelord] \cap \stateSet_{2}}$ such that $\hyperWeakCombinedTransition{\sd_{2}}{\hidden}{\bar{\gd}_{2}}$ \textbf{then}}	
							\IF{Condition~\ref{def:weakCostBisimMinorCost:stepConditionOnBorderStates} of Definition~\ref{def:weakCostBisimMinorCost} is not satisfied \textbf{then} $\costrelord' = \costrelord' \setminus \setnocond{(s_{2},s_{1})}$}
							\STATE{\vskip-1.2em}
							\ENDIF
						\ELSE
							\IF{Condition~\ref{def:weakCostBisimMinorCost:stepConditionOnInternalStates} of Definition~\ref{def:weakCostBisimMinorCost} is not satisfied \textbf{then} $\costrelord' = \costrelord' \setminus \setnocond{(s_{2},s_{1})}$}
							\STATE{\vskip-1.2em}
							\ENDIF
						\ENDIF
					\ENDFOR
				\UNTIL{$\costrelord' \neq \costrelord$}
			\RETURN{$(\wbrelord, \costrelord)$}
			\end{algorithmic}
		\end{minipage}
		}; 
		\coordinate (ENDMC) at ($(headingMC.north west)!33.5ex!(headingMC.south west)$); 
		\draw (headingMC.south west) to (ENDMC) {}; 
		\draw (ENDMC) to ($(ENDMC)+(2em,0)$) {};

		\node[draw, text height=2ex, minimum width=55mm, inner sep=0pt, minimum height=3ex] (headingQ) at (-4.45,-6) {\progHeader{$\QuotientProc(\aut_{1}, \aut_{2})$}};
		\node[below=-0.1ex of headingQ] {%
		\hskip-4mm
		\begin{minipage}[h]{55mm}
			\begin{algorithmic}
				\STATE{$\wbrelord = \{\stateSet_{1} \uplus \stateSet_{2}\}$;}
				\STATE{$(\equivclass,a,\sd) = \proc{FindSplit}(\wbrelord)$;}
				\WHILE{$\equivclass \neq \emptyset$}
					\STATE{$\wbrelord = \proc{Refine}(\wbrelord,(\equivclass,a,\sd))$;}
					\STATE{$(\equivclass,a,\sd) = \proc{FindSplit}(\wbrelord)$;}
				\ENDWHILE
				\RETURN{$\wbrelord$}
			\end{algorithmic}
		\end{minipage}
		}; 
		\coordinate (ENDQ) at ($(headingQ.north west)!20ex!(headingQ.south west)$); 
		\draw (headingQ.south west) to (ENDQ) {}; 
		\draw (ENDQ) to ($(ENDQ)+(2em,0)$) {};

		\node[draw, text height=2ex, minimum width=80mm, inner sep = 0pt, minimum height=3ex] (headingFS) at (3.2,-6) {\progHeader{$\FindSplitProc(\wbrelord)$}}; 
		\node[below=0.5ex of headingFS] {%
		\hskip-4mm
		\begin{minipage}[h]{80mm}
			\begin{algorithmic}
				\FORALL{$(s,a,\sd) \in \transitionRelation = \transitionRelation_{1} \uplus \transitionRelation_{2}$}
					\FORALL{$t \in \relclass{s}{\wbrelord}$}
						\IF{$\LPproblemTBetaMuRel{t}{a}{\sd}{\wbrelord}$ has no solution \textbf{then}} 
							\RETURN{$(\relclass{s}{\wbrelord},a,\sd)$}
						\ENDIF
					\ENDFOR
				\ENDFOR
				\RETURN{$(\emptyset,\hidden,\dirac{\startState})$}
			\end{algorithmic}
		\end{minipage}
		}; 
		\coordinate (ENDFS) at ($(headingFS.north west)!18ex!(headingFS.south west)$); 
		\draw (headingFS.south west) to (ENDFS) {}; 
		\draw (ENDFS) to ($(ENDFS)+(2em,0)$) {};
	\end{tikzpicture}
	\caption{Minor cost weak bisimulation decision procedure}
	\label{fig:minorCostWeakBisimDecisionProcedure}
\end{figure}

In order to algorithmically decide whether $\costAutPed{\costf}{1} \weakCostBisimMinorCost \costAutPed{\costf}{2}$, we extend the polynomial decision procedure $\QuotientProc$ that establishes whether $\aut_{1} \weakBisim \aut_{2}$ holds~\cite{HT12}, to the $\MinorCostProc$ algorithm depicted in Figure~\ref{fig:minorCostWeakBisimDecisionProcedure} that computes $(\wbrelord, \costrelord)$ justifying $\costAutPed{\costf}{1} \weakCostBisimMinorCost \costAutPed{\costf}{2}$: 
We first compute $\wbrelord = \QuotientProc(\aut_{1},\aut_{2})$ and then we consider as candidate cost relation $\costrelord = \costrelord'$ all pairs $s_{2} \wbrel s_{1}$ with $s_{2} \in \stateSet_{2}$ and $s_{1} \in \stateSet_{1}$.
In the main loop of $\MinorCostProc$ we repeatedly refine $\costrelord$ by removing all pairs that do not satisfy the conditions of Definition~\ref{def:weakCostBisimMinorCost}: 
If a check fails, we remove the offending pair $(s_{2}, s_{1})$ from $\costrelord'$.

On termination of the loop, $\costrelord$ contains only pairs satisfying Definition~\ref{def:weakCostBisimMinorCost}, so deciding whether $\costAutPed{\costf}{1} \weakCostBisimMinorCost \costAutPed{\costf}{2}$ reduces to check whether $\startState_{2} \costrel \startState_{1}$ and whether for each $s_{2} \in \stateSet_{2}$ there exists $s_{1} \in \stateSet_{1}$ such that $s_{2} \costrel s_{1}$.

Given two \CPA{}s $\costAutPed{\costf}{1}$ and $\costAutPed{\costf}{2}$, let $N = \max \setnocond{\setcardinality{\stateSet_{1} \uplus \stateSet_{2}},\setcardinality{\transitionRelation_{1} \uplus \transitionRelation_{2}}}$.
Computing $\wbrelord = \QuotientProc(\aut_{1},\aut_{2})$ is polynomial in $N$ (cf.~\cite[Theorem~11]{HT12}), say $P(N)$;
in the worst case, that occurs when we remove all pairs from $\costrelord$, the main loop of $\MinorCostProc$ is performed at most $N^{2}$ times;
according to Theorem~\ref{thm:minCostLPequivalentToCostWeakTransitionLifting} and its corollaries, finding $\gd_{2} \in \Disc{\borderStateSetOrd[\wbrelord] \cap \stateSet_{2}}$ such that $\hyperWeakCombinedTransition{\sd_{2}}{\hidden}{\gd_{2}}$ and $\cost[2]{\hyperWeakCombinedTransition{\sd_{2}}{\hidden}{\gd_{2}}} = \min \setcond{\cost[2]{\hyperWeakCombinedTransition{\sd_{2}}{\hidden}{\gd}}}{\gd \in \Disc{\borderStateSet[\wbrelord] \cap \stateSet_{2}}}$ is polynomial in $N$, say $R(N)$, by solving the $\minCostLPproblemTBetaMuRel{\sd_{2}}{\hidden}{\dirac{b_{2}}}{\relord[B]}$ LP problem where $b_{2} \in \borderStateSet[\wbrelord] \cap \stateSet_{2}$ and $\relord[B]$ is the reflexive, symmetric, and transitive closure of $\borderStateSet[\wbrelord]$. 
Similarly, $R(N)$ is also the complexity of either finding $\gd_{1} \in \Disc{\borderStateSetOrd[\wbrelord] \cap \stateSet_{1}}$ such that $\weakCombinedTransition{s_{1}}{a}{\gd_{1}}$, $\gd_{2} \liftrel[\costrelord] \gd_{1}$, and $\cost[1]{\weakCombinedTransition{s_{1}}{a}{\gd_{1}}} \leq \cost[2]{\weakCombinedTransition{\strongTransition{s_{2}}{a}{\sd_{2}}}{\hidden}{\gd_{2}}}$, or finding $\sd_{1} \in \Disc{\stateSet_{1}}$ such that $\weakCombinedTransition{s_{1}}{a}{\sd_{1}}$, $\sd_{2} \liftrel[\costrelord] \sd_{1}$, and $\cost[1]{\weakCombinedTransition{s_{1}}{a}{\sd_{1}}} \leq \cost[2]{\strongTransition{s_{2}}{a}{\sd_{2}}}$.
This implies that the total complexity of $\MinorCostProc$ is $P(N) + N^{2} \cdot 2R(N)$.

\begin{theorem}
	Given two \CPA{}s $\costAutPed{\costf}{1}$ and $\costAutPed{\costf}{2}$, checking $\costAutPed{\costf}{1} \weakCostBisimMinorCost \costAutPed{\costf}{2}$ is polynomial in $N = \max \setnocond{\setcardinality{\stateSet_{1} \uplus \stateSet_{2}}, \setcardinality{\transitionRelation_{1} \uplus \transitionRelation_{2}}}$.
\end{theorem}

\subsubsection{Deciding Weak Probabilistic Cost-Preserving Bisimulation}
\label{ssec:decisionProcedureWPCPB}

Regarding weak probabilistic cost-preserving bisimulation, the algorithm is actually simpler, since in order to check for the existence of weak combined transitions with a given cost $\givenCost$, it is enough to add the new constraint $\sum_{(x,y) \in E} \cost[f]{(x,y)} \cdot f_{x,y} = \givenCost$ to the $\minCostLPproblemTBetaMuRel{t}{a}{\sd}{\relord}$ LP problem.
This allows us to check in polynomial time whether two \CPA{}s are weak probabilistic cost-preserving bisimilar: 
We compute $\QuotientProc$ on the two \CPA{}s where we have replaced in procedure $\FindSplitProc$ the test for feasibility of $\LPproblemTBetaMuRel{t}{a}{\sd}{\relord[W]}$ with the test for feasibility of $\minCostLPproblemTBetaMuRel{t}{a}{\sd}{\relord}$ extended with the constraint $\sum_{(x,y) \in E} \cost[f]{(x,y)} \cdot f_{x,y} = \cost{\strongTransition{s}{a}{\sd_{s}}}$.
\begin{theorem}
	Given two \CPA{}s $\costAutPed{\costf}{1}$ and $\costAutPed{\costf}{2}$, checking $\costAutPed{\costf}{1} \weakCostBisim \costAutPed{\costf}{2}$ is polynomial in $N = \max \setnocond{\setcardinality{\stateSet_{1} \uplus \stateSet_{2}}, \setcardinality{\transitionRelation_{1} \uplus \transitionRelation_{2}}}$.
\end{theorem}

\subsubsection{Deciding Strong Probabilistic Bisimulations on \CPA{}s}
\label{ssec:decisionProcedureSPB}

We now consider the decision algorithm for both minor cost strong probabilistic bisimulation and strong probabilistic cost-preserving bisimulation for the two \CPA{}s $\costAutPed{\costf}{1}$ and $\costAutPed{\costf}{2}$; 
by definition, the only difference between the two bisimulations is that the former requires that one cost is at most the other, while the latter requires that the two costs are the same.  
The remaining requirements are precisely those of strong probabilistic bisimulation on \PA{}s, so we use its decision procedure as base: 
The main procedure is again $\QuotientProc$ where we have replaced in procedure $\FindSplitProc$ the test for feasibility of $\LPproblemTBetaMuRel{t}{a}{\sd}{\relord[W]}$ with the test for feasibility of the following LP problem~\cite{Zha08}: 
Let $\stateSet$ be the disjoint union of the states of the two \PA{}s; $\strongTransition{t}{a}{\sd_{1}}$, $\strongTransition{t}{a}{\sd_{2}}$, \dots, $\strongTransition{t}{a}{\sd_{k}}$ be an enumeration of the transitions enabled by $t$ with label $a$ (we assume that $k > 0$, otherwise the test trivially fails); 
$\relord[W](u) = \setcond{u \in \stateSet}{u \rel[W] v}$; 
and $\relord[W]^{-1}(v) = \setcond{u \in \stateSet}{u \rel[W] v}$.
\[
	\begin{array}{ll}
		\sum_{i = 1}^{k} p_{i} = 1 \\
		0 \leq p_{i} \leq 1 & \text{for each $i \in \setnocond{1, \dots, k}$}\\
		0 \leq f_{u,v} \leq 1 & \text{for each $(u,v) \in \relord[W]$}\\
		\probeval{\sd}{u} = \sum_{v \in \relord[W](u)} f_{u,v} & \text{for each $u \in \stateSet$} \\
		\sum_{u \in \relord[W]^{-1}(v)} f_{u,v} = \sum_{i = 1}^{k} p_{i} \cdot \probeval{\sd_{i}}{v} & \text{for each $t \in \stateSet$}
	\end{array}
\]
By~\cite[Lemma 5.3.1]{Zha08}, we have that the above LP problem is feasible if and only if $t$ enables a strong combined transition $\strongCombinedTransition{t}{a}{\sd_{t}}$ such that $\sd \liftrel[{\relord[W]}] \sd_{t}$;
such strong combined transition $\strongCombinedTransition{t}{a}{\sd_{t}}$ is actually obtained by combining each transition $\strongTransition{t}{a}{\sd_{i}}$ with weight $p_{i}$ if $p_{i} > 0$.
It is immediate to see that the size of the above LP problem is polynomial in $N = \max \setnocond{\setcardinality{\stateSet_{1} \uplus \stateSet_{2}}, \setcardinality{\transitionRelation_{1} \uplus \transitionRelation_{2}}}$, so it can be solved in polynomial time.

Extending this approach to the cost bisimulations we have presented is now trivial: 
it is enough to add the constraint $\sum_{i = i}^{k} p_{i} \cdot \cost[d]{\strongTransition{t}{a}{\sd_{i}}} = \cost[c]{\strongTransition{s}{a}{\sd}}$ when we decide the strong probabilistic cost-preserving bisimulation, or the constraint $\sum_{i = i}^{k} p_{i} \cdot \cost[d]{\strongTransition{t}{a}{\sd_{i}}} \leq \cost[c]{\strongTransition{s}{a}{\sd}}$ when we decide the minor cost strong probabilistic bisimulation.
Obviously these two constraints do not change the complexity class of the LP problem, so we can check in polynomial time the strong probabilistic bisimulations on \CPA{}s.

\begin{theorem}
	Given two \CPA{}s $\costAutPed{\costf}{1}$ and $\costAutPed{\costf}{2}$, checking $\costAutPed{\costf}{1} \strongCostProbBisim \costAutPed{\costf}{2}$ and $\costAutPed{\costf}{1} \strongCostProbBisimMinorCost \costAutPed{\costf}{2}$ is polynomial in $N = \max \setnocond{\setcardinality{\stateSet_{1} \uplus \stateSet_{2}}, \setcardinality{\transitionRelation_{1} \uplus \transitionRelation_{2}}}$.
\end{theorem}

\subsubsection{Deciding Strong Bisimulations on \CPA{}s}
\label{ssec:decisionProcedureSB}

The last decision algorithm we propose allows us to check the minor cost strong bisimulation and the strong cost-preserving bisimulation for the two \CPA{}s $\costAutPed{\costf}{1}$ and $\costAutPed{\costf}{2}$.  
As for the probabilistic case, the only difference between the two bisimulations is that the former asks that one cost is at most the other, while the latter asks them to be the same.  
The remaining requirements again coincide with strong bisimulation on \PA{}s. 
This makes the algorithm straightforward, the main procedure is, as before, $\QuotientProc$ where we replace in procedure $\FindSplitProc$ the test for feasibility of $\LPproblemTBetaMuRel{t}{a}{\sd}{\relord[W]}$ with a test iterating over all transitions enabled by $t$ with label $a$; 
if among them we find a transition $\strongTransition{t}{a}{\sd_{t}}$ such that $\sd \liftrel[{\relord[W]}] \sd_{t}$, then the test succeeds, otherwise it fails.  
Since deciding $\sd \liftrel[{\relord[W]}] \sd_{t}$ is equivalent to solving a network flow problem~\cite{BEMC00} whose number of arcs is quadratic in $\setcardinality{\stateSet_{1} \uplus \stateSet_{2}}$, this iteration has a complexity that is in $\bigO{\setcardinality{\transitionRelation_{t}} \cdot \setcardinality{\stateSet_{1} \uplus \stateSet_{2}}}$ (where $\transitionRelation_{t}$ is the set of transitions with label $a$ enabled by $t$). 
The cost condition can be incorporated in the obvious and straightforward manner: 
Before verifying $\sd \liftrel[{\relord[W]}] \sd_{t}$ for the current $\strongTransition{t}{a}{\sd_{t}}$, we check that $\cost[d]{\strongTransition{t}{a}{\sd_{t}}} \leq \cost[c]{\strongTransition{s}{a}{\sd}}$ provided we are computing the minor cost strong bisimulation, or that $\cost[d]{\strongTransition{t}{a}{\sd_{t}}} = \cost[c]{\strongTransition{s}{a}{\sd}}$ in case we are computing the strong cost-preserving bisimulation.  
This additional check does not increase the overall complexity, so also deciding the strong bisimulations on \CPA{}s can be done in polynomial time.

\begin{theorem}
	Given two \CPA{}s $\costAutPed{\costf}{1}$ and $\costAutPed{\costf}{2}$, checking $\costAutPed{\costf}{1} \strongCostBisim \costAutPed{\costf}{2}$ and $\costAutPed{\costf}{1} \strongCostBisimMinorCost \costAutPed{\costf}{2}$ is polynomial in $N = \max \setnocond{\setcardinality{\stateSet_{1} \uplus \stateSet_{2}}, \setcardinality{\transitionRelation_{1} \uplus \transitionRelation_{2}}}$.
\end{theorem}

In summary we have devised polynomial time algorithms for all the cost related relations presented in this paper. 
For the strong relations, these are essentially echoing the strong (probabilistic) bisimulation decision algorithms~\cite{Zha08,BEMC00} for \PA{}s, though presented in our setting. 
The central innovation and contribution of our algorithmic work arguably lies in the treatment of weak transition costs, embodied in weak cost-preserving probabilistic bisimulation and minor cost weak probabilistic bisimulation.

\section{Discussion}
\label{sec:discussion}

This section puts our work in the greater context of related work and also discusses several other options to follow.

Givan, Dean and Greig~\cite{GDG03} have introduced the idea of strong bisimilarity for \MDP{}s with state and transition costs, together with algorithms for minimisation to the quotient model. 
The minimisation with respect to weak probabilistic bisimulation on \PA{} has lately been discussed~\cite{EHSTZ13}, and it remains to be investigated how the minimisation can be applied for the minor cost approach meaningfully. 
For the cost-preserving bisimilarity, the adaptations are straightforward, so we can indeed minimise with respect to weak transition costs.

In the present work we focus our attention on the minimisation of the cost, which is quite natural. 
The converse, i.e., maximisation of the cost, is hardly definable whenever the model contains cycles possibly occurring with probability $1$. 
In such cases, the maximum cost is infinite, obtained by performing such cycles forever. 
Fairness arguments might be added to enforce to eventually leave such cycles almost surely, but still do not help since it does not change the cost suprema. 
This seems to indicate that we need to intertwine arguments about almost sure cycle termination with finite expectation in order to overcome this problem.

It is  also interesting to discuss how state costs can be handled. 
Indeed it is possible to turn state costs to transition costs by moving them on incoming or outgoing transitions. 
The concrete choice makes a difference, because the labels of incoming and outgoing transitions generally differ.  
If already transitions costs were present prior to the move, we end up with a second cost structure. 
Multiple cost structures can indeed also be integrated into our setting rather easily, one just needs to take the minor cost for all structures in the decision problem.  

For \MDP{}s, multiple reward structures have been investigated~\cite{EKVY08} in the context of model checking, and our approach naturally combines with that. 
Chatterjee, Majumdar, and Henzinger~\cite{CMH06} investigated them in a setting with discounting.  
In fact, our polynomial time LP approach can be extended to compute the minimum cost of discounted weak combined transitions, if we can assume a polynomially bounded number of internal steps. 
Conversely, one can compute an upper bound on discounted but non-polynomially bounded weak combined transitions in polynomial time.

If discounting is integrated into the weak bisimulation definitions we propose, this however induces difficult-to-grasp equalities. 
This is because sequences of internal transitions of different length are abstracted away by weak bisimilarity, but they would imply different discounts.  
For similar reasons, our cost model does by itself not talk about traces.  
As long as internal transitions carry nonzero costs, the definition of the cost of a weak trace is not obvious.
Even if two executions fragments have the same trace, i.e., the same sequence of visible actions, different execution fragments usually have different costs when they involve different internal transitions, in particular after the last external action of the trace.  
Moreover, even if the execution fragment does not involve internal transitions, it can have different costs as resulting by the resolution of probabilistic and nondeterministic choices, the latter performed by the scheduler.

Still, cost-preserving bisimilarity implies equal trace costs, and if $\costAutPed{\costf}{1}$ is in minor cost weak probabilistic bisimilarity with $\costAutPed{\costf}{2}$, then the trace costs of $\costAutPed{\costf}{1}$ are bounded from above by $\costAutPed{\costf}{2}$. 
Trace costs appear central in many cost related formalisms not involving probabilities, such as weighted timed and energy automata~\cite{Qua09,BFLMS08}, though without (internal) actions playing a dedicated role here, so it is worth to investigate trace costs in the \CPA{} model as well. 

While minor cost weak bisimilarity is implicitly asymmetric, we have still formulated it as an equivalence relation. 
The wireless communication channel example has demonstrated that this approach is undoubtedly useful. 
Yet, it seems worthwhile to also take inspiration from simulation and simulation distance approaches~\cite{AK12,CHR12} in this matter. 
Another avenue that is worth exploring is to consider branching instead of weak bisimulation~\cite{vGW96}, not only because our notion of border states fits particularly well to that setting~\cite{FPP06}. 
We conjecture that requiring the underlying weak bisimulations to be branching is enough to achieve the corresponding transitivity and compositional properties, as well as polynomial decision algorithms.

\section{Concluding Remarks}
\label{sec:conclusion}

In this paper we have presented an extension of Probabilistic Automata to Cost Probabilistic Automata and we have proposed cost related strong and weak probabilistic bisimulations on these models. 
We have distinguished cost-preserving and cost-bounding variations. 
The latter is based on the idea that the defender matches a transition with a cost that is bounded by at most the cost of the challenger. 
We have exemplified the use of this idea with a power sensitive wireless sensing example.

The bisimilarities introduced are precongruences, respectively congruences, with respect to parallel composition of \CPA{}s, which sets the foundational ground for compositional construction and minimization approaches of \CPA{}s. 
Especially in the minor cost weak setting the proofs to establish transitivity and precongruence are all but straightforward.

Moreover we have shown how to compute in polynomial time the minimum cost for each transition, and hence to decide each of the relations. 
With this, it is practically possible to roll out compositional construction and minimisation techniques to operations research, automated planning, and decision support applications.  
This is because, the \CPA{} model encompasses \MDP{}s, so the results apply readily to these models as well.

\noindent{\textbf{Acknowledgements.}} 
This work is supported by the DFG/NWO bilateral research programme ROCKS, by the DFG as part of the SFB/TR~14~AVACS, by the EU FP7 Programme under grant agreement no.\@ 295261 (MEALS) and 318490 (SENSATION), and by the CAS/SAFEA International Partnership Program for Creative Research Teams. 
Part of this work has been done while Andrea Turrini was at Saarland University supported by the Cluster of Excellence ``Multimodal Computing and Interaction'' (MMCI), part of the German Excellence Initiative.

\newpage 

\appendix

\section{Proofs of the Results Enunciated in the Paper}
\label{app:proofs}

In this appendix we provide the detailed non-trivial proofs for the results enunciated in the main part of the paper.

\begin{result}[Proposition~\ref{pro:costAsExpectedRewardInMDP}]
\label{res:costAsExpectedRewardInMDP}
	Given an \MDP{} $\aut[M] = (\stateSet, \startState, \actionSet, \transitionProbability)$ and a policy $\policy$, 
	let $\aut$ be the \PA{} $(\stateSet, \startState, \actionSet, \transitionRelation)$ where $\transitionRelation = \setcond{(s, a, \probeval{\transitionProbability}{s,a})}{s \in \stateSet, a \in \actionSet(s)}$. 
	For each $N \in \nat$, $\tr = (s, a, \sd) \in \transitionRelation$ and $\alpha \in \finiteFrags{\aut[M]}$, let $\cost{\tr} = \functioneval{\rewardFunction}{s,a}$, 
	$\probeval{\schedeval{\sched}{\alpha}}{\tr} =\probeval{\policyeval{\policy}{\alpha}}{a}$ if $\length{\alpha} < N$, $0$ otherwise, 
	and $\weakCombinedTransition{\startState}{\hidden}{\sd}$ be the weak combined transition of $\aut$ induced by the scheduler $\sched$ when all actions are considered as internal. 
	Then, it holds that for each  $N \in \nat$, 
	\[
		\expectedReward[N]{\policy} = \cost[\sched]{\weakCombinedTransition{\startState}{\hidden}{\sd}}\text{.}
	\]
\end{result}
\begin{myproof}
	Given $\alpha = s_{1} a_{1} s_{2} a_{2} \dots$, let $\finiteFrags{\aut[M],N}$ be $\setcond{\alpha \in \finiteFrags{\aut[M]}}{\length{\alpha} = N}$, $\alpha^{i}$ be the action $a_{i}$, and $\alpha_{i}$ be the state $s_{i}$.
	\begin{align*}
		\expectedReward[N]{\policy} 
		& {} = \sum_{\alpha \in \finiteFrags{\aut[M],N}} \functioneval{\rewardFunction}{\alpha} \cdot \probPolicyFrag{\alpha} \\
		& {} = \sum_{\alpha \in \finiteFrags{\aut[M],N}} \left( \sum_{i=1}^{N} \functioneval{\rewardFunction}{\alpha_{i},\alpha^{i}} \right) \cdot \left( \probeval{\dirac{\startState}}{\alpha_{1}} \cdot \prod_{i=1}^{N} \probeval{\policyeval{\policy}{\fragRestriction{\alpha}{i}}}{\alpha^{i}} \cdot \probeval{\probeval{\transitionProbability}{\alpha_{i},\alpha^{i}}}{\alpha_{i+1}} \right) \\
		& {} = \sum_{\alpha \in \finiteFrags{\aut[M],N}} \probeval{\dirac{\startState}}{\alpha_{1}} \cdot \left( \sum_{i=1}^{N} \functioneval{\rewardFunction}{\alpha_{i},\alpha^{i}} \right) \cdot \prod_{i=1}^{N} \probeval{\policyeval{\policy}{\fragRestriction{\alpha}{i}}}{\alpha^{i}} \cdot \probeval{\probeval{\transitionProbability}{\alpha_{i},\alpha^{i}}}{\alpha_{i+1}} \\
		& {} = \sum_{s \in \stateSet} \sum_{\alpha \in \setcond{\alpha \in \finiteFrags{\aut[M],N}}{\alpha_{1} = s}} \probeval{\dirac{\startState}}{s} \cdot \left( \sum_{i=1}^{N} \functioneval{\rewardFunction}{\alpha_{i},\alpha^{i}} \right) \cdot \prod_{i=1}^{N} \probeval{\policyeval{\policy}{\fragRestriction{\alpha}{i}}}{\alpha^{i}} \cdot \probeval{\probeval{\transitionProbability}{\alpha_{i},\alpha^{i}}}{\alpha_{i+1}} \\
		& {} = \sum_{s \in \stateSet} \probeval{\dirac{\startState}}{s} \cdot \sum_{\alpha \in \setcond{\alpha \in \finiteFrags{\aut[M],N}}{\alpha_{1} = s}} \left( \sum_{i=1}^{N} \functioneval{\rewardFunction}{\alpha_{i},\alpha^{i}} \right) \cdot \prod_{i=1}^{N} \probeval{\policyeval{\policy}{\fragRestriction{\alpha}{i}}}{\alpha^{i}} \cdot \probeval{\probeval{\transitionProbability}{\alpha_{i},\alpha^{i}}}{\alpha_{i+1}} \\
		& {} = \sum_{\alpha \in \setcond{\alpha \in \finiteFrags{\aut[M],N}}{\alpha_{1} = \startState}} \left( \sum_{i=1}^{N} \functioneval{\rewardFunction}{\alpha_{i},\alpha^{i}} \right) \cdot \prod_{i=1}^{N} \probeval{\schedeval{\sched}{\fragRestriction{\alpha}{i}}}{\tr_{i}} \cdot \probeval{\sd_{\tr_{i}}}{\alpha_{i+1}} \\
		\intertext{where $\tr_{i} = (\alpha_{i},\alpha^{i},\probeval{\transitionProbability}{\alpha_{i},\alpha^{i}})$}
		& {} = \sum_{\alpha \in \setcond{\alpha \in \finiteFrags{\aut[M],N}}{\alpha_{1} = \startState}} \left( \sum_{i=1}^{N} \cost{\tr_{i}} \right) \cdot \prod_{i=1}^{N} \probeval{\schedeval{\sched}{\fragRestriction{\alpha}{i}}}{\tr_{i}} \cdot \probeval{\sd_{\tr_{i}}}{\alpha_{i+1}} \\
		& \stackrel{\dag}{=} \sum_{\alpha \in \setcond{\alpha \in \finiteFrags{\aut[M],N}}{\alpha_{1} = \startState}} \cost[\sched]{\alpha} \cdot \prod_{i=1}^{N} \probeval{\schedeval{\sched}{\fragRestriction{\alpha}{i}}}{\tr_{i}} \cdot \probeval{\sd_{\tr_{i}}}{\alpha_{i+1}} \\
		& \stackrel{\ddag}{=} \sum_{\alpha \in \setcond{\alpha \in \finiteFrags{\aut[M],N}}{\alpha_{1} = \startState}} \cost[\sched]{\alpha} \cdot \probeval{\sd_{\sched,\startState}}{\alpha} \\
		& \stackrel{\S}{=} \sum_{\alpha \in \setcond{\alpha \in \finiteFrags{\aut[M],N}}{\alpha_{1} = \startState}} \cost[\sched]{\alpha} \cdot \probeval{\sd_{\sched,\startState}}{\alpha} + \sum_{\alpha \in \setcond{\alpha \in \finiteFrags{\aut[M]}}{\length{\alpha} \neq N \land \alpha_{1} = \startState}} \cost[\sched]{\alpha} \cdot \probeval{\sd_{\sched,\startState}}{\alpha}\\
		& {} = \sum_{\alpha \in \setcond{\alpha \in \finiteFrags{\aut[M]}}{\alpha_{1} = \startState}} \cost[\sched]{\alpha} \cdot \probeval{\sd_{\sched,\startState}}{\alpha}\\
		& {} = \sum_{\alpha \in \setcond{\alpha \in \finiteFrags{\aut[M]}}{\alpha_{1} = \startState}} \cost[\sched]{\alpha} \cdot \probeval{\sd_{\sched,\startState}}{\alpha} + \sum_{\alpha \in \setcond{\alpha \in \finiteFrags{\aut[M]}}{\alpha_{1} \neq \startState}} \cost[\sched]{\alpha} \cdot \probeval{\sd_{\sched,\startState}}{\alpha}\\
		& {} = \sum_{\alpha \in \finiteFrags{\aut[M]}} \cost[\sched]{\alpha} \cdot \probeval{\sd_{\sched,\startState}}{\alpha}\\
		& {} = \cost[\sched]{\weakCombinedTransition{s}{\hidden}{\sd}}\text{,}
	\end{align*}
	where equalities marked by $\dag$, $\ddag$, and $\S$ are justified as follows:
	\begin{itemize}
		\item \textbf{Equality $\dag$:}
			we prove that $\probeval{\sd_{\sched,\startState}}{\cone{\alpha}} = \prod_{i=1}^{n} \probeval{\schedeval{\sched}{\fragRestriction{\alpha}{i}}}{\tr_{i}} \cdot \probeval{\sd_{\tr_{i}}}{\alpha_{i+1}}$ by induction on the length $n \leq N$ of $\alpha$.
			If $n = 0$, then $\probeval{\sd_{\sched,\startState}}{\cone{\alpha}} = \probeval{\sd_{\sched,\startState}}{\cone{s}} = 1 = \prod_{i=1}^{0} \probeval{\schedeval{\sched}{\fragRestriction{\alpha}{i}}}{\tr_{i}} \cdot \probeval{\sd_{\tr_{i}}}{\alpha_{i+1}} = \prod_{i=1}^{n} \probeval{\schedeval{\sched}{\fragRestriction{\alpha}{i}}}{\tr_{i}} \cdot \probeval{\sd_{\tr_{i}}}{\alpha_{i+1}}$;
			if $n > 0$, then $\alpha = \alpha'at$ for some action $a$ and state $t$, therefore
			\begin{align*}
				\probeval{\sd_{\sched,\startState}}{\cone{\alpha}}
				& {} = \probeval{\sd_{\sched,\startState}}{\cone{\alpha'at}} \\
				& {} = \probeval{\sd_{\sched,\startState}}{\cone{\alpha'}} \cdot \sum_{\tr \in \transitionsWithLabel{a}} \probeval{\schedeval{\sched}{\alpha'}}{\tr} \cdot  \probeval{\sd_{\tr}}{t}\\
				& {} = \left( \prod_{i=1}^{n-1} \probeval{\schedeval{\sched}{\fragRestriction{\alpha}{i}}}{\tr_{i}} \cdot \probeval{\sd_{\tr_{i}}}{\alpha_{i+1}} \right) \cdot \sum_{\tr \in \transitionsWithLabel{a}} \probeval{\schedeval{\sched}{\alpha'}}{\tr} \cdot  \probeval{\sd_{\tr}}{t}\\
				\intertext{by inductive hypothesis}
				& {} = \left( \prod_{i=1}^{n-1} \probeval{\schedeval{\sched}{\fragRestriction{\alpha}{i}}}{\tr_{i}} \cdot \probeval{\sd_{\tr_{i}}}{\alpha_{i+1}} \right) \cdot \probeval{\schedeval{\sched}{\fragRestriction{\alpha}{n}}}{\tr_{n}} \cdot \probeval{\sd_{\tr_{n}}}{t}\\
				\intertext{where $\tr_{n} = (\fragRestriction{\alpha'}{n},a,\probeval{\transitionProbability}{\fragRestriction{\alpha'}{n},a})$}
				& {} = \left( \prod_{i=1}^{n-1} \probeval{\schedeval{\sched}{\fragRestriction{\alpha}{i}}}{\tr_{i}} \cdot \probeval{\sd_{\tr_{i}}}{\alpha_{i+1}} \right) \cdot \probeval{\schedeval{\sched}{\fragRestriction{\alpha}{n}}}{\tr_{n}} \cdot \probeval{\sd_{\tr_{n}}}{\alpha_{n+1}}\\
				& {} = \prod_{i=1}^{n} \probeval{\schedeval{\sched}{\fragRestriction{\alpha}{i}}}{\tr_{i}} \cdot \probeval{\sd_{\tr_{i}}}{\alpha_{i+1}}
			\end{align*}
			where $\sum_{\tr \in \transitionsWithLabel{a}} \probeval{\schedeval{\sched}{\alpha'}}{\tr} \cdot  \probeval{\sd_{\tr}}{t}$ reduces to $\probeval{\schedeval{\sched}{\fragRestriction{\alpha}{n}}}{\tr_{n}} \cdot \probeval{\sd_{\tr_{n}}}{t}$ with $\fragRestriction{\alpha}{n} = \alpha'$since by definition of \MDP{}, there is only the transition $\tr_{n}$ with label $a$ from state $\fragRestriction{\alpha}{n}$, thus for each other $\tr \in \transitionsWithLabel{a}$, $\probeval{\schedeval{\sched}{\alpha'}}{\tr} = 0$.
		\item \textbf{Equality $\ddag$:}
			we assume, without loss of generality, that $\probeval{\sd_{\sched,\startState}}{\cone{\alpha}} > 0$ and we prove that $\cost[\sched]{\alpha} = \sum_{i=1}^{N} \cost{\tr_{i}}$ by induction on the length $N$ of $\alpha$.
			If $N = 0$, then $\cost[\sched]{\alpha} = 0 = \sum_{i=1}^{0} \cost{\tr_{i}} = \sum_{i=1}^{N} \cost{\tr_{i}}$;
			if $N > 0$, then $\alpha = \alpha'at$ for some action $a$ and state $t$, therefore
			\begin{align*}
				\cost[\sched]{\alpha}
				& {} = \cost[\sched]{\alpha'} + \sum_{\tr_{N} \in \transitionsWithLabel{a}} \cost{\tr_{N}} \cdot \schedeval{\widehat{\sched}}{\alpha', t, a, \tr_{N}}\\
				& {} = \left( \sum_{i=1}^{N-1} \cost{\tr_{i}} \right) + \sum_{\tr \in \transitionsWithLabel{a}} \cost{\tr} \cdot \schedeval{\widehat{\sched}}{\alpha', t, a, \tr}\\
				\intertext{by inductive hypothesis}
				& {} = \sum_{i=1}^{N-1} \cost{\tr_{i}} + \cost{\tr_{N}} \cdot 1 \\
				\intertext{where $\tr_{N} = (\fragRestriction{\alpha'}{N},a,\probeval{\transitionProbability}{\fragRestriction{\alpha'}{N},a})$}
				& {} = \sum_{i=1}^{N} \cost{\tr_{i}}
			\end{align*}
			since $\schedeval{\widehat{\sched}}{\alpha', t, a, \tr}$ is $0$ for each $\tr \in \transitionsWithLabel{a}$ that is different from $\tr_{N}$, as $\source{\tr} \neq \fragRestriction{\alpha'}{N}$, thus $\probeval{\schedeval{\sched}{\alpha'}}{\tr} = 0$. 
			When we consider the transition $\tr_{N}$, we
                        have
                        that \[\schedeval{\widehat{\sched}}{\alpha',
                          t, a, \tr_{N}} =
                        \dfrac{\probeval{\schedeval{\sched}{\alpha'}}{\tr_{N}}
                          \cdot \probeval{\sd_{\tr_{N}}}{t}}{\sum_{\tr
                            \in \transitionsWithLabel{a}}
                          \probeval{\schedeval{\sched}{\alpha'}}{\tr}
                          \cdot \probeval{\sd_{\tr}}{t}} =
                        \dfrac{\probeval{\schedeval{\sched}{\alpha'}}{\tr_{N}}
                          \cdot
                          \probeval{\sd_{\tr_{N}}}{t}}{\probeval{\schedeval{\sched}{\alpha'}}{\tr_{N}}
                          \cdot \probeval{\sd_{\tr_{N}}}{t}} = 1\text{,}
\]
 thus $\sum_{\tr \in \transitionsWithLabel{a}} \cost{\tr} \cdot \schedeval{\widehat{\sched}}{\alpha', t, a, \tr} = \cost{\tr_{N}}$ holds.
			Note that $\probeval{\sd_{\tr_{N}}}{t} = \probeval{\probeval{\transitionProbability}{\fragRestriction{\alpha'}{N},a}}{t} > 0$ follows by definition of finite execution fragment for $\alpha = \alpha'at$.
		\item \textbf{Equality $\S$:} 
			when $\length{\alpha} < N$, by definition of $\sched$ it follows that $\probeval{\schedeval{\sched}{\alpha}}{\bot} = 0$, hence $\probeval{\sd_{\sched,\startState}}{\alpha} = \probeval{\sd_{\sched,\startState}}{\cone{\alpha}} \cdot \probeval{\schedeval{\sched}{\alpha}}{\bot} = 0$.
			When $\length{\alpha} > N$, by definition of $\sched$ it follows that $\probeval{\schedeval{\sched}{\fragRestriction{\alpha}{N}}}{\bot} = 1$, thus $\probeval{\sd_{\sched,\startState}}{\cone{\alpha}} = 0$, hence $\probeval{\sd_{\sched,\startState}}{\alpha} = \probeval{\sd_{\sched,\startState}}{\cone{\alpha}} \cdot \probeval{\schedeval{\sched}{\alpha}}{\bot} = 0$.
	\end{itemize}
	This concludes the proof that $\expectedReward[N]{\policy} = \cost[\sched]{\weakCombinedTransition{s}{\hidden}{\sd}}$.
\end{myproof}

\begin{result}[Proposition~\ref{pro:equivalenceOfCostDefinitions}]
\label{res:equivalenceOfCostDefinitions}
	Given a \CPA{} $\costAut{\costf}$, a state $s$, an action $a$, a probability distribution $\sd$, and a scheduler $\sched$ inducing the weak combined transition $\weakCombinedTransition{s}{a}{\sd}$, it holds that 
	\[
		\costf^{\mathit{Ray}}_{\sched}(\weakCombinedTransition{s}{a}{\sd}) = \costf^{\mathit{Ball}}_{\sched}(\weakCombinedTransition{s}{a}{\sd})
	\]
	where costs $\costf^{\mathit{Ray}}_{\sched}$ and $\costf^{\mathit{Ball}}_{\sched}$ are defined according to Definition~\ref{def:costAsUsualDefinition} and~\ref{def:costAsWorkingForTheProof}, respectively.
\end{result}
\begin{myproof}
	As preliminary result, we show by induction that for each execution fragment $\alpha \in \finiteFrags{\aut}$, if $\probeval{\sd_{\sched,s}}{\cone{\alpha}} > 0$, then $\costf^{\mathit{Ray}}_{\sched}(\alpha) = \sum_{\alpha' \prefix \alpha} \dfrac{\costf^{\mathit{Ball}}_{\sched,s}(\alpha')}{\probeval{\sd_{\sched,s}}{\cone{\alpha'}}}$.
	Note that $\probeval{\sd_{\sched,s}}{\cone{\alpha}} > 0$ implies that for each execution fragment $\alpha'$ such that $\alpha' \prefix \alpha$, $\probeval{\sd_{\sched,s}}{\cone{\alpha'}} > 0$ as well.
	
	Fix a generic execution fragment $\alpha\in \finiteFrags{\aut}$ such that $\probeval{\sd_{\sched,s}}{\cone{\alpha}} > 0$; 
	if $\length{\alpha} = 0$, then $\costf^{\mathit{Ray}}_{\sched}(\alpha) = 0 = \costf^{\mathit{Ball}}_{\sched,s}(\alpha) = \dfrac{\costf^{\mathit{Ball}}_{\sched,s}(\alpha)}{\probeval{\sd_{\sched,s}}{\cone{\alpha}}} = \sum_{\phi \prefix \alpha} \dfrac{\costf^{\mathit{Ball}}_{\sched,s}(\phi)}{\probeval{\sd_{\sched,s}}{\cone{\phi}}}$.
	
	Now, suppose that $\length{\alpha} > 0$;
	this implies that there exists $\alpha'$, $a$, $t$ such that $\alpha = \alpha' a t$.
	Thus, 
	\begin{align*}
		\costf^{\mathit{Ray}}_{\sched}(\alpha) & {} = \costf^{\mathit{Ray}}_{\sched}(\alpha') + \sum_{\tr \in \transitionsWithLabel{a}} \cost{\tr} \cdot \schedeval{\widehat{\sched}}{\alpha', t, a, \tr} \\
		& {} = \costf^{\mathit{Ray}}_{\sched}(\alpha') + \sum_{\tr \in \transitionsWithLabel{a}} \cost{\tr} \cdot \dfrac{\probeval{\schedeval{\sched}{\alpha'}}{\tr} \cdot \probeval{\sd_{\tr}}{t}}{\sum_{\tr' \in \transitionsWithLabel{a}} \probeval{\schedeval{\sched}{\alpha'}}{\tr'} \cdot \probeval{\sd_{\tr'}}{t}} \\
		& {} = \costf^{\mathit{Ray}}_{\sched}(\alpha') + \sum_{\tr \in \transitionsWithLabel{a}} \cost{\tr} \cdot \dfrac{\probeval{\schedeval{\sched}{\alpha'}}{\tr} \cdot \probeval{\sd_{\tr}}{t}}{\sum_{\tr' \in \transitionsWithLabel{a}} \probeval{\schedeval{\sched}{\alpha'}}{\tr'} \cdot \probeval{\sd_{\tr'}}{t}} \\
		& {} = \costf^{\mathit{Ray}}_{\sched}(\alpha') + \sum_{\tr \in \transitionsWithLabel{a}} \dfrac{\probeval{\sd_{\sched,s}}{\cone{\alpha'}}}{\probeval{\sd_{\sched,s}}{\cone{\alpha'}}} \cdot \cost{\tr} \cdot \dfrac{\probeval{\schedeval{\sched}{\alpha'}}{\tr} \cdot \probeval{\sd_{\tr}}{t}}{\sum_{\tr' \in \transitionsWithLabel{a}} \probeval{\schedeval{\sched}{\alpha'}}{\tr'} \cdot \probeval{\sd_{\tr'}}{t}} \\
		& {} = \costf^{\mathit{Ray}}_{\sched}(\alpha') + \probeval{\sd_{\sched,s}}{\cone{\alpha'}} \cdot \sum_{\tr \in \transitionsWithLabel{a}} \dfrac{\cost{\tr} \cdot \probeval{\schedeval{\sched}{\alpha'}}{\tr} \cdot \probeval{\sd_{\tr}}{t}}{\probeval{\sd_{\sched,s}}{\cone{\alpha'}} \cdot \sum_{\tr' \in \transitionsWithLabel{a}} \probeval{\schedeval{\sched}{\alpha'}}{\tr'} \cdot \probeval{\sd_{\tr'}}{t}} \\
		& {} = \costf^{\mathit{Ray}}_{\sched}(\alpha') + \dfrac{\probeval{\sd_{\sched,s}}{\cone{\alpha'}} \cdot \sum_{\tr \in \transitionsWithLabel{a}} \cost{\tr} \cdot \probeval{\schedeval{\sched}{\alpha'}}{\tr} \cdot \probeval{\sd_{\tr}}{t}}{\probeval{\sd_{\sched,s}}{\cone{\alpha'}} \cdot \sum_{\tr' \in \transitionsWithLabel{a}} \probeval{\schedeval{\sched}{\alpha'}}{\tr'} \cdot \probeval{\sd_{\tr'}}{t}} \\
		& {} = \costf^{\mathit{Ray}}_{\sched}(\alpha') + \dfrac{\costf^{\mathit{Ball}}_{\sched,s}(\alpha)}{\probeval{\sd_{\sched,s}}{\cone{\alpha}}} \\
		& {} = \dfrac{\costf^{\mathit{Ball}}_{\sched,s}(\alpha)}{\probeval{\sd_{\sched,s}}{\cone{\alpha}}} + \sum_{\phi \prefix \alpha'} \dfrac{\costf^{\mathit{Ball}}_{\sched,s}(\phi)}{\probeval{\sd_{\sched,s}}{\cone{\phi}}}\\
		& {} = \sum_{\phi \prefix \alpha} \dfrac{\costf^{\mathit{Ball}}_{\sched,s}(\phi)}{\probeval{\sd_{\sched,s}}{\cone{\phi}}}\text{.}
	\end{align*}
	It is interesting to note that whenever $\probeval{\sd_{\sched,s}}{\cone{\alpha}} = 0$, then $\costf^{\mathit{Ball}}_{\sched,s}(\alpha) = 0$ by definition as well as $\costf^{\mathit{Ray}}_{\sched}(\alpha) \cdot \probeval{\sd_{\sched,s}}{\cone{\alpha}}$.
	
	Now we are ready to prove the statement of the proposition:
	\begin{align*}
		\costf^{\mathit{Ray}}_{\sched}(\weakCombinedTransition{s}{a}{\sd}) & {} = \sum_{\alpha \in \finiteFrags{\aut}} \costf^{\mathit{Ray}}_{\sched}(\alpha) \cdot \probeval{\sd_{\sched,s}}{\alpha} \\
		& {} = \sum_{\alpha \in \finiteFrags{\aut}} \probeval{\sd_{\sched,s}}{\cone{\alpha}} \cdot \probeval{\schedeval{\sched}{\alpha}}{\bot} \cdot \costf^{\mathit{Ray}}_{\sched}(\alpha) \\
		& {} = \sum_{\alpha \in \finiteFrags{\aut}} \probeval{\sd_{\sched,s}}{\cone{\alpha}} \cdot \probeval{\schedeval{\sched}{\alpha}}{\bot} \cdot \sum_{\phi \prefix \alpha} \dfrac{\costf^{\mathit{Ball}}_{\sched,s}(\phi)}{\probeval{\sd_{\sched,s}}{\cone{\phi}}} \\
		& {} = \sum_{\alpha \in \finiteFrags{\aut}} \sum_{\phi \prefix \alpha} \dfrac{\costf^{\mathit{Ball}}_{\sched,s}(\phi)}{\probeval{\sd_{\sched,s}}{\cone{\phi}}} \cdot \probeval{\sd_{\sched,s}}{\cone{\alpha}} \cdot \probeval{\schedeval{\sched}{\alpha}}{\bot} \\
		& {} = \sum_{\phi \in \finiteFrags{\aut}} \sum_{\alpha \in \cone{\phi}} \dfrac{\costf^{\mathit{Ball}}_{\sched,s}(\phi)}{\probeval{\sd_{\sched,s}}{\cone{\phi}}} \cdot \probeval{\sd_{\sched,s}}{\cone{\alpha}} \cdot \probeval{\schedeval{\sched}{\alpha}}{\bot} \\
		& {} = \sum_{\phi \in \finiteFrags{\aut}} \dfrac{\costf^{\mathit{Ball}}_{\sched,s}(\phi)}{\probeval{\sd_{\sched,s}}{\cone{\phi}}} \cdot \sum_{\alpha \in \cone{\phi}} \probeval{\sd_{\sched,s}}{\cone{\alpha}} \cdot \probeval{\schedeval{\sched}{\alpha}}{\bot} \\
		& \stackrel{\dag}{=} \sum_{\phi \in \finiteFrags{\aut}} \dfrac{\costf^{\mathit{Ball}}_{\sched,s}(\phi)}{\probeval{\sd_{\sched,s}}{\cone{\phi}}} \cdot \probeval{\sd_{\sched,s}}{\cone{\phi}} \\
		& {} = \sum_{\phi \in \finiteFrags{\aut}} \costf^{\mathit{Ball}}_{\sched,s}(\phi) \\
		& {} = \costf^{\mathit{Ball}}_{\sched,s}(\weakCombinedTransition{s}{a}{\sd})\text{,}
	\end{align*}
	where $\stackrel{\dag}{=}$ is justified by the definition of the probability of a cone and the fact that $\sched$ induces a weak transition. 
\end{myproof}

\begin{result}[Proposition~\ref{pro:strongCostBisimEqIsEquivalenceRelation}]
\label{res:strongCostBisimEqIsEquivalenceRelation}
	Strong and strong probabilistic cost-preserving bisimilarities are equivalence relations on the set of \CPA{}s.
\end{result}
\begin{myproof}
	It is trivial to show that both strong and strong probabilistic cost-preserving bisimulations are reflexive and symmetric, since the identity relation $\idrelord$ suffices for reflexivity and the symmetry of the underlying equivalence relation is the base for the symmetry.
	Transitivity is more interesting, that is, given three \CPA{}s $\costAutPed{\costf}{1}$, $\costAutPed{\costf}{2}$, and $\costAutPed{\costf}{3}$ such that $\costAutPed{\costf}{1} \strongCostProbBisim \costAutPed{\costf}{2}$ and $\costAutPed{\costf}{2} \strongCostProbBisim \costAutPed{\costf}{3}$, then $\costAutPed{\costf}{1} \strongCostProbBisim \costAutPed{\costf}{3}$ (and similarly for $\strongCostBisim$);
	we provide the proof for strong probabilistic cost-preserving bisimulation since the proof for strong cost-preserving bisimulation is essentially the same, where the involved combined transitions are just ordinary transitions and the families are just singletons.
	
	Since $\costAutPed{\costf}{1} \strongCostProbBisim \costAutPed{\costf}{2}$ and $\costAutPed{\costf}{2} \strongCostProbBisim \costAutPed{\costf}{3}$, it follows that $\aut_{1} \strongProbBisim \aut_{2}$ and $\aut_{2} \strongProbBisim \aut_{3}$, since the step condition of strong probabilistic cost-preserving bisimulation is the step condition of strong probabilistic bisimulation extended with a constraint on the cost of the involved transitions.
	Let $\relord_{12}$ and $\relord_{23}$ be the corresponding relations.
	By transitivity of strong probabilistic bisimulation on \PA{}s~\cite{Seg95}, we have that $\aut_{1} \strongProbBisim \aut_{3}$ and this is justified by $\relord_{13} = \relord_{12} \relationComposition \relord_{23}$. 
	We claim that $\relord_{13}$ is also a strong probabilistic cost-preserving bisimulation;
	to show this claim, we need to check that for each $(s_{1},s_{3}) \in \relord_{13}$ and $\strongTransition{s_{1}}{a}{\sd_{1}}$, there exists $\strongCombinedTransition{s_{3}}{a}{\sd_{3}}$ such that $\sd_{1} \liftrel[\relord_{13}] \sd_{3}$ and $\cost[3]{\strongCombinedTransition{s_{3}}{a}{\sd_{3}}} = \cost[1]{\strongTransition{s_{1}}{a}{\sd_{1}}}$.
	
	Let $(s_{1},s_{3}) \in \relord_{13}$ and $\strongTransition{s_{1}}{a}{\sd_{1}}$. 
	Suppose that $s_{1} \in \stateSet_{1}$ and $s_{3} \in \stateSet_{3}$; 
	the symmetric case is analogous while the case where both states belong to the same automaton is trivial.
	By definition of $\relord_{13}$, we know that there exists $s_{2} \in \stateSet_{2}$ such that $s_{1} \rel_{12} s_{2} \rel_{23} s_{3}$;
	moreover, by $\costAutPed{\costf}{1} \strongCostProbBisim \costAutPed{\costf}{2}$, there exists $\strongCombinedTransition{s_{2}}{a}{\sd_{2}}$ such that $\sd_{1} \liftrel[\relord_{12}] \sd_{2}$ and $\cost[2]{\strongCombinedTransition{s_{2}}{a}{\sd_{2}}} = \cost[1]{\strongTransition{s_{1}}{a}{\sd_{1}}}$.
	Let $\family{\strongTransition{s_{2}}{a}{\sd_{2,i}}}{i \in I}$ and $\family{p_{i}}{i \in I}$ be the families of transitions and weights generating $\strongCombinedTransition{s_{2}}{a}{\sd_{2}}$.
	Since $s_{2} \rel_{23} s_{3}$ and $\costAutPed{\costf}{2} \strongCostProbBisim \costAutPed{\costf}{3}$, for each $i \in I$ there exists $\strongCombinedTransition{s_{3}}{a}{\sd_{3,i}}$ such that $\sd_{2,i} \liftrel[\relord_{23}] \sd_{3,i}$ and $\cost[3]{\strongCombinedTransition{s_{3}}{a}{\sd_{3,i}}} = \cost[2]{\strongTransition{s_{2}}{a}{\sd_{2,i}}}$.
	Let $\strongCombinedTransition{s_{3}}{a}{\sd_{3}}$ be the strong combined transition such that $\sd_{3} = \sum_{i \in I} p_{i} \cdot \sd_{3,i}$.
	By properties of the lifting $\liftrel[\functionGenericArgument]$, it is immediate to see that $\sd_{1} \liftrel[\relord_{13}] \sd_{3}$; 
	for the cost of the strong combined transition, we have:
        $\cost[3]{\strongCombinedTransition{s_{3}}{a}{\sd_{3}}} =
        \sum_{i \in I} p_{i} \cdot
        \cost[3]{\strongCombinedTransition{s_{3}}{a}{\sd_{3,i}}} =
        \sum_{i \in I} p_{i} \cdot
        \cost[2]{\strongTransition{s_{2}}{a}{\sd_{2,i}}} =
        \cost[2]{\strongCombinedTransition{s_{2}}{a}{\sd_{2}}} =
        \cost[1]{\strongTransition{s_{1}}{a}{\sd_{1}}}$, as
        required. 

	This completes the proof that $\relord_{13}$ is a strong probabilistic cost-preserving bisimulation, thus $\costAutPed{\costf}{1} \strongCostProbBisim \costAutPed{\costf}{3}$.
\end{myproof}

\begin{result}[Proposition~\ref{pro:strongCostBisimEqIsCompositional}]
\label{res:strongCostBisimEqIsCompositional}
	Given two \CPA{}s $\costAutPed{\costf}{1}$ and $\costAutPed{\costf}{2}$, if $\costAutPed{\costf}{1} \strongCostProbBisim \costAutPed{\costf}{2}$, then for each $\costAutPed{\costf}{3}$ compatible with both $\costAutPed{\costf}{1}$ and $\costAutPed{\costf}{2}$ and each pair of cost-preserving functions $\costf_{l}$ and $\costf_{r}$ with $\gen{\costf_{l}} = \gen{\costf_{r}}$, $\costAutPed{\costf}{1} \parComp_{\costf_{l}} \costAutPed{\costf}{3} \strongCostProbBisim \costAutPed{\costf}{2} \parComp_{\costf_{r}} \costAutPed{\costf}{3}$, and similarly for $\strongCostBisim$.
\end{result}
\begin{myproof}
	We detail here the proof for strong probabilistic cost-preserving
	bisimilarity. The one for strong cost-preserving bisimilarity is a
	simplification.
	Denoting by $\stateSet_{12}$ the set $\stateSet_{1} \uplus \stateSet_{2}$, let $\relord$ be the equivalence relation on $\stateSet_{12}$ justifying $\costAutPed{\costf}{1} \strongCostProbBisim \costAutPed{\costf}{2}$ and $\relord_{p}$ be the strong probabilistic bisimulation justifying $\aut_{1} \parComp \aut_{3} \strongProbBisim \aut_{2} \parComp \aut_{3}$.
	Note that $\relord_{p} = \relord \times \idrelord$, where $\idrelord$ is the identity relation on $\stateSet_{3}$.
	The existence of $\relord_{p}$ is ensured by the fact that $\aut_{1} \strongProbBisim \aut_{2}$ and strong probabilistic bisimilarity on probabilistic automata is preserved by parallel composition.
	We now show that $\relord_{p}$ is a strong probabilistic cost-preserving bisimulation between $\costAutPed{\costf}{1} \parComp_{\costf_{l}} \costAutPed{\costf}{3}$ and $\costAutPed{\costf}{2} \parComp_{\costf_{r}} \costAutPed{\costf}{3}$.
	The fact that $\relord_{p}$ is an equivalence relation follows directly from being a strong probabilistic bisimulation; 
	moreover, this implies $(\startState_{1}, \startState_{3}) \rel_{p} (\startState_{2}, \startState_{3})$ as well.
	
	So, consider a pair of states $(s_{1}, s_{3}) \rel_{p} (s_{2}, s_{3})$ and suppose that $\strongTransition{(s_{1}, s_{3})}{a}{\sd_{1} \times \sd_{3}}$. 
	Just for simplicity, assume that $s_{1} \in \stateSet_{1}$ and $s_{2} \in \stateSet_{2}$; the remaining cases are essentially the same.
	There are three cases:
	\begin{description}
	\item[Case $a \in \actionSet_{3} \setminus \actionSet_{1}$]
		In this case, $\sd_{1} = \dirac{s_{1}}$ and the transition $\strongTransition{(s_{1}, s_{3})}{a}{\sd_{1} \times \sd_{3}}$ can be matched by $(s_{2},s_{3})$ via the transition $\strongTransition{(s_{2},s_{3})}{a}{\dirac{s_{2}} \times \sd_{3}}$ that is trivially also a strong combined transition;
		it is immediate to see that $\dirac{s_{1}} \times \sd_{3} \liftrel[\relord_{p}] \dirac{s_{2}} \times \sd_{3}$. 
		For the cost, we have that 
		\begin{align*}
		\cost[r]{\strongTransition{(s_{2},s_{3})}{a}{\dirac{s_{2}} \times \sd_{3}}} 
		& {} = \functioneval{\gen{\costf_{r}}}{\cost[2]{\strongTransition{s_{2}}{\apparent{a}}{\dirac{s_{2}}}}, \cost[3]{\strongTransition{s_{3}}{a}{\sd_{3}}}} \\
		& {} = \functioneval{\gen{\costf_{r}}}{0, \cost[3]{\strongTransition{s_{3}}{a}{\sd_{3}}}} \\
		& {} = \functioneval{\gen{\costf_{l}}}{\cost[1]{\strongTransition{s_{1}}{\apparent{a}}{\dirac{s_{1}}}}, \cost[3]{\strongTransition{s_{3}}{a}{\sd_{3}}}} \\
		& {} = \cost[l]{\strongTransition{(s_{1},s_{3})}{a}{\dirac{s_{1}} \times \sd_{3}}}\text{,}
		\end{align*}
		as required.
		
	\item[Case $a \in \actionSet_{1} \setminus \actionSet_{3}$]
		In this case, $\sd_{3} = \dirac{s_{3}}$.
		Since $\costAutPed{\costf}{1} \strongCostProbBisim \costAutPed{\costf}{2}$, it follows that there exists $\strongCombinedTransition{s_{2}}{a}{\sd_{2}}$ such that $\sd_{1} \liftrel \sd_{2}$, and $\cost[2]{\strongCombinedTransition{s_{2}}{a}{\sd_{2}}} = \cost[1]{\strongTransition{s_{1}}{a}{\sd_{1}}}$.
		This implies that there exists the strong combined transition $\strongCombinedTransition{(s_{2}, s_{3})}{a}{\sd_{2} \times \dirac{s_{3}}}$ such that $\sd_{1} \times \dirac{s_{3}} \liftrel[\relord_{p}] \sd_{2} \times \dirac{s_{3}}$.
		For the cost of such transition, we have that 
		\begin{align*}
		\cost[r]{\strongCombinedTransition{(s_{2},s_{3})}{a}{\sd_{2} \times \dirac{s_{3}}}} 
		& {} = \functioneval{\gen{\costf_{r}}}{\cost[2]{\strongCombinedTransition{s_{2}}{a}{\sd_{2}}}, \cost[3]{\strongCombinedTransition{s_{3}}{\apparent{a}}{\dirac{s_{3}}}}} \\
		& {} = \functioneval{\gen{\costf_{r}}}{\cost[2]{\strongCombinedTransition{s_{2}}{a}{\sd_{2}}}, 0} \\
		& {} = \functioneval{\gen{\costf_{r}}}{\cost[1]{\strongTransition{s_{1}}{a}{\sd_{1}}}, 0} \\
		& {} = \functioneval{\gen{\costf_{l}}}{\cost[1]{\strongTransition{s_{1}}{a}{\sd_{1}}}, 0} \\
		& {} = \functioneval{\gen{\costf_{l}}}{\cost[1]{\strongTransition{s_{1}}{a}{\sd_{1}}}, \cost[3]{\strongTransition{s_{3}}{\apparent{a}}{\dirac{s_{3}}}}} \\
		& {} = \cost[l]{\strongTransition{(s_{1},s_{3})}{a}{\sd_{1} \times \dirac{s_{3}}}}\text{,}
		\end{align*} 
		as required.
		
	\item[Case $a \in \actionSet_{3} \cap \actionSet_{1}$]
		In this case, we have that $\strongTransition{(s_{1}, s_{3})}{a}{\sd_{1} \times \sd_{3}}$ is generated in the parallel composition by the two transitions $\strongTransition{s_{2}}{a}{\sd_{2}}$ and $\strongTransition{s_{3}}{a}{\sd_{3}}$.
		Since by hypothesis we have that $\costAutPed{\costf}{1} \strongCostProbBisim \costAutPed{\costf}{2}$, it follows that there exists $\strongCombinedTransition{s_{2}}{a}{\sd_{2}}$ such that $\sd_{1} \liftrel \sd_{2}$, and $\cost[2]{\strongCombinedTransition{s_{2}}{a}{\sd_{2}}} = \cost[1]{\strongTransition{s_{1}}{a}{\sd_{1}}}$.
		Let $\family{\strongTransition{s_{2}}{a}{\sd_{2,i}}}{i \in I}$ and $\family{p_{i}}{i \in I}$ be such that $\sum_{i \in I} p_{i} \cdot \sd_{2,i} = \sd_{2}$.
		Let $\strongCombinedTransition{(s_{2}, s_{3})}{a}{\sd_{2} \times \sd_{3}}$ be the strong combined transition for $\aut_{2} \times \aut_{3}$ obtained from $\family{\strongTransition{(s_{2}, s_{3})}{a}{\sd_{2,i} \times \sd_{3}}}{i \in I}$ and $\family{p_{i}}{i \in I}$, i.e., $\sd_{2} \times \sd_{3} = \sum_{i \in I} p_{i} \cdot \sd_{2,i} \times \sd_{3}$.
		
		It is immediate to see that $\sd_{1} \times \sd_{3} \liftrel[\relord_{p}] \sd_{2} \times \sd_{3}$;
		for the cost of such strong combined transition, we have that 
		\begin{align*}
		\cost[r]{\strongCombinedTransition{(s_{2},s_{3})}{a}{\sd_{2} \times \sd_{3}}} 
		& {} = \sum_{i \in I} p_{i} \cdot \cost[r]{\strongTransition{(s_{2},s_{3})}{a}{\sd_{2,i} \times \sd_{3}}} \\
		& {} = \sum_{i \in I} p_{i} \cdot \functioneval{\gen{\costf_{r}}}{\cost[2]{\strongCombinedTransition{s_{2}}{a}{\sd_{2,i}}}, \cost[3]{\strongTransition{s_{3}}{a}{\sd_{3}}}} \\
		& {} = \functioneval{\gen{\costf_{r}}}{\sum_{i \in I} p_{i} \cdot \cost[2]{\strongTransition{s_{2}}{a}{\sd_{2,i}}}, \cost[3]{\strongTransition{s_{3}}{a}{\sd_{3}}}} \\
		& {} = \functioneval{\gen{\costf_{r}}}{\cost[2]{\strongCombinedTransition{s_{2}}{a}{\sd_{2}}}, \cost[3]{\strongTransition{s_{3}}{a}{\sd_{3}}}} \\
		& {} = \functioneval{\gen{\costf_{l}}}{\cost[2]{\strongCombinedTransition{s_{2}}{a}{\sd_{2}}}, \cost[3]{\strongTransition{s_{3}}{a}{\sd_{3}}}} \\
		& {} = \functioneval{\gen{\costf_{l}}}{\cost[1]{\strongTransition{s_{1}}{a}{\sd_{1}}}, \cost[3]{\strongTransition{s_{3}}{a}{\sd_{3}}}} \\
		& {} = \cost[l]{\strongTransition{(s_{1},s_{3})}{a}{\sd_{1} \times \sd_{3}}}\text{,}
		\end{align*} 
		as required.
	\end{description}
	This completes the proof that $\relord_{p}$ is a strong probabilistic cost-preserving bisimulation, thus $\costAutPed{\costf}{1} \parComp_{\costf_{l}} \costAutPed{\costf}{3} \strongCostProbBisim \costAutPed{\costf}{2} \parComp_{\costf_{r}} \costAutPed{\costf}{3}$.
\end{myproof}

\begin{result}[Proposition~\ref{pro:weakCostBisimEqIsEquivalenceRelation}]
\label{res:weakCostBisimEqIsEquivalenceRelation}
	Weak probabilistic cost-preserving bisimilarity is an equivalence relation on the set of \CPA{}s.
\end{result}
\begin{myproof}
	Reflexivity and symmetry are straightforward, because $\idrelord \subseteq \mathord{\weakCostBisim}$ and because the underlying equivalence
	relation is symmetric.
	Transitivity needs a more detailed account: 
	Given three \CPA{}s $\costAutPed{\costf}{1}$, $\costAutPed{\costf}{2}$, and $\costAutPed{\costf}{3}$, if $\costAutPed{\costf}{1} \weakCostBisim \costAutPed{\costf}{2}$ and $\costAutPed{\costf}{2} \weakCostBisim \costAutPed{\costf}{3}$, then $\costAutPed{\costf}{1} \weakCostBisim \costAutPed{\costf}{3}$.

	Since $\costAutPed{\costf}{1} \weakCostBisim \costAutPed{\costf}{2}$ and $\costAutPed{\costf}{2} \weakCostBisim \costAutPed{\costf}{3}$, it follows that $\aut_{1} \weakBisim \aut_{2}$ and $\aut_{2} \weakBisim \aut_{3}$, thus, by transitivity of weak probabilistic bisimulation on \PA{}s~\cite{Seg95}, we have that $\aut_{1} \weakBisim \aut_{3}$. 
	Let $\relord$ be the corresponding equivalence relation.
	We claim that $\relord$ is also a weak probabilistic cost-preserving bisimulation;
	the only remaining thing we need to check is that for each $(s_{1},s_{3}) \in \relord$ and $\strongTransition{s_{1}}{a}{\sd_{1}}$, there exists $\weakCombinedTransition{s_{3}}{a}{\sd_{3}}$ such that $\sd_{1} \liftrel \sd_{3}$ and $\cost[3]{\weakCombinedTransition{s_{3}}{a}{\sd_{3}}} = \cost[1]{\strongTransition{s_{1}}{a}{\sd_{1}}}$.
	The existence of $\weakCombinedTransition{s_{3}}{a}{\sd_{3}}$ such that $\sd_{1} \liftrel \sd_{3}$ is ensured again by the fact that $\relord$ justifies $\aut_{1} \weakBisim \aut_{3}$;
	moreover, $\weakCombinedTransition{s_{3}}{a}{\sd_{3}}$ is essentially obtained from $\weakCombinedTransition{s_{2}}{a}{\sd_{2}}$ (induced by some scheduler $\sched_{2}$ and used to match $\strongTransition{s_{1}}{a}{\sd_{1}}$ when checking the step condition for the pair $(s_{1},s_{2})$ in $\aut_{1} \weakBisim \aut_{2}$) by replacing each transition $\strongTransition{s'_{2}}{b}{\sd'_{2}}$ chosen by the scheduler $\sched_{2}$ by the corresponding matching $\weakCombinedTransition{s'_{3}}{b}{\sd'_{3}}$ where $b \in \setnocond{a,\hidden}$.
	Since by hypothesis $\cost[2]{\strongTransition{s'_{2}}{b}{\sd'_{2}}} = \cost[3]{\weakCombinedTransition{s'_{3}}{b}{\sd'_{3}}}$, it is immediate to derive that indeed $\cost[3]{\weakCombinedTransition{s_{3}}{a}{\sd_{3}}} = \cost[1]{\strongTransition{s_{1}}{a}{\sd_{1}}}$ by using the definition of cost provided in Definition~\ref{def:costAsWorkingForTheProof}.

	This completes the proof that $\relord_{p}$ is a weak probabilistic cost-preserving bisimulation, thus $\costAutPed{\costf}{1} \parComp_{\costf_{l}} \costAutPed{\costf}{3} \weakCostBisim \costAutPed{\costf}{2} \parComp_{\costf_{r}} \costAutPed{\costf}{3}$.
\end{myproof}

\begin{result}[Lemma~\ref{lem:costWeakTransPCfunctionOfComponentWTs}]
\label{res:costWeakTransPCfunctionOfComponentWTs}
	Given two compatible \CPA{}s $\costAutPed{\costf}{1}$ and $\costAutPed{\costf}{2}$ and a cost preserving\break function $\costf$, let $\weakCombinedTransition{(s_{1}, s_{2})}{a}{\sd_{1} \times \sd_{2}}$ be a weak combined transition of $\costAutPed{\costf}{1} \parComp_{\costf} \costAutPed{\costf}{2}$.
	Then, $\cost{\weakCombinedTransition{(s_{1}, s_{2})}{a}{\sd_{1} \times \sd_{2}}} = \functioneval{\gen{\costf}}{\cost[1]{\weakCombinedTransition{s_{1}}{a}{\sd_{1}}}, \cost[2]{\weakCombinedTransition{s_{2}}{a}{\sd_{2}}}}$, where for $i=1,2$, $\weakCombinedTransition{s_{i}}{a}{\sd_{i}} = \project{i}{\weakCombinedTransition{(s_{1}, s_{2})}{a}{\sd_{1} \times \sd_{2}}}$.
\end{result}
\begin{myproof}
	Let $\aut = \aut_{1} \parComp \aut_{2}$ and $\sched$ be the scheduler inducing $\weakCombinedTransition{(s_{1}, s_{2})}{a}{\sd_{1} \times \sd_{2}}$;
	by definition~\ref{def:costAsWorkingForTheProof}, 
	$\cost{\weakCombinedTransition{(s_{1}, s_{2})}{a}{\sd_{1} \times \sd_{2}}} = \sum_{\alpha \in \finiteFrags{\aut}} \cost[\sched, (s_{1}, s_{2})]{\alpha}$ where
	\[
		\cost[\sched, (s_{1}, s_{2})]{\alpha} =
		\begin{cases}
			\probeval{\sd_{\sched, (s_{1}, s_{2})}}{\cone{\alpha'}} \cdot \sum_{\tr \in \transitionsWithLabel{b}} \cost{\tr} \cdot \probeval{\schedeval{\sched}{\alpha'}}{\tr} \cdot \probeval{\sd_{\tr}}{t_{1}, t_{2}} & \text{if $\alpha = \alpha' b (t_{1}, t_{2})$,} \\
			0 & \text{otherwise.}
		\end{cases}
	\]
	Suppose that for each $\alpha \in \finiteFrags{\aut}$, $\cost[\sched, (s_{1}, s_{2})]{\alpha} = \functioneval{\gen{\costf}}{\cost[1_{\sched, (s_{1}, s_{2})}]{\alpha}, \cost[2_{\sched, (s_{1}, s_{2})}]{\alpha}}$; 
	this implies that $\sum_{\alpha \in \finiteFrags{\aut}} \cost[\sched, (s_{1}, s_{2})]{\alpha} = \sum_{\alpha \in \finiteFrags{\aut}} \functioneval{\gen{\costf}}{\cost[1_{\sched, (s_{1}, s_{2})}]{\alpha}, \cost[2_{\sched, (s_{1}, s_{2})}]{\alpha}}$. 
	Since $\gen{\costf}$ is distributive by hypothesis, we can move the summation inside $\gen{\costf}$, that is, $\sum_{\alpha \in \finiteFrags{\aut}} \cost[\sched, (s_{1}, s_{2})]{\alpha} = \functioneval{\gen{\costf}}{\sum_{\alpha \in \finiteFrags{\aut}} \cost[1_{\sched, (s_{1}, s_{2})}]{\alpha}, \sum_{\alpha \in \finiteFrags{\aut}} \cost[2_{\sched, (s_{1}, s_{2})}]{\alpha}}$, i.e., $\cost{\weakCombinedTransition{(s_{1}, s_{2})}{a}{\sd_{1} \times \sd_{2}}} = \functioneval{\gen{\costf}}{\cost[1]{\weakCombinedTransition{s_{1}}{a}{\sd_{1}}}, \cost[2]{\weakCombinedTransition{s_{2}}{a}{\sd_{2}}}}$, as required.
	
	We now show that $\cost[\sched, (s_{1}, s_{2})]{\alpha} = \functioneval{\gen{\costf}}{\cost[1_{\sched, (s_{1}, s_{2})}]{\alpha}, \cost[2_{\sched, (s_{1}, s_{2})}]{\alpha}}$ holds for each $\alpha \in \finiteFrags{\aut}$.
	Suppose that $\alpha = t$; this case is obvious, since $\cost[\sched, (s_{1}, s_{2})]{t} = 0 = \functioneval{\gen{\costf}}{0, 0} = \functioneval{\gen{\costf}}{\cost[1_{\sched, (s_{1}, s_{2})}]{t}, \cost[2_{\sched, (s_{1}, s_{2})}]{t}}$;
	the equality $0 = \functioneval{\gen{\costf}}{0,0}$ follows by the property of being zero-preserving.
	
	Suppose that $\alpha = \alpha' b (t_{1}, t_{2})$ for some $\alpha' \in \finiteFrags{\aut}$, $b \in \actionSet$, and $(t_{1}, t_{2}) \in \stateSet$.
	There are three cases:
	\begin{description}
	\item[Case $b \in \actionSet_{1} \setminus \actionSet_{2}$]
		In this case, each transition $\tr = \strongTransition{(v_{1}, v_{2})}{b}{\sd_{1} \times \sd_{2}} \in \transitionsWithLabel{b}$ has been obtained by combining $\strongTransition{v_{1}}{b}{\sd_{1}}$ and the apparent transition $\strongTransition{v_{2}}{\apparent{b}}{\dirac{v_{2}}}$ where $\sd_{2} = \dirac{v_{2}}$; 
		its cost is by definition $\cost{\tr} = \functioneval{\gen{\costf}}{\cost[1]{\strongTransition{v_{1}}{b}{\sd_{1}}}, 0}$. 
		This means that
		\begin{align*}
		& \phantom{{}={}} \cost[\sched, (s_{1}, s_{2})]{\alpha} \\
		& {}= \probeval{\sd_{\sched, (s_{1}, s_{2})}}{\cone{\alpha'}} \cdot \sum_{\tr \in \transitionsWithLabel{b}} \cost{\tr} \cdot \probeval{\schedeval{\sched}{\alpha'}}{\tr} \cdot \probeval{\sd_{\tr}}{t_{1}, t_{2}} \\
		& {}= \sum_{\tr \in \transitionsWithLabel{b}} \probeval{\sd_{\sched, (s_{1}, s_{2})}}{\cone{\alpha'}} \cdot \probeval{\schedeval{\sched}{\alpha'}}{\tr} \cdot \probeval{\sd_{\tr}}{t_{1}, t_{2}} \cdot \cost{\tr} \\
		\intertext{by reordering of summations and products,}
		& {}= \sum_{\tr \in \transitionsWithLabel{b}} \probeval{\sd_{\sched, (s_{1}, s_{2})}}{\cone{\alpha'}} \cdot \probeval{\schedeval{\sched}{\alpha'}}{\tr} \cdot \probeval{\sd_{\tr}}{t_{1}, t_{2}} \cdot \functioneval{\gen{\costf}}{\cost[1]{\strongTransition{v_{1}}{b}{\sd_{1}}}, \cost[2]{\strongTransition{v_{2}}{\apparent{b}}{\dirac{v_{2}}}}} \\
		\intertext{by definition of $\costf$,}
		& {}= \functioneval{\gen{\costf}}{\sum_{\tr \in \transitionsWithLabel{b}} \probeval{\sd_{\sched, (s_{1}, s_{2})}}{\cone{\alpha'}} \cdot \probeval{\schedeval{\sched}{\alpha'}}{\tr} \cdot \probeval{\sd_{\tr}}{t_{1}, t_{2}} \cdot \cost[1]{\strongTransition{v_{1}}{b}{\sd_{1}}}, 0} \\
		\intertext{by distributivity of $\gen{\costf}$ and $\cost[2]{\strongTransition{v_{2}}{\apparent{b}}{\dirac{v_{2}}}} = 0$,}
		& {}= \functioneval{\gen{\costf}}{\probeval{\sd_{\sched, (s_{1}, s_{2})}}{\cone{\alpha'}} \cdot \sum_{\tr \in \transitionsWithLabel{b}} \cost[1]{\strongTransition{v_{1}}{b}{\sd_{1}}} \cdot \probeval{\schedeval{\sched}{\alpha'}}{\tr} \cdot \probeval{\sd_{\tr}}{t_{1}, t_{2}}, 0} \\
		\intertext{by reordering of summations and products,}
		& {}= \functioneval{\gen{\costf}}{\cost[1_{\sched, (s_{1}, s_{2})}]{\alpha}, \cost[2_{\sched, (s_{1}, s_{2})}]{\alpha}}\text{.}
		\end{align*}
		The last equality comes from the fact that by
                definition,
                $\cost[2]{\strongTransition{v_{2}}{\apparent{b}}{\dirac{v_{2}}}}
                = 0$, thus 
\[\eqalign{
  0 
&=0\cdot \probeval{\sd_{\sched, (s_{1}, s_{2})}}{\cone{\alpha'}}
   \cdot \sum_{\tr \in \transitionsWithLabel{b}}
   \probeval{\schedeval{\sched}{\alpha'}}{\tr} \cdot
   \probeval{\sd_{\tr}}{t_{1}, t_{2}}\cr
&=\cost[2]{\strongTransition{v_{2}}{\apparent{b}}{\dirac{v_{2}}}} \cdot
  \probeval{\sd_{\sched, (s_{1}, s_{2})}}{\cone{\alpha'}} \cdot
  \sum_{\tr \in \transitionsWithLabel{b}}
  \probeval{\schedeval{\sched}{\alpha'}}{\tr} \cdot
  \probeval{\sd_{\tr}}{t_{1}, t_{2}}\cr
&=\probeval{\sd_{\sched, 
  (s_{1},s_{2})}}{\cone{\alpha'}} \cdot \sum_{\tr \in
  \transitionsWithLabel{b}}
  \cost[2]{\strongTransition{v_{2}}{\apparent{b}}{\dirac{v_{2}}}} \cdot
  \probeval{\schedeval{\sched}{\alpha'}}{\tr} \cdot
  \probeval{\sd_{\tr}}{t_{1}, t_{2}}\cr
&=\cost[2_{\sched, (s_{1},s_{2})}]{\alpha}\text{.}
  }
\]
		
	\item[Case $b \in \actionSet_{2} \setminus \actionSet_{1}$]
		This case is symmetric to the previous case.
		
	\item[Case $b \in \actionSet_{1} \cap \actionSet_{2}$]
		In this case, every transition $\tr = \strongTransition{(v_{1}, v_{2})}{b}{\sd_{1} \times \sd_{2}} \in \transitionsWithLabel{b}$ has\break been obtained by combining $\strongTransition{v_{1}}{b}{\sd_{1}}$ and $\strongTransition{v_{2}}{b}{\sd_{2}}$; 
		its cost, by definition, is $\cost{\tr} = \functioneval{\gen{\costf}}{\cost[1]{\strongTransition{v_{1}}{b}{\sd_{1}}}, \cost[2]{\strongTransition{v_{2}}{b}{\sd_{2}}}}$. 
		This means that
		\begin{align*}
		& \phantom{{}={}} \cost[\sched, (s_{1}, s_{2})]{\alpha} \\
		& {}= \probeval{\sd_{\sched, (s_{1}, s_{2})}}{\cone{\alpha'}} \cdot \sum_{\tr \in \transitionsWithLabel{b}} \cost{\tr} \cdot \probeval{\schedeval{\sched}{\alpha'}}{\tr} \cdot \probeval{\sd_{\tr}}{t_{1}, t_{2}} \\
		& {}= \sum_{\tr \in \transitionsWithLabel{b}} \probeval{\sd_{\sched, (s_{1}, s_{2})}}{\cone{\alpha'}} \cdot \probeval{\schedeval{\sched}{\alpha'}}{\tr} \cdot \probeval{\sd_{\tr}}{t_{1}, t_{2}} \cdot \cost{\tr} \\
		\intertext{by reordering of summations and products,}
		& {}= \sum_{\tr \in \transitionsWithLabel{b}} \probeval{\sd_{\sched, (s_{1}, s_{2})}}{\cone{\alpha'}} \cdot \probeval{\schedeval{\sched}{\alpha'}}{\tr} \cdot \probeval{\sd_{\tr}}{t_{1}, t_{2}} \cdot \functioneval{\gen{\costf}}{\cost[1]{\strongTransition{v_{1}}{b}{\sd_{1}}}, \cost[2]{\strongTransition{v_{2}}{b}{\sd_{2}}}} \\
		\intertext{by definition of $\costf$,}
		& {}= \gen{\costf}(\sum_{\tr \in \transitionsWithLabel{b}} \probeval{\sd_{\sched, (s_{1}, s_{2})}}{\cone{\alpha'}} \cdot \probeval{\schedeval{\sched}{\alpha'}}{\tr} \cdot \probeval{\sd_{\tr}}{t_{1}, t_{2}} \cdot \cost[1]{\strongTransition{v_{1}}{b}{\sd_{1}}}, \\
		& \phantom{{}= \gen{\costf}(} \sum_{\tr \in \transitionsWithLabel{b}} \probeval{\sd_{\sched, (s_{1}, s_{2})}}{\cone{\alpha'}} \cdot \probeval{\schedeval{\sched}{\alpha'}}{\tr} \cdot \probeval{\sd_{\tr}}{t_{1}, t_{2}} \cdot \cost[2]{\strongTransition{v_{2}}{b}{\sd_{2}}}) \\
		\intertext{by distributivity of $\gen{\costf}$,}
		& {}= \gen{\costf}(\probeval{\sd_{\sched, (s_{1}, s_{2})}}{\cone{\alpha'}} \cdot \sum_{\tr \in \transitionsWithLabel{b}} \cost[1]{\strongTransition{v_{1}}{b}{\sd_{1}}} \cdot \probeval{\schedeval{\sched}{\alpha'}}{\tr} \cdot \probeval{\sd_{\tr}}{t_{1}, t_{2}}, \\
		& \phantom{{}= \gen{\costf}(} \probeval{\sd_{\sched, (s_{1}, s_{2})}}{\cone{\alpha'}} \cdot \sum_{\tr \in \transitionsWithLabel{b}} \cost[2]{\strongTransition{v_{2}}{b}{\sd_{2}}} \cdot \probeval{\schedeval{\sched}{\alpha'}}{\tr} \cdot \probeval{\sd_{\tr}}{t_{1}, t_{2}}) \\
		\intertext{by reordering of summations and products,}
		& {}= \functioneval{\gen{\costf}}{\cost[1_{\sched, (s_{1}, s_{2})}]{\alpha}, \cost[2_{\sched, (s_{1}, s_{2})}]{\alpha}} \\
		\intertext{by definition of $\cost[1_{\sched, (s_{1}, s_{2})}]{\alpha}$ and $\cost[2_{\sched, (s_{1}, s_{2})}]{\alpha}$.}
		\end{align*}
	\end{description}
	This completes the proof that $\cost[\sched, (s_{1}, s_{2})]{\alpha} = \functioneval{\gen{\costf}}{\cost[1_{\sched, (s_{1}, s_{2})}]{\alpha}, \cost[2_{\sched, (s_{1}, s_{2})}]{\alpha}}$ holds for each $\alpha \in \finiteFrags{\aut}$, hence $\cost{\weakCombinedTransition{(s_{1}, s_{2})}{a}{\sd_{1} \times \sd_{2}}} = \functioneval{\gen{\costf}}{\cost[1]{\weakCombinedTransition{s_{1}}{a}{\sd_{1}}}, \cost[2]{\weakCombinedTransition{s_{2}}{a}{\sd_{2}}}}$.
\end{myproof}

\begin{result}[Proposition~\ref{pro:weakCostBisimEqIsCompositional}]
\label{res:weakCostBisimEqIsCompositional}
	Given two \CPA{}s $\costAutPed{\costf}{1}$ and $\costAutPed{\costf}{2}$, if $\costAutPed{\costf}{1} \weakCostBisim \costAutPed{\costf}{2}$, then for each $\costAutPed{\costf}{3}$ compatible with both $\costAutPed{\costf}{1}$ and $\costAutPed{\costf}{2}$ and each pair of cost-preserving functions $\costf_{l}$ and $\costf_{r}$ with $\gen{\costf_{l}} = \gen{\costf_{r}}$, $\costAutPed{\costf}{1} \parComp_{\costf_{l}} \costAutPed{\costf}{3} \weakCostBisim \costAutPed{\costf}{2} \parComp_{\costf_{r}} \costAutPed{\costf}{3}$.
\end{result}
\begin{myproof}
	Denoted by $\stateSet_{12}$ the set $\stateSet_{1} \uplus \stateSet_{2}$, let $\relord$ be the equivalence relation on $\stateSet_{12}$ justifying $\costAutPed{\costf}{1} \weakCostBisim \costAutPed{\costf}{2}$ and $\relord_{p}$ be a weak probabilistic bisimulation justifying $\aut_{1} \parComp \aut_{3} \weakBisim \aut_{2} \parComp \aut_{3}$.
	Note that $\relord_{p} = \relord \times \idrelord$, where $\idrelord$ is the identity relation on $\stateSet_{3}$.
	The existence of $\relord_{p}$ is ensured by the fact that $\aut_{1} \weakBisim \aut_{2}$ and that weak probabilistic bisimilarity on probabilistic automata is preserved by parallel composition.
	We now show that $\relord_{p}$ is a weak probabilistic cost-preserving bisimulation between $\costAutPed{\costf}{1} \parComp_{\costf_{l}} \costAutPed{\costf}{3}$ and $\costAutPed{\costf}{2} \parComp_{\costf_{r}} \costAutPed{\costf}{3}$.
	The fact that $\relord_{p}$ is an equivalence relation follows directly from being a weak probabilistic bisimulation; 
	moreover, this implies $(\startState_{1}, \startState_{3}) \rel_{p} (\startState_{2}, \startState_{3})$ as well.
	
	So, consider a pair of states $(s_{1}, s_{3}) \rel_{p} (s_{2}, s_{3})$ and suppose that $\strongTransition{(s_{1}, s_{3})}{a}{\sd_{1} \times \sd_{3}}$. 
	Just for simplicity, assume that $s_{1} \in \stateSet_{1}$ and $s_{2} \in \stateSet_{2}$; the remaining cases are essentially the same.
	There are three cases:
	\begin{description}
	\item[Case $a \in \actionSet_{3} \setminus \actionSet_{1}$]
		In this case, $\sd_{1} = \dirac{s_{1}}$ and the transition $\strongTransition{(s_{1}, s_{3})}{a}{\sd_{1} \times \sd_{3}}$ can be matched by $(s_{2},s_{3})$ via the transition $\strongTransition{(s_{2},s_{3})}{a}{\dirac{s_{2}} \times \sd_{3}}$ that is trivially also a weak combined transition;
		it is immediate to see that $\dirac{s_{1}} \times \sd_{3} \liftrel[\relord_{p}] \dirac{s_{2}} \times \sd_{3}$. 
		For the cost, we have that 
		\begin{align*}
		\cost[r]{\strongTransition{(s_{2},s_{3})}{a}{\dirac{s_{2}} \times \sd_{3}}} 
		& {} = \functioneval{\gen{\costf_{r}}}{\cost[2]{\strongTransition{s_{2}}{\apparent{a}}{\dirac{s_{2}}}}, \cost[3]{\strongTransition{s_{3}}{a}{\sd_{3}}}} \\
		& {} = \functioneval{\gen{\costf_{r}}}{0, \cost[3]{\strongTransition{s_{3}}{a}{\sd_{3}}}} \\
		& {} = \functioneval{\gen{\costf_{l}}}{0, \cost[3]{\strongTransition{s_{3}}{a}{\sd_{3}}}} \\
		& {} = \functioneval{\gen{\costf_{l}}}{\cost[1]{\strongTransition{s_{1}}{\apparent{a}}{\dirac{s_{1}}}}, \cost[3]{\strongTransition{s_{3}}{a}{\sd_{3}}}} \\
		& {} = \cost[l]{\strongTransition{(s_{1},s_{3})}{a}{\dirac{s_{1}} \times \sd_{3}}}\text{,} 
		\end{align*}
		as required.
		
	\item[Case $a \in \actionSet_{1} \setminus \actionSet_{3}$]
		In this case, $\sd_{3} = \dirac{s_{3}}$.
		Since $\costAutPed{\costf}{1} \weakCostBisim \costAutPed{\costf}{2}$, it follows that there exists $\weakCombinedTransition{s_{2}}{a}{\sd_{2}}$ such that $\sd_{1} \liftrel \sd_{2}$, and $\cost[2]{\weakCombinedTransition{s_{2}}{a}{\sd_{2}}} = \cost[1]{\strongTransition{s_{1}}{a}{\sd_{1}}}$.
		This implies that there exists a weak combined transition $\weakCombinedTransition{(s_{2}, s_{3})}{a}{\sd_{2} \times \dirac{s_{3}}}$ such that $\sd_{1} \times \dirac{s_{3}} \liftrel[\relord_{p}] \sd_{2} \times \dirac{s_{3}}$.
		By Lemma~\ref{lem:costWeakTransPCfunctionOfComponentWTs}, we have that 
		\begin{align*}
		\cost[r]{\weakCombinedTransition{(s_{2},s_{3})}{a}{\sd_{2} \times \dirac{s_{3}}}} 
		& {} = \functioneval{\gen{\costf_{r}}}{\cost[2]{\weakCombinedTransition{s_{2}}{a}{\sd_{2}}}, \cost[3]{\weakCombinedTransition{s_{3}}{\hidden}{\dirac{s_{3}}}}} \\
		& {} = \functioneval{\gen{\costf_{r}}}{\cost[2]{\weakCombinedTransition{s_{2}}{a}{\sd_{2}}}, 0} \\
		& {} = \functioneval{\gen{\costf_{r}}}{\cost[1]{\strongTransition{s_{1}}{a}{\sd_{1}}}, 0} \\
		& {} = \functioneval{\gen{\costf_{l}}}{\cost[1]{\strongTransition{s_{1}}{a}{\sd_{1}}}, 0} \\
		& {} = \functioneval{\gen{\costf_{l}}}{\cost[1]{\strongTransition{s_{1}}{a}{\sd_{1}}}, \cost[3]{\strongTransition{s_{3}}{\apparent{a}}{\dirac{s_{3}}}}} \\
		& {} = \cost[l]{\strongTransition{(s_{1},s_{3})}{a}{\sd_{1} \times \dirac{s_{3}}}}\text{,}
		\end{align*} 
		as required.
		
	\item[Case $a \in \actionSet_{3} \cap \actionSet_{1}$]
		In this case, we have that $\strongTransition{(s_{1}, s_{3})}{a}{\sd_{1} \times \sd_{3}}$ is a transition of the composed automaton obtained by combining the two transitions $\strongTransition{s_{2}}{a}{\sd_{2}}$ and $\strongTransition{s_{3}}{a}{\sd_{3}}$.
		Since $\costAutPed{\costf}{1} \weakCostBisim \costAutPed{\costf}{2}$, it follows that there exists $\weakCombinedTransition{s_{2}}{a}{\sd_{2}}$ induced by the scheduler $\sched$ such that $\sd_{1} \liftrel \sd_{2}$, and $\cost[2]{\weakCombinedTransition{s_{2}}{a}{\sd_{2}}} = \cost[1]{\strongTransition{s_{1}}{a}{\sd_{1}}}$.
		Let $\weakCombinedTransition{(s_{2}, s_{3})}{a}{\sd_{2} \times \sd_{3}}$ be the weak combined transition for $\aut_{2} \times \aut_{3}$ obtained by the scheduler $\sched_{p}$ that mimics $\sched$ for all internal transitions of $\aut_{2}$ used in $\weakCombinedTransition{s_{2}}{a}{\sd_{2}}$ and schedules $\strongTransition{(t_{2},s_{3})}{a}{\gd_{2} \times \sd_{3}}$ on an execution fragment $\alpha$ with the same probability of scheduling $\strongTransition{t_{2}}{a}{\gd_{2}}$ given by $\sched$ on $\project{2}{\alpha}$.
		It is worthwhile to note that $\project{3}{\weakCombinedTransition{(s_{2}, s_{3})}{a}{\sd_{2} \times \sd_{3}}} = \weakCombinedTransition{s_{3}}{a}{\sd_{3}} = \strongTransition{s_{3}}{a}{\sd_{3}}$.
		It is immediate to see that $\sd_{1} \times \sd_{3} \liftrel[\relord_{p}] \sd_{2} \times \sd_{3}$;
		for the cost of such weak combined transition, by Lemma~\ref{lem:costWeakTransPCfunctionOfComponentWTs}, we have that 
		\begin{align*}
		\cost[r]{\weakCombinedTransition{(s_{2},s_{3})}{a}{\sd_{2} \times \sd_{3}}} 
		& {} = \functioneval{\gen{\costf_{r}}}{\cost[2]{\weakCombinedTransition{s_{2}}{a}{\sd_{2}}}, \cost[3]{\weakCombinedTransition{s_{3}}{a}{\sd_{3}}}} \\
		& {} = \functioneval{\gen{\costf_{r}}}{\cost[2]{\weakCombinedTransition{s_{2}}{a}{\sd_{2}}}, \cost[3]{\strongTransition{s_{3}}{a}{\sd_{3}}}} \\
		& {} = \functioneval{\gen{\costf_{r}}}{\cost[1]{\strongTransition{s_{1}}{a}{\sd_{1}}}, \cost[3]{\strongTransition{s_{3}}{a}{\sd_{3}}}} \\
		& {} = \functioneval{\gen{\costf_{l}}}{\cost[1]{\strongTransition{s_{1}}{a}{\sd_{1}}}, \cost[3]{\strongTransition{s_{3}}{a}{\sd_{3}}}} \\
		& {} = \cost[l]{\strongCombinedTransition{(s_{1},s_{3})}{a}{\sd_{1} \times \sd_{3}}}\text{,}
		\end{align*} 
		as required.
	\end{description}
	This completes the proof that $\relord_{p}$ is a weak probabilistic cost-preserving bisimulation, thus $\costAutPed{\costf}{1} \parComp_{\costf_{l}} \costAutPed{\costf}{3} \weakCostBisim \costAutPed{\costf}{2} \parComp_{\costf_{r}} \costAutPed{\costf}{3}$.
\end{myproof}

\begin{result}[Proposition~\ref{pro:strongCostBisimMinorCostIsPreorder}]
\label{res:strongCostBisimMinorCostIsPreorder}
	Minor cost strong and strong probabilistic bisimilarities are preorders on the set of \CPA{}s.
\end{result}
\begin{myproof}
	It is trivial to show that both bisimulations are reflexive, so we concentrate on transitivity, that is, given three \CPA{}s $\costAutPed{\costf}{1}$, $\costAutPed{\costf}{2}$, and $\costAutPed{\costf}{3}$, if $\costAutPed{\costf}{1} \strongCostProbBisimMinorCost \costAutPed{\costf}{2}$ and $\costAutPed{\costf}{2} \strongCostProbBisimMinorCost \costAutPed{\costf}{3}$, then $\costAutPed{\costf}{1} \strongCostProbBisimMinorCost \costAutPed{\costf}{3}$, and similarly for $\strongCostBisimMinorCost$.
	As for Proposition~\ref{pro:strongCostBisimEqIsEquivalenceRelation}, we provide the proof only for minor cost strong probabilistic bisimulation; 
	the proof for minor cost strong bisimulation is essentially the same, except that the involved combined transitions are just ordinary transitions and that families are just singletons.
	
	Since $\costAutPed{\costf}{1} \strongCostProbBisimMinorCost \costAutPed{\costf}{2}$ and $\costAutPed{\costf}{2} \strongCostProbBisimMinorCost \costAutPed{\costf}{3}$, it follows by definition that $\aut_{2} \strongProbBisim \aut_{1}$ and $\aut_{3} \strongProbBisim \aut_{2}$.
	Let $\relord_{21}$ and $\relord_{32}$ be the corresponding relations.
	By transitivity of strong probabilistic bisimulation on \PA{}s~\cite{Seg95}, we have that $\aut_{3} \strongProbBisim \aut_{1}$ and this is justified by $\relord_{31} = \relord_{32} \relationComposition \relord_{21}$. 
	We claim that $\relord_{31}$ is also a minor cost strong probabilistic bisimulation;
	to show this claim, we need to check that for each $\strongTransition{s_{3}}{a}{\sd_{3}} \in \transitionRelation_{3}$ and $s_{1} \in \stateSet_{1}$ such that $s_{3} \rel_{31} s_{1}$, there exists $\strongCombinedTransition{s_{1}}{a}{\sd_{1}}$ such that $\sd_{3} \liftrel[\relord_{31}] \sd_{1}$ and $\cost[1]{\strongCombinedTransition{s_{1}}{a}{\sd_{1}}} \leq \cost[3]{\strongTransition{s_{3}}{a}{\sd_{3}}}$.
	
	Let $\strongTransition{s_{3}}{a}{\sd_{3}}$ and $s_{3} \rel_{31} s_{1}$ with $s_{1} \in \stateSet_{1}$.
	By definition of $\relord_{31}$, we know that there exists $s_{2} \in \stateSet_{2}$ such that $s_{3} \rel_{32} s_{2} \rel_{21} s_{1}$;
	moreover, by $\costAutPed{\costf}{2} \strongCostProbBisimMinorCost \costAutPed{\costf}{3}$, there exists $\strongCombinedTransition{s_{2}}{a}{\sd_{2}}$ such that $\sd_{3} \liftrel[\relord_{32}] \sd_{2}$ and $\cost[2]{\strongCombinedTransition{s_{2}}{a}{\sd_{2}}} \leq \cost[3]{\strongTransition{s_{3}}{a}{\sd_{3}}}$.
	Let $\family{\strongTransition{s_{2}}{a}{\sd_{2,i}}}{i \in I}$ and $\family{p_{i}}{i \in I}$ be the families of transitions and weights generating $\strongCombinedTransition{s_{2}}{a}{\sd_{2}}$.
	Since $s_{2} \rel_{21} s_{1}$ and $\costAutPed{\costf}{1} \strongCostProbBisimMinorCost \costAutPed{\costf}{2}$, for each $i \in I$ there exists $\strongCombinedTransition{s_{1}}{a}{\sd_{1,i}}$ such that $\sd_{2,i} \liftrel[\relord_{21}] \sd_{1,i}$ and $\cost[1]{\strongCombinedTransition{s_{1}}{a}{\sd_{1,i}}} \leq \cost[2]{\strongTransition{s_{2}}{a}{\sd_{2,i}}}$.
	Let $\strongCombinedTransition{s_{1}}{a}{\sd_{1}}$ be the strong combined transition such that $\sd_{1} = \sum_{i \in I} p_{i} \cdot \sd_{1,i}$.
	By properties of the lifting $\liftrel[\functionGenericArgument]$, it is immediate to see that $\sd_{3} \liftrel[\relord_{31}] \sd_{1}$; 
	for the cost of the strong combined transition, we have: $\cost[1]{\strongCombinedTransition{s_{1}}{a}{\sd_{1}}} = \sum_{i \in I} p_{i} \cdot \cost[1]{\strongCombinedTransition{s_{1}}{a}{\sd_{1,i}}} \leq \sum_{i \in I} p_{i} \cdot \cost[2]{\strongTransition{s_{2}}{a}{\sd_{2,i}}} = \cost[2]{\strongCombinedTransition{s_{2}}{a}{\sd_{2}}} \leq \cost[3]{\strongTransition{s_{3}}{a}{\sd_{3}}}$, as required.

	This completes the proof that $\relord_{31}$ is a minor cost strong probabilistic bisimulation, thus $\costAutPed{\costf}{1} \strongCostProbBisimMinorCost \costAutPed{\costf}{3}$.
\end{myproof}

\begin{result}[Proposition~\ref{pro:strongMinCostSimIsCompositional}]
\label{res:strongMinCostSimIsCompositional}
	Given two \CPA{}s $\costAutPed{\costf}{1}$ and $\costAutPed{\costf}{2}$, if $\costAutPed{\costf}{1} \strongCostProbBisimMinorCost \costAutPed{\costf}{2}$, then for each $\costAutPed{\costf}{3}$ compatible with both $\costAutPed{\costf}{1}$ and $\costAutPed{\costf}{2}$ and each pair of cost-preserving functions $\costf_{l}$ and $\costf_{r}$ with $\gen{\costf_{l}} = \gen{\costf_{r}}$, $\costAutPed{\costf}{1} \parComp_{\costf_{l}} \costAutPed{\costf}{3} \strongCostProbBisimMinorCost \costAutPed{\costf}{2} \parComp_{\costf_{r}} \costAutPed{\costf}{3}$, and similarly for $\strongCostBisimMinorCost$.
\end{result}
\begin{myproof}
	We provide the proof for minor cost strong probabilistic bisimulation, the one for minor cost strong bisimulation is again a simplification thereof.
	Denoted by $\stateSet_{12}$ the set $\stateSet_{1} \uplus \stateSet_{2}$, let $\relord$ be the equivalence relation on $\stateSet_{12}$ justifying $\costAutPed{\costf}{1} \strongCostProbBisimMinorCost \costAutPed{\costf}{2}$ and $\relord_{p}$ be the strong probabilistic bisimulation justifying $\aut_{1} \parComp \aut_{3} \strongProbBisim \aut_{2} \parComp \aut_{3}$.
	Note that $\relord_{p} = \relord \times \idrelord$, where $\idrelord$ is the identity relation on $\stateSet_{3}$.
	The existence of $\relord_{p}$ is ensured by the fact that $\aut_{1} \strongProbBisim \aut_{2}$ and that strong probabilistic bisimilarity on probabilistic automata is preserved by parallel composition.
	We now show that $\relord_{p}$ is a minor cost strong probabilistic bisimulation from $\costAutPed{\costf}{1} \parComp_{\costf_{l}} \costAutPed{\costf}{3}$ to $\costAutPed{\costf}{2} \parComp_{\costf_{r}} \costAutPed{\costf}{3}$.
	The fact that $\relord_{p}$ is an equivalence relation follows directly from being a strong probabilistic bisimulation; 
	moreover, this implies $(\startState_{1}, \startState_{3}) \rel_{p} (\startState_{2}, \startState_{3})$ as well.
	
	So, consider $\strongTransition{(s_{2}, s_{3})}{a}{\sd_{2} \times \sd_{3}} \in \transitionRelation_{2,3}$ and $(s_{1}, s_{3}) \in \stateSet_{1,3}$ with $(s_{2}, s_{3}) \rel_{p} (s_{1}, s_{3})$. 
	There are three cases:
	\begin{description}
	\item[Case $a \in \actionSet_{3} \setminus \actionSet_{2}$]
		In this case, $\sd_{2} = \dirac{s_{2}}$ and the transition $\strongTransition{(s_{2}, s_{3})}{a}{\sd_{2} \times \sd_{3}}$ can be matched by $(s_{1},s_{3})$ via the transition $\strongTransition{(s_{1},s_{3})}{a}{\dirac{s_{1}} \times \sd_{3}}$ that is trivially also a strong combined transition;
		it is immediate to see that $\dirac{s_{2}} \times \sd_{3} \liftrel[\relord_{p}] \dirac{s_{1}} \times \sd_{3}$. 
		For the cost, we have that 
		\begin{align*}
		\cost[r]{\strongTransition{(s_{1},s_{3})}{a}{\dirac{s_{1}} \times \sd_{3}}} 
		& {} = \functioneval{\gen{\costf_{r}}}{\cost[1]{\strongTransition{s_{1}}{\apparent{a}}{\dirac{s_{1}}}}, \cost[3]{\strongTransition{s_{3}}{a}{\sd_{3}}}} \\
		& {} = \functioneval{\gen{\costf_{r}}}{0, \cost[3]{\strongTransition{s_{3}}{a}{\sd_{3}}}} \\
		& {} = \functioneval{\gen{\costf_{l}}}{0, \cost[3]{\strongTransition{s_{3}}{a}{\sd_{3}}}} \\
		& {} = \functioneval{\gen{\costf_{l}}}{\cost[2]{\strongTransition{s_{2}}{\apparent{a}}{\dirac{s_{2}}}}, \cost[3]{\strongTransition{s_{3}}{a}{\sd_{3}}}} \\
		& {} = \cost[l]{\strongTransition{(s_{2},s_{3})}{a}{\dirac{s_{2}} \times \sd_{3}}}\text{,}
		\end{align*}
		as required.
		
	\item[Case $a \in \actionSet_{2} \setminus \actionSet_{3}$]
		In this case, $\sd_{3} = \dirac{s_{3}}$.
		Since $\costAutPed{\costf}{1} \strongCostProbBisimMinorCost \costAutPed{\costf}{2}$, it follows that there exists $\strongCombinedTransition{s_{1}}{a}{\sd_{1}}$ such that $\sd_{2} \liftrel \sd_{1}$, and $\cost[1]{\strongCombinedTransition{s_{1}}{a}{\sd_{1}}} \leq \cost[2]{\strongTransition{s_{2}}{a}{\sd_{2}}}$.
		This implies that there exists a strong combined transition $\strongCombinedTransition{(s_{1}, s_{3})}{a}{\sd_{1} \times \dirac{s_{3}}}$ such that $\sd_{2} \times \dirac{s_{3}} \liftrel[\relord_{p}] \sd_{1} \times \dirac{s_{3}}$.
		For the cost of such transition, we have that 
		\begin{align*}
		\cost[r]{\strongCombinedTransition{(s_{1},s_{3})}{a}{\sd_{1} \times \dirac{s_{3}}}} 
		& {} = \functioneval{\gen{\costf_{r}}}{\cost[1]{\strongCombinedTransition{s_{1}}{a}{\sd_{1}}}, \cost[3]{\strongCombinedTransition{s_{3}}{\apparent{a}}{\dirac{s_{3}}}}} \\
		& {} = \functioneval{\gen{\costf_{r}}}{\cost[1]{\strongCombinedTransition{s_{1}}{a}{\sd_{1}}}, 0} \\
		& {} \leq \functioneval{\gen{\costf_{r}}}{\cost[2]{\strongTransition{s_{2}}{a}{\sd_{2}}}, 0} \\
		& {} = \functioneval{\gen{\costf_{l}}}{\cost[2]{\strongTransition{s_{2}}{a}{\sd_{2}}}, 0} \\
		& {} = \functioneval{\gen{\costf_{l}}}{\cost[2]{\strongTransition{s_{2}}{a}{\sd_{2}}}, \cost[3]{\strongTransition{s_{3}}{\apparent{a}}{\dirac{s_{3}}}}} \\
		& {} = \cost[l]{\strongTransition{(s_{2},s_{3})}{a}{\sd_{2} \times \dirac{s_{3}}}}\text{,}
		\end{align*} 
		as required.
		
	\item[Case $a \in \actionSet_{3} \cap \actionSet_{2}$]
		In this case, we have that $\strongTransition{(s_{2}, s_{3})}{a}{\sd_{2} \times \sd_{3}}$ is generated in the parallel composition by the two transitions $\strongTransition{s_{1}}{a}{\sd_{1}}$ and $\strongTransition{s_{3}}{a}{\sd_{3}}$.
		Since by hypothesis we have that $\costAutPed{\costf}{1} \strongCostProbBisimMinorCost \costAutPed{\costf}{2}$, it follows that there exists $\strongCombinedTransition{s_{1}}{a}{\sd_{1}}$ such that $\sd_{2} \liftrel \sd_{1}$, and $\cost[1]{\strongCombinedTransition{s_{1}}{a}{\sd_{1}}} \leq \cost[2]{\strongTransition{s_{2}}{a}{\sd_{2}}}$.
		Let $\family{\strongTransition{s_{1}}{a}{\sd_{1,i}}}{i \in I}$ and $\family{p_{i}}{i \in I}$ be such that $\sum_{i \in I} p_{i} \cdot \sd_{1,i} = \sd_{1}$.
		Let $\strongCombinedTransition{(s_{1}, s_{3})}{a}{\sd_{1} \times \sd_{3}}$ be the strong combined transition for $\aut_{1} \times \aut_{3}$ obtained from $\family{\strongTransition{(s_{1}, s_{3})}{a}{\sd_{1,i} \times \sd_{3}}}{i \in I}$ and $\family{p_{i}}{i \in I}$, i.e., $\sd_{1} \times \sd_{3} = \sum_{i \in I} p_{i} \cdot \sd_{1,i} \times \sd_{3}$.
		
		It is immediate to see that $\sd_{2} \times \sd_{3} \liftrel[\relord_{p}] \sd_{1} \times \sd_{3}$;
		for the cost of such strong combined transition, we have that 
		\begin{align*}
		\cost[r]{\strongCombinedTransition{(s_{1},s_{3})}{a}{\sd_{1} \times \sd_{3}}} 
		& {} = \sum_{i \in I} p_{i} \cdot \cost[r]{\strongTransition{(s_{1},s_{3})}{a}{\sd_{1,i} \times \sd_{3}}} \\
		& {} = \sum_{i \in I} p_{i} \cdot \functioneval{\gen{\costf_{r}}}{\cost[1]{\strongCombinedTransition{s_{1}}{a}{\sd_{1,i}}}, \cost[3]{\strongTransition{s_{3}}{a}{\sd_{3}}}} \\
		& {} = \functioneval{\gen{\costf_{r}}}{\sum_{i \in I} p_{i} \cdot \cost[1]{\strongTransition{s_{1}}{a}{\sd_{1,i}}}, \cost[3]{\strongTransition{s_{3}}{a}{\sd_{3}}}} \\
		& {} = \functioneval{\gen{\costf_{r}}}{\cost[1]{\strongCombinedTransition{s_{1}}{a}{\sd_{1}}}, \cost[3]{\strongTransition{s_{3}}{a}{\sd_{3}}}} \\
		& {} = \functioneval{\gen{\costf_{l}}}{\cost[1]{\strongCombinedTransition{s_{1}}{a}{\sd_{1}}}, \cost[3]{\strongTransition{s_{3}}{a}{\sd_{3}}}} \\
		& {} \leq \functioneval{\gen{\costf_{l}}}{\cost[2]{\strongTransition{s_{2}}{a}{\sd_{2}}}, \cost[3]{\strongTransition{s_{3}}{a}{\sd_{3}}}} \\
		& {} = \cost[l]{\strongTransition{(s_{2},s_{3})}{a}{\sd_{2} \times \sd_{3}}}\text{,}
		\end{align*} 
		as required.
	\end{description}
	This completes the proof that $\relord_{p}$ is a minor cost strong probabilistic bisimulation, thus $\costAutPed{\costf}{1} \parComp_{\costf_{l}} \costAutPed{\costf}{3} \strongCostProbBisimMinorCost \costAutPed{\costf}{2} \parComp_{\costf_{r}} \costAutPed{\costf}{3}$.
\end{myproof}

\begin{result}[Proposition~\ref{pro:strongMinCostProbBisimImpliesWeakMinCostBisim}]
\label{res:strongMinCostProbBisimImpliesWeakMinCostBisim}
	Given two \CPA{}s $\costAutPed{\costf}{1}$ and $\costAutPed{\costf}{2}$, if $\costAutPed{\costf}{1} \strongCostProbBisimMinorCost \costAutPed{\costf}{2}$, then $\costAutPed{\costf}{1} \weakCostBisimMinorCost \costAutPed{\costf}{2}$.
\end{result}
\begin{myproof}
	Let $\wbrelord$ be the equivalence relation justifying $\costAutPed{\costf}{1} \strongCostProbBisimMinorCost \costAutPed{\costf}{2}$; 
	by construction it is also a strong probabilistic bisimulation between $\aut_{1}$ and $\aut_{2}$, thus it is also a weak probabilistic bisimulation between $\aut_{1}$ and $\aut_{2}$.
	Let $\costrelord$ be $\wbrelord \cap \stateSet_{2} \times \stateSet_{1}$. 
	Obviously we have that $\costrelord \subseteq \wbrelord \cap \stateSet_{2} \times \stateSet_{1}$, $\startState_{2} \costrel \startState_{1}$, and for each $s_{2} \in \stateSet_{2}$ there exists $s_{1} \in \stateSet_{1}$ such that $s_{2} \costrel s_{1}$. 
	(This is ensured by the fact that every state in both automata is reachable from the corresponding start state.)
	To complete the proof, we need to show the step condition: 
	Let $\strongTransition{s_{2}}{a}{\sd_{2}}$ and $s_{1} \in \stateSet_{1}$ be such that $s_{2} \costrel s_{1}$.
	Suppose that there does not exist $\gd_{2} \in \Disc{\borderStateSetOrd[\wbrelord] \cap \stateSet_{2}}$ such that $\hyperWeakCombinedTransition{\sd_{2}}{\hidden}{\gd_{2}}$, then by hypothesis, there exists $\sd_{1} \in \Disc{\stateSet_{1}}$ such that $\strongCombinedTransition{s_{1}}{a}{\sd_{1}}$, $\sd_{2} \liftrel[\wbrelord] \sd_{1}$, and $\cost[1]{\strongCombinedTransition{s_{1}}{a}{\sd_{1}}} \leq \cost[2]{\strongTransition{s_{2}}{a}{\sd_{2}}}$.
	Since $\costrelord = \wbrelord \cap \stateSet_{2} \times \stateSet_{1}$ and the definition of lifting only involves pairs belonging to $\stateSet_{2} \times \stateSet_{1}$, we have also $\sd_{2} \liftrel[\costrelord] \sd_{1}$, as required.
	Suppose that there exists $\gd_{2} \in \Disc{\borderStateSetOrd[\wbrelord] \cap \stateSet_{2}}$ such that $\hyperWeakCombinedTransition{\sd_{2}}{\hidden}{\gd_{2}}$ and $\min \setcond{\cost[2]{\hyperWeakCombinedTransition{\sd_{2}}{\hidden}{\gd}}}{\gd \in \Disc{\borderStateSet[\wbrelord] \cap \stateSet_{2}}} = \cost[2]{\hyperWeakCombinedTransition{\sd_{2}}{\hidden}{\gd_{2}}}$.
	In order to find the matching transition from $s_{1}$, we replace each ordinary transition $\strongTransition{t_{2}}{b}{\gamma_{2}}$ inside $\weakCombinedTransition{s_{2}}{a}{\gd_{2}}$ (where $b \in \setnocond{a, \hidden}$) with the matching transitions $\strongCombinedTransition{t_{1}}{b}{\gamma_{1}}$ with the corresponding probabilities.
	It is routine to verify that the result of this replacement is indeed a weak combined transition $\weakCombinedTransition{s_{1}}{a}{\gd_{1}}$ for some $\gd_{1} \in \Disc{\stateSet_{1}}$ such that $\gd_{2} \liftrel[\wbrelord] \gd_{1}$, thus $\gd_{2} \liftrel[\costrelord] \gd_{1}$ as before.
	Moreover, since by hypothesis each ordinary transition $\strongTransition{t_{2}}{b}{\gamma_{2}}$ inside $\weakCombinedTransition{s_{2}}{a}{\gd_{2}}$ has been matched by $\strongCombinedTransition{t_{1}}{b}{\gamma_{1}}$ such that $\cost[1]{\strongCombinedTransition{t_{1}}{b}{\gamma_{1}}} \leq \cost[2]{\strongTransition{t_{2}}{b}{\gamma_{2}}}$, it is trivial to derive that indeed $\cost[1]{\weakCombinedTransition{s_{1}}{a}{\gd_{1}}} \leq \cost[2]{\weakCombinedTransition{\strongTransition{s_{2}}{a}{\sd_{2}}}{\hidden}{\gd_{2}}}$.
	The last thing we have to check is that $\gd_{1} \in \Disc{\borderStateSet[\wbrelord] \cap \stateSet_{1}}$. 
	This trivially holds since each $v_{2} \in \Supp{\gd_{2}}$ is a border state, thus $v_{2}$ enables a transition $\strongTransition{v_{2}}{b}{\theta_{2}}$ such that $b \in \externalActionSet_{2}$ or $\probeval{\theta_{2}}{\relclass{v_{2}}{\wbrelord}} < 1$.
	Since $\gd_{2} \liftrel[\costrelord] \gd_{1}$, we have that each $v_{1} \in \Supp{\gd_{1}}$ is related to some $u_{2} \in \Supp{\gd_{2}}$, thus also $v_{1}$ enables a transition $\strongTransition{v_{1}}{b}{\theta_{1}}$ (with $\theta_{2} \liftrel[\wbrelord] \theta_{1}$ by $\costAutPed{\costf}{1} \strongCostProbBisimMinorCost \costAutPed{\costf}{2}$) such that $b \in \externalActionSet_{1}$ or $\probeval{\theta_{1}}{\relclass{v_{1}}{\wbrelord}} < 1$, respectively, i.e., $v_{1}$ is a border state, as required.
	
	This completes the proof that $(\wbrelord, \costrelord)$ is a minor cost weak probabilistic bisimulation, thus $\costAutPed{\costf}{1} \weakCostBisimMinorCost \costAutPed{\costf}{2}$.
\end{myproof}

\begin{result}[Proposition~\ref{pro:weakMinCostSimIsPreorder}]
\label{res:weakMinCostSimIsPreorder}
	Minor cost weak probabilistic bisimilarity is a preorder on the set of \CPA{}s.
\end{result}
\begin{myproof}
	Reflexivity is straightforward and we omit it, so let us consider transitivity, that is, given three \CPA{}s $\costAutPed{\costf}{1}$, $\costAutPed{\costf}{2}$, and $\costAutPed{\costf}{3}$, if $\costAutPed{\costf}{1} \weakCostBisimMinorCost \costAutPed{\costf}{2}$ and $\costAutPed{\costf}{2} \weakCostBisimMinorCost \costAutPed{\costf}{3}$, then $\costAutPed{\costf}{1} \weakCostBisimMinorCost \costAutPed{\costf}{3}$.

	Let $(\wbrelord_{21}, \costrelord_{21})$ and $(\wbrelord_{32},
        \costrelord_{32})$ be a minor cost weak bisimulations justifying $\costAutPed{\costf}{1} \weakCostBisimMinorCost \costAutPed{\costf}{2}$ and $\costAutPed{\costf}{2} \weakCostBisimMinorCost \costAutPed{\costf}{3}$, respectively.
	Let $\wbrelord_{31}$ be $\wbrelord_{32} \relationComposition \wbrelord_{21}$.
	It is known~\cite{Seg95} that $\wbrelord_{31}$ is a weak probabilistic bisimulation between $\aut_{1}$ and $\aut_{3}$.
	Let $\costrelord_{31}$ be $\costrelord_{32} \relationComposition \costrelord_{21}$.
	We claim that $(\wbrelord_{31}, \costrelord_{31})$ is a minor cost weak bisimulation from $\costAutPed{\costf}{1}$ to $\costAutPed{\costf}{3}$.
	
	$\startState_{3} \costrel_{31} \startState_{1}$ is immediate since by hypothesis and by definition of $\costrelord_{32} \relationComposition \costrelord_{21}$, $\startState_{3} \costrel_{32} \startState_{2} \costrel_{21} \startState_{1}$.
	
	It is immediate to see that for each $s_{3} \in \stateSet_{3}$ there exists $s_{1} \in \stateSet_{1}$ such that $s_{3} \costrel_{31} s_{1}$:
	Let $s_{3} \in \stateSet_{3}$; 
	by definition of $\costrelord_{32}$, there exists $s_{2} \in \stateSet_{2}$ such that $s_{3} \costrel_{32} s_{2}$ and by definition of $\costrelord_{21}$, there exists $s_{1} \in \stateSet_{1}$ such that $s_{2} \costrel_{21} s_{1}$, hence $s_{3} \costrel_{31} s_{1}$, as required.
	The fact that $s_{3} \wbrel_{31} s_{1}$ is immediate by the way $\wbrelord_{31}$ is constructed and the fact that $\costrelord_{32} \subseteq \wbrelord_{32}$ and $\costrelord_{21} \subseteq \wbrelord_{21}$.
	
	Before continuing with the proof, consider the set $\borderStateSetOrd[\wbrelord_{31}]$: 
	It is immediate to see that $\borderStateSetOrd[\wbrelord_{31}] \cap \stateSet_{3} = \borderStateSetOrd[\wbrelord_{32}] \cap \stateSet_{3}$.
	In fact, by definition of $\borderStateSetOrd[\functionGenericArgument]$, a state $t_{3} \in \stateSet_{3}$ belongs to $\borderStateSetOrd[\wbrelord_{31}]$ since $t_{3}$ enables a transition $\strongTransition{t_{3}}{b}{\gd_{3}}$ such that $b \in \externalActionSet_{3}$ (but this is independent from the equivalence relation, thus $t_{3} \in \borderStateSetOrd[\wbrelord_{32}] \cap \stateSet_{3}$) or $\probeval{\gd_{3}}{\relclass{t_{3}}{\wbrelord_{31}}} < 1$, i.e., there exists $t'_{3} \in \Supp{\gd_{3}}$ such that $(t_{3}, t'_{3}) \notin \wbrelord_{31}$.
	By definition of $\wbrelord_{31}$, it follows that $(t_{3}, t'_{3}) \notin \wbrelord_{32}$ holds as well (otherwise $(t_{3}, t'_{3}) \in \wbrelord_{31}$ would hold), hence $t_{3} \in \borderStateSetOrd[\wbrelord_{32}] \cap \stateSet_{3}$, as required.
	On the other hand, consider $t_{3} \in \stateSet_{3}$ such that $t_{3} \notin \borderStateSetOrd[\wbrelord_{31}]$.
	This implies that for each $\strongTransition{t_{3}}{b}{\gd_{3}}$, $b \in \internalActionSet$ and $\probeval{\gd_{3}}{\relclass{t_{3}}{\wbrelord_{31}}} = 1$, i.e., for each $t'_{3} \in \Supp{\gd_{3}}$, $t'_{3} \wbrel_{31} t_{3}$.
	By definition of $\wbrelord_{31}$, it follows that $t'_{3} \wbrel_{31} t_{3}$ holds as well, hence $\probeval{\gd_{3}}{\relclass{t_{3}}{\wbrelord_{32}}} = 1$, thus $t_{3} \notin \borderStateSetOrd[\wbrelord_{32}]$, as required.
	
	Now, assume $\strongTransition{s_{3}}{a}{\sd_{3}}$ and $s_{3} \costrel_{31} s_{1}$.
	Moreover, assume that $a \in \externalActionSet$ (the case $a \in \internalActionSet$ is just a simplification of this case).
	Let $s_{2} \in \stateSet_{2}$ be a state such that $s_{3} \costrel_{32} s_{2} \costrel_{21} s_{1}$.
	There are two cases:
	\begin{enumerate}
	\item 
	\label{item:mcwsTransitiveCaseBorder}
		There exists $\gd_{3} \in \Disc{\borderStateSetOrd[\wbrelord_{31}] \cap \stateSet_{3}}$ such that $\hyperWeakCombinedTransition{\sd_{3}}{\hidden}{\gd_{3}}$ and $\min \setcond{\cost[3]{\hyperWeakCombinedTransition{\sd_{3}}{\hidden}{\gd}}}{\gd \in \Disc{\borderStateSet[\wbrelord_{31}] \cap \stateSet_{3}}} = \cost[3]{\hyperWeakCombinedTransition{\sd_{3}}{\hidden}{\gd_{3}}}$.
		Since $\borderStateSetOrd[\wbrelord_{31}] \cap \stateSet_{3} = \borderStateSetOrd[\wbrelord_{32}] \cap \stateSet_{3}$ and $\costAutPed{\costf}{2} \weakCostBisimMinorCost \costAutPed{\costf}{3}$, it follows that there exists $\gd_{2} \in \Disc{\borderStateSetOrd[\wbrelord_{32}] \cap \stateSet_{2}}$ such that $\weakCombinedTransition{s_{2}}{a}{\gd_{2}}$, $\gd_{3} \liftrel[\costrelord_{32}] \gd_{2}$, and $\cost[2]{\weakCombinedTransition{s_{2}}{a}{\gd_{2}}} \leq \cost[3]{\weakCombinedTransition{\strongTransition{s_{3}}{a}{\sd_{3}}}{\hidden}{\gd_{3}}}$.
		Let $\gamma_{2} \in \Disc{\stateSet_{2}}$ be such that $\strongCombinedTransition{\weakCombinedTransition{s_{2}}{\hidden}{\gamma_{2}}}{a}{\gd_{2}}$, that is, $\gamma_{2}$ is the probability distribution reached exactly before the $a$ action along $\weakCombinedTransition{s_{2}}{a}{\gd_{2}}$.
		Note that $\Supp{\gamma_{2}} \subseteq \borderStateSetOrd[\wbrelord_{32}] \cap \stateSet_{2}$.
		This implies that there exists $\gamma_{1} \in \Disc{\borderStateSetOrd[\wbrelord_{21}] \cap \stateSet_{1}}$ such that $\weakCombinedTransition{s_{1}}{\hidden}{\gamma_{1}}$, $\gamma_{2} \liftrel[\costrelord_{21}] \gamma_{1}$, and $\cost[1]{\weakCombinedTransition{s_{1}}{\hidden}{\gamma_{1}}} \leq \cost[2]{\weakCombinedTransition{s_{2}}{\hidden}{\gamma_{2}}}$.
		Now, from each $t_{2} \in \Supp{\gamma_{2}}$, let $\weakCombinedTransition{t_{2}}{a}{\theta_{t_{2}}}$ be the weak combined transition enabled by $t_{2}$ such that $\sum_{t_{2} \in \Supp{\gamma_{2}}} \probeval{\gamma_{2}}{t_{2}} \cdot \theta_{t_{2}} = \gd_{2}$.
		By construction, the external action is performed immediately, so $\weakCombinedTransition{t_{2}}{a}{\theta_{t_{2}}}$ is actually $\hyperWeakCombinedTransition{\strongCombinedTransition{t_{2}}{a}{\phi_{t_{2}}}}{\hidden}{\theta_{t_{2}}}$ for some distribution $\phi_{t_{2}}$.
		Let $\hyperWeakCombinedTransition{\strongTransition{t_{2}}{a}{\varphi_{t_{2}}}}{\hidden}{\psi_{t_{2}}}$ be a component of such weak combined transition, i.e., $\sum_{\strongTransition{t_{2}}{a}{\varphi_{t_{2}}} \in \transitionRelation_{2}} \probeval{\schedeval{\sched}{t_{2}}}{\strongTransition{t_{2}}{a}{\varphi_{t_{2}}}} \cdot \varphi_{t_{2}} = \phi_{t_{2}}$ and $\sum_{\strongTransition{t_{2}}{a}{\varphi_{t_{2}}} \in \transitionRelation_{2}} \probeval{\schedeval{\sched}{t_{2}}}{\strongTransition{t_{2}}{a}{\varphi_{t_{2}}}} \cdot \psi_{t_{2}} = \theta_{t_{2}}$ where $\sched$ is the scheduler inducing $\weakCombinedTransition{t_{2}}{a}{\theta_{t_{2}}}$.
		Since $\gd_{2} \in \Disc{\borderStateSetOrd[\wbrelord_{32}] \cap \stateSet_{2}}$, it is immediate to see that also $\psi_{t_{2}} \in \Disc{\borderStateSetOrd[\wbrelord_{32}] \cap \stateSet_{2}}$ for each $t_{2} \in \Supp{\gamma_{2}}$, thus for each $t_{1} \in \Supp{\gamma_{1}}$ such that $t_{2} \costrel_{21} t_{1}$, there exists $\psi_{t_{1}} \in \Disc{\borderStateSetOrd[\wbrelord_{21}] \cap \stateSet_{1}}$ such that $\weakCombinedTransition{t_{1}}{a}{\psi_{t_{1}}}$, $\psi_{t_{2}} \liftrel[\costrelord_{21}] \psi_{t_{1}}$, and $\cost[1]{\weakCombinedTransition{t_{1}}{a}{\psi_{t_{1}}}} \leq \cost[2]{\weakCombinedTransition{t_{2}}{a}{\psi_{t_{2}}}}$.
		By combining these transitions to obtain $\weakCombinedTransition{t_{1}}{a}{\theta_{t_{1}}}$ as in the construction of $\weakCombinedTransition{t_{2}}{a}{\theta_{t_{2}}}$, we obtain that $\theta_{t_{2}} \liftrel[\costrelord_{21}] \theta_{t_{1}}$, and $\cost[1]{\weakCombinedTransition{t_{1}}{a}{\theta_{t_{1}}}} \leq \cost[2]{\weakCombinedTransition{t_{2}}{a}{\theta_{t_{2}}}}$.
		By extending $\weakCombinedTransition{s_{1}}{\hidden}{\gamma_{1}}$ with $\hyperWeakCombinedTransition{\gamma_{1}}{a}{\gd_{1}}$ with $\gd_{1} = \sum_{t_{1} \in \Supp{\gamma_{1}}} \probeval{\gamma_{1}}{t_{1}} \cdot \theta_{t_{1}}$, we have that the resulting weak combined transition $\weakCombinedTransition{s_{1}}{a}{\gd_{1}}$ satisfies $\gd_{1} \in \Disc{\borderStateSetOrd[\wbrelord_{31}] \cap \stateSet_{1}}$, $\gd_{3} \liftrel[\costrelord_{31}] \gd_{1}$, and $\cost[1]{\weakCombinedTransition{s_{1}}{a}{\gd_{1}}} \leq \cost[3]{\weakCombinedTransition{\strongTransition{s_{3}}{a}{\sd_{3}}}{\hidden}{\gd_{3}}}$, as required.
	\item 
		There does not exist $\gd_{3} \in \Disc{\borderStateSetOrd[\wbrelord_{31}] \cap \stateSet_{3}}$ such that $\hyperWeakCombinedTransition{\sd_{3}}{\hidden}{\gd_{3}}$.
		Since $\borderStateSetOrd[\wbrelord_{31}] \cap \stateSet_{3} = \borderStateSetOrd[\wbrelord_{32}] \cap \stateSet_{3}$ and $\costAutPed{\costf}{2} \weakCostBisimMinorCost \costAutPed{\costf}{3}$, it follows that there exists $\gd_{2} \in \Disc{\stateSet_{2}}$ such that $\weakCombinedTransition{s_{2}}{a}{\gd_{2}}$, $\gd_{3} \liftrel[\costrelord_{32}] \gd_{2}$, and $\cost[2]{\weakCombinedTransition{s_{2}}{a}{\gd_{2}}} \leq \cost[3]{\strongTransition{s_{3}}{a}{\sd_{3}}}$.
		As in the case (\ref{item:mcwsTransitiveCaseBorder}), let $\gamma_{2} \in \Disc{\stateSet_{2}}$ be such that $\strongCombinedTransition{\weakCombinedTransition{s_{2}}{\hidden}{\gamma_{2}}}{a}{\gd_{2}}$, that is, $\gamma_{2}$ are the probability distribution reached exactly before the $a$ action along $\weakCombinedTransition{s_{2}}{a}{\gd_{2}}$.
		Note that $\Supp{\gamma_{2}} \subseteq \borderStateSetOrd[\wbrelord_{32}] \cap \stateSet_{2}$.
		This implies that there exists $\gamma_{1} \in \Disc{\borderStateSetOrd[\wbrelord_{21}] \cap \stateSet_{1}}$ such that $\weakCombinedTransition{s_{1}}{\hidden}{\gamma_{1}}$, $\gamma_{2} \liftrel[\costrelord_{21}] \gamma_{1}$, and $\cost[1]{\weakCombinedTransition{s_{1}}{\hidden}{\gamma_{1}}} \leq \cost[2]{\weakCombinedTransition{s_{2}}{\hidden}{\gamma_{2}}}$.
		Now, from each $t_{2} \in \Supp{\gamma_{2}}$, let $\weakCombinedTransition{t_{2}}{a}{\theta_{t_{2}}}$ be the weak combined transition enabled by $t_{2}$ such that $\sum_{t_{2} \in \Supp{\gamma_{2}}} \probeval{\gamma_{2}}{t_{2}} \cdot \theta_{t_{2}} = \gd_{2}$.
		For each $t_{2} \in \Supp{\gamma_{2}}$, if $\Supp{\theta_{t_{2}}} \subseteq \borderStateSetOrd[\wbrelord_{21}] \cap \stateSet_{2}$, then we are in the same situation as in the case (\ref{item:mcwsTransitiveCaseBorder}), that is, we are able to construct $\weakCombinedTransition{t_{1}}{a}{\theta_{t_{1}}}$ such that $\theta_{t_{2}} \liftrel[\costrelord_{21}] \theta_{t_{1}}$ and $\cost[1]{\weakCombinedTransition{t_{1}}{a}{\theta_{t_{1}}}} \leq \cost[2]{\weakCombinedTransition{t_{2}}{a}{\theta_{t_{2}}}}$.
		Now, suppose that $\Supp{\theta_{t_{2}}} \not\subseteq \borderStateSetOrd[\wbrelord_{21}] \cap \stateSet_{2}$.
		Let $\phi_{t_{2}}$ be the distribution such that $\hyperWeakCombinedTransition{\strongCombinedTransition{t_{2}}{a}{\phi_{t_{2}}}}{\hidden}{\theta_{t_{2}}}$. 
		Let $\hyperWeakCombinedTransition{\strongTransition{t_{2}}{a}{\varphi_{t_{2}}}}{\hidden}{\psi_{t_{2}}}$ be a component of such weak combined transition, i.e., $\sum_{\strongTransition{t_{2}}{a}{\varphi_{t_{2}}} \in \transitionRelation_{2}} \probeval{\schedeval{\sched}{t_{2}}}{\strongTransition{t_{2}}{a}{\varphi_{t_{2}}}} \cdot \varphi_{t_{2}} = \phi_{t_{2}}$ and $\sum_{\strongTransition{t_{2}}{a}{\varphi_{t_{2}}} \in \transitionRelation_{2}} \probeval{\schedeval{\sched}{t_{2}}}{\strongTransition{t_{2}}{a}{\varphi_{t_{2}}}} \cdot \psi_{t_{2}} = \theta_{t_{2}}$ where $\sched$ is the scheduler inducing $\weakCombinedTransition{t_{2}}{a}{\theta_{t_{2}}}$.
		If $\Supp{\psi_{t_{2}}} \subseteq \borderStateSetOrd[\wbrelord_{21}] \cap \stateSet_{2}$, then we are in the same situation as in the case (\ref{item:mcwsTransitiveCaseBorder}), otherwise from each $u_{2} \in \Supp{\varphi_{2}}$, let $\weakCombinedTransition{u_{2}}{\hidden}{\kappa_{u_{2}}}$ be the weak combined transition enabled by $u_{2}$ such that $\sum_{u_{2} \in \Supp{\varphi_{2}}} \probeval{\varphi_{2}}{u_{2}} \cdot \kappa_{u_{2}} = \psi_{2}$.
		For each $u_{2} \in \Supp{\varphi_{2}}$, if $\Supp{\kappa_{u_{2}}} \subseteq \borderStateSetOrd[\wbrelord_{21}] \cap \stateSet_{2}$, then we are in the same situation as in the case (\ref{item:mcwsTransitiveCaseBorder}), so suppose that $\Supp{\kappa_{u_{2}}} \not\subseteq \borderStateSetOrd[\wbrelord_{21}] \cap \stateSet_{2}$.
		Let $\lambda_{u_{2}}$ be the distribution such that $\hyperWeakCombinedTransition{\strongCombinedTransition{u_{2}}{\hidden}{\lambda_{u_{2}}}}{\hidden}{\kappa_{u_{2}}}$. 		
		(Note that here we are assuming that $u_{2}$ does not stop immediately with non-zero probability. This is not an issue since if $u_{2}$ needs to stop immediately, then this is matched by $u_{1}$ such that $u_{2} \costrel_{21} u_{1}$ by stopping immediately with the same probability, and in both cases the cost is $0$.)
		Let $\hyperWeakCombinedTransition{\strongTransition{u_{2}}{\hidden}{\xi_{u_{2}}}}{\hidden}{\chi_{u_{2}}}$ be a component of such weak combined transition, i.e., $\sum_{\strongTransition{u_{2}}{\hidden}{\xi_{u_{2}}} \in \transitionRelation_{2}} \probeval{\schedeval{\sched}{u_{2}}}{\strongTransition{u_{2}}{\hidden}{\xi_{u_{2}}}} \cdot \xi_{u_{2}} = \lambda_{u_{2}}$ and $\sum_{\strongTransition{u_{2}}{\hidden}{\xi_{u_{2}}} \in \transitionRelation_{2}} \probeval{\schedeval{\sched}{u_{2}}}{\strongTransition{u_{2}}{\hidden}{\xi_{u_{2}}}} \cdot \chi_{u_{2}} = \kappa_{u_{2}}$ where $\sched_{u}$ is the scheduler inducing $\weakCombinedTransition{u_{2}}{\hidden}{\kappa_{u_{2}}}$.
		If $\Supp{\chi_{u_{2}}} \subseteq \borderStateSetOrd[\wbrelord_{21}] \cap \stateSet_{2}$, then we are in the same situation as in the case (\ref{item:mcwsTransitiveCaseBorder}), otherwise we can apply the same technique until we obtain a strong transition $\strongTransition{v_{2}}{\hidden}{\omega_{v_{2}}}$ such that there does not exist $\varpi_{v_{2}} \in \Disc{\borderStateSetOrd[\wbrelord_{32}] \cap \stateSet_{2}}$ such that $\hyperWeakCombinedTransition{\omega_{v_{2}}}{\hidden}{\varpi_{v_{2}}}$ (actually, we usually obtain a strong combined transition, but we just focus on its components).
		For this transition, since $\costAutPed{\costf}{1} \weakCostBisimMinorCost \costAutPed{\costf}{2}$, we have that for each $v_{1} \in \stateSet_{1}$ such that $v_{2} \costrel_{21} v_{1}$, there exists $\omega_{v_{1}} \in \Disc{\stateSet_{1}}$ such that $\weakCombinedTransition{v_{1}}{\hidden}{\omega_{v_{1}}}$, $\omega_{2} \liftrel[\costrelord_{21}] \omega_{1}$ and $\cost[1]{\weakCombinedTransition{v_{1}}{\hidden}{\omega_{1}}} \leq \cost[2]{\strongTransition{v_{2}}{\hidden}{\omega_{2}}}$.
		By combining all these weak combined transitions according to the combinations used for constructing $\weakCombinedTransition{s_{2}}{a}{\gd_{2}}$, we obtain a weak combined transition $\weakCombinedTransition{s_{1}}{a}{\gd_{1}}$ such that $\gd_{1} \in \Disc{\stateSet_{1}}$, $\gd_{3} \liftrel[\costrelord_{31}] \gd_{1}$, and $\cost[1]{\weakCombinedTransition{s_{1}}{a}{\gd_{1}}} \leq \cost[3]{\strongTransition{s_{3}}{\hidden}{\sd_{3}}}$, as required.
	\end{enumerate}
	This completes the proof that $(\wbrelord_{31}, \costrelord_{31})$ is a minor cost weak bisimulation from $\costAutPed{\costf}{1}$ to $\costAutPed{\costf}{3}$, thus $\costAutPed{\costf}{1} \weakCostBisimMinorCost \costAutPed{\costf}{3}$.
\end{myproof}

\begin{result}[Proposition~\ref{pro:weakMinCostSimIsCompositional}]
\label{res:weakMinCostSimIsCompositional}
	Given two \CPA{}s $\costAutPed{\costf}{1}$ and $\costAutPed{\costf}{2}$, if $\costAutPed{\costf}{1} \weakCostBisimMinorCost \costAutPed{\costf}{2}$, then for each $\costAutPed{\costf}{3}$ compatible with both $\costAutPed{\costf}{1}$ and $\costAutPed{\costf}{2}$ and each pair of cost-preserving functions $\costf_{l}$ and $\costf_{r}$ with $\gen{\costf_{l}} = \gen{\costf_{r}}$, $\costAutPed{\costf}{1} \parComp_{\costf_{l}} \costAutPed{\costf}{3} \weakCostBisimMinorCost \costAutPed{\costf}{2} \parComp_{\costf_{r}} \costAutPed{\costf}{3}$.
\end{result}
\begin{myproof}
	Denoting by $\stateSet_{12}$ the set $\stateSet_{1} \uplus \stateSet_{2}$, let $(\wbrelord, \costrelord)$ be a minor cost weak probabilistic bisimulation justifying $\costAutPed{\costf}{1} \weakCostBisimMinorCost \costAutPed{\costf}{2}$ and $(\wbrelord_{p}, \costrelord_{p})$ be defined as follows:
	\begin{itemize}
	\item 
		$\wbrelord_{p}$ is the weak probabilistic bisimulation between $\aut_{1} \parComp \aut_{3}$ and $\aut_{2} \parComp \aut_{3}$. 
		Note that $\wbrelord_{p} = \wbrelord \times \idrelord$, where $\idrelord$ is the identity relation on $\stateSet_{3}$.
		Its existence is ensured by the fact that the $\aut_{1} \weakBisim \aut_{2}$ and the weak probabilistic bisimulation on probabilistic automata is preserved by parallel composition;
	\item
		$\costrelord_{p} = \setcond{((s_{2},s_{3}), (s_{1},s_{3}))}{(s_{2},s_{1}) \in \costrelord, s_{3} \in \stateSet_{3}}$.
		Essentially, $\costrelord_{p}$ is the product of $\costrelord$ with the identity relation $\idrelord$ on $\stateSet_{3}$.
	\end{itemize}
	
	\noindent By the way $\costrelord_{p}$ is defined, it is immediate to see that $\costrelord_{p} \subseteq \wbrelord_{p} \cap (\stateSet_{2} \times \stateSet_{3}) \times (\stateSet_{1} \times \stateSet_{3})$ such that for each $(s_{2}, s_{3}) \in \stateSet_{2} \times \stateSet_{3}$ there exists $(s_{1},s_{3}) \in \stateSet_{1} \times \stateSet_{3}$ such that $(s_{2}, s_{3}) \costrel_{p} (s_{1}, s_{3})$. 
	
	We now show that $(\wbrelord_{p}, \costrelord_{p})$ is actually a minor cost weak probabilistic bisimulation from $\costAutPed{\costf}{1} \parComp_{\costf_{l}} \costAutPed{\costf}{3}$ to $\costAutPed{\costf}{2} \parComp_{\costf_{r}} \costAutPed{\costf}{3}$.
	
	The fact that $\wbrelord_{p}$ is a weak probabilistic bisimulation between $\aut_{1} \parComp \aut_{3}$ and $\aut_{2} \parComp \aut_{3}$ is immediate by definition.
	Since by hypothesis, $\startState_{2} \costrel \startState_{1}$, it is immediate to see that $(\startState_{2}, \startState_{3}) \costrel_{p} (\startState_{1}, \startState_{3})$.
	So, consider a pair of states $(s_{2},s_{3})$ and $(s_{1},s_{3})$ such that $(s_{2},s_{3}) \costrel_{p} (s_{1},s_{3})$ and a transition $\strongTransition{(s_{2},s_{3})}{a}{\sd_{2} \times \sd_{3}}$.
	Now, there are three cases:
	\begin{description}
	\item[Case $a \in \actionSet_{3} \setminus \actionSet_{2}$]
		In this case, $\sd_{2} = \dirac{s_{2}}$.
		Suppose that there exists the distribution $\gd_{2} \times \gd_{3} \in \Disc{\borderStateSetOrd[\wbrelord_{p}] \cap (\stateSet_{2} \times \stateSet_{3})}$ such that $\hyperWeakCombinedTransition{\sd_{2} \times \sd_{3}}{\hidden}{\gd_{2} \times \gd_{3}}$.
		Since $\hyperWeakCombinedTransition{\sd_{2} \times \sd_{3}}{\hidden}{\gd_{2} \times \gd_{3}}$ is an internal weak combined transition (we remark that $\hidden$ is used as symbol for any internal action, it is not a specific action), each transition chosen by the scheduler inducing $\hyperWeakCombinedTransition{\sd_{2} \times \sd_{3}}{\hidden}{\gd_{2} \times \gd_{3}}$ either corresponds to a transition from $\aut_{2}$ or from $\aut_{3}$, but none of them is the result of the synchronisation between transitions of the two automata.
		Moreover, when a transition from $\aut_{2}$ is performed from the product state $(t_{2},t_{3})$, the reached states are of the form $(t'_{2},t_{3})$, and similarly for transitions from $\aut_{3}$.
		Since each state $(t_{2},t_{3})$ is a border state either because $t_{2}$ or $t_{3}$ is a border state, the minimum cost is obtained only by choosing transitions from either $\aut_{2}$ or $\aut_{3}$, but not from both.
		The only exception is when transitions have cost $0$, since they do not affect the resulting cost but they also do not affect whether it is cheaper to reach the border with only transitions from either $\aut_{2}$ or $\aut_{3}$.

		Now, suppose that only transitions from $\aut_{2}$ are used, hence we have $\gd_{3} = \sd_{3}$:
		Let $\family{\strongTransition{s_{2}}{\hidden}{\theta_{i}}}{i \in I}$ be the set of transitions chosen by the scheduler $\sched$ for the execution fragment $\alpha = s_{2}$ during the construction of $\hyperWeakCombinedTransition{\sd_{2}}{\hidden}{\gd_{2}}$.
		Since $s_{2} \costrel s_{1}$ and for each $\theta_{i}$ there exists $\phi_{i} \in \Disc{\borderStateSetOrd[\wbrelord] \cap \stateSet_{2}}$ such that $\hyperWeakCombinedTransition{\theta_{i}}{\hidden}{\phi_{i}}$ and $\sum_{i \in I} \probeval{\schedeval{\sched}{s_{2}}}{\strongTransition{s_{2}}{\hidden}{\theta_{i}}} \cdot \phi_{i} = \gd_{2}$, by $\costAutPed{\costf}{1} \weakCostBisimMinorCost \costAutPed{\costf}{2}$ it follows that there exists $\chi_{i} \in \Disc{\borderStateSetOrd[\wbrelord] \cap \stateSet_{1}}$ such that $\weakCombinedTransition{s_{1}}{\hidden}{\chi_{i}}$, $\phi_{i} \liftrel[\costrelord] \chi_{i}$, and $\cost[1]{\weakCombinedTransition{s_{1}}{\hidden}{\chi_{i}}} \leq \cost[2]{\weakCombinedTransition{\strongTransition{s_{2}}{\hidden}{\sd_{2}}}{\hidden}{\phi_{i}}}$.
		It is easy to see that the convex combination $\sum_{i \in I} \probeval{\schedeval{\sched}{s_{2}}}{\strongTransition{s_{2}}{\hidden}{\theta_{i}}} \cdot \weakCombinedTransition{s_{1}}{\hidden}{\chi_{i}}$ results in a transition $\weakCombinedTransition{s_{1}}{\hidden}{\chi}$ such that $\chi \in \Disc{\borderStateSetOrd[\wbrelord] \cap \stateSet_{1}}$,  $\gd_{2} \liftrel[\costrelord] \chi$, and $\cost[1]{\weakCombinedTransition{s_{1}}{\hidden}{\chi}} \leq \cost[2]{\weakCombinedTransition{\strongTransition{s_{2}}{\hidden}{\sd_{2}}}{\hidden}{\gd_{2}}}$.
		Since $\gd_{2} \liftrel[\costrelord] \chi$, we have $\gd_{2} \times \gd_{3} \liftrel[\costrelord_{p}] \chi \times \gd_{3}$ and $\chi \times \gd_{3} \in \Disc{\borderStateSetOrd[\wbrelord_{p}] \cap \stateSet_{1} \times \stateSet_{3}}$.
		To complete the proof for this case, we need to show that $\cost[l]{\weakCombinedTransition{(s_{1},s_{3})}{a}{\chi \times \sd_{3}}} \leq \cost[r]{\weakCombinedTransition{(s_{2},s_{3})}{a}{\gd_{2} \times \sd_{3}}}$: 
		\begin{align*}
		\cost[l]{\weakCombinedTransition{(s_{1},s_{3})}{a}{\chi \times \sd_{3}}} 
		& {} = \functioneval{\gen{\costf_{l}}}{\cost[1]{\weakCombinedTransition{s_{1}}{\hidden}{\chi}}, \cost[3]{\weakCombinedTransition{s_{3}}{a}{\sd_{3}}}} \\
		& \leq \functioneval{\gen{\costf_{l}}}{\cost[2]{\weakCombinedTransition{\strongTransition{s_{2}}{\hidden}{\sd_{2}}}{\hidden}{\gd_{2}}}, \cost[3]{\weakCombinedTransition{s_{3}}{a}{\sd_{3}}}} \\
		& {} = \functioneval{\gen{\costf_{r}}}{\cost[2]{\weakCombinedTransition{\strongTransition{s_{2}}{\hidden}{\sd_{2}}}{\hidden}{\gd_{2}}}, \cost[3]{\weakCombinedTransition{s_{3}}{a}{\sd_{3}}}} \\
		& {} = \cost[r]{\weakCombinedTransition{(s_{2},s_{3})}{a}{\gd_{2} \times \sd_{3}}}\text{,}
		\end{align*} 
		hence $\cost[l]{\weakCombinedTransition{(s_{1},s_{3})}{a}{\chi \times \sd_{3}}} \leq \cost[r]{\weakCombinedTransition{(s_{2},s_{3})}{a}{\gd_{2} \times \sd_{3}}}$ as required.
		
		Instead, if only transitions from $\aut_{3}$ are used, hence $\gd_{2} = \dirac{s_{2}}$, then we simply consider the weak combined transition $\hyperWeakCombinedTransition{\strongTransition{(s_{1},s_{3})}{a}{\dirac{s_{1}} \times \sd_{3}}}{\hidden}{\dirac{s_{1}} \times \gd_{3}}$ obtained by performing only transitions from $\aut_{3}$ that fulfils the required properties: 
		$\dirac{s_{2}} \times \gd_{3} \liftrel[\costrelord_{p}] \dirac{s_{1}} \times \gd_{3}$, \begin{align*}
		\cost[l]{\weakCombinedTransition{(s_{1},s_{3})}{a}{\dirac{s_{1}} \times \gd_{3}}} 
		& {} = \functioneval{\gen{\costf_{l}}}{\cost[1]{\strongTransition{s_{1}}{\apparent{a}}{\dirac{s_{1}}}}, \cost[3]{\weakCombinedTransition{\strongTransition{s_{3}}{a}{\sd_{3}}}{\hidden}{\gd_{3}}}} \\
		& {} = \functioneval{\gen{\costf_{l}}}{0, \cost[3]{\weakCombinedTransition{\strongTransition{s_{3}}{a}{\sd_{3}}}{\hidden}{\gd_{3}}}} \\
		& {} = \functioneval{\gen{\costf_{r}}}{0, \cost[3]{\weakCombinedTransition{\strongTransition{s_{3}}{a}{\sd_{3}}}{\hidden}{\gd_{3}}}} \\
		& {} = \functioneval{\gen{\costf_{r}}}{\cost[2]{\strongTransition{s_{2}}{\apparent{a}}{\dirac{s_{2}}}}, \cost[3]{\weakCombinedTransition{\strongTransition{s_{3}}{a}{\sd_{3}}}{\hidden}{\gd_{3}}}} \\
		& {} = \cost[r]{\weakCombinedTransition{(s_{2},s_{3})}{a}{\dirac{s_{2}} \times \gd_{3}}}\text{,} 
		\end{align*}
		thus we obtain that $\cost[l]{\weakCombinedTransition{(s_{1},s_{3})}{a}{\dirac{s_{1}} \times \gd_{3}}} \leq \cost[r]{\weakCombinedTransition{(s_{2},s_{3})}{a}{\dirac{s_{2}} \times \gd_{3}}}$, as expected.
		
		Suppose that there does not exist $\gd_{2} \times \gd_{3} \in \Disc{\borderStateSetOrd[\wbrelord_{p}] \cap (\stateSet_{2} \times \stateSet_{3})}$ such that $\hyperWeakCombinedTransition{\sd_{2} \times \sd_{3}}{\hidden}{\gd_{2} \times \gd_{3}}$:
		The step condition is trivially satisfied by taking $\sd_{1} = \dirac{s_{1}}$ and the weak combined transition $\weakCombinedTransition{(s_{1},s_{3})}{a}{\dirac{s_{1}} \times \sd_{3}} = \strongTransition{(s_{1}, s_{3})}{a}{\dirac{s_{1}} \times \sd_{3}}$.
		The condition $\dirac{s_{2}} \times \sd_{3} \liftrel[\costrelord_{p}] \dirac{s_{1}} \times \sd_{3}$ trivially holds since $s_{2} \costrel s_{1}$ and thus, for each $t_{3} \in \Supp{\sd_{3}}$, $(s_{2},t_{3}) \costrel_{p} (s_{1},t_{3})$;
		\begin{align*}
		\cost[l]{\weakCombinedTransition{(s_{1},s_{3})}{a}{\dirac{s_{1}} \times \sd_{3}}} 
		& {} = \cost[l]{\strongTransition{(s_{1}, s_{3})}{a}{\dirac{s_{1}} \times \sd_{3}}} \\
		& {} = \functioneval{\gen{\costf_{l}}}{\cost[1]{\strongTransition{s_{1}}{\apparent{a}}{\dirac{s_{1}}}}, \cost[3]{\strongTransition{s_{3}}{a}{\sd_{3}}}} \\
		& {} = \functioneval{\gen{\costf_{l}}}{0, \cost[3]{\strongTransition{s_{3}}{a}{\sd_{3}}}} \\
		& {} = \functioneval{\gen{\costf_{r}}}{0, \cost[3]{\strongTransition{s_{3}}{a}{\sd_{3}}}} \\
		& {} = \functioneval{\gen{\costf_{r}}}{\cost[2]{\strongTransition{s_{2}}{\apparent{a}}{\dirac{s_{2}}}}, \cost[3]{\strongTransition{s_{3}}{a}{\sd_{3}}}} \\
		& {} = \cost[r]{\strongTransition{(s_{2}, s_{3})}{a}{\dirac{s_{2}} \times \sd_{3}}}\text{,}
		\end{align*}
		thus $\cost[l]{\weakCombinedTransition{(s_{1},s_{3})}{a}{\dirac{s_{1}} \times \sd_{3}}} \leq \cost[r]{\strongTransition{(s_{2}, s_{3})}{a}{\dirac{s_{2}} \times \sd_{3}}}$, as required.
		
	\item[Case $a \in \actionSet_{2} \setminus \actionSet_{3}$]
		It is essentially the same as the previous case, where the roles of $\aut_{2}$/$\aut_{1}$ and $\aut_{3}$ are exchanged.
		For instance, we have $\sd_{3} = \dirac{s_{3}}$ and consider the case that there exists the distribution $\gd_{2} \times \gd_{3} \in \Disc{\borderStateSetOrd[\wbrelord_{p}] \cap (\stateSet_{2} \times \stateSet_{3})}$ such that $\hyperWeakCombinedTransition{\sd_{2} \times \sd_{3}}{\hidden}{\gd_{2} \times \gd_{3}}$ and suppose that only transitions from $\aut_{2}$ are used in such transition.
		Let $\family{\strongTransition{s_{2}}{\hidden}{\theta_{i}}}{i \in I}$ be the set of transitions chosen by the scheduler $\sched$ for the execution fragment $\alpha = s_{2}$ during the construction of $\hyperWeakCombinedTransition{\sd_{2}}{\hidden}{\gd_{2}}$.
		Since $\costAutPed{1}{\costf} \weakCostBisimMinorCost \costAutPed{\costf}{2}$, $s_{2} \costrel s_{1}$, $\strongTransition{s_{2}}{a}{\sd_{2}}$, and for each $\theta_{i}$ there exists $\phi_{i} \in \Disc{\borderStateSetOrd[\wbrelord] \cap \stateSet_{2}}$ such that $\hyperWeakCombinedTransition{\theta_{i}}{\hidden}{\phi_{i}}$ and $\sum_{i \in I} \probeval{\schedeval{\sched}{s_{2}}}{\strongTransition{s_{2}}{\hidden}{\theta_{i}}} \cdot \phi_{i} = \gd_{2}$, by $\costAutPed{\costf}{1} \weakCostBisimMinorCost \costAutPed{\costf}{2}$ it follows that there exists $\chi_{i} \in \Disc{\borderStateSetOrd[\wbrelord] \cap \stateSet_{1}}$ such that $\weakCombinedTransition{s_{1}}{a}{\chi_{i}}$, $\phi_{i} \liftrel[\costrelord] \chi_{i}$, and $\cost[1]{\weakCombinedTransition{s_{1}}{a}{\chi_{i}}} \leq \cost[2]{\weakCombinedTransition{\strongTransition{s_{2}}{\hidden}{\sd_{2}}}{\hidden}{\phi_{i}}}$.
		It is easy to see that the convex combination $\sum_{i \in I} \probeval{\schedeval{\sched}{s_{2}}}{\strongTransition{s_{2}}{\hidden}{\theta_{i}}} \cdot \weakCombinedTransition{s_{1}}{a}{\chi_{i}}$ results in a transition $\weakCombinedTransition{s_{1}}{a}{\chi}$ such that $\chi \in \Disc{\borderStateSetOrd[\wbrelord] \cap \stateSet_{1}}$,  $\gd_{2} \liftrel[\costrelord] \chi$, and $\cost[1]{\weakCombinedTransition{s_{1}}{a}{\chi}} \leq \cost[2]{\weakCombinedTransition{\strongTransition{s_{2}}{a}{\sd_{2}}}{\hidden}{\gd_{2}}}$.
		Since $\gd_{2} \liftrel[\costrelord] \chi$ and trivially $\gd_{3} \liftrel[\costrelord] \gd_{3}$, we have $\gd_{2} \times \gd_{3} \liftrel[\costrelord_{p}] \chi \times \gd_{3}$ and $\chi \times \gd_{3} \in \Disc{\borderStateSetOrd[\wbrelord_{p}] \cap \stateSet_{1} \times \gd_{3}}$.
		To complete the proof for this case, we need to show that $\cost[l]{\weakCombinedTransition{(s_{1},s_{3})}{a}{\chi \times \sd_{3}}} \leq \cost[r]{\weakCombinedTransition{(s_{2},s_{3})}{a}{\gd_{2} \times \sd_{3}}}$: 
		\begin{align*}
		\cost[l]{\weakCombinedTransition{(s_{1},s_{3})}{a}{\chi \times \sd_{3}}} 
		& {} = \functioneval{\gen{\costf_{r}}}{\cost[1]{\weakCombinedTransition{s_{1}}{a}{\chi}}, \cost[3]{\strongTransition{s_{3}}{\apparent{a}}{\dirac{s_{3}}}}} \\
		& {} = \functioneval{\gen{\costf_{r}}}{\cost[1]{\weakCombinedTransition{s_{1}}{a}{\chi}}, 0} \\
		& \leq \functioneval{\gen{\costf_{r}}}{\cost[2]{\weakCombinedTransition{\strongTransition{s_{2}}{\hidden}{\sd_{2}}}{\hidden}{\gd_{2}}}, 0} \\
		& {} = \functioneval{\gen{\costf_{l}}}{\cost[2]{\weakCombinedTransition{\strongTransition{s_{2}}{\hidden}{\sd_{2}}}{\hidden}{\gd_{2}}}, 0} \\
		& {} = \functioneval{\gen{\costf_{l}}}{\cost[2]{\weakCombinedTransition{\strongTransition{s_{2}}{\hidden}{\sd_{2}}}{\hidden}{\gd_{2}}}, \cost[3]{\strongTransition{s_{3}}{\apparent{a}}{\dirac{s_{3}}}}} \\
		& {} = \cost[r]{\weakCombinedTransition{(s_{2},s_{3})}{a}{\gd_{2} \times \sd_{3}}}\text{,}
		\end{align*} 
		hence $\cost[l]{\weakCombinedTransition{(s_{1},s_{3})}{a}{\chi \times \sd_{3}}} \leq \cost[r]{\weakCombinedTransition{(s_{2},s_{3})}{a}{\gd_{2} \times \sd_{3}}}$ as required.

	\item[Case $a \in \actionSet_{3} \cap \actionSet_{2}$]
		the definition of parallel composition implies that
                one obtains the transition $\strongTransition{(s_{2},s_{3})}{a}{\sd_{2} \times \sd_{3}}$ by combining the transitions $\strongTransition{s_{2}}{a}{\sd_{2}} \in \transitionRelation_{2}$ and $\strongTransition{s_{3}}{a}{\sd_{3}} \in \transitionRelation_{3}$.
		
		The remainder of the proof for this case is just the expected combination of the above two cases.
	\end{description}
	This completes the proof that $(\wbrelord_{p}, \costrelord_{p})$ is a minor cost weak probabilistic bisimulation, thus $\costAutPed{\costf}{1} \parComp_{\costf_{l}} \costAutPed{\costf}{3} \weakCostBisimMinorCost \costAutPed{\costf}{2} \parComp_{\costf_{r}} \costAutPed{\costf}{3}$.
\end{myproof}

\begin{result}[Theorem~\ref{thm:minCostLPequivalentToCostWeakTransitionLifting}]
\label{res:minCostLPequivalentToCostWeakTransitionLifting}
	Given a \CPA{} $\costAut{\costf}$, $\relord \subseteq \stateSet \times \stateSet$, $a \in \actionSet$, $\sd \in \Disc{\stateSet}$, and $t \in \stateSet$, consider the $\minCostLPproblemTBetaMuRel{t}{a}{\sd}{\relord}$ LP problem.
	The following implications hold:
	\begin{enumerate}
		\item
			If there exists a scheduler $\sched$ for $\aut$ that induces $\weakCombinedTransition{t}{a}{\sd_{t}}$ such that $\sd \liftrel \sd_{t}$, then $\minCostLPproblemTBetaMuRel{t}{a}{\sd}{\relord}$ has an optimal solution $f^{o}$ such that $\minCostLPValue \leq \cost{\weakCombinedTransition{t}{a}{\sd_{t}}}$.
		\item 
			If $\minCostLPproblemTBetaMuRel{t}{a}{\sd}{\relord}$ has an optimal solution $f^{o}$, then there exists a scheduler $\sched$ for $\aut$ that induces $\weakCombinedTransition{t}{a}{\sd_{t}}$ such that $\sd \liftrel \sd_{t}$ and $\cost{\weakCombinedTransition{t}{a}{\sd_{t}}} = \minCostLPValue$.
	\end{enumerate}
\end{result}
\begin{myproof}
	The proof is mainly based on the proof of~\cite[Theorem~8]{HT12}. 
	We recall that the min cost $\minCostLPValue$ is defined as $\minCostLPValue = \sum_{(x,y) \in E} \cost[f]{(x,y)} \cdot f^{o}_{x,y}$.

	Let $F(b)$ be the set $\setcond{\alpha \in \finiteFrags{\aut}}{\trace{\alpha} = \trace{b}}$ and $F(b, q)$ be the set $\setcond{\alpha \in F(b)}{\last{\alpha} = q}$; denote by $E_{\hidden}$, $E^{a}$, and $E^{a}_{\hidden}$ the sets $\setcond{(q,q^{\tr})}{\tr = \strongTransition{q}{\hidden}{\sd} \in \transitionRelation}$, $\setcond{(q,q_{a}^{\tr})}{\tr = \strongTransition{q}{a}{\sd} \in \transitionRelation}$, and $\setcond{(q_{a},q^{\tr}_{a})}{\tr = \strongTransition{q}{\hidden}{\sd} \in \transitionRelation}$, respectively.
	
	We prove the theorem for $a \in \externalActionSet$; the case $a \in \internalActionSet$ is similar.
	\begin{enumerate}
	\item 
		Suppose that there exists a scheduler $\sched$ for $\aut$ that induces $\weakCombinedTransition{t}{a}{\sd_{t}}$ such that $\sd \liftrel \sd_{t}$.
		This implies, by~\cite[Theorem~8]{HT12}, that $\LPproblemTBetaMuRel{t}{a}{\sd}{\rel}$ has a solution $f^{*}$ such that for each transition $\tr = \strongTransition{q}{b}{\sd}$, 
		\begin{enumerate}
		\item 
			$f^{*}_{q,q^{\tr}} = \sum_{\alpha \in F(\hidden, q)} \probeval{\sd_{\sched,t}}{\cone{\alpha}} \cdot \probeval{\schedeval{\sched}{\alpha}}{\tr}$ if $b = \hidden$, 
		\item
			$f^{*}_{q_{a},q^{\tr}_{a}} = \sum_{\alpha \in F(a, q)} \probeval{\sd_{\sched,t}}{\cone{\alpha}} \cdot \probeval{\schedeval{\sched}{\alpha}}{\tr}$ if $b = \hidden$, and
		\item 
			$f^{*}_{q,q^{\tr}_{a}} = \sum_{\alpha \in F(\hidden, q)} \probeval{\sd_{\sched,t}}{\cone{\alpha}} \cdot \probeval{\schedeval{\sched}{\alpha}}{\tr}$ if $b \neq \hidden$, 
		\end{enumerate}
		This implies that $\sum_{(x,y) \in E} \cost[f]{(x,y)} f^{*}_{x,y} = \cost[\sched]{\weakCombinedTransition{t}{a}{\sd_{t}}}$. 
		In fact, it holds that
		\begin{align*}
			& \phantom{{}={}} \sum_{(x,y) \in E} \cost[f]{(x,y)} \cdot f^{*}_{x,y} \\
			& {} = \sum_{(x,y) \in E_{\hidden}} \cost[f]{(x,y)} \cdot f^{*}_{x,y} + \sum_{(x,y) \in E^{a}_{\hidden}} \cost[f]{(x,y)} \cdot f^{*}_{x,y} \\
			& \phantom{{}={}} + \sum_{(x,y) \in E^{a}} \cost[f]{(x,y)} \cdot f^{*}_{x,y} + \sum_{(x,y) \in E \setminus (E_{\hidden} \cup E^{a} \cup E^{a}_{\hidden})} \cost[f]{(x,y)} \cdot f^{*}_{x,y} \\
			& {} = \sum_{\tr = \strongTransition{q}{\hidden}{\sd} \in \transitionRelation} \cost{\tr} \cdot f^{*}_{q,q^{\tr}} + \sum_{\tr = \strongTransition{q}{\hidden}{\sd} \in \transitionRelation} \cost{\tr} \cdot f^{*}_{q_{a},q^{\tr}_{a}} + \sum_{\tr = \strongTransition{q}{a}{\sd} \in \transitionRelation} \cost{\tr} \cdot f^{*}_{q,q^{\tr}_{a}} \\ 
			& {} = \sum_{\tr = \strongTransition{q}{\hidden}{\sd} \in \transitionRelation} \cost{\tr} \cdot \sum_{\alpha \in F(\hidden, q)} \probeval{\sd_{\sched,t}}{\cone{\alpha}} \cdot \probeval{\schedeval{\sched}{\alpha}}{\tr} \\ 
			& \phantom{{}={}} + \sum_{\tr = \strongTransition{q}{\hidden}{\sd} \in \transitionRelation} \cost{\tr} \cdot \sum_{\alpha \in F(a, q)} \probeval{\sd_{\sched,t}}{\cone{\alpha}} \cdot \probeval{\schedeval{\sched}{\alpha}}{\tr} \\
			& \phantom{{}={}} + \sum_{\tr = \strongTransition{q}{a}{\sd} \in \transitionRelation} \cost{\tr} \cdot \sum_{\alpha \in F(\hidden, q)} \probeval{\sd_{\sched,t}}{\cone{\alpha}} \cdot \probeval{\schedeval{\sched}{\alpha}}{\tr} \\ 
			& {} = \sum_{\tr = \strongTransition{q}{\hidden}{\sd} \in \transitionRelation} \sum_{\alpha \in F(\hidden, q)} \cost{\tr} \cdot \probeval{\sd_{\sched,t}}{\cone{\alpha}} \cdot \probeval{\schedeval{\sched}{\alpha}}{\tr} \\ 
			& \phantom{{}={}} + \sum_{\tr = \strongTransition{q}{\hidden}{\sd} \in \transitionRelation} \sum_{\alpha \in F(a, q)} \cost{\tr} \cdot \probeval{\sd_{\sched,t}}{\cone{\alpha}} \cdot \probeval{\schedeval{\sched}{\alpha}}{\tr} \\
			& \phantom{{}={}} + \sum_{\tr = \strongTransition{q}{a}{\sd} \in \transitionRelation} \sum_{\alpha \in F(\hidden, q)} \cost{\tr} \cdot \probeval{\sd_{\sched,t}}{\cone{\alpha}} \cdot \probeval{\schedeval{\sched}{\alpha}}{\tr} \\ 
			& {} = \sum_{\alpha \in F(\hidden)} \sum_{\tr \in \transitionsWithLabel{\hidden}} \cost{\tr} \cdot \probeval{\sd_{\sched,t}}{\cone{\alpha}} \cdot \probeval{\schedeval{\sched}{\alpha}}{\tr} \\ 
			& \phantom{{}={}} + \sum_{\alpha \in F(a)} \sum_{\tr \in \transitionsWithLabel{\hidden}} \cost{\tr} \cdot \probeval{\sd_{\sched,t}}{\cone{\alpha}} \cdot \probeval{\schedeval{\sched}{\alpha}}{\tr} \\
			& \phantom{{}={}} + \sum_{\alpha \in F(\hidden)} \sum_{\tr \in \transitionsWithLabel{a}} \cost{\tr} \cdot \probeval{\sd_{\sched,t}}{\cone{\alpha}} \cdot \probeval{\schedeval{\sched}{\alpha}}{\tr} \\ 
			\intertext{since $\probeval{\schedeval{\sched}{\alpha}}{\tr} = 0$ when $\source{\tr} \neq \last{\alpha}$}
			& {} = \sum_{\alpha \in F(\hidden)} \sum_{\tr \in \transitionsWithLabel{\hidden}} \cost{\tr} \cdot \probeval{\sd_{\sched,t}}{\cone{\alpha}} \cdot \probeval{\schedeval{\sched}{\alpha}}{\tr} \sum_{r \in \stateSet} \probeval{\sd_{\tr}}{r} \\ 
			& \phantom{{}={}} + \sum_{\alpha \in F(a)} \sum_{\tr \in \transitionsWithLabel{\hidden}} \cost{\tr} \cdot \probeval{\sd_{\sched,t}}{\cone{\alpha}} \cdot \probeval{\schedeval{\sched}{\alpha}}{\tr} \sum_{r \in \stateSet} \probeval{\sd_{\tr}}{r} \\
			& \phantom{{}={}} + \sum_{\alpha \in F(\hidden)} \sum_{\tr \in \transitionsWithLabel{a}} \cost{\tr} \cdot \probeval{\sd_{\sched,t}}{\cone{\alpha}} \cdot \probeval{\schedeval{\sched}{\alpha}}{\tr} \sum_{r \in \stateSet} \probeval{\sd_{\tr}}{r} \\ 
			& {} = \sum_{\alpha \in F(\hidden)} \sum_{r \in \stateSet} \sum_{\tr \in \transitionsWithLabel{\hidden}} \cost{\tr} \cdot \probeval{\sd_{\sched,t}}{\cone{\alpha}} \cdot \probeval{\schedeval{\sched}{\alpha}}{\tr} \cdot \probeval{\sd_{\tr}}{r} \\ 
			& \phantom{{}={}} + \sum_{\alpha \in F(a)} \sum_{r \in \stateSet} \sum_{\tr \in \transitionsWithLabel{\hidden}} \cost{\tr} \cdot \probeval{\sd_{\sched,t}}{\cone{\alpha}} \cdot \probeval{\schedeval{\sched}{\alpha}}{\tr} \cdot \probeval{\sd_{\tr}}{r} \\
			& \phantom{{}={}} + \sum_{\alpha \in F(\hidden)} \sum_{r \in \stateSet} \sum_{\tr \in \transitionsWithLabel{a}} \cost{\tr} \cdot \probeval{\sd_{\sched,t}}{\cone{\alpha}} \cdot \probeval{\schedeval{\sched}{\alpha}}{\tr} \cdot \probeval{\sd_{\tr}}{r} \\ 
			& {} = \sum_{\alpha \in F(\hidden)} \sum_{r \in \stateSet} \probeval{\sd_{\sched,t}}{\cone{\alpha}} \cdot \sum_{\tr \in \transitionsWithLabel{\hidden}} \cost{\tr} \cdot \probeval{\schedeval{\sched}{\alpha}}{\tr} \cdot \probeval{\sd_{\tr}}{r} \\ 
			& \phantom{{}={}} + \sum_{\alpha \in F(a)} \sum_{r \in \stateSet} \probeval{\sd_{\sched,t}}{\cone{\alpha}} \cdot \sum_{\tr \in \transitionsWithLabel{\hidden}} \cost{\tr} \cdot \probeval{\schedeval{\sched}{\alpha}}{\tr} \cdot \probeval{\sd_{\tr}}{r} \\
			& \phantom{{}={}} + \sum_{\alpha \in F(\hidden)} \sum_{r \in \stateSet} \probeval{\sd_{\sched,t}}{\cone{\alpha}} \cdot \sum_{\tr \in \transitionsWithLabel{a}} \cost{\tr} \cdot \probeval{\schedeval{\sched}{\alpha}}{\tr} \cdot \probeval{\sd_{\tr}}{r} \\ 
			& {} = \sum_{\alpha \hidden r \in \setcond{\beta \hidden r}{\beta \in F(\hidden)}} \probeval{\sd_{\sched,t}}{\cone{\alpha}} \cdot \sum_{\tr \in \transitionsWithLabel{\hidden}} \cost{\tr} \cdot \probeval{\schedeval{\sched}{\alpha}}{\tr} \cdot \probeval{\sd_{\tr}}{r} \\ 
			& \phantom{{}={}} + \sum_{\alpha \hidden r \in \setcond{\beta \hidden r}{\beta \in F(a)}} \probeval{\sd_{\sched,t}}{\cone{\alpha}} \cdot \sum_{\tr \in \transitionsWithLabel{\hidden}} \cost{\tr} \cdot \probeval{\schedeval{\sched}{\alpha}}{\tr} \cdot \probeval{\sd_{\tr}}{r} \\
			& \phantom{{}={}} + \sum_{\alpha a r \in \setcond{\beta a r}{\beta \in F(\hidden)}} \probeval{\sd_{\sched,t}}{\cone{\alpha}} \cdot \sum_{\tr \in \transitionsWithLabel{a}} \cost{\tr} \cdot \probeval{\schedeval{\sched}{\alpha}}{\tr} \cdot \probeval{\sd_{\tr}}{r} \\ 
			& {} = \sum_{\alpha b r \in \setcond{\beta b r}{\beta \in F(\hidden) \cup F(a)}} \probeval{\sd_{\sched,t}}{\cone{\alpha}} \cdot \sum_{\tr \in \transitionsWithLabel{b}} \cost{\tr} \cdot \probeval{\schedeval{\sched}{\alpha}}{\tr} \cdot \probeval{\sd_{\tr}}{r} \\ 
			\intertext{since $\probeval{\schedeval{\sched}{\alpha}}{\tr} = 0$ when $\trace{\alpha b r} \notin \setnocond{\emptytrace, \trace{a}}$}
			& {} = \sum_{\alpha b r \in \setcond{\beta b r}{\beta \in \finiteFrags{\aut}}} \probeval{\sd_{\sched,t}}{\cone{\alpha}} \cdot \sum_{\tr \in \transitionsWithLabel{b}} \cost{\tr} \cdot \probeval{\schedeval{\sched}{\alpha}}{\tr} \cdot \probeval{\sd_{\tr}}{r} \\ 
			& {} = 0 + \sum_{\alpha b r \in \setcond{\beta b r}{\beta \in \finiteFrags{\aut}}} \cost[\sched, t]{\alpha b r} \\
			& {} = \sum_{\alpha \in \setcond{\beta \in \finiteFrags{\aut}}{\beta = q \in \stateSet}} \cost[\sched, t]{\alpha} + \sum_{\alpha \in \setcond{\beta b r}{\beta \in \finiteFrags{\aut}}} \cost[\sched, t]{\alpha} \\
			& {} = \sum_{\alpha \in \finiteFrags{\aut}} \cost[\sched, t]{\alpha} \\
			& {} = \cost[\sched]{\weakCombinedTransition{t}{a}{\sd_{t}}}\text{.}
		\end{align*}
		Since $f^{*}$ is a feasible solution of $\LPproblemTBetaMuRel{t}{a}{\sd}{\rel}$, Proposition~\ref{pro:minCostLPProblemIsFeasibleIFFLPProblemIsFeasible} implies that $f^{*}$ is also a feasible solution of $\minCostLPproblemTBetaMuRel{t}{a}{\sd}{\rel}$.
		This implies that there exists a (possibly different) optimal solution $f^{o}$ such that $\minCostLPValue = \sum_{(x,y) \in E} \cost[f]{(x,y)} \cdot f^{o}_{x,y} \leq \sum_{(x,y) \in E} \cost[f]{(x,y)} \cdot f^{*}_{x,y} = \cost[\sched]{\weakCombinedTransition{t}{a}{\sd_{t}}}$, as required.

	\item 
		Suppose that $\minCostLPproblemTBetaMuRel{t}{a}{\sd}{\rel}$ has an optimal solution $f^{o}$;
		Proposition~\ref{pro:minCostLPProblemIsFeasibleIFFLPProblemIsFeasible} implies that $f^{o}$ is also a feasible solution of $\LPproblemTBetaMuRel{t}{a}{\sd}{\rel}$ and thus, by~\cite[Theorem~8]{HT12}, there exists a scheduler $\sched$ for $\aut$ that induces $\weakCombinedTransition{t}{a}{\sd_{t}}$ such that $\sd \liftrel \sd_{t}$.
		In particular, from the proof of~\cite[Theorem~8]{HT12} we know that such scheduler is defined as follows: 
		For each execution fragment $\phi \in \finiteFrags{\aut}$,
		\[
			\probeval{\schedeval{\sched}{\phi}}{x} =
				\begin{cases}
					f^{o}_{v,v^{\tr}}/\incomingflow^{o}_{v} & \text{if $\incomingflow^{o}_{v} \neq 0$, $\trace{\phi} = \emptytrace$, and $x = \tr = \strongTransition{v}{\hidden}{\rho} \in \transitionRelation$;} \\
					f^{o}_{v,v^{\tr}_{a}}/\incomingflow^{o}_{v} & \text{if $\incomingflow^{o}_{v} \neq 0$, $\trace{\phi} = \emptytrace$, $a \neq \hidden$, and $x = \tr = \strongTransition{v}{a}{\rho} \in \transitionRelation$;} \\
					f^{o}_{v_{a},v^{\tr}_{a}}/\incomingflow^{o}_{v_{a}} & \text{if $\incomingflow^{o}_{v_{a}} \neq 0$, $\trace{\phi} = a \neq \hidden$, and $x = \tr = \strongTransition{v}{\hidden}{\rho} \in \transitionRelation$;} \\
					f^{o}_{v,\relimage{v}{\rel}}/\incomingflow^{o}_{v} & \text{if $\incomingflow^{o}_{v} \neq 0$, $\trace{\phi} = \emptytrace$, $a = \hidden$, and $x = \bot$;} \\
					f^{o}_{v_{a},\relimage{v}{\rel}}/\incomingflow^{o}_{v_{a}} & \text{if $\incomingflow^{o}_{v_{a}} \neq 0$, $\trace{\phi} = a \neq \hidden$, and $x = \bot$;} \\
					1 & \text{if $\trace{\phi} \notin \setnocond{\emptytrace, \trace{a}}$ and $x = \bot$;} \\
					1 & \text{if $\incomingflow^{o}_{v} = 0$, $\trace{\phi} = \emptytrace$ and $x = \bot$;} \\
					1 & \text{if $\incomingflow^{o}_{v_{a}} = 0$, $\trace{\phi} = a \neq \hidden$ and $x = \bot$;} \\
					0 & \text{otherwise}
				\end{cases}
		\]
		where $v = \last{\phi}$, $f^{o}_{v,\relimage{v}{\rel}}$ is the total flow from the vertex $v$ to the vertices $u_{\rel}$ such that $v \rel u$, and $\incomingflow^{o}_{v}$ is the total incoming flow in the vertex $v$.
		As pointed out in the proof of~\cite[Theorem~8]{HT12}, $\sched$ is a determinate scheduler, i.e., for each pair of $\phi, \phi' \in \finiteFrags{\aut}$ such that $\last{\phi} = \last{\phi'}$ and $\trace{\phi} = \trace{\phi'}$, we have that $\schedeval{\sched}{\phi} = \schedeval{\sched}{\phi'}$.
		
		\begin{align*}
			& \phantom{{}={}} \cost[\sched]{\weakCombinedTransition{t}{a}{\sd_{t}}} \\
			& {} = \sum_{\alpha \in \finiteFrags{\aut}} \cost[\sched, t]{\alpha} \\
			& {} = \sum_{\alpha \in \setcond{\beta \in \finiteFrags{\aut}}{\beta = q \in \stateSet}} \cost[\sched, t]{\alpha} + \sum_{\alpha \in \setcond{\beta b r}{\beta \in \finiteFrags{\aut}}} \cost[\sched, t]{\alpha} \\
			& {} = 0 + \sum_{\alpha b r \in \setcond{\beta b r}{\beta \in \finiteFrags{\aut}}} \cost[\sched, t]{\alpha b r} \\
			& {} = \sum_{\alpha b r \in \setcond{\beta b r}{\beta \in \finiteFrags{\aut}}} \probeval{\sd_{\sched,t}}{\cone{\alpha}} \cdot \sum_{\tr \in \transitionsWithLabel{b}} \cost{\tr} \cdot \probeval{\schedeval{\sched}{\alpha}}{\tr} \cdot \probeval{\sd_{\tr}}{r} \\ 
			& {} = \sum_{\alpha \in \finiteFrags{\aut}} \sum_{b \in \actionSet} \sum_{r \in \stateSet} \probeval{\sd_{\sched,t}}{\cone{\alpha}} \cdot \sum_{\tr \in \transitionsWithLabel{b}} \cost{\tr} \cdot \probeval{\schedeval{\sched}{\alpha}}{\tr} \cdot \probeval{\sd_{\tr}}{r} \\ 
			& {} = \sum_{\alpha \in \finiteFrags{\aut}} \sum_{b \in \actionSet} \probeval{\sd_{\sched,t}}{\cone{\alpha}} \cdot \sum_{\tr \in \transitionsWithLabel{b}} \cost{\tr} \cdot \probeval{\schedeval{\sched}{\alpha}}{\tr} \cdot \sum_{r \in \stateSet} \probeval{\sd_{\tr}}{r} \\ 
			& {} = \sum_{\alpha \in \finiteFrags{\aut}} \sum_{b \in \actionSet} \probeval{\sd_{\sched,t}}{\cone{\alpha}} \cdot \sum_{\tr \in \transitionsWithLabel{b}} \cost{\tr} \cdot \probeval{\schedeval{\sched}{\alpha}}{\tr} \\ 
			& {} = \sum_{\alpha \in F(\hidden)} \sum_{b \in \setnocond{a, \hidden}} \probeval{\sd_{\sched,t}}{\cone{\alpha}} \cdot \sum_{\tr \in \transitionsWithLabel{b}} \cost{\tr} \cdot \probeval{\schedeval{\sched}{\alpha}}{\tr} \\ 
			& \phantom{{}={}} + \sum_{\alpha \in F(a)} \probeval{\sd_{\sched,t}}{\cone{\alpha}} \cdot \sum_{\tr \in \transitionsWithLabel{\hidden}} \cost{\tr} \cdot \probeval{\schedeval{\sched}{\alpha}}{\tr} \\ 
			& {} = \sum_{\alpha \in F(\hidden)} \probeval{\sd_{\sched,t}}{\cone{\alpha}} \cdot \sum_{\tr \in \transitionsWithLabel{\hidden}} \cost{\tr} \cdot \probeval{\schedeval{\sched}{\alpha}}{\tr} \\ 
			& \phantom{{}={}} + \sum_{\alpha \in F(\hidden)} \probeval{\sd_{\sched,t}}{\cone{\alpha}} \cdot \sum_{\tr \in \transitionsWithLabel{a}} \cost{\tr} \cdot \probeval{\schedeval{\sched}{\alpha}}{\tr} \\ 
			& \phantom{{}={}} + \sum_{\alpha \in F(a)} \probeval{\sd_{\sched,t}}{\cone{\alpha}} \cdot \sum_{\tr \in \transitionsWithLabel{\hidden}} \cost{\tr} \cdot \probeval{\schedeval{\sched}{\alpha}}{\tr} \\ 
			& {} = \sum_{v \in \stateSet} \sum_{\alpha \in F(\hidden, v)} \probeval{\sd_{\sched,t}}{\cone{\alpha}} \cdot \sum_{\tr \in \transitionsWithLabel{\hidden}} \cost{\tr} \cdot \probeval{\schedeval{\sched}{\alpha}}{\tr} \\ 
			& \phantom{{}={}} + \sum_{v \in \stateSet} \sum_{\alpha \in F(\hidden, v)} \probeval{\sd_{\sched,t}}{\cone{\alpha}} \cdot \sum_{\tr \in \transitionsWithLabel{a}} \cost{\tr} \cdot \probeval{\schedeval{\sched}{\alpha}}{\tr} \\ 
			& \phantom{{}={}} + \sum_{v \in \stateSet} \sum_{\alpha \in F(a, v)} \probeval{\sd_{\sched,t}}{\cone{\alpha}} \cdot \sum_{\tr \in \transitionsWithLabel{\hidden}} \cost{\tr} \cdot \probeval{\schedeval{\sched}{\alpha}}{\tr} \\ 
			& {} = \sum_{v \in \stateSet} \sum_{\alpha \in F(\hidden, v)} \probeval{\sd_{\sched,t}}{\cone{\alpha}} \cdot \sum_{\tr = \strongTransition{v}{\hidden}{\sd} \in \transitionsWithLabel{\hidden}} \cost{\tr} \cdot \probeval{\schedeval{\sched}{\alpha}}{\tr} \\ 
			& \phantom{{}={}} + \sum_{v \in \stateSet} \sum_{\alpha \in F(\hidden, v)} \probeval{\sd_{\sched,t}}{\cone{\alpha}} \cdot \sum_{\tr = \strongTransition{v}{a}{\sd} \in \transitionsWithLabel{a}} \cost{\tr} \cdot \probeval{\schedeval{\sched}{\alpha}}{\tr} \\ 
			& \phantom{{}={}} + \sum_{v \in \stateSet} \sum_{\alpha \in F(a, v)} \probeval{\sd_{\sched,t}}{\cone{\alpha}} \cdot \sum_{\tr = \strongTransition{v}{\hidden}{\sd} \in \transitionsWithLabel{\hidden}} \cost{\tr} \cdot \probeval{\schedeval{\sched}{\alpha}}{\tr} \\ 
			& {} = \sum_{v \in \stateSet} \sum_{\alpha \in F(\hidden, v)} \probeval{\sd_{\sched,t}}{\cone{\alpha}} \cdot \sum_{\tr = \strongTransition{v}{\hidden}{\sd} \in \transitionRelation} \cost{\tr} \cdot \probeval{\schedeval{\sched}{\alpha}}{\tr} \\ 
			& \phantom{{}={}} + \sum_{v \in \stateSet} \sum_{\alpha \in F(\hidden, v)} \probeval{\sd_{\sched,t}}{\cone{\alpha}} \cdot \sum_{\tr = \strongTransition{v}{a}{\sd} \in \transitionRelation} \cost{\tr} \cdot \probeval{\schedeval{\sched}{\alpha}}{\tr} \\ 
			& \phantom{{}={}} + \sum_{v \in \stateSet} \sum_{\alpha \in F(a, v)} \probeval{\sd_{\sched,t}}{\cone{\alpha}} \cdot \sum_{\tr = \strongTransition{v}{\hidden}{\sd} \in \transitionRelation} \cost{\tr} \cdot \probeval{\schedeval{\sched}{\alpha}}{\tr} \\ 
			& {} = \sum_{v \in \stateSet} \sum_{\alpha \in F(\hidden, v)} \probeval{\sd_{\sched,t}}{\cone{\alpha}} \cdot \sum_{\tr = \strongTransition{v}{\hidden}{\sd} \in \transitionRelation} \cost{\tr} \cdot \frac{f^{o}_{v,v^{\tr}}}{\incomingflow^{o}_{v}} \\ 
			& \phantom{{}={}} + \sum_{v \in \stateSet} \sum_{\alpha \in F(\hidden, v)} \probeval{\sd_{\sched,t}}{\cone{\alpha}} \cdot \sum_{\tr = \strongTransition{v}{a}{\sd} \in \transitionRelation} \cost{\tr} \cdot \frac{f^{o}_{v,v^{\tr}_{a}}}{\incomingflow^{o}_{v}} \\ 
			& \phantom{{}={}} + \sum_{v \in \stateSet} \sum_{\alpha \in F(a, v)} \probeval{\sd_{\sched,t}}{\cone{\alpha}} \cdot \sum_{\tr = \strongTransition{v}{\hidden}{\sd} \in \transitionRelation} \cost{\tr} \cdot \frac{f^{o}_{v_{a},v^{\tr}_{a}}}{\incomingflow^{o}_{v_{a}}} \\
			& {} = \sum_{v \in \stateSet} \sum_{\alpha \in F(\hidden, v)} \sum_{\tr = \strongTransition{v}{\hidden}{\sd} \in \transitionRelation} \cost{\tr} \cdot \frac{f^{o}_{v,v^{\tr}}}{\incomingflow^{o}_{v}} \cdot \probeval{\sd_{\sched,t}}{\cone{\alpha}} \\ 
			& \phantom{{}={}} + \sum_{v \in \stateSet} \sum_{\alpha \in F(\hidden, v)} \sum_{\tr = \strongTransition{v}{a}{\sd} \in \transitionRelation} \cost{\tr} \cdot \frac{f^{o}_{v,v^{\tr}_{a}}}{\incomingflow^{o}_{v}} \cdot \probeval{\sd_{\sched,t}}{\cone{\alpha}} \\ 
			& \phantom{{}={}} + \sum_{v \in \stateSet} \sum_{\alpha \in F(a, v)} \sum_{\tr = \strongTransition{v}{\hidden}{\sd} \in \transitionRelation} \cost{\tr} \cdot \frac{f^{o}_{v_{a},v^{\tr}_{a}}}{\incomingflow^{o}_{v_{a}}} \cdot \probeval{\sd_{\sched,t}}{\cone{\alpha}} \\ 
			& {} = \sum_{v \in \stateSet} \sum_{\tr = \strongTransition{v}{\hidden}{\sd} \in \transitionRelation} \cost{\tr} \cdot \frac{f^{o}_{v,v^{\tr}}}{\incomingflow^{o}_{v}} \cdot \sum_{\alpha \in F(\hidden, v)} \probeval{\sd_{\sched,t}}{\cone{\alpha}} \\ 
			& \phantom{{}={}} + \sum_{v \in \stateSet} \sum_{\tr = \strongTransition{v}{a}{\sd} \in \transitionRelation} \cost{\tr} \cdot \frac{f^{o}_{v,v^{\tr}_{a}}}{\incomingflow^{o}_{v}} \cdot \sum_{\alpha \in F(\hidden, v)} \probeval{\sd_{\sched,t}}{\cone{\alpha}} \\ 
			& \phantom{{}={}} + \sum_{v \in \stateSet} \sum_{\tr = \strongTransition{v}{\hidden}{\sd} \in \transitionRelation} \cost{\tr} \cdot \frac{f^{o}_{v_{a},v^{\tr}_{a}}}{\incomingflow^{o}_{v_{a}}} \cdot \sum_{\alpha \in F(a, v)} \probeval{\sd_{\sched,t}}{\cone{\alpha}} \\ 
			& {} = \sum_{v \in \stateSet} \sum_{\tr = \strongTransition{v}{\hidden}{\sd} \in \transitionRelation} \cost{\tr} \cdot \frac{f^{o}_{v,v^{\tr}}}{\incomingflow^{o}_{v}} \cdot \incomingflow^{o}_{v} + \sum_{v \in \stateSet} \sum_{\tr = \strongTransition{v}{a}{\sd} \in \transitionRelation} \cost{\tr} \cdot \frac{f^{o}_{v,v^{\tr}_{a}}}{\incomingflow^{o}_{v}} \cdot \incomingflow^{o}_{v} \\
			& \phantom{{}={}} + \sum_{v \in \stateSet} \sum_{\tr = \strongTransition{v}{\hidden}{\sd} \in \transitionRelation} \cost{\tr} \cdot \frac{f^{o}_{v_{a},v^{\tr}_{a}}}{\incomingflow^{o}_{v_{a}}} \cdot \incomingflow^{o}_{v_{a}} \\ 
			\intertext{by~\cite[Long version, Corollary~2]{HT12}}
			& {} = \sum_{v \in \stateSet} \sum_{\tr = \strongTransition{v}{\hidden}{\sd} \in \transitionRelation} \cost{\tr} \cdot f^{o}_{v,v^{\tr}} + \sum_{v \in \stateSet} \sum_{\tr = \strongTransition{v}{a}{\sd} \in \transitionRelation} \cost{\tr} \cdot f^{o}_{v,v^{\tr}_{a}} \\
			& \phantom{{}={}} + \sum_{v \in \stateSet} \sum_{\tr = \strongTransition{v}{\hidden}{\sd} \in \transitionRelation} \cost{\tr} \cdot f^{o}_{v_{a},v^{\tr}_{a}} \\ 
			& {} = \sum_{\tr = \strongTransition{v}{\hidden}{\sd} \in \transitionRelation} \cost{\tr} \cdot f^{o}_{v,v^{\tr}} + \sum_{\tr = \strongTransition{v}{a}{\sd} \in \transitionRelation} \cost{\tr} \cdot f^{o}_{v,v^{\tr}_{a}} + \sum_{\tr = \strongTransition{v}{\hidden}{\sd} \in \transitionRelation} \cost{\tr} \cdot f^{o}_{v_{a},v^{\tr}_{a}} \\ 
			& {} = \sum_{\tr = \strongTransition{v}{\hidden}{\sd} \in \transitionRelation} \cost[f]{(v,v^{\tr})} \cdot f^{o}_{v,v^{\tr}} + \sum_{\tr = \strongTransition{v}{a}{\sd} \in \transitionRelation} \cost[f]{(v,v^{\tr}_{a})} \cdot f^{o}_{v,v^{\tr}_{a}} \\
			& \phantom{{}={}} + \sum_{\tr = \strongTransition{v}{\hidden}{\sd} \in \transitionRelation} \cost[f]{(v_{a},v^{\tr}_{a})} \cdot f^{o}_{v_{a},v^{\tr}_{a}} + 0 \\
			& {} = \sum_{(v, v^{\tr}) \in E_{\hidden}} \cost[f]{(v,v^{\tr})} \cdot f^{o}_{v,v^{\tr}} + \sum_{(v,v^{\tr}_{a}) \in E^{a}} \cost[f]{(v,v^{\tr}_{a})} \cdot f^{o}_{v,v^{\tr}_{a}} \\ 
			& \phantom{{}={}} + \sum_{(v_{a},v^{\tr}_{a}) \in E^{a}_{\hidden}} \cost[f]{(v_{a},v^{\tr}_{a})} \cdot f^{o}_{v_{a},v^{\tr}_{a}} + \sum_{(x,y) \in E \setminus (E_{\hidden} \cup E^{a} \cup E^{a}_{\hidden})} 0 \cdot f^{o}_{x,y} \\
			& {} = \sum_{(v, v^{\tr}) \in E_{\hidden}} \cost[f]{(v,v^{\tr})} \cdot f^{o}_{v,v^{\tr}} + \sum_{(v,v^{\tr}_{a}) \in E^{a}} \cost[f]{(v,v^{\tr}_{a})} \cdot f^{o}_{v,v^{\tr}_{a}} \\ 
			& \phantom{{}={}} + \sum_{(v_{a},v^{\tr}_{a}) \in E^{a}_{\hidden}} \cost[f]{(v_{a},v^{\tr}_{a})} \cdot f^{o}_{v_{a},v^{\tr}_{a}} + \sum_{(x,y) \in E \setminus (E_{\hidden} \cup E^{a} \cup E^{a}_{\hidden})} \cost[f]{(x,y)} \cdot f^{o}_{x,y} \\
			& {} = \sum_{(x,y) \in E} \cost[f]{(x,y)} \cdot f^{o}_{x,y} \\
			& {} = \minCostLPValue\text{.}
		\end{align*}
	\end{enumerate}
	This completes the proof that if $\minCostLPproblemTBetaMuRel{t}{a}{\sd}{\relord}$ has an optimal solution $f^{o}$, then there exists a scheduler $\sched$ for $\aut$ that induces $\weakCombinedTransition{t}{a}{\sd_{t}}$ such that $\sd \liftrel \sd_{t}$ and $\cost{\weakCombinedTransition{t}{a}{\sd_{t}}} = \minCostLPValue$, and the proof of the theorem.
\end{myproof}

\end{document}